# Structural Insights and Advanced Spectroscopic Characterization of Thiazolothiazoles: Unveiling Potential for Optoelectronic and Sensing Applications

Karolina Gutmańska[a]*, Agnieszka Podborska[b], Andrzej Sławek[b], Ramesh Sivasamy[b], Lulu Alluhaibi[c], Alexey Maximenko[c], Anna Ordyszewska[a], Konrad Szaciłowski[b,d]*, Anna Dołęga[a], Tomasz Mazur[b]*

[a] *Gdansk University of Technology, Chemical Faculty, Department of Inorganic Chemistry, Narutowicza 11/12, 80-233 Gdańsk, Poland*

[b] *AGH University of Krakow, Academic Centre of Materials and Technology, al. Mickiewicza 30, 30-059 Kraków, Poland*

[c] *National Synchrotron Radiation Centre SOLARIS, Jagiellonian University, ul. Czerwone Maki 98, Kraków 30-392, Poland*

[d] *University of the West of England, Unconventional Computing Lab, Bristol BS16 1QY, United Kingdom*

*Corresponding authors: karolina.gutmanska@pg.edu.pl, szacilow@agh.edu.pl, tmazur@agh.edu.pl*

## Abstract

**Thiazolothiazoles (TzTz)** represent a class of compounds with distinctive structural motifs and exceptional optical properties, positioning them as promising candidates for breakthroughs in optoelectronic and sensing technologies. X-ray crystallographic analyses of TzTz units symmetrically substituted with functional groups such as imidazole, *o*-vanillin, *p*-vanillin, phenyl, thiazole, cinnamate, and bistrifluoromethylphenyl have revealed complex structural features, including π-π stacking interactions, hydrogen-bond networks, and specific chalcogen and halogen interactions. These interactions collectively enhance the stability and define the unique spectroscopic profiles of these compounds.

Beyond classical spectral fingerprints (FTIR, NMR, and UV-Vis spectra), fluorescence studies at various temperatures, complemented by XANES synchrotron analyses, have underscored their remarkable thermal and electronic properties. The findings presented here offer a comprehensive framework for the characterization and analysis of TzTz compounds, emphasizing their potential as components in smart electronic and optoelectronic devices.

## Introduction

In recent years, there has been a significant increase in interest in the study of modern organic materials with exceptional optical and electronic properties,[1] which have the potential for wide application in advanced technologies such as optoelectronics,[2] photovoltaics,[3] and biosensing.[4] A particularly promising class of compounds in this field are heterocyclic compounds, known for their ability to absorb and emit light, as well as their structural tunability, which enables the tailoring of their properties to specific application needs.[5]

One such compound of interest is thiazolo[5,4-d]thiazole, a heterocyclic structure composed of fused thiazole rings.[6] Its structure contains both sulfur and nitrogen atoms, which contribute to its unique physicochemical and optical properties, including intense fluorescence and strong ultraviolet absorption.[6a, 7] Since their discovery in 1960, thiazolothiazoles attract attention due to their unusual properties as well as synthetic accessibility[8] and even prospective medicinal applications.[9] The planarity of the TzTz core and its strong π–π stacking tendency render TzTz derivatives effective semiconductors with excellent charge mobilities, making them ideal candidates for use in OFETs and organic solar cells.[6a] Thiazolo[5,4-d]thiazole derivatives became important electron acceptors in donor-acceptor π-conjugated systems due to their electron-deficient nature.[10]

The complexity of the absorption and emission mechanisms in these compounds arises from the π-electron conjugation within the ring, which enables efficient charge transfer and stabilization of excited states.[2b] The optical properties of thiazolothiazoles are highly sensitive to the presence of specific functional groups,[11] environmental changes, such as solvent polarity,[10, 12] the degree of protonation or deprotonation[12c, 13] and presence of metal cations,[14] allowing for flexible tuning to meet the demands of various applications (for viologen-type compounds see works of Li and Woodward).[15] Consequently, studies on photoluminescence and angular changes in optical dipoles in excited states are crucial for fully understanding their potential and require advanced spectroscopic analyses.

From a practical perspective, the ability to modify the thiazolothiazole ring facilitates tailoring their optical properties. Most of optoelectronic applications of thiazolothiazoles (solar cells, OFETs, etc.) are based either on polymeric structures with thiazolothiazole moiety as one of the building blocks,[16] or asymmetric (usually of push-pull architecture) derivatives.[17] In this context, symmetrical derivatives have attracted much less attention. On the other hand, symmetrical derivatives pose much smaller synthetic obstacles and should also offer significant tenability of their electronic and optical properties. This could lead to the development of new dyes with high resistance to fading, precisely controlled emission parameters, and selectivity for specific biomolecules or metal ions, as well as n–type organic semiconductors,[18] which are much less common and less stable that their *p*-type counterparts. Given the increasing demand for optical materials with precisely controlled parameters, detailed studies on the properties of thiazolothiazoles especially their optical and spectroscopic properties, may inspire the development of innovative technologies, particularly in fields such as chemical detection, advanced display technologies, and light energy storage.

In this work we report structural, spectroscopic and electronic properties of a series of thiazolothiazoles with simple aromatic substituents as well as several diphenyl derivatives substituted with electron withdrawing (trifluoromethyl) and electron donating (hydroxyl, methoxyl) substituents. Some of them (o-vanilin, 2-imidazoyl and 2-thiazoyl derivatives) are also potential ligands for transition metal ions. Here we will mostly focus on structure – property relationships in the context of applications in optoelectronics, especially in memristive devices. Weak electron acceptor character of the core thiazolothiazoyl unit suggests possibility of reversible resistive switching due to redox-controlled delocalization of electrons in molecular materials.[19]

## Results and discussion

### Synthesis and Molecular Structure of Thiazolothiazoles

The synthesis of thiazolothiazoles (TzTzs), including both known compounds and several new species, was carried out using the classical Hantzsch reaction.[20] In this case, the reaction involves the condensation of dithiooxamide with an appropriate aldehyde.[8a, 21] This method is relatively straightforward but requires high-boiling solvents and poses challenges in isolating and crystallizing pure products. The reaction is conducted at temperatures ranging from 120°C to 150°C for 3–8 hours. The formation of the TzTz ring system is quite easily confirmed by the photoluminescence observable under UV light. Gradual cooling of the reaction mixture to room temperature promotes the precipitation of crystalline product, facilitating characterization and subsequent purification. Immediate purification is recommended due to the very low solubility of the final compound in common solvents.

The thiazolothiazoles (TzTzs) studied in this work were categorized into three groups based on the different substituents at the TzTz unit: Type A, Type B, and Type C, as shown in Scheme I. Type A consists of diverse phenyl groups with electron-withdrawing and electron-donating substituents. Type B includes 5-membered heterocycles as substituents. Finally, Type C, represented by a cinnamyl-substituted TzTz, can be summarized as a highly conjugated species.

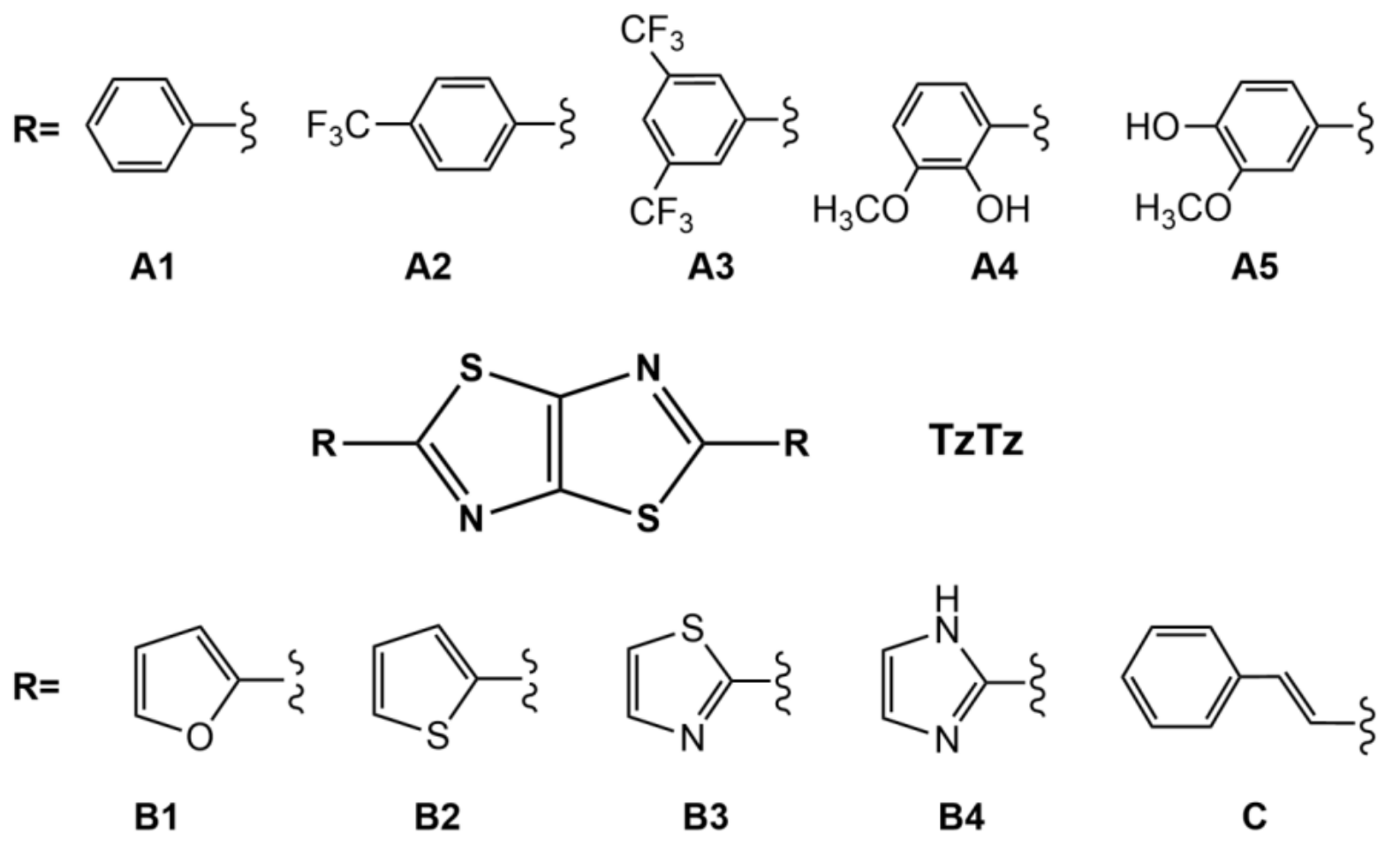

***Scheme I.*** *Chemical structures of* ***TzTzs*** *under study.*

The characterization of TzTzs should always include a detailed description of their crystal structure or, at a minimum, the identification of a phase using powder XRD. It has been observed multiple times that crystal packing significantly influences the physical properties of these materials. Therefore, obtaining a compound in a defined polymorphic form is essential to ensure the reproducibility of its properties, particularly its luminescence.[22]

Typically, simple TzTz derivatives are flat, aromatic molecules. They adopt an anti-conformation of the nitrogen and sulfur atoms around the central double bond within the TzTz rings, along with an all-trans arrangement of their substituents. The crystal structure of the parent thiazolo(5,4-d)thiazole was first reported in 1987 by Bolognesi et al.[23] As illustrated in Figure 1, the fused thiazole rings are symmetrically arranged around an inversion center located at the midpoint of the C1-C1B bond (as designated in Bolognesi's original work). These rings are nearly perfectly coplanar, showing a maximum displacement of only 0.002 Å—a distinguishing characteristic of all derivatives of the simplest TzTz molecule.

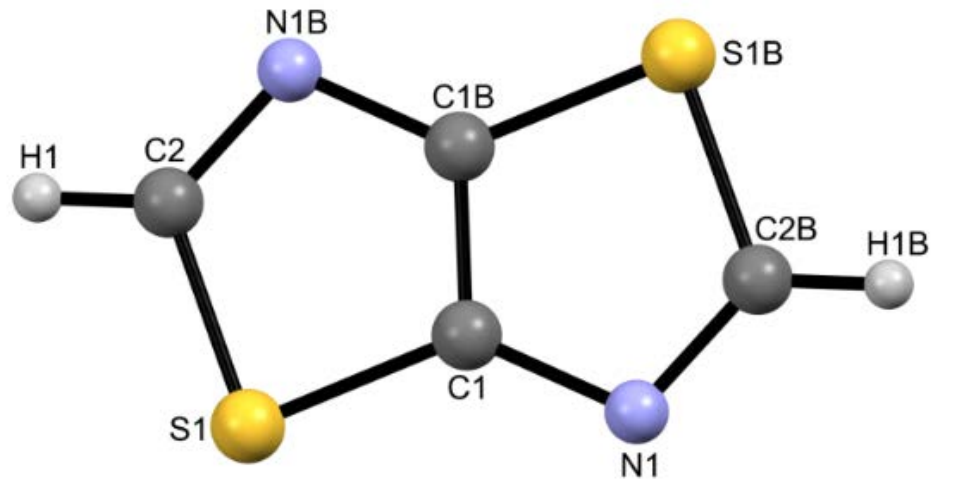

***Figure 1.*** *Molecular structure of thiazolo(5,4-d)thiazole.[23]*

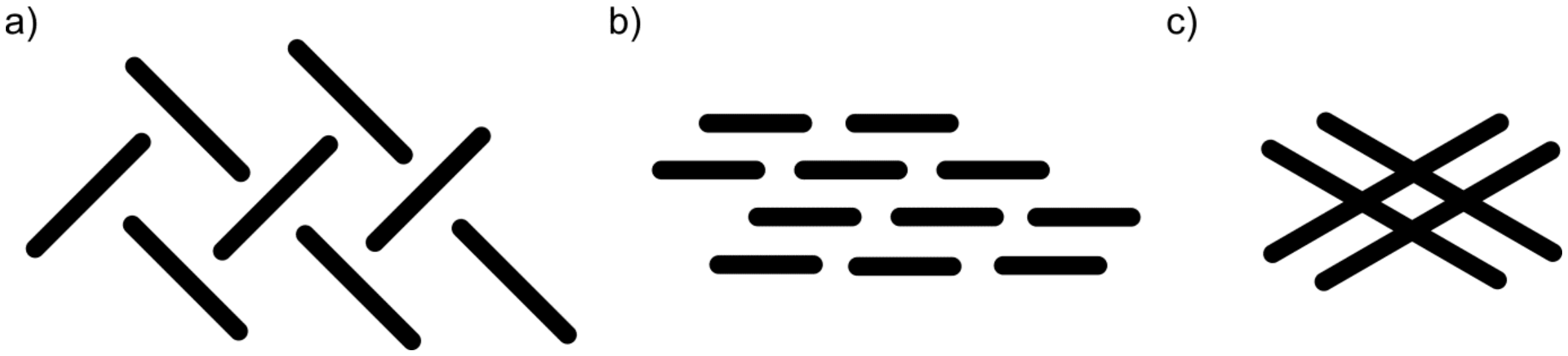

***Figure 2****. Typical crystal packing of thiazolo(5,4-d)thiazole derivatives: a) herringbone; b) parallel; c) grid. More than one two dimensional pattern may be observed in various directions.[24]*

Similar to the aromatic hydrocarbons the crystal packing of aromatic TzTzs may be categorized into the typical patterns such as herringbone or parallel illustrated in Figure 2a and b.[24] In the case of larger TzTz molecules or less usual intermolecular interactions the grid arrangement of molecules is possible shown in Figure 2c (for example see the crystal structure of 2,5-bis(2-fluorophenyl)[1,3]thiazolo[5,4-d][1,3]thiazole).[25] We distinguish this grid type to emphasize the interactions between the long edge of one molecule and aromatic system or heteroatoms of the other molecule, for example CH···π interactions in 2,5-bis(pyridin-4-yl)[1,3]thiazolo[5,4-d][1,3]thiazole.[22b] On the other hand the π-stacking interactions in the crystals of TzTzs may be realized in the form of approximately vertical or shifted stacks (Figure 3).

The crystal data for the new structures of **A3**–**A5**, **B1**, **B4**, and **C** are presented in Table S1 of the Supplementary Materials. The geometrical parameters of the molecules, including bond lengths and angles, are summarized in Tables S2 and S3. Selected intermolecular interactions are listed in Tables 1 and 2. Corresponding data for the remaining compounds can be found in the following references: **A1,** [293 K, yellow?[26], 100 K, colorless?[27]] **A2,**[28] **B2,**[29] and **B3**.[30]

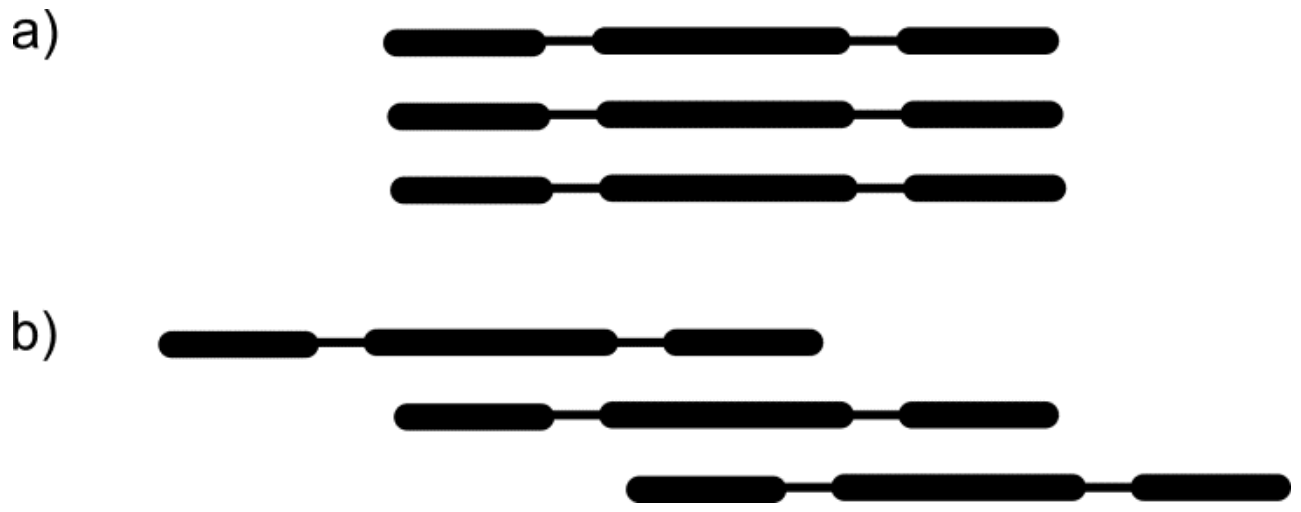

***Figure 3***. *π-stacking interactions in the crystal packing of thiazolo(5,4-d)thiazole derivatives: a) vertical; b) shifted. The longer bar represents the central TzTz unit, while the shorter bars indicate the substituents attached to the central unit.*

The molecular architectures of all TzTzs depicted in Figure 4 exhibit a characteristic coplanar arrangement of flat aromatic systems. Even in compound C, which uniquely includes an additional linker between its aromatic rings, this coplanarity is maintained. Consistently, the molecules are centrosymmetric, as previously discussed in this chapter, with an inversion center positioned at the midpoint of the C–C bond within the TzTz core. In compounds **A4**, **A5**, and **B1–B3**, this inherent centrosymmetry dictates a *trans* configuration for the asymmetrically substituted phenyl rings or heterocycles. In **A4**, the conformation observed through diffraction measurements is stabilized by an intramolecular hydrogen bond, O1–H1···N1. While the geometrical parameters of molecules, which are collected in Table S2, do not vary significantly between the compounds, the principal distinctions among all these compounds emerge upon examining their crystal packing arrangements, necessitating a separate analysis for each structure type. The statement is further supported by the analysis of the radial distribution of sulfur-sulfur pairs, which serves as a reliable measure of crystal packing in the case of TzTzs and highlights significant differences among the compounds (Figure S1). We focused on the sulfur-sulfur distance distribution to support our analysis of sulfur K-edge XAS for the studied thiazolothiazoles.

Among the studied TzTzs there are 6 typical herringbone packings: **A1**, **A4**, **B1**-**B3**, **C**, only one parallel **A3** and three examples of grid: **A2**, **A5** and **B4**.

The compounds **A1** and **B1**-**B3** realize very similar type of intermolecular interactions, which involve CH···π, CH···heteroatom, chalcogen···chalcogen or chalcogen···pnictogen and shifted π···π stacking interactions shown in Figure 5 for **B1**. Tables 1 and 2 characterize some of these intermolecular contacts for all TzTzs.

The only example of parallel packing among the studied compounds, **A3**, crystallizes in the orthorhombic system with the space group *Ibam*, in contrast to the remaining compounds, which crystallize in the monoclinic crystal system (Table S1). The centrosymmetric molecule is planar, except for the fluorine atoms in the trifluoromethyl groups. The molecules form layers along the *a* and *b* axes, which stack along the *c* axis. Within each layer, the molecules of compound **A3** are stabilized by C–H···F hydrogen bonds which also contribute to interlayer stabilization.

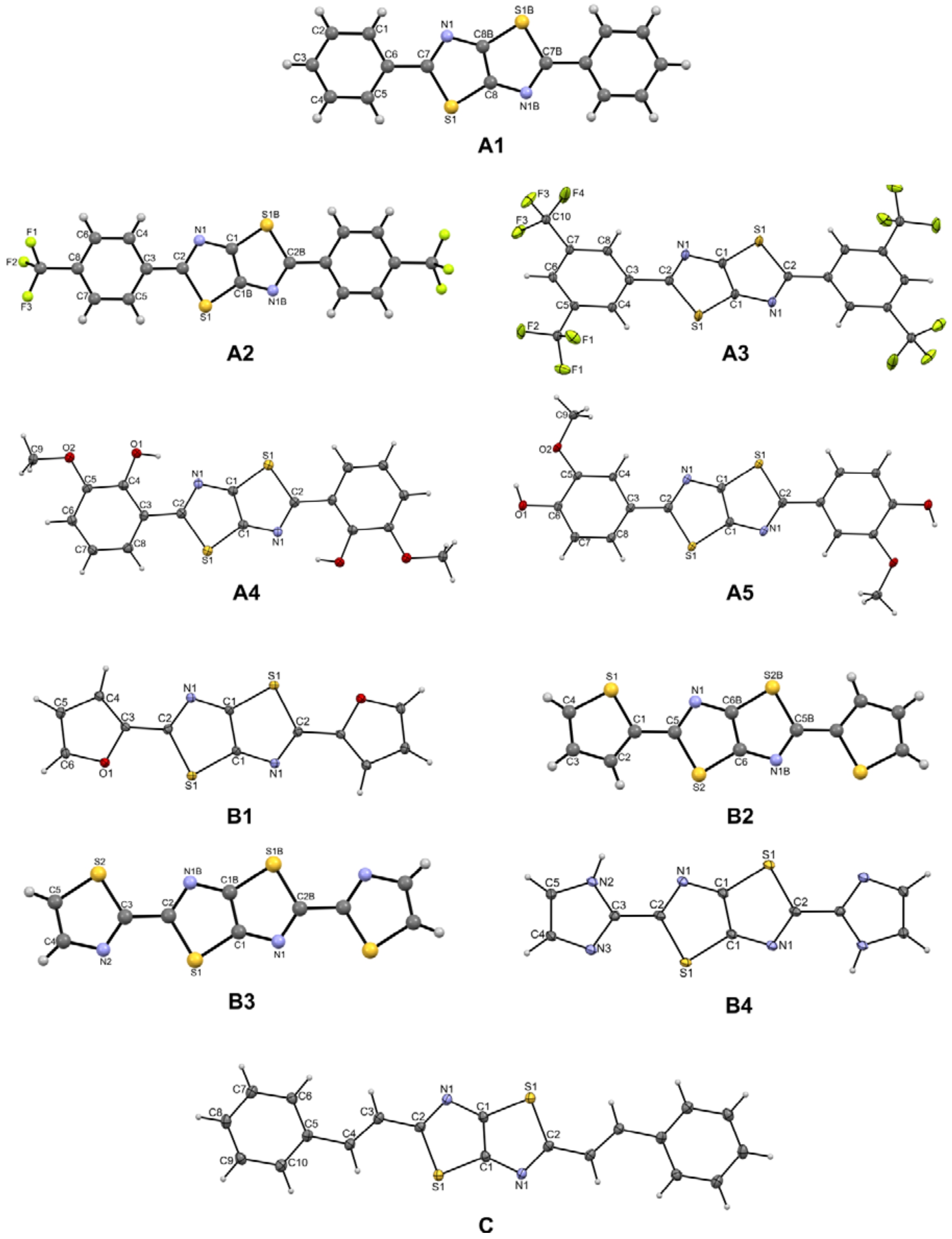

***Figure 4.*** *Molecular structures of* ***A1***–***A5****,* ***B1***–***B4****, and* ***C****, with the numbering schemes. The structures determined in this work (****A3***–***A5****,* ***B1****,* ***B4****,* ***C****) are depicted with 50% probability ellipsoids, while the remaining structures, sourced from the literature, are presented as ball-and-stick models. The source data for these figures were obtained from the Cambridge Structural Database (CSD)****.***

**Table 1** Chalcogen and chalcogen-pnictogen interactions in the studied compounds.

| Compound | A1a | A1b | A2 | A3 | A4 | A5 | B1 | B2 | B3 | B4 | C |
|---|---|---|---|---|---|---|---|---|---|---|---|
| S1/S2⋯S1$^{iii/iv/vi}$/ S1B$^{i/ii}$/S2B$^{v}$ [Å] | 3.566 | 3.532 | - | - | 3.775 | 3.964 | - | 3.579 | 3.507 | 3.744 | 3.360 |
| S1⋯N1$^{vii}$ [Å] | - | - | - | - | - | - | 3.343 | - | - | - | - |

$^{i}$: 1-x, 2-y, 2-z; $^{ii}$: 2-x,2-y,1-z; $^{iii}$: 1-x,-y,-z; $^{iv}$: x,1+y,z; $^{v}$: -x,1-y,-z; $^{vi}$: 1+x,y,z; $^{vii}$: -x,2-y,1-z; $^{viii}$: x,-1+y,z, ix: 1/2-x,-1/2+y,1.5-z

***Table 2** Shortest π⋯π stacking interactions in the studied compounds characterized by the distances between the centroids of the rings, slippage angle and slippage value.[]*

| Compound | π-π stacking interactions | Distances between Cg-Cg [Å] | Slippage angle [°]/ slippage value [Å] | Cg definition |
|---|---|---|---|---|
| **A1a** | $Cg_{Ph1}$⋯$Cg_{TzTz1}$ | 3.787 | 16.17/1.329 | $Cg_{Ph1}$: C1–C2–C3–C4–C5–C6<br>$Cg_{TzTz1}$: S1$^{i}$–C7$^{i}$–N1$^{i}$–C8B$^{ii}$–C8$^{i}$ |
| | $Cg_{TzTz2}$⋯$Cg_{TzTz3}$ | 4.005 | 19.92/1.699 | $Cg_{TzTz2}$: S1–C7–N1–C8B$^{iii}$–C8<br>$Cg_{TzTz3}$: S1$^{ii}$–C7$^{ii}$–N1$^{ii}$–C8B$^{ii}$–C8$^{i}$ |
| **A1b** | $Cg_{Ph2}$⋯$Cg_{TzTz4}$ | 3.715 | 16.65/1.337 | $Cg_{Ph2}$: C3–C4–C5–C6–C7–C8<br>$Cg_{TzTz4}$: C1B$^{iv}$–C1$^{i}$–N1$^{i}$–C2$^{i}$–S1B$^{iv}$ |
| | $Cg_{TzTz5}$⋯$Cg_{TzTz6}$ | 3.935 | 19.86/1.662 | $Cg_{TzTz5}$: C1B$^{v}$ – C1– N1– C2– S1B$^{v}$<br>$Cg_{TzTz6}$: C1B$^{iv}$–C1$^{i}$–N1B$^{iv}$–C2B$^{iv}$–S1$^{i}$ |
| **A2** | $Cg_{Ph3}$⋯$Cg_{TzTz7}$ | 3.766 | 15.32/1.242 | $Cg_{Ph3}$: C3$^{vi}$–C4$^{vi}$–C5$^{iv}$–C6$^{iv}$–C7$^{vi}$–C8$^{vi}$<br>$Cg_{TzTz7}$: C1–C1B$^{vii}$–N1B$^{vii}$–C2B$^{vii}$–S1B$^{vii}$ |
| | $Cg_{Ph3}$⋯$Cg_{TzTz8}$ | 3.949 | 20.72/1.716 | $Cg_{Ph3}$: C3$^{vi}$–C4$^{vi}$–C5$^{iv}$–C6$^{iv}$–C7$^{vi}$–C8$^{vi}$<br>$Cg_{TzTz8}$: C1B$^{vii}$–C1–N1B–C2–S1 |
| **A3** | $Cg_{TzTz9}$⋯$Cg_{TzTz10}$ | 3.764 | 8.93/0.584 | $Cg_{TzTz9}$: C1$^{ix}$–C1–N1–C2–S1$^{ix}$<br>$Cg_{TzTz10}$:C1$^{xi}$–C1$^{x}$–N1$^{x}$–C2$^{x}$–S1$^{xi}$ |
| | $Cg_{Ph4}$⋯$Cg_{Ph5}$ | 4.058 | 23.56/1.621 | $Cg_{Ph4}$: C3–C4–C5–C6–C7–C8<br>$Cg_{Ph5}$: C3$^{viii}$–C4$^{viii}$–C5$^{viii}$–C6$^{viii}$–C7$^{viii}$–C8$^{viii}$ |
| **A4** | $Cg_{TzTz12}$⋯$Cg_{Ph6}$ | 3.642 | 12.07/1.050 | $Cg_{TzTz12}$: C1– C1$^{xii}$ – N1$^{xii}$ – C2$^{xii}$ – S1$^{xii}$<br>$Cg_{Ph6}$: C3$^{xiii}$–C4$^{xiii}$–C5$^{xiii}$–C6$^{xiii}$–C7$^{xiii}$–C8$^{xiii}$ |
| | $Cg_{TzTz11}$⋯$Cg_{Ph6}$ | 3.659 | 12.70/1.107 | $Cg_{TzTz11}$: C1$^{xii}$–C1–N1–C2–S1<br>$Cg_{Ph6}$: C3$^{xiii}$–C4$^{xiii}$–C5$^{xiii}$–C6$^{xiii}$–C7$^{xiii}$–C8$^{xiii}$ |
| **A5** | $Cg_{Ph7}$⋯$Cg_{Ph8}$ | 3.964 | 20.09/1.776 | $Cg_{Ph7}$: C3–C4–C5–C6–C7–C8<br>$Cg_{Ph8}$: C3$^{xiii}$–C4$^{xiii}$–C5$^{xiii}$–C6$^{xiii}$–C7$^{xiii}$–C8$^{xiii}$ |
| | $Cg_{TzTz13}$⋯$Cg_{TzTz14}$ | | | $Cg_{TzTz13}$: C1–C1–N1–C2–S1$^{xiv}$<br>$Cg_{TzTz14}$: C1$^{xv}$–C1$^{xv}$–N1$^{xv}$–C2$^{xv}$–S1$^{xiv}$ |
| **B1** | $Cg_{Fh1}$⋯$Cg_{TzTz16}$ | 3.537 | 12.91/0.942 | $Cg_{Fh1}$: C3$^{iv}$–C2$^{iv}$–C1$^{iv}$–S1$^{iv}$–C4$^{iv}$<br>$Cg_{TzTz16}$: C1–C1$^{v}$–N1$^{v}$–C2$^{v}$–S1 |
| | $Cg_{Fh1}$⋯$Cg_{TzTz15}$ | 3.584 | 15.14/1.112 | $Cg_{Fh1}$: C3$^{iv}$–C2$^{iv}$–C1$^{iv}$–S1$^{iv}$–C4$^{iv}$<br>$Cg_{TzTz15}$: C1$^{v}$–C1–N1–C2–S1$^{v}$ |
| **B2** | $Cg_{TzTz17}$⋯$Cg_{Th1}$ | 3.981 | 21.40/1.628 | $Cg_{Th1}$: C3$^{xx}$–C2$^{xx}$–C1$^{xx}$–S1$^{xx}$–C4$^{xx}$<br>$Cg_{TzTz17}$: C6B$^{xvi}$–C6–N1–C5–S2 |
| **B3** | $Cg_{Tz1}$⋯$Cg_{TzTz18}$ | 3.595 | 10.41/0.766 | $Cg_{Tz1}$: C4$^{i}$–C5$^{i}$ –N2$^{i}$ –C3$^{i}$–S1$^{i}$<br>$Cg_{TzTz18}$: C1–S1–C2–N1B$^{xx}$–C1B$^{xx}$ |
| | $Cg_{Tz1}$⋯$Cg_{TzTz19}$ | 3.717 | 16.15/1.224 | $Cg_{Tz1}$: C4$^{i}$–C5$^{i}$ –N2$^{i}$ –C3$^{i}$–S1$^{i}$<br>$Cg_{TzTz19}$: C1B$^{xx}$–S1B$^{xx}$–C2B$^{xx}$–N1–C1 |
| **B4** | $Cg_{TzTz20}$⋯$Cg_{TzTz21}$ | 3.540 | 10.07/0.861 | $Cg_{TzTz20}$: C1$^{xvii}$ –C1$^{xix}$–N1$^{xix}$–C2$^{xix}$–S1$^{xvii}$<br>$Cg_{TzTz21}$: C1–C1–N1–C2$^{iv}$–S1$^{iv}$ |
| | $Cg_{TzTz22}$⋯$Cg_{TzTz21}$ | 3.744 | 17.14/1.494 | $Cg_{TzTz22}$: C1$^{xix}$ –C $^{xvii}$ –N1$^{xvii}$ –C2$^{xvii}$ –S1$^{xix}$<br>$Cg_{TzTz21}$: C1–C1–N1–C2$^{iv}$–S1$^{iv}$ |
| | $Cg_{Im1}$⋯$Cg_{Im2}$ | 3.744 | 17.14/1.534 | $Cg_{Im1}$: N2–C3–N3–C4–C5<br>$Cg_{Im2}$: N2$^{xvii}$–C3$^{xvii}$–N3$^{xvii}$–C4$^{xvii}$–C5$^{xvii}$ |
| **C** | $Cg1_{C=C}$ ⋯$Cg_{TzTz25}$ | 3.471 | 1.73/0.133 | $Cg1_{C=C}$: C3$^{iv}$–C4$^{iv}$<br>$Cg_{TzTz25}$: C1–C1$^{v}$–N1$^{v}$–C2$^{v}$–S1 |
| | $Cg2_{C=C}$ ⋯$Cg_{Ph9}$ | 3.475 | 1.47/0.113 | $Cg2_{C=C}$: C3$^{v}$–C4$^{v}$<br>$Cg_{Ph9}$: C5$^{iv}$–C6$^{iv}$–C7$^{iv}$–C8$^{iv}$–C9$^{iv}$–C10$^{iv}$ |
| | $Cg_{TzTz23}$⋯$Cg_{TzTz24}$ | 3.711 | 15.21/1.211 | $Cg_{TzTz23}$: C1$^{v}$–C1–N1–C2–S1$^{v}$<br>$Cg_{TzTz24}$: C1$^{i}$–C1$^{iv}$–N1$^{iv}$–C2$^{iv}$–S1$^{i}$ |

$Cg_{Ph}$ - centroid of the phenyl ring; $Cg_{Tz}$ - centroid of the thiazole ring; $Cg_{TzTz}$ - centroid of the thiazole ring from the TzTz system; $Cg_{C=C}$ - centroid of the C=C bond determined for the compound C; $Cg_{Th}$ - centroid of the thiophene ring; $Cg_{Im}$ - centroid of the imidazole ring; $Cg_{Fh}$- centroid of the furyl ring. Symmetry operations: $^{i}$ x,-1+y,z; $^{ii}$ 2-x,1-y,2-z; $^{iii}$ 2-x,2-y,2-z; $^{iv}$ 1-x,1-y,1-z; $^{v}$ 1-x,2-y,1-z; $^{vi}$ x,-1+y,z; $^{vii}$ -x,-y,-z; $^{xiii}$ 1-x,y,1/2-z; $^{ix}$ 1-x,1-y,z; $^{x}$ 1-x,y,1/2-z; $^{xi}$ x,1-y,1/2-z; $^{xii}$ 2-x,-y,1-z; $^{xiv}$ 2-x,1-y,1-z; $^{xv}$ x,1+y,z; $^{xvi}$ 1-x,1-y,-z; $^{xvii}$ -1+x,y,z; $^{xviii}$ 1-x,3-y,1-z; $^{xix}$ -x, 1-y, 1-z; $^{xx}$ 1-x, -y, -z

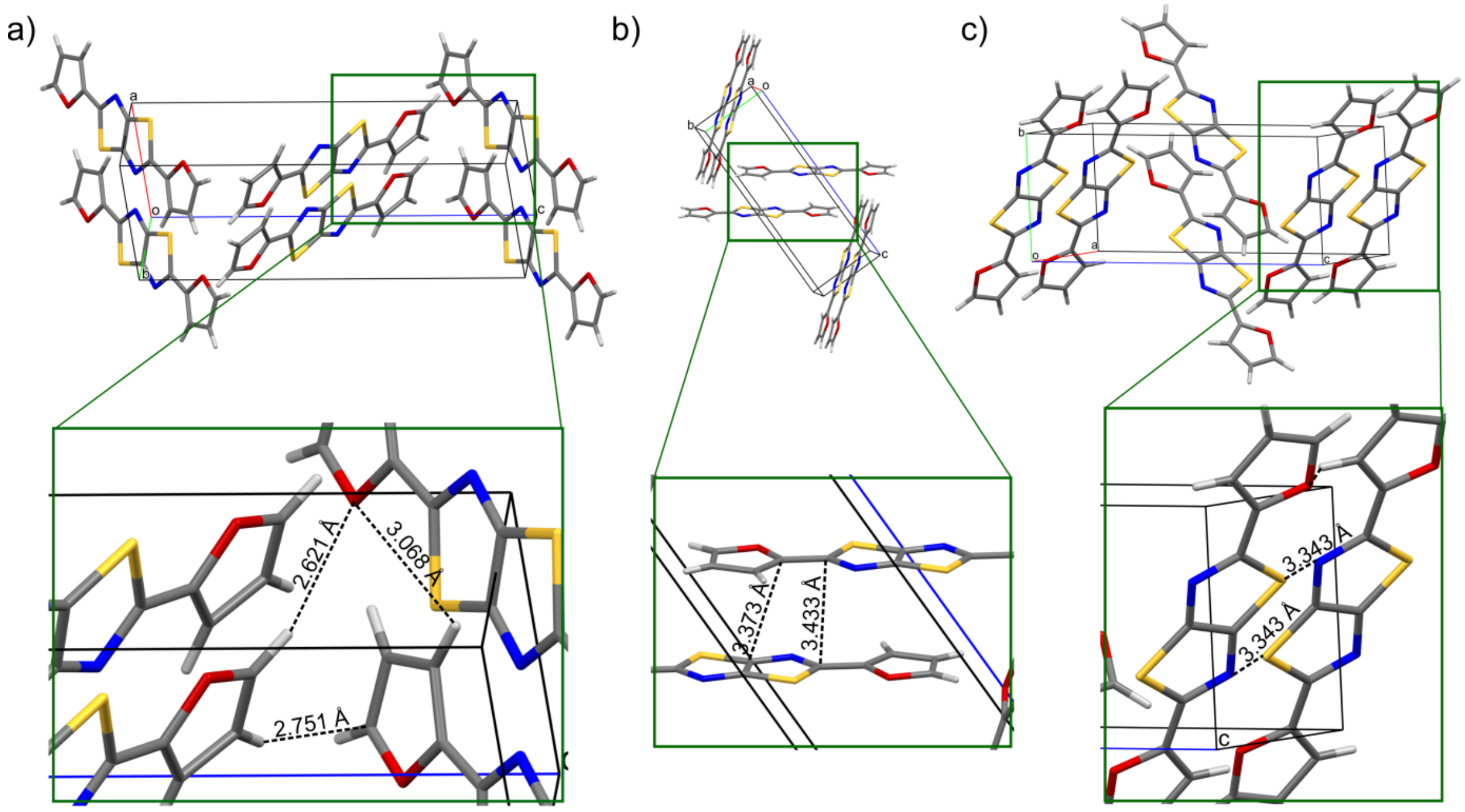

***Figure 5.*** *Crystal packing of **B1**: a) along axis b, the shortest intermolecular contacts include two CH···O interactions and one CH···π; b) along axis a, the shortest π···π stacking interactions are shown; c) along a, the shortest contact between sulfur and nitrogen atoms are visualized.*

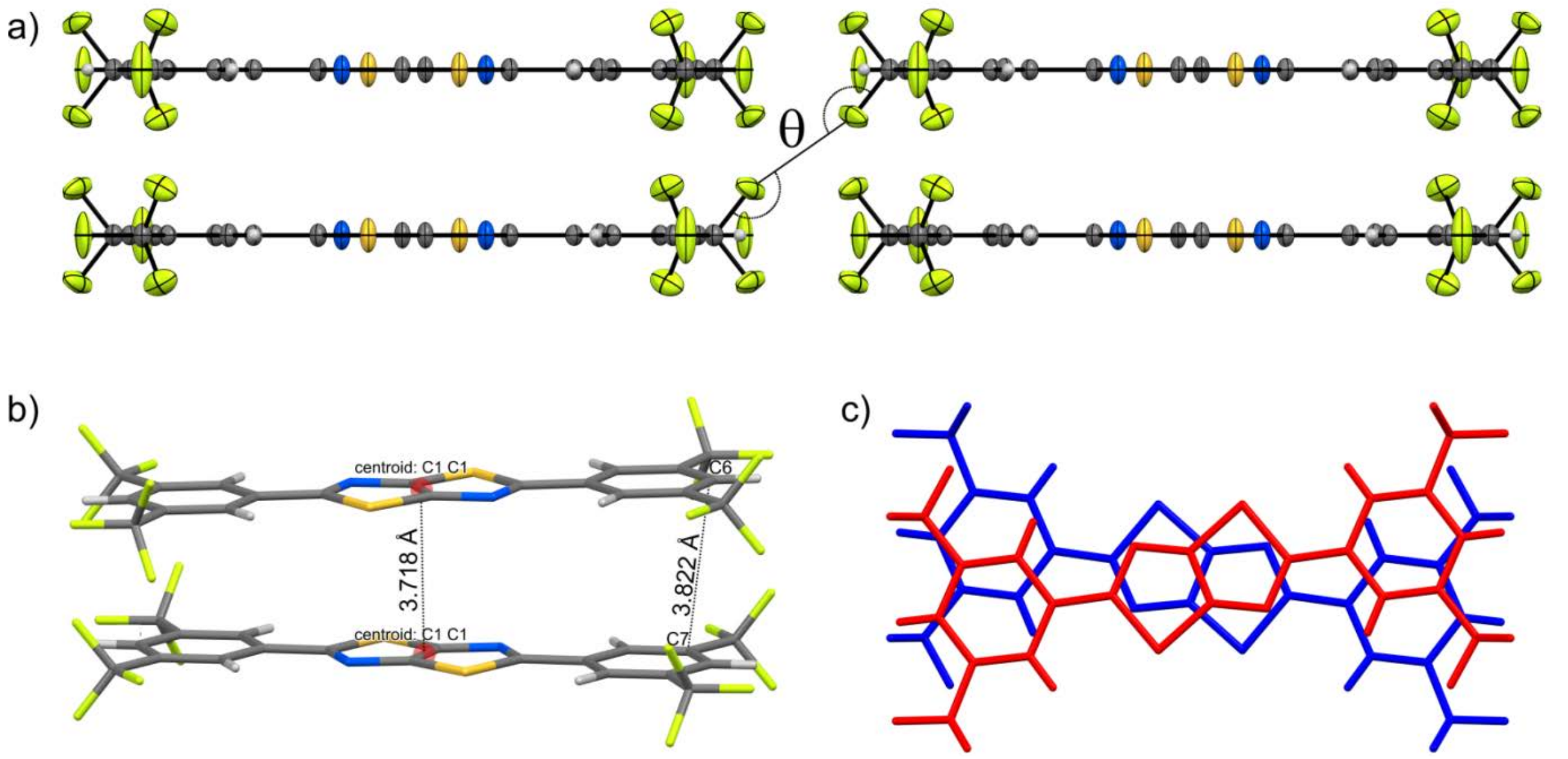

***Figure 6.*** *Crystal packing of **A3**: a) visualized along axis a, the torsion angle ϑ for C–F···F–C, bonds is shown; b) visualized approximately along axis a, the shortest π···π stacking interactions are shown; c) visualized along c to illustrate the rotation of the molecules between the layers.*

Furthermore, the presence of as many as 12 fluorine atoms per molecule facilitates the formation of interlayer halogen interactions, which play a key role in layer stabilization. The torsion angle θ for C–F···F–C, defined in Figure 6a, is 162.86°, and the fluorine-fluorine distance is 2.826 Å—typical values for the first type of F···F interactions in organic compounds.[31] The crystal lattice is further stabilized by vertical π···π stacking interactions as defined in Figure 3b and illustrated in Figure 6b. Interestingly, the parallel-aligned molecules exhibit a slight rotation, leading to variations in π···π stacking interaction lengths (Figure 6c). The closest interactions occur between the TzTz systems of neighboring molecules, while the more distant ones are observed between the 3,5-trifluoromethylphenyl substituents. The shortest distances shown in Figure 6b are as follows: between the centroids of C1-C1$^{i}$··· C1$^{ii}$-C1$^{iii}$ bonds – 3.718 Å, and C6(phenyl ring)···C7$^{ii}$(phenyl ring) – 3.822 Å ($^{i}$ 1-x, 1-y,z; $^{ii}$ 1-x, y, ½-z; $^{iii}$ x, 1-y, ½-z).

The typical S···S or S···N interactions commonly observed in thiazole systems are absent in this structure. These interactions were present in all previously studied compounds and played a significant role in stabilization. The absence of S···S/N interactions, combined with the presence of halogen interactions, highlights how substituent modifications influence the stabilization mechanisms.

The final type of crystal packing, observed in compounds **A2**, **A5**, and **B4**, is a grid arrangement, as illustrated in Figure 2c. In this arrangement, the predominant intermolecular interactions occur along the longer edge of the approximately planar molecule. In Figure 7 we provide compound **B4** as an example, where the molecular orientation is dictated by the formation of N–H···N hydrogen bonds between the imidazole substituents and the TzTz system (Figure 7a). Chains of hydrogen bonds extend along the *c*-axis, while in another direction, the crystal structure is stabilized by shifted π···π stacking interactions (Figure 7b).

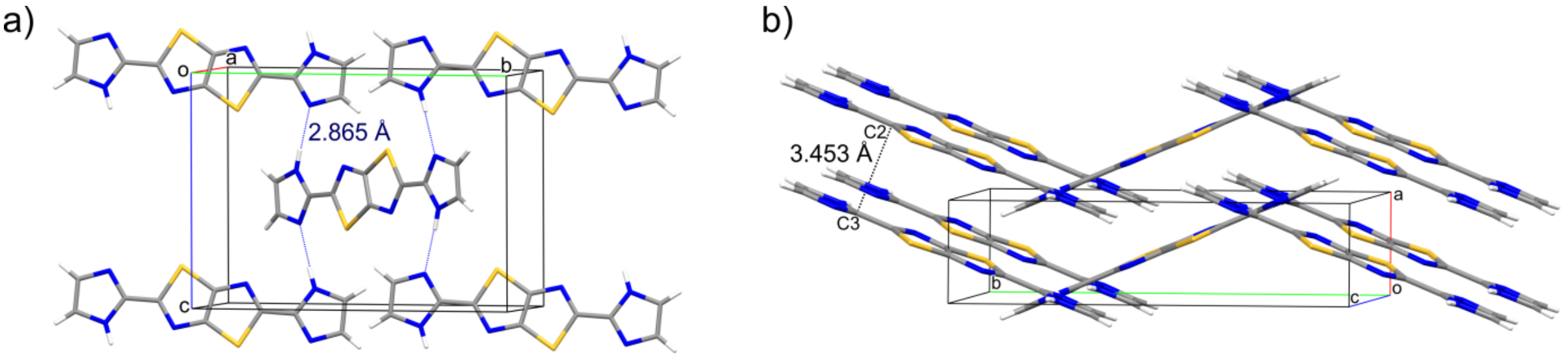

***Figure 7.*** *Crystal packing of **B4**: a) along axis a, the hydrogen bonding chains are visualized; b) along axis c, to illustrate the π···π stacking interactions.*

The crystal packings of the remaining compounds **A1**, **A2**, **A4**, **A5**, **B2**, **B3**, **C** are illustrated in Figure S2 and S3 of supplementary materials. The selected intermolecular interactions are additionally listed in Tables 1 and 2 and Figure 8, which shows the contributions of various short contacts on Hirshfeld surfaces of the studied molecules. We would like to point to the interactions summarized for two molecular structures of **A1** deposited in CSD – **A1**a and **A1**b.[26-27] Though the unit cells were selected in a different way for this two measurements, the interaction diagram generated with the Crystal Explorer[32] indicates the same crystal packing in both cases. Moreover, as expected, for **A1** and **C** we observe very large percentage of hydrophobic H···H interactions, for **A2** and **A3** very large percentage of diverse F···X contacts, **A4**, **A5** and **B1** O···X contacts, while **B4** is distinguished by a substantial contribution of NH···N hydrogen bonds.

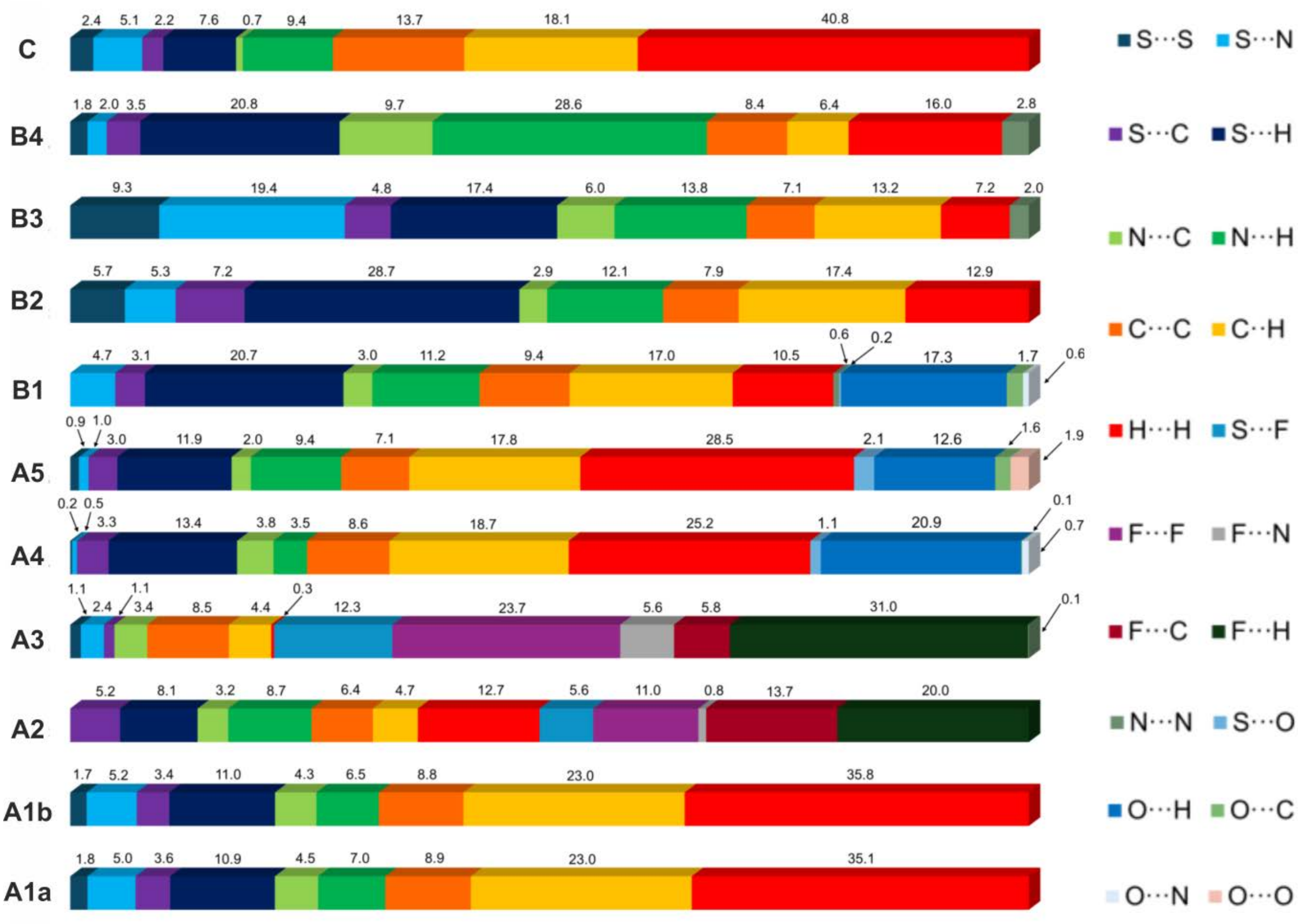

***Figure 8*** *The contributions of various types of contacts on the Hirshfeld surfaces of the studied TzTzs. Analysis performed with the use of CrystalExplorer.*[32]

**Electronic Properties of Thiazolothiazoles**

The thiazolo[5,4-d]thiazole core is a bicyclic heterocyclic aromatic moiety with 10 π electrons, thus is isoelectronic with naphthalene and thieno[3,2-b]thiophene. The presence of heteroatoms significantly changes their geometry as well as local and global electronic properties. All three molecules are planar, but in the case of naphthalene introduction of any substituent generates a significant steric hindrance due to hydrogen atoms in vicinal positions. In the case of thienothiophene it is highly reduced at there are only two hydrogen atoms at each ring. In the case of thiazolothiazole no steric hindrance should be present, which may render formation of extended planar conjugated systems much easier. Globally, thienothiophene has slightly higher energies of frontier orbitals, which emphasizes its character as an electron donor. This feature of thienothiophene is exploited in the design and fabrication of high mobility organic *p*-type semiconductors.[33] In contrast to thienothiophene, thiazolothiazole exhibits the characteristics of a moderate electron acceptor (Figure 9), with a LUMO level that is 0.46 eV lower compared to naphthalene. HOMO-LUMO separations increase on introduction of heteroatoms from 4.79 eV for naphthalene to 5.16 eV for thiazolothiazole.

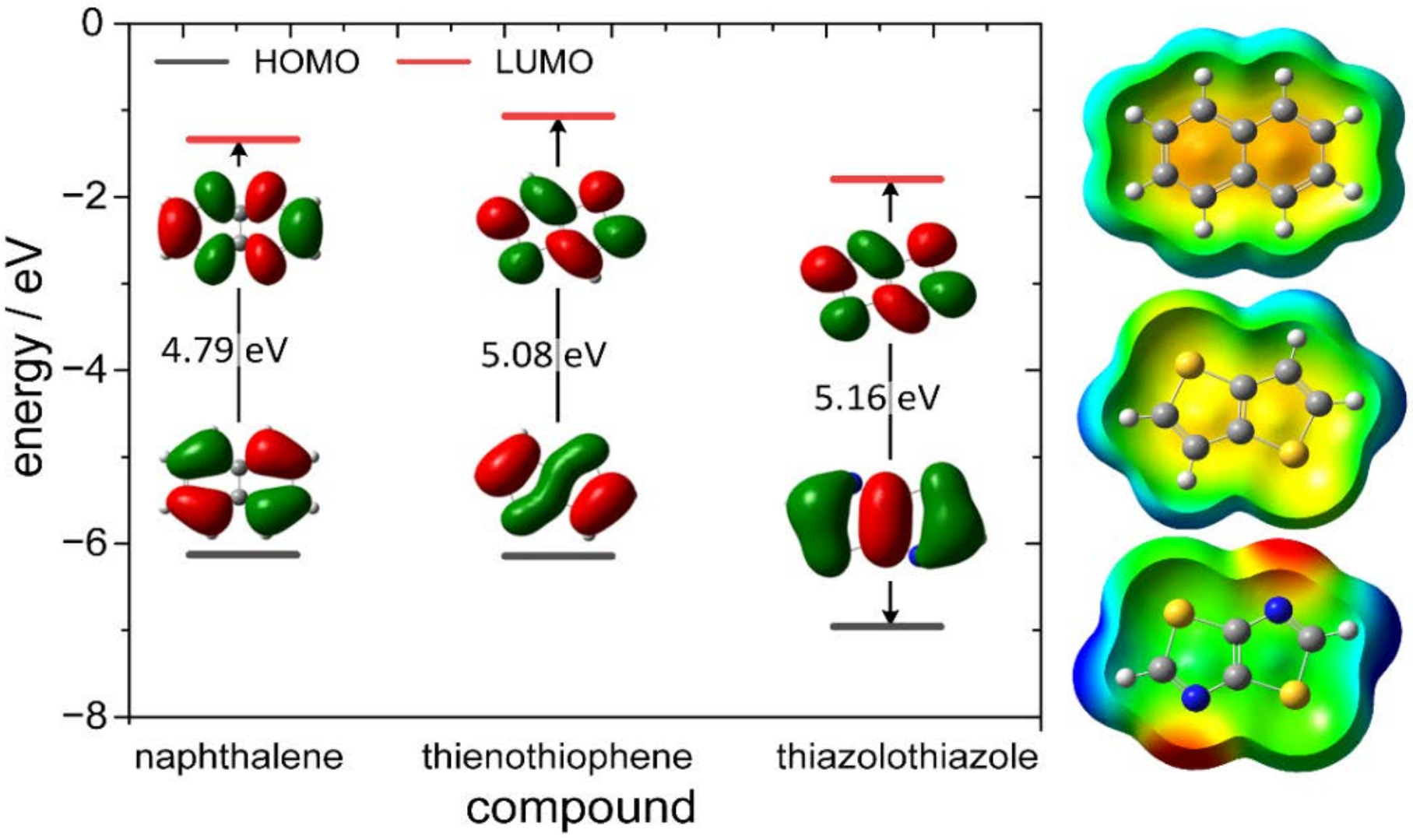

***Figure 9****. Energy and contours of frontiers orbitals of naphthalene, thienothiophene and thiazolothiazole along with electrostatic potentials mapper onto electron density isosurface as calculated using DFT approach at the B3LYP/TZVP level of theory. Electrostatic potential colour scale spans from −0.04 a.u. (red) to +0.04 a.u. (dark blue) and is identical for all three molecules.*

Introduction of heteroatoms results in local changes of electron density. In the case of naphthalene, the whole ring system bears local negative charge due to symmetrical delocalized cloud of π electrons. Introduction of sulfur slightly decreases the negative charge in the π-electron cloud, which also extends to lone electron pair at sulfur atoms. In the case of thiazolothiazole, the changes are striking: the nitrogen atoms carry the largest negative charge, while the peripheral protons are positively charged (Figure 9). This change may play a crucial role in molecular arrangement and intermolecular interactions in the solid phase.

***Electron structure***

DFT geometry optimization confirms planarity of all studied molecules. As a consequence, electronic π conjugation extends over whole molecules including aromatic or unsaturated substituents. Frontier orbital delocalization clearly indicates this process - see Table 3. Introduction of moderately electron withdrawing substituents like in **B3**, or electron donating groups like in **B2** does not change the charge distribution in the thiazolothiazole core. Stronger electron acceptors, which is the case of **A2** and **A3**, or the presence of electron donors as in **A4** and **A5** significantly change the charge distribution in the thiazolothiazole core. The strong negative character of nitrogen atoms vanishes and the distribution of charge becomes more uniform (**A2**, **A3**, **A5**) or becomes dominated by the substituent (**A3**, **A4**). This behaviour may have further consequence on transport phenomena in these materials when applied as semiconductors, as well as may affect their luminescence and other optical properties.

***Table 3.*** *Contours of frontier orbitals and maps of electrostatic potential distribution for studied thiazolothiazoles as calculated by the DFT method at the b3lyp/tzvp level of theory. Electrostatic potential colour scale spans from −0.04 a.u. (red) to +0.04 a.u. (dark blue) and is identical for all three molecules.*

| | **HOMO** | **LUMO** | **ESP map** |
|---|---|---|---|
| **A1** | 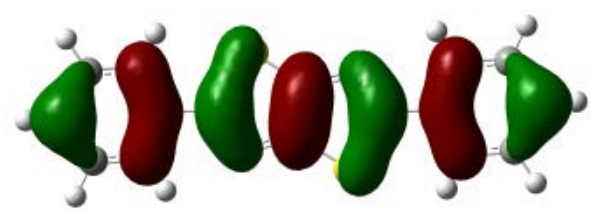 | 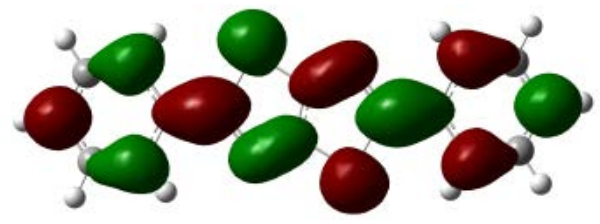 | 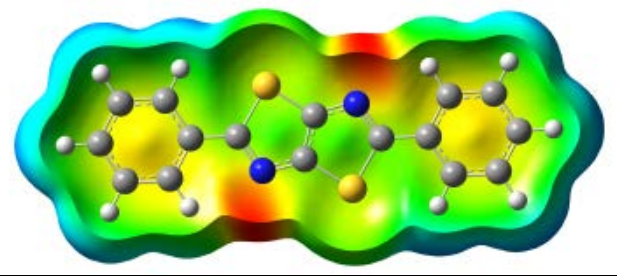 |
| **A2** | 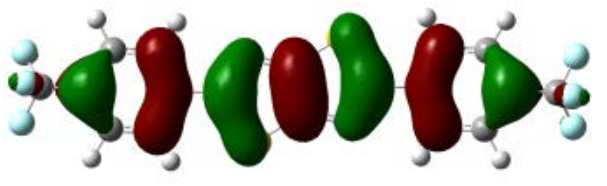 | 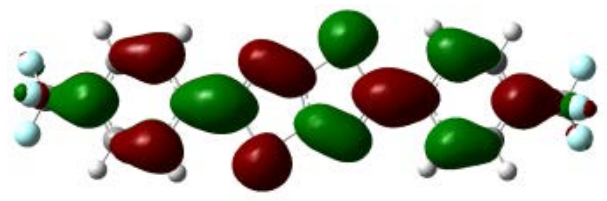 | 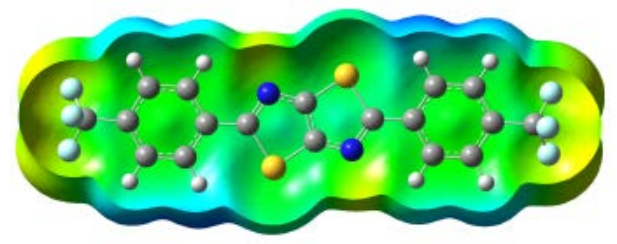 |
| **A3** | 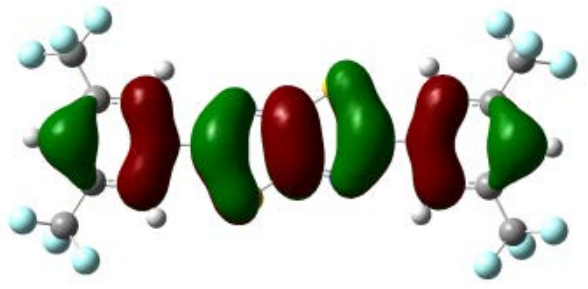 | 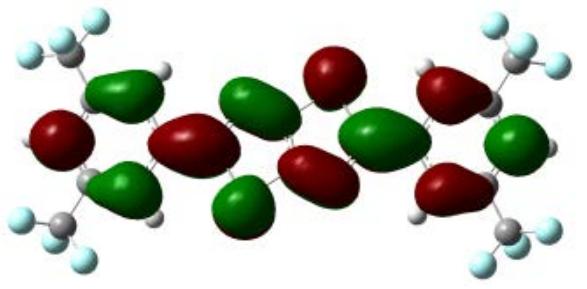 | 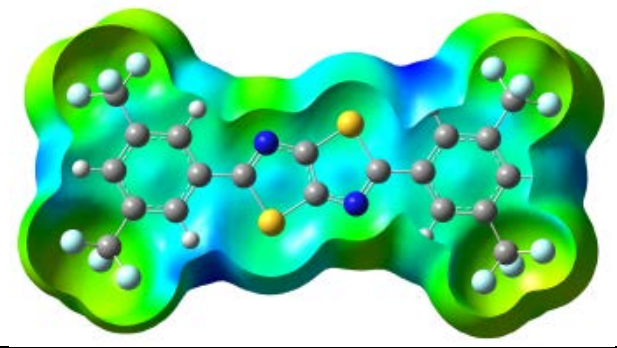 |
| **A4** | 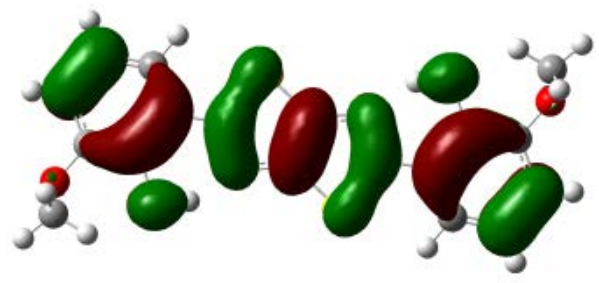 | 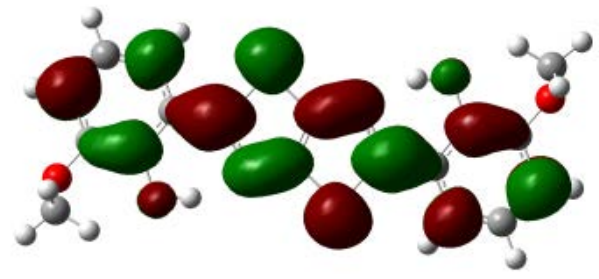 | 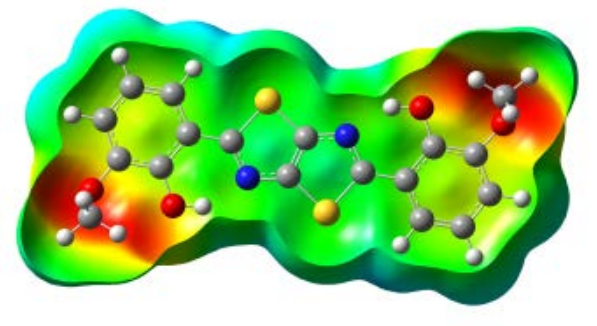 |
| **A5** | 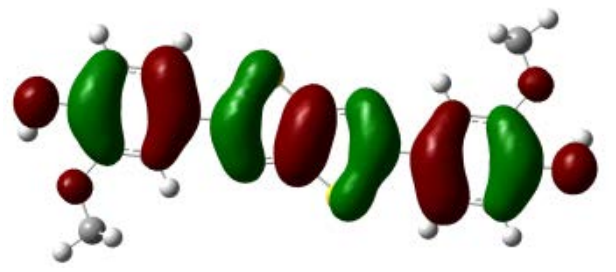 | 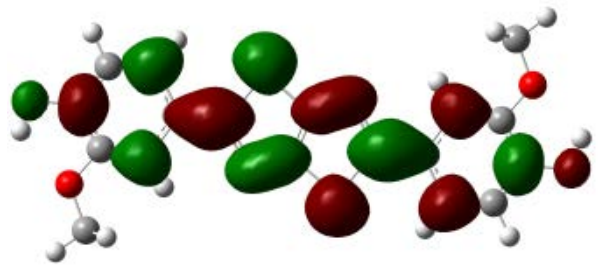 | 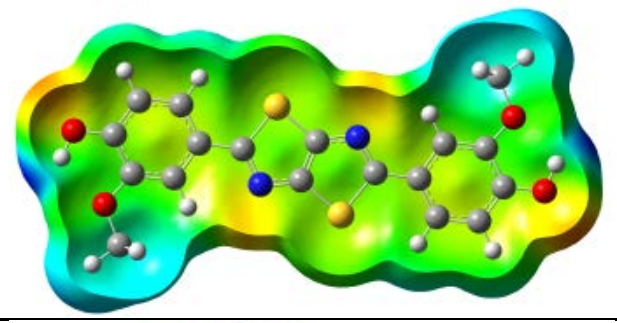 |
| **B1** | 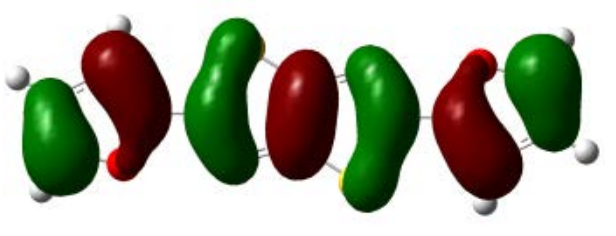 | 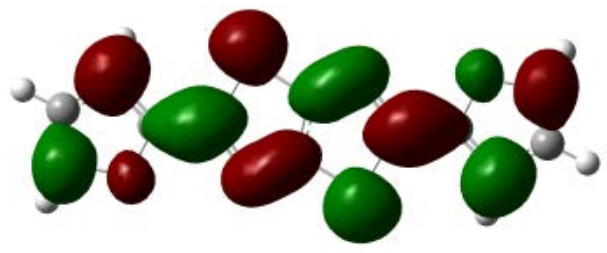 | 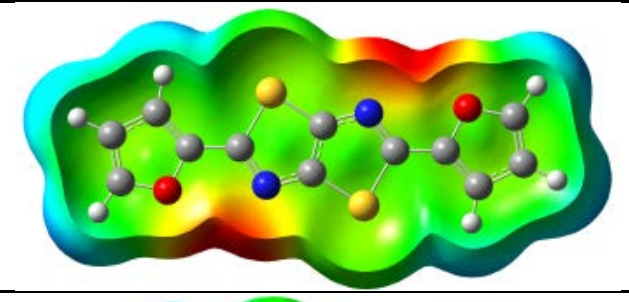 |
| **B2** | 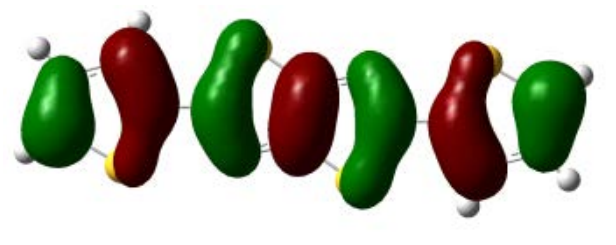 | 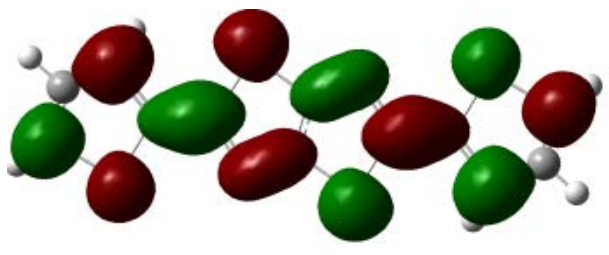 | 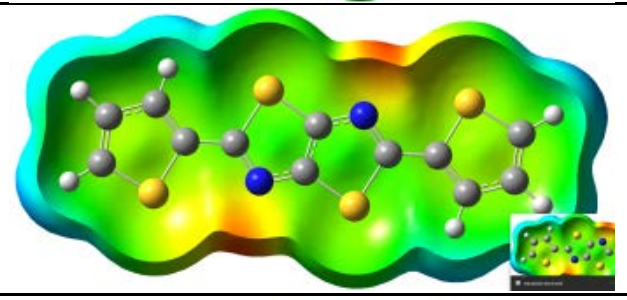 |
| **B3** | 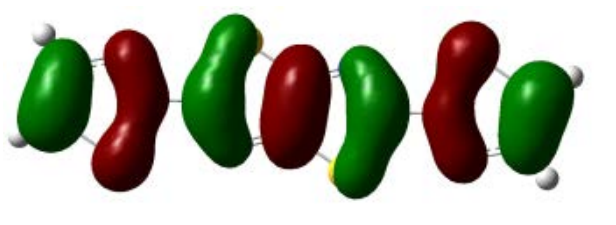 | 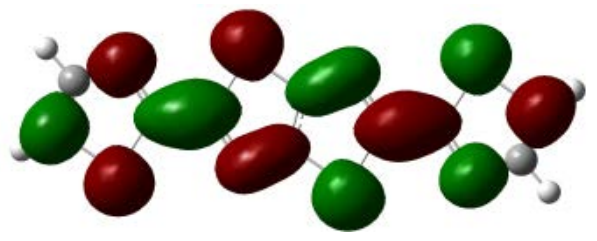 | 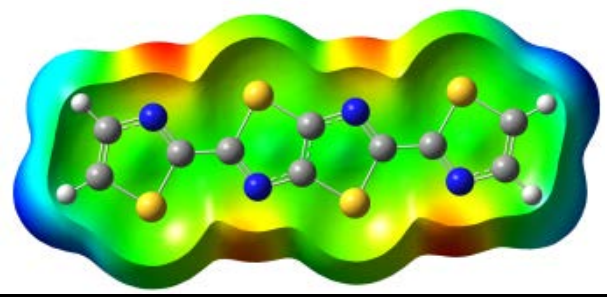 |
| **B4** | 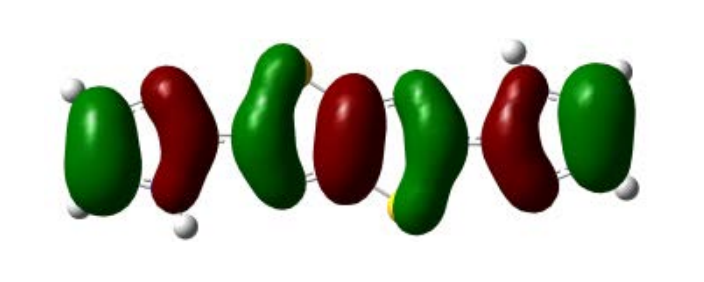 | 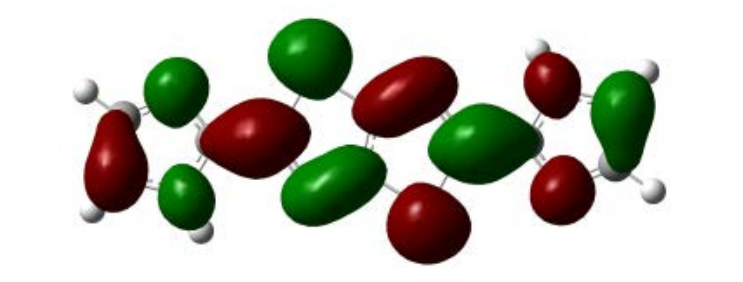 | 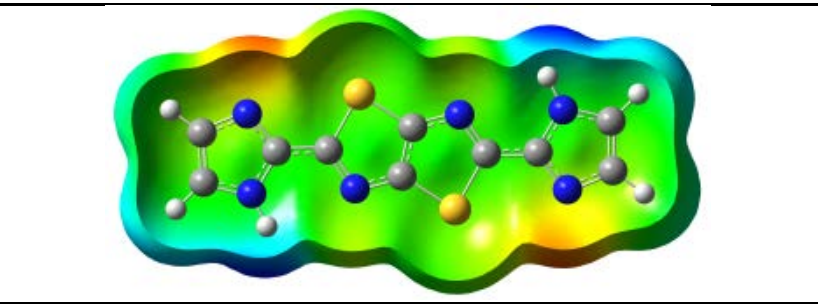 |
| **C** | 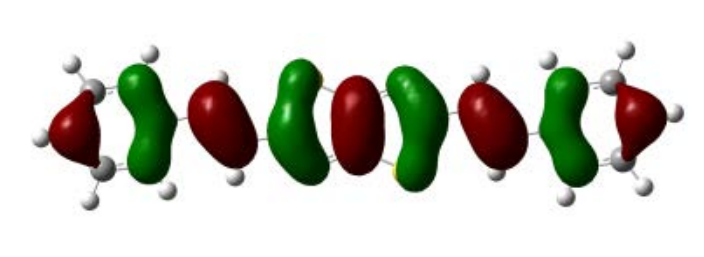 | 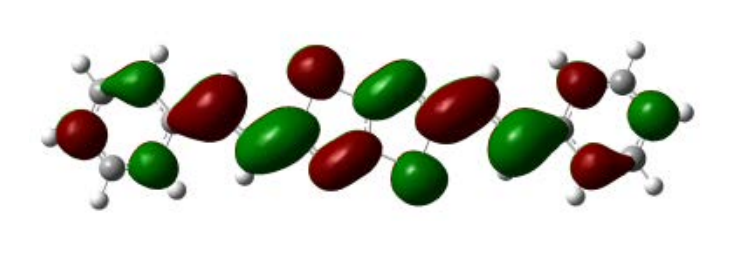 | 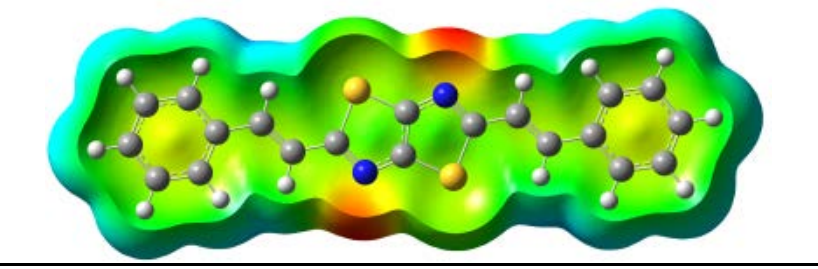 |

Along with rather minor changes in local electrostatic charge distribution, introduction of electron donors or electron acceptors significantly changes energies of frontier orbitals, with the HOMO level being the most sensitive one (Figure 10a). Interestingly, just rearrangement of substituents in vanillin derivatives results in a transition from mild electron acceptor to strong electron donor. As the sensitivity of HOMO and LUMO orbitals to the substituents are different, also the HOMO-LUMO energy separation significantly changes with substituents (Figure 10b). This offers an easy tool for engineering of the excited states and other photophysical properties of thiazolothiazoles, *e.g.* ability to engage in photoinduced electron transfer reactions, which may have severe consequences for their applications as fluorescent labels or homogeneous photocatalysts. In may be also important from the point of view of optoelectronic applications.

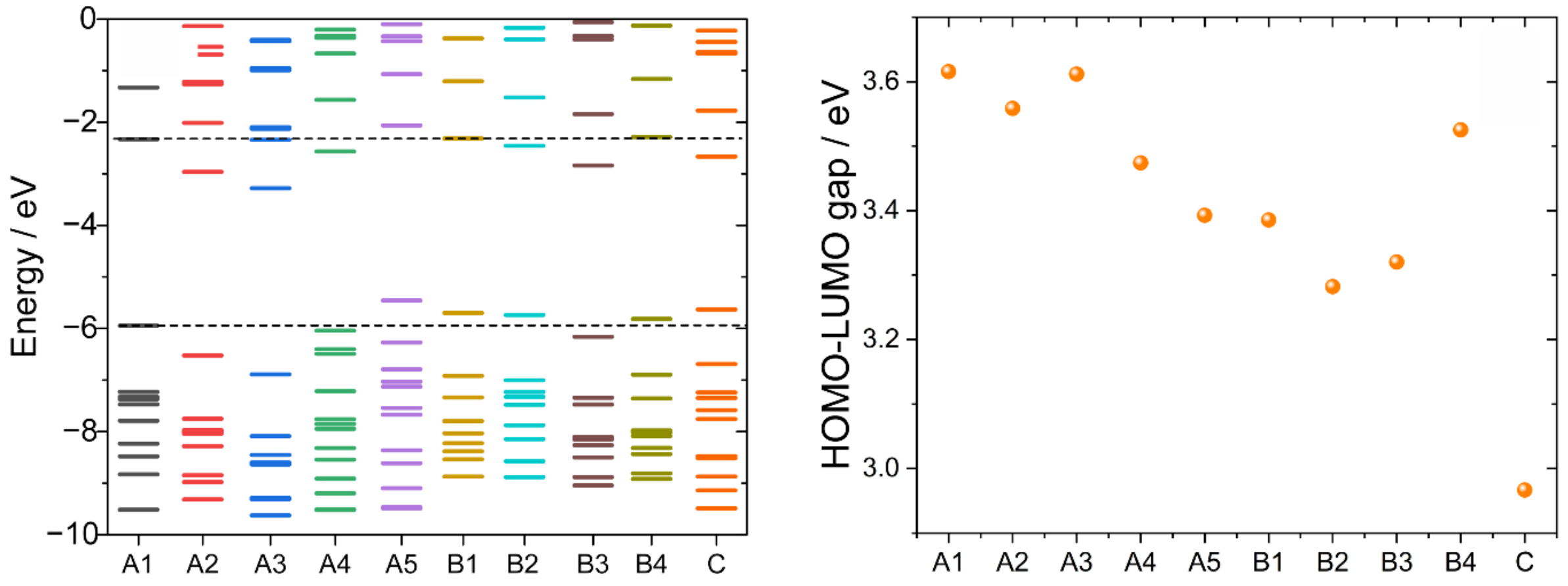

***Figure 10.*** *Orbital Energy diagrams (left side of the figure) and HOMO-LUMO gaps (right side of the figure) for all studied thiazolothiazoles derived from DFT calculations for single molecules at the B3LYP/TZVP level of theory. Dashed lined are reference levels for diphenyl derivative,* ***A1****.*

It has been already noted that various thiazolothiazole derivatives assume various special arrangements in the solid phase. It turns out, that these arrangements also affect electronic structure of solid phases. The most striking difference is observed in the case of **A2** and **A3** molecules. The first one crystallizes in parallel stacks without any shift between neighbouring molecules, whereas the latter crystallizes in herringbone structure. If was found that slipping of two adjacent thiazolothiazole molecules results in significant delocalization of π orbital over neighbouring molecules (Figure 11), whereas perfect face-to-face arrangement prevents such interaction. It may be related to the long-distance sulphur-nitrogen attractive interaction due to a significant difference in electronegativity: 2.5 and 3.00 for sulphur and nitrogen, respectively, and their local donor-acceptor character. It is also possible, that the presence of larger number of strong electron acceptors such as trifluoromethyl moieties in **A3** results in electron density depletion in the core. The latter effect is however less probable, at the local Mulliken charges of atoms are not much affected by the substituents. The HOMO contour for **A3** clearly suggests some antibonding interaction between adjacent molecules. These interactions are directly related to the transport properties of solid phases. One can assume that the principal charge transport mechanism will be electron hopping. The probability of electron transfer between neighbouring molecules in solid thiazolothiazoles can be described using Marcus formalism (Equation 1):[34]

$$k_{et} = \frac{2\pi}{\hbar}|H_{DA}|^2 \frac{1}{\sqrt{4\pi\lambda k_B T}} \exp\left(\frac{-(\lambda+\Delta G^o)^2}{4\pi\lambda k_B T}\right) \qquad \text{(Equation 1)}$$

where $|H_{DA}|$ is the electronic coupling matrix element between adjacent molecules, $\lambda$ is the reorganizational energy, $k_B$ is the Boltzmann constant, $T$ is the absolute temperature and $\Delta G^o$ is the free enthalpy of the electron transfer reaction. Thus, the electronic conductivity of the molecular material is directly related to the electron transfer rate between closest neighbours in the solid.[35] In the case of molecular materials like thiazolothiazoles, where electron donor and electron acceptor are identical molecules involved in an electron exchange process, $\Delta G^o = 0$. When the process takes place in the molecular solid, the external reorganization energy, resulting from the rearrangement of the solvent molecules, can be neglected and the internal reorganization energy should be similar for all compounds.

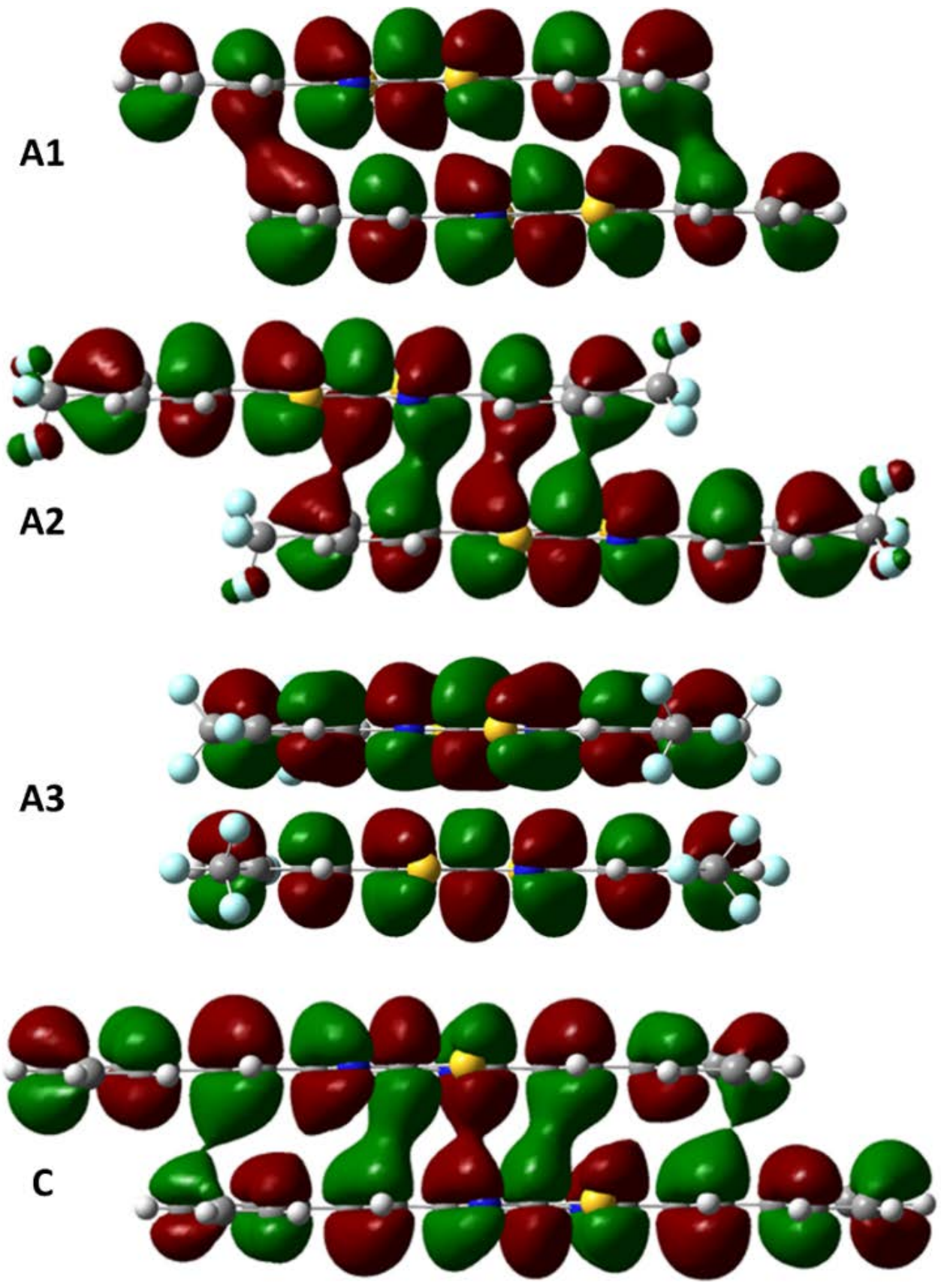

***Figure 11***. *Frontier orbitals of molecular dimers **A1···A1**, **A2···A2**, **A3···A3** and **C···C** as calculated for frozen, XRD-derived geometries, using DFT methodology at the B3LYP/TZVP level of theory.*

The electronic coupling element can be calculated based on HOMO and LUMO orbitals splitting due to the interaction of the molecule of interest with its nearest neighbour in the lattice as (Equations 2, 3):[36]

$$|H_{DA}|_{hole} = \frac{1}{2}(E_{HOMO} - E_{HOMO-1}) \quad \text{(Equation 2)}$$

$$|H_{DA}|_{electron} = \frac{1}{2}(E_{LUMO+1} - E_{LUMO}) \quad \text{(Equation 3)}$$

Electronic coupling elements are collected in Table 4. Quite unexpectedly the **A3** molecule, which does not show any overlap of HOMO orbitals of individual molecules shows the highest coupling matrix element of all studied compounds. *p*-vanilin- **A5**, furyl- **B1** and thiazoyl- **B3** derivatives are also promising candidates for higher hole mobilities. Matrix coupling elements for electron mobility are much lower, therefore, despite electron acceptor properties of thiazolothiazoles they may show good performance as *p*-type semiconductors. This is also substantiated by electron coupling matrix elements calculated for negatively charged dimers – charging increases the probability of electron transfer by ca. 2 orders of magnitude (hole transfer) compared to much smaller, by 1 order of magnitude, increase of electron transfer probability. This feature may not be very suitable for applications in organic field effect transistors (OFETs), but is essential for memristive applications.[19, 37]

***Table 4.*** *Electron coupling elements for electron and hole transfer for thiazolothiazole derivatives.*

| compound | neutral dimer | | negatively charged dimer | |
|---|---|---|---|---|
| | $\lvert H_{DA}\rvert_{hole}$ [meV] | $\lvert H_{DA}\rvert_{electron}$ [meV] | $\lvert H_{DA}\rvert_{hole}$ [meV] | $\lvert H_{DA}\rvert_{electron}$ [meV] |
| **A1** | 63.0 | 25.0 | 1411.6 | 449.3 |
| **A2** | 0.8 | 49.7 | 1446.0 | 429.1 |
| **A3** | 132.0 | 56.2 | 1411.3 | 364.8 |
| **A4** | 84.1 | 27.1 | 1407.9 | 511.8 |
| **A5** | 122.6 | 15.5 | 1345.9 | 475.0 |
| **B1** | 120.4 | 64.2 | 1270.1 | 532.5 |
| **B2** | 26.8 | 11.8 | 1320.6 | 477.0 |
| **B3** | 125.9 | 31.6 | 1319.8 | 461.2 |
| **B4** | 90.9 | 55.5 | 1374.4 | 463.1 |
| **C** | 16.9 | 41.0 | 1254.6 | 321.8 |

More information on the electronic structure of these materials is delivered by the periodic DFT calculations. The calculated crystal structure details and the calculated band gaps of the samples are listed in the Table S4-S6. The optimized geometries for all the compounds are presented in Figure S4 and Figure S5. The fully relaxed, geometry-optimized structures of the TzTz crystals were used to calculate their electronic structures that are shown in Figure 12 for **A1**, **A3** and **A4** and Figure S6 for rest of the compounds. Paths through the Brillouin zone for the crystal structure calculation are shown in Figure S7.

All investigated crystals exhibit the Valance Band Maximum (VBM) near the Fermi level, which suggests a *p*-type semiconductor. The density of states shows the dense charge distribution, indicating the presence of multiple electronic states. From the PDOS plots, the CBM is predominately influenced by the electron-deficient TzTz group which behaves as electron acceptor. In contrast the VBM is dominated by the electron-rich groups: 2-thiazoyl, phenyl, phenylethenyl, trifluoromethylphenyl, 2-imidazoyl, 2-methoxyphenyl, 2-furanyl, 2-thienyl and 3-methoxyphenyl, which behave as electron donors. These results indicate the formation of a donor-acceptor (D-A) framework, which facilitates inter- and intra-molecular electronic transitions, a critical factor for efficient charge transport in semiconducting materials. Below and above the Fermi level in the PDOS, the p orbitals of the TzTz and the tail molecules are overlapping within in the same energy ranges indicating the covalent-like interaction between them. These interactions often result in delocalized π-electrons that enhance electronic conductivity and can significantly influence the VBM and CBM. Among all the studied ten crystal structures, the four of them: **A2**, **A3**, **A4** and **B3** are showing the direct-band gap *p*-type semiconductor where the VBM and CBM are at the same position and the remaining six compounds: **A1**, **A5**, **B1**, **B2**, **B4**, **C** feature indirect band gap with VBM and CBM at the different locations.

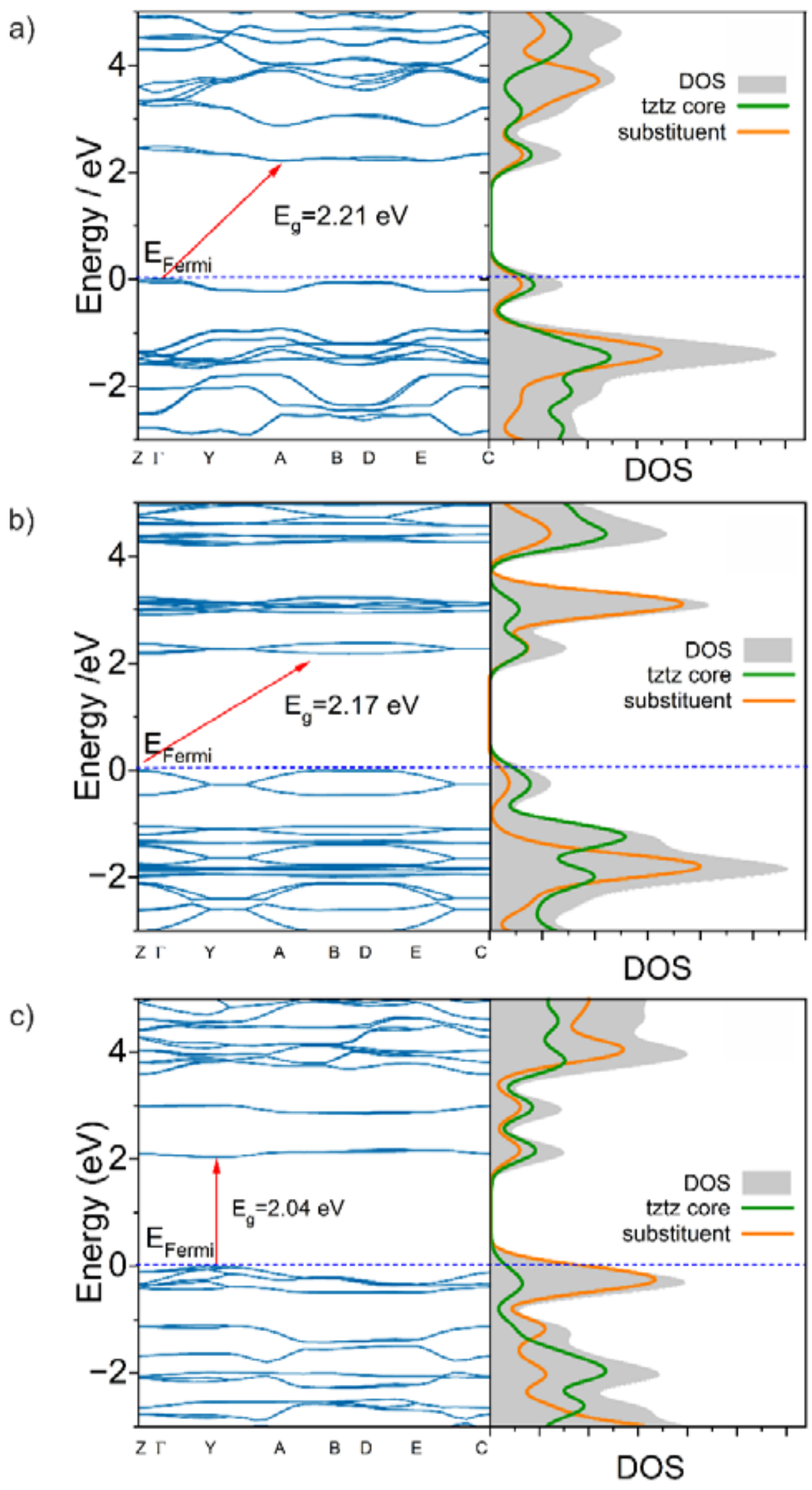

***Figure 12.*** *Theoretical band structure and partial density of states (pDOS, division between TzTz core and substituent) for the most representative TzTz derivatives: a)* ***A1****,"basic" compound with unsubstituted phenyl ring, b)* ***A3****, electron acceptor functional groups –$CF_3$ and c)* ***A4****, electron donor functional group –OMe.*

## Spectral analysis

### FTIR and NMR studies

The solution NMR spectra of TzTzs are technically challenging to obtain due to the compounds' limited solubility in most common deuterated solvents. Nevertheless, we attempted to measure the spectra for all compounds and assigned all proton and carbon signals in their molecules. The NMR spectra, including $^{1}H$, $^{13}C$, and $^{19}F$ (where applicable) assignments, are provided in the Supplementary Information as Table S7.

The FTIR spectra of compounds containing the TzTz system were analyzed by comparing the experimental spectra with theoretical spectra calculated using DFT approach at the B3LYP/TZVP level of theory. This approach enables a more precise assignment of characteristic bands and a deeper understanding of differences arising from intermolecular interactions and specific chemical structural features. Again characterizing these compounds is challenging due to the presence of numerous overlapping and complex bands in their spectra, which result from the coexistence of conjugated aromatic and heteroaromatic rings.[38] Despite these difficulties, it is possible to distinguish the characteristic bands associated with the TzTz system and the substituents. For the FTIR experimental and theoretical spectra, please refer to Table S8, and for the assignments of the characteristic absorptions, please refer to Tables S9 and S10.

As with other aromatic compounds, the TzTz ring exhibits bands corresponding to the asymmetric and symmetric stretching vibrations of C–H bonds. These bands are observed in the range of 3113–3006 $cm^{-1}$ in experimental spectra and 3252–3134 $cm^{-1}$ in calculated spectra.[39] In compounds **A3** and **A4**, which contain –OH groups the $\nu_{aromC\text{–}H}$ overlap with the stretching vibration mode of the O–H group (Table S8). For compound **A4**, it was not possible to isolate the $\nu_{O\text{–}H}$, which is frequently true for hydroxyl groups involved in intramolecular hydrogen bonds.[40] For compound **A5**, the broad and intense $\nu_{O\text{–}H}$ mode is centered at 3267 $cm^{-1}$ ,[41] which confirms the formation of a chain of hydrogen bonds with high proton polarizability along crystallographic axis *b*.[42] In the theoretical spectra, the O–H stretching vibration band appears at 3335 $cm^{-1}$ for **A4** and 3749 $cm^{-1}$ for **A5**. For compounds **A3** and **A4**, symmetric and asymmetric vibrations of methyl groups (from –$OCH_3$) are also observed, appearing in the range of 2968–2835 $cm^{-1}$ in the experimental spectra.[43] In theoretical spectra, these bands shift to higher frequencies, appearing in the range of 3200–3019 $cm^{-1}$.

Among compounds containing a five-membered, heteroaromatic rings, compound **B4** is particular and its stretching modes corresponding to N–H and C–H aromatic vibrations are significantly broadened and shifted, covering the wide range of 3060–2537 $cm^{-1}$. In contrast, the theoretical spectra exhibit only two distinct bands: N–H stretching vibrations at 3637 $cm^{-1}$ and C–H at 3253 $cm^{-1}$.[38, 44] These discrepancies arise due to the neglect of intermolecular interactions by computational methods. The broadening of the $\nu_{N\text{–}H}$ modes is attributed to the properties of the imidazole fragment, which, similar to the hydroxyl group in **A5**, participates in a chain of N–H···N hydrogen bonds along the crystallographic *c* axis (see also Figure 7). When the imidazoyl substituent is replaced with furyl, thiazoyl or tienyl, a significant decrease in the intensity of these bands is observed, emphasizing the crucial role of hydrogen bonding in shaping the spectrum.

The most challenging to identify and characterize are the vibrational bands of the TzTz ring, particularly the stretching vibrations of C=C and C=N, which occur in overlapping regions. C=C vibrations are located in the ranges 1679–1549 $cm^{-1}$ and 1450–1410 $cm^{-1}$,[45] while C=N vibrations for most compounds are observed in the range of 1558–1496 $cm^{-1}$ which is much lower than reported in the literature.[46] In the experimental spectra of compounds **A3** and **B3**, it is not possible to unambiguously assign the band corresponding to C=N vibrations. In compound **B4**, these vibrations occur at lower frequencies, between 1485–1431 $cm^{-1}$. Comparison of experimental and theoretical spectra played a key role in the correct assignment of these bands – most importantly we have assumed that the order of the bands was not reversed in the results of the calculations and we have compared the intensity of calculated and experimental bands. Another characteristic feature of the FTIR spectrum is the strong stretching vibration of the C–N bond, appearing in the range of 1397–1219 $cm^{-1}$.[45c, 46] Interestingly, in this case, the band is exceptionally distinct, whereas in most organic compounds, it is typically less noticeable. Moreover, the intensity of this band increases with the number of nitrogen atoms in the molecule.

In the range of 1220–1070 $cm^{-1}$, FTIR spectra exhibit C–H deformation vibrations, predominantly of the scissoring type, characteristic of aromatic rings and specific substituted phenol groups. In compounds **A2** and **A3**, a very intense C–F stretching vibration band, originating from the $CF_3$ group, is also observed. In experimental spectra, this band appears in the range of 1109–1084 $cm^{-1}$, while in computed spectra, it is found in the range of 1119–1131 $cm^{-1}$.[47] The C–F bond is very strong, and fluorine groups are characterized by low polarizability and bond rigidity, limiting their impact on the activation of infrared bands, especially for $–CF_3$ groups.[48] For compounds **A4**, **A5** and furyl derivative **B1**, strong C–O and C–O–C stretching vibrations are observed in the ranges of 1238–1200 $cm^{-1}$ and 1021 $cm^{-1}$, associated with the presence of $–OCH_3$, –OH and C-O-C groups.[41a, 49] The presence of these modes is easily recognized due to their strong intensity and band broadening. In the range of approximately 1000 to 650 $cm^{-1}$, numerous C-H deformation vibrations of aromatic rings dominate.[50] Additionally, one of the few shared features includes weak-intensity bands attributed to C–S vibrations in thiazole or thiophene rings, observed in the range of 710–529 $cm^{-1}$.[39a, 45a, 51]

**UV-Vis range: absorption and photoluminescence studies**

The analysis of UV-Vis spectra in solution for all compounds reveals absorption maxima ranging from 3.10 eV (400 nm) for **C** to 3.50 eV (354 nm) for **A1** and **A3**, while in the solid state, the bandgaps range from 2.34 eV (530 nm) for **A4** to 2.91 eV (426 nm) for **A3**. Emission maxima in solution span from 2.61 eV (474 nm) for **C** to 3.03 eV (409 nm) for **A1**. The TzTz compounds were also emissive in the solid state, except for the **B4** compound. The correlation between absorption and emission maxima for an exemplary TzTz derivative, **A2**, along with other UV-Vis spectral features, is shown in Figure 13. Similar spectra for other compounds are provided in Figure S7 (A-type compounds) and Figure S8 (B- and C-type compounds). In all cases the lowest energy transition has a $\pi$–$\pi^*$ (HOMO→LUMO) transition and are in a good agreement with theoretically predicted spectra (Figure S10, Table S11).

Comparison of the Stokes shift for the same compound measured in MeCN and in solid state revealed a consistent trend of increased values for solid state measurements. Substantial increases in Stokes shift were observed for **A1**, **A2**, **A5**, **B1**, **B2**, and **C**. This phenomenon occurs due to the increased intermolecular interactions or structural distortions in solid state[52] especially when it comes to planar structures, which can experience π-conjugation in solid state,[53] which results in emission maxima redshift (towards lower energies). At the same time presence of specific functional groups can affect this shift, leading to more efficient non-emissive energy dissipation form the excited states. High Stokes shifts are advantageous for minimizing reabsorption of emitted photons, which is beneficial for optoelectronic applications.[54] Here Stokes shifts of ca. 0.4 eV in solution and up to 1 eV in the solid phase can be achieved despite low dipole moment difference between ground and excited states of the chromophore (fully symmetrical HOMO and LUMO orbitals, cf. Table 3), which contradicts the Lippert-Mataga rule.[55]

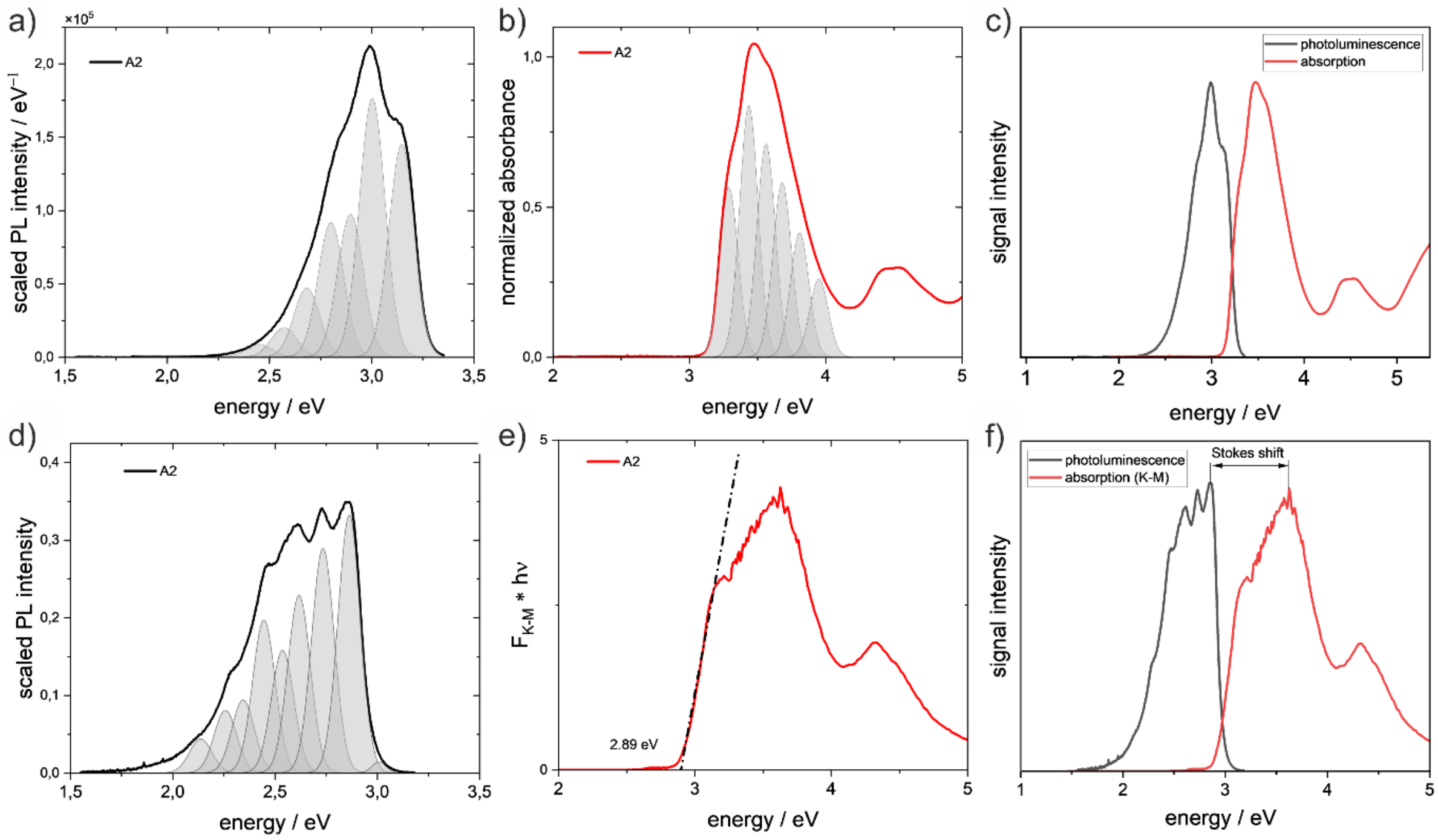

***Figure 13.*** *Spectroscopic UV-Vis range overview of the **A2** compound. Upper row presents results for MeCN solution of the compound: a) emission spectrum b) absorption spectrum c) Stokes shift. Bottom row presents results for powdered sample: d) emission spectrum e) Kubelka-Munk plot f) Stokes shift.*

Spectra in solution (MeCN) and in the solid state exhibit visible broadening, which may be attributed to oscillation broadening or vibrational progression. To analyse this effect, each spectrum was deconvoluted into Gaussian functions. The energy differences between peaks constituting the main spectral features were measured to assess the oscillation broadening of the bands for each type of spectrum: absorption spectrum in MeCN and emission spectrum in MeCN and the solid state. All calculated mean values are presented in Table 5.

Similarly, strongly expressed vibrational progression has been reported for the **A1** and also TzTz molecule itself.[56] Broadening of absorption peak in the spectrum measured in MeCN range from 0.11 eV to 0.16 eV, which is equivalent to 869.8 $cm^{-1}$ and 1258.3 $cm^{-1}$ and can be identified in the IR spectrum as oscillations related to deformation of the TzTz core unit. It should be noted that the TzTz core usually show a series similar distorting vibrations, resulting from coupling with different vibrational modes of substituents. This is the reason for slightly different vibronic progression patterns for absorption and emission spectra. Electronic excitation thus does not initiate any strong distortion in molecular geometry. These values are more high-energetic than the one reported for TzTz unit, which is due to the additional molecular mass in form of functional groups in symmetrical configuration. The same outcome happens for each analysed of the compounds. This fact shows that TzTz core acts as main chromophore, which can be also concluded from previously presented pDOS analysis. At the same time, a good correlation between predicted (DFT) and registered (FT-IR) frequencies has been observed. Vibrational progression is similar for the same compounds in solution for absorption and emission with highest difference over 200 $cm^{-1}$ reported for **A2**, **A4**, **B1**, and **B4**. It indicates that the same oscillation is influencing electronic state transitions. Comparison of emission data for solution and solid state shows that most significant differences in vibrational modes happen for **B1** (591 $cm^{-1}$), **A3**, **A4**, and **A5** (more than 340 $cm^{-1}$ difference). The dataset describing spectrum broadening is presented in a form of Table 5 and can be compared with Figure S11 and Figure S12 where deconvoluted spectra are shown.

More detailed analysis of diffuse reflectance spectra has been performed on the basis of Tauc method. First, diffuse reflectance spectra were converted to the Kubelka-Munk function, defined as follows (Equation 4):

$$F(R) = \frac{(1-R)^2}{2R} \qquad \text{(Equation 4)}$$

where $R$ is the reflectance. For powder samples dispersed in scattering media it is commonly assumed that F($R$) is proportional to the absorption coefficient α.[57] Then the Tauc function has been applied to fit the linear fragment of the spectrum (Equation 5):[58]

$$(\alpha \cdot h\nu)^{\frac{1}{r}} = A(h\nu - E_g) \qquad \text{(Equation 5)}$$

where $A$ is the proportionality constant (independent of the photon energy), $h$ is the Planck constant, $v$ is the photon frequency, $E_g$ is the band gap, $\alpha$ is the absorption coefficient, and $r$ is the exponent describing the nature of the band gap: $r$ = ½ for direct and $r$ = 2 for indirect transitions, respectively.[58]

***Table 5.*** *Parameters describing spectrum broadening and vibrational progression.*

| | code | **A1** | **A2** | **A3** | **A4** | **A5** |
|---|---|---|---|---|---|---|
| abs MeCN | Energy difference [eV] | 0.12 | 0.14 | 0.12 | 0.14 | 0.13 |
| | Energy difference [$cm^{-1}$] | 957 | 1161 | 1007 | 1114 | 1071 |
| | Vibration mode (DFT) | 974 | 1247 | 1028 | 1124 | 1047 |
| em MeCN | Energy difference [eV] | 0.11 | 0.12 | 0.12 | 0.11 | 0.13 |
| | Energy difference [cm⁻¹] | 882 | 938 | 929 | 898 | 1009 |
| | Vibration mode (DFT) | 880 | 977 | 919 | 930 | 1023 |
| em solid state | Energy difference [eV] | 0.11 | 0.10 | 0.16 | 0.16 | 0.17 |
| | Energy difference [cm⁻¹] | 875 | 804 | 1276 | 1291 | 1400 |
| | Vibration mode (exp) | 880 | 809 | 1283 | 1248 | 1423 |
| | Code | **B1** | **B2** | **B3** | **B4** | **C** |
| abs MeCN | Energy difference [eV] | 0.13 | 0.12 | 0.11 | 0.16 | 0.14 |
| | Energy difference [cm⁻¹] | 1082 | 967 | 870 | 1258 | 1129 |
| | Vibration mode (DFT) | 1020 | 973 | 880 | 1294 | 1200 |
| em MeCN | Energy difference [eV] | 0.10 | 0.10 | 0.10 | 0.13 | 0.13 |
| | Energy difference [cm⁻¹] | 832 | 846 | 823 | 1035 | 1083 |
| | Vibration mode (DFT) | 851 | 854 | 835 | 1041 | 1198 |
| em solid state | Energy difference [eV] | 0.18 | 0.13 | 0.15 | - | 0.12 |
| | Energy difference [cm⁻¹] | 1424 | 1036 | 1200 | - | 944 |
| | Vibration mode (exp) | 1436 | 1054 | 1240 | - | 943 |

In the case of amorphous (or almost amorphous) material it is assumed that $r$ = 1 seems to be the most reasonable choice.[59] The same approximation is commonly used for molecular crystalline materials as well as for ionic crystal with only a week covalent interaction between ionic species.[60] Despite the fact that DFT models predict both direct and indirect band gaps, the $r$ = 1 case provides the best fit for the spectra, which is consistent with relatively weak intermolecular interactions in covalent crystals. Therefore, the final equation that allows the determination of the band gap is derived as follows (Equation 6):

$$F(R) \cdot h\nu = A(h\nu - E_g) \qquad \text{(Equation 6)}$$

Optical band gaps, along with other spectroscopic parameters are collated in Table 6. Fitted parameters for bandgap determination are presented in Table S12.

**Lifetime measurements**

For photoluminescence lifetime measurements in solid state in RT, in almost each case two-exponential fit was calculated, with $\chi^2$ parameters being below 1.3 (Table 6). An exception was observed for compound **C** in MeCN, where the photoluminescence lifetime was described by single-exponential decay. Analysis of $\tau_1$ across compound types (**A1**-**A5**, **B1**-**B4**, and **C**) revealed variability between solution and solid state. Some of the compounds (**A3, B4**) exhibit substantial increases in photoluminescence lifetimes, while others (**A2, A4**) show slight decreases or only minimal changes, suggesting limited variation in their emissive properties between solution and solid-state forms. In contrast, $\tau_2$ consistently increases for solid state across nearly all examples. Specifically, **A3** and **B3** showed the most pronounced increases in both $\tau_1$ and $\tau_2$, into long-lived fluorescence lifetimes of the order of ns.

***Table 6.*** *UV-Vis spectroscopic parameters alongside with optical bandgaps and lifetimes*

| code | substituent | MeCN solution | | | | Solid state | | | | |
|---|---|---|---|---|---|---|---|---|---|---|
| | | Max abs [eV] | Max em [eV] | Stokes shift [eV] | Life time [ns] | Band gap [eV] | Max abs [eV] | Max em [eV] | Stokes shift [eV] | Life time [ns] |
| **A1** | 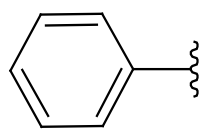 | 3.50 | 3.03 | 0.47 | $\tau_1$=0.2428<br>$\tau_2$=0.5205<br>$\chi^2$=1.124 | 2.89 | 3.48 | 2.67 | 0.81 | $\tau_1$=0.7844<br>$\tau_2$=1.9463<br>$\chi^2$=1.093 |
| **A2** | 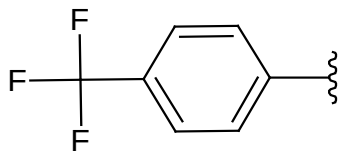 | 3.47 | 2.99 | 0.48 | $\tau_1$= 0.266<br>$\tau_2$= 0.581<br>$\chi^2$= 2.641 | 2.89 | 3.63 | 2.85 | 0.78 | $\tau_1$= 0.246<br>$\tau_2$= 0.721<br>$\chi^2$= 1.14 |
| **A3** | 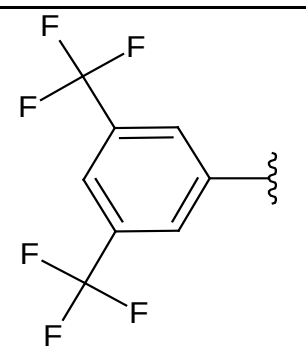 | 3.50 | 3.01 | 0.49 | $\tau_1$= 0.241<br>$\tau_2$= 0.457<br>$\chi^2$= 1.017 | 2.91 | 3.32 | 2.70 | 0.62 | $\tau_1$= 1.803<br>$\tau_2$= 4.488<br>$\chi^2$= 1.290 |
| **A4** | 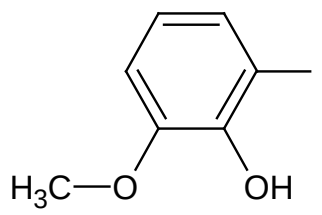 | 3.32 | 2.98 | 0.34 | $\tau_1$= 0.269<br>$\tau_2$= 0.489<br>$\chi^2$= 1.174 | 2.34 | 3.11 | 2.00 | 1.11 | $\tau_1$=0.257<br>$\tau_2$= 1.125<br>$\chi^2$= 1.005 |
| **A5** | 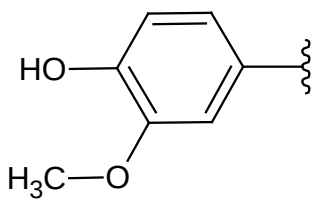 | 3.31 | 2.80 | 0.51 | $\tau_1$= 0.204<br>$\tau_2$= 0.652<br>$\chi^2$= 1.084 | 2.61 | 3.24 | 2.43 | 0.81 | $\tau_1$= 0.246<br>$\tau_2$= 1.857<br>$\chi^2$= 1.323 |
| **B1** | 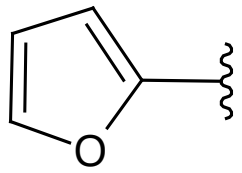 | 3.34 | 2.88 | 0.46 | $\tau_1$= 0.303<br>$\tau_2$=0.566<br>$\chi^2$= 1.310 | 2.72 | 2.93 | 2.13 | 0.80 | $\tau_1$= 0.719<br>$\tau_2$= 1.856<br>$\chi^2$= 1.019 |
| **B2** | 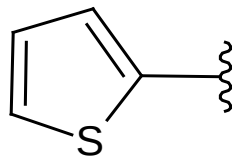 | 3.23 | 2.77 | 0.46 | $\tau_1$= 0.282<br>$\tau_2$= 0.542<br>$\chi^2$= 1.171 | 2.57 | 3.04 | 2.33 | 0.71 | $\tau_1$= 0.507<br>$\tau_2$= 1.200<br>$\chi^2$= 1.078 |
| **B3** | 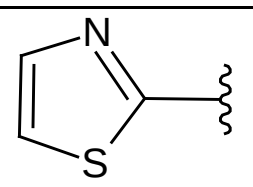 | 3.21 | 2.79 | 0.42 | $\tau$1= 0.344<br>$\tau$2= 0.659<br>$\chi$2= 1.514 | 2.61 | 3.00 | 2.35 | 0.65 | $\tau$1= 1.419<br>$\tau$2= 2.842<br>$\chi$2= 1.115 |
| **B4** | 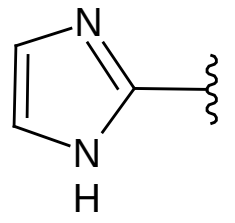 | 3.33 | 2.86 | 0.47 | $\tau$1= 1.052<br>$\tau$2= 2.123<br>$\chi$2= 1.323 | 2.62 | 3.15 | - | - | - |
| **C** | 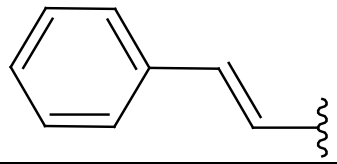 | 3.10 | 2.61 | 0.49 | $\tau_1$=0.818<br>$\chi^2$=1.651 | 2.40 | 1.99 | 3.04 | 1.05 | $\tau_1$= 0.588<br>$\tau_2$= 1.748<br>$\chi^2$= 1.161 |

These observations align with the photoluminescence phenomena principles, where the solid-state environment often promotes extended lifetimes due to decreased rate of non-radiative decay. The results underline the importance of both molecular structure and neighbouring environment in determining photoluminescence – which must be taken into account upon development of materials for solid-state optoelectronic applications.

It is worth to notice that the powder samples exhibited much longer lifetimes than their solution counterparts. For most compounds, the lifetimes determined for solutions are twice as short as for solid samples. Only for samples **A2**, **A4** and **A5**, the lifetimes are comparable. The longest time for solid samples is expected since bulk organic aggregates require higher reorganization energy than molecular monomers in dilute solutions[11a]. This also explains Stoke shifts, which are significantly larger in the case of solid samples.

***XAS analysis***

Sulphur exhibits various oxidation states, from -2 to +6, forming chemical bonds with elements of different electronegativity. This greatly influences the local electronic structure of sulphur, and spectroscopic methods are an excellent tool for studying these phenomena.

X-ray absorption spectroscopy (XAS), often exploited combined with synchrotron radiation, is an element-specific technique that probes the electronic structure and local coordination. XAS probes electronic transitions of a core electron (1s for K-edge) to an empty state above Fermi energy. X-ray absorption near edge structure (XANES) part of XAS, typically 50-100 eV above the edge, arises from transitions that may include bound states due to resonant excitations. Extended X-ray absorption fine structure (EXAFS) is a higher-energy oscillatory part of XAS, which may reveal a few coordination spheres around an absorber by analysing the possible scattering path on adjacent atoms. It is worth noting that XAS in soft and tender energy ranges, as for S K-edge, is often influenced by self-absorption, a phenomenon whereby the fluorescence detector signal is distorted due to reabsorption of the emitted fluorescence photons by the sample itself. The XAS measurements presented in this work were performed in transmission mode, which eliminates this effect.

Figure S13 in Supporting Information shows the full XAS spectra, *i.e.* XANES and EXAFS, measured for the studied thiazolothiazole derivatives. The spectra differ from each other, which is more evident in the XANES shown in Figure 14. The shape of XANES reflects a complex relationship between charge screening and orbital hybridisation effects.[61] Charge screening describes how the local chemical environment of an atom and the surrounding electron density influence the effective nuclear charge felt by the core and valence electrons. Orbital hybridization refers to the mixing of atomic orbitals of the absorber with orbitals from the neighbouring atoms. This results in the creation of hybridized orbitals which may disrupt the electron density of unoccupied states available for the excited core electrons.

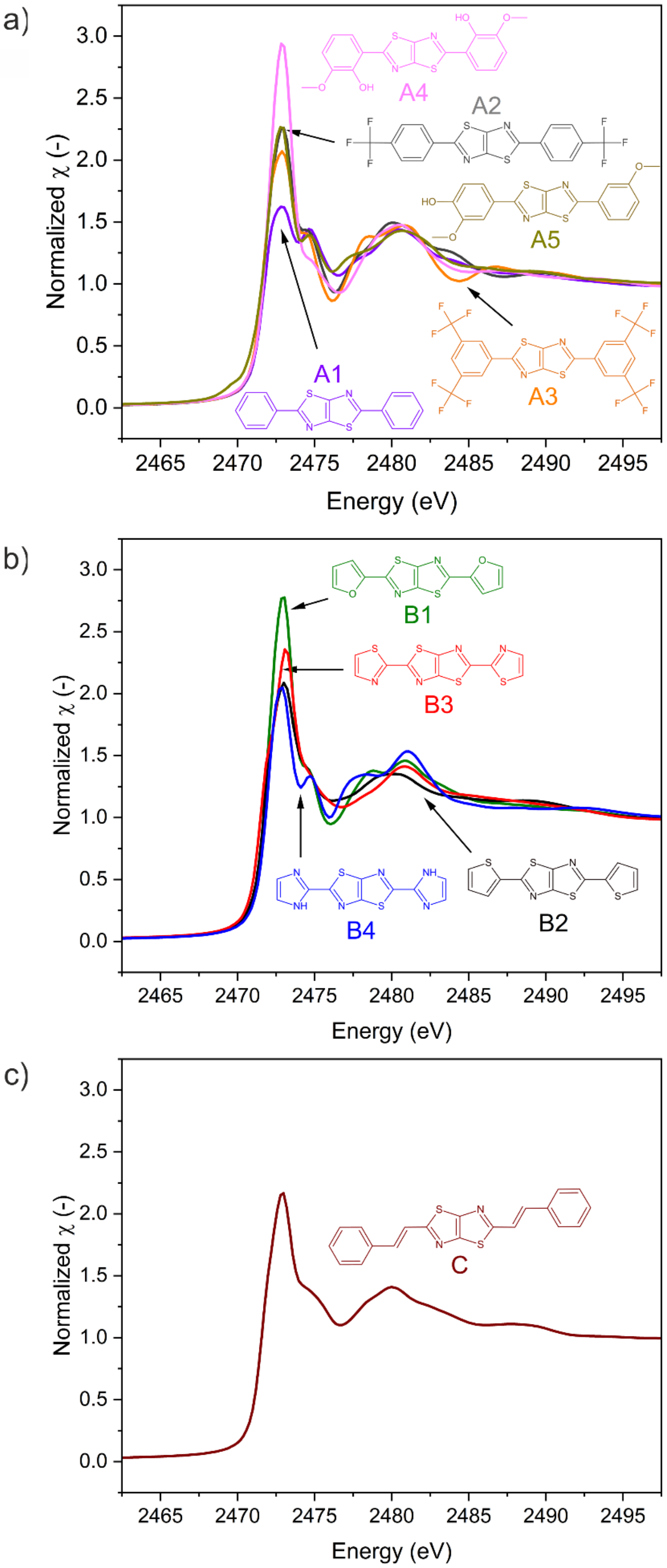

***Figure 14.*** *Sulphur K-edge XANES for the studied thiazolothiazoles measured in transmission mode: a)* ***A1****-****A5****; b)* ***B1****-****B4****; c)* ***C****.*

Because of low electron density, the K-edge XANES of sulphur may be very sensitive to both charge screening and hybridization effects. Due to weak charge screening, position of the sulphur K-edge strongly depends on the oxidation state of this element[62], as it varies from 2470.41 eV for $CuFeS_2$ (-2 oxidation state), up to 2482.85 eV for $Na_2SO_4$ (+6 oxidation state).[62-63] For the studied thiazolothiazole derivatives, sulphur occurs in a formal -2 oxidation state, resulting in absorption K-edges values between 2471.46 and 2472.08 eV (Table 7). These values may confirm sulphur −2, as even for simple sulfides with a formal oxidation state of -2, the absorption edge value may vary, e.g. 2471.02 eV for $Fe_{1-x}S$, 2471.67 eV for HgS, 2471.7 eV for $As_2S_3$, 2473.22 eV for CaS, and 2473.8 eV for ZnS. The negligible difference in electronegativity between S (2.58) and C (2.55) in thiazolothiazoles implies that the bond between these two elements exhibit largely covalent character with minimal polarity. Therefore, the formal oxidation state may not be sufficient to understand the bonding environment and electron distribution in these compounds.

**Table 7.** XAS S K-edge values for parameters for the studied thiazolothiazole derivatives.

| | $E_{edge}$ [eV] | $E_{white\ line}$ [eV] |
|---|---|---|
| **A1** | 2471.78 | 2472.81 |
| **A2** | 2471.78 | 2472.99 |
| **A3** | 2471.67 | 2472.88 |
| **A4** | 2472.04 | 2472.83 |
| **A5** | 2471.98 | 2472.77 |
| **B1** | 2472.08 | 2473.04 |
| **B2** | 2471.70 | 2472.99 |
| **B3** | 2471.46 | 2473.06 |
| **B4** | 2472.00 | 2472.91 |
| **C** | 2471.59 | 2472.98 |

In S K-edge the main electronic transitions occur according to the Laporte rule allowing for the dipole transitions with orbital quantum number $\Delta l = \pm 1$. The basic electron configuration of sulphur is [Ne] $3s^2\ 3p^4$, but the one expected for thiazolothiazoles with -2 oxidation state is closed shell [Ne] $3s^2\ 3p^6$, as for Ar. Herein, the most probable transition is from 1s sulphur orbital to σ* and π* anti-bonding molecular orbitals formed with the neighbouring atoms. Therefore, orbital hybridization should play an important role in shaping the near-edge structure. To study the orbital hybridization in detail, the XAS spectra of the studied thiazolothiazoles were also modelled using Finite Difference Method Near Edge Structure (FDMNES) *ab initio* code[64]. The experimental and theoretical data are shown for each of the compounds Supporting Information in Figure S14.

### *Electric measurements*

Representatives **A4**, **A5**, **B3**, **B4**, and **C** were selected based on their superior solubility in DMF compared to the remaining compounds, which are almost insoluble in DMF or other solvents suitable for spin-coating. Layered devices were prepared, and their current – voltage (I-V) characteristics were measured across varying temperatures, as shown in Figure 15. The I-V responses remained stable for all compounds between −20°C and 120°C. The forward-to-reverse bias current amounts to 1.07 ± 0.08 (average value for all compounds), the highest for **A5** (1.18) and the lowest for **C** (1.0). Furthermore the current-voltage dependencies are almost linear, with slopes almost identical for both low and high temperature. It indicates the lack of Schottky barrier at the copper/thiazolothiazole interface. On the other hand formation of metallic filaments can be excluded due to small resistivity change during switching. Therefore the only reasonable mechanistic explanation will be electron injection-enhanced electron hopping mechanism. This hypothesis is supported by much higher electron coupling matrix elements $|H_{DA}|$ (cf. Equation 1) for negatively charged dimers as compared with neutral ones (cf. Table 4). This observation is consistent with other molecular crystal organic[37] and organometallic[19] semiconductors. Basic endurance and retention tests, which are presented in the Supplementary Information, revealed that while three out of five (**A5**, **B3** and **C**) compounds maintained distinct high-resistance (HRS) and low-resistance (LRS) states, their endurance and stability varied significantly for each of them.

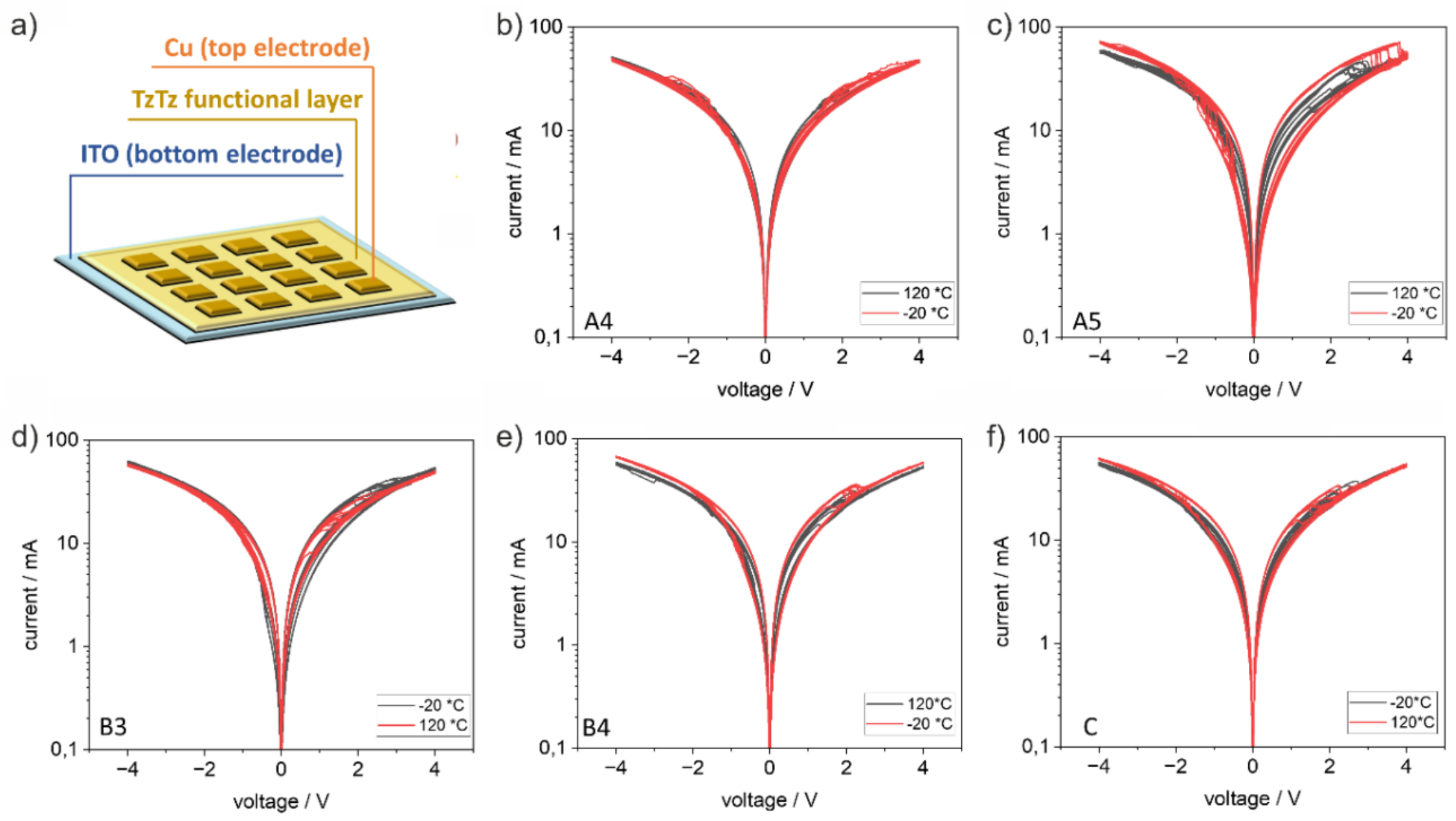

***Figure 15.*** *Temperature–dependent current-voltage I-V characteristics for the selected TzTzs: a) Scheme of the device; b)* ***A4****; c)* ***A5****; d)* ***B3****, e)* ***B4****; f)* ***C.***

For **A4**, the resistive states remained separated during 750 on-off cycles but diminished toward convergence into a single state, indicating instability during both cycling and retention evaluations. In contrast, **A5** exhibited consistent separation of HRS and LRS throughout the 750 cycles, with no degradation in performance upon switching, thus demonstrating sufficient stability.

Compound **B3** displayed diminishing state separation after 750 cycles, aligning with the narrow hysteresis curve observed in Figure S15. State retention for **B3** was reliable over the 2-hour test. Compound **B4** showed indistinguishable resistive states after 100 cycles and lost retention within 2 hours, suggesting that it may be better suited for volatile memory or dynamic signal applications. Similarly, compound **C** maintained distinct resistive states only up to 300 cycles, highlighting its potential for low-duty-cycle applications. These results, only for exemplary compounds from each group show diversity in charge transport mechanisms. As such TzTzs are offering a range of switching characteristics that can be fine-tuned for specific technological requirements. However, other parameters, such as on-off ratio need to be refined before introducing compound for industrial applications (Figure S16).

From a neuromorphic perspective, **A4** demonstrated basic plasticity behaviours such as depression and potentiation at threshold pulse values of 1.6 V, suggesting its potential use in synaptic plasticity change-based learning models. Results for **A4** and other compounds are presented in Figure 16. **A5** required higher pulse voltages 2.4 V of higher for switching, reflecting a more energy-intensive but highly stable device suitable for neural networks.

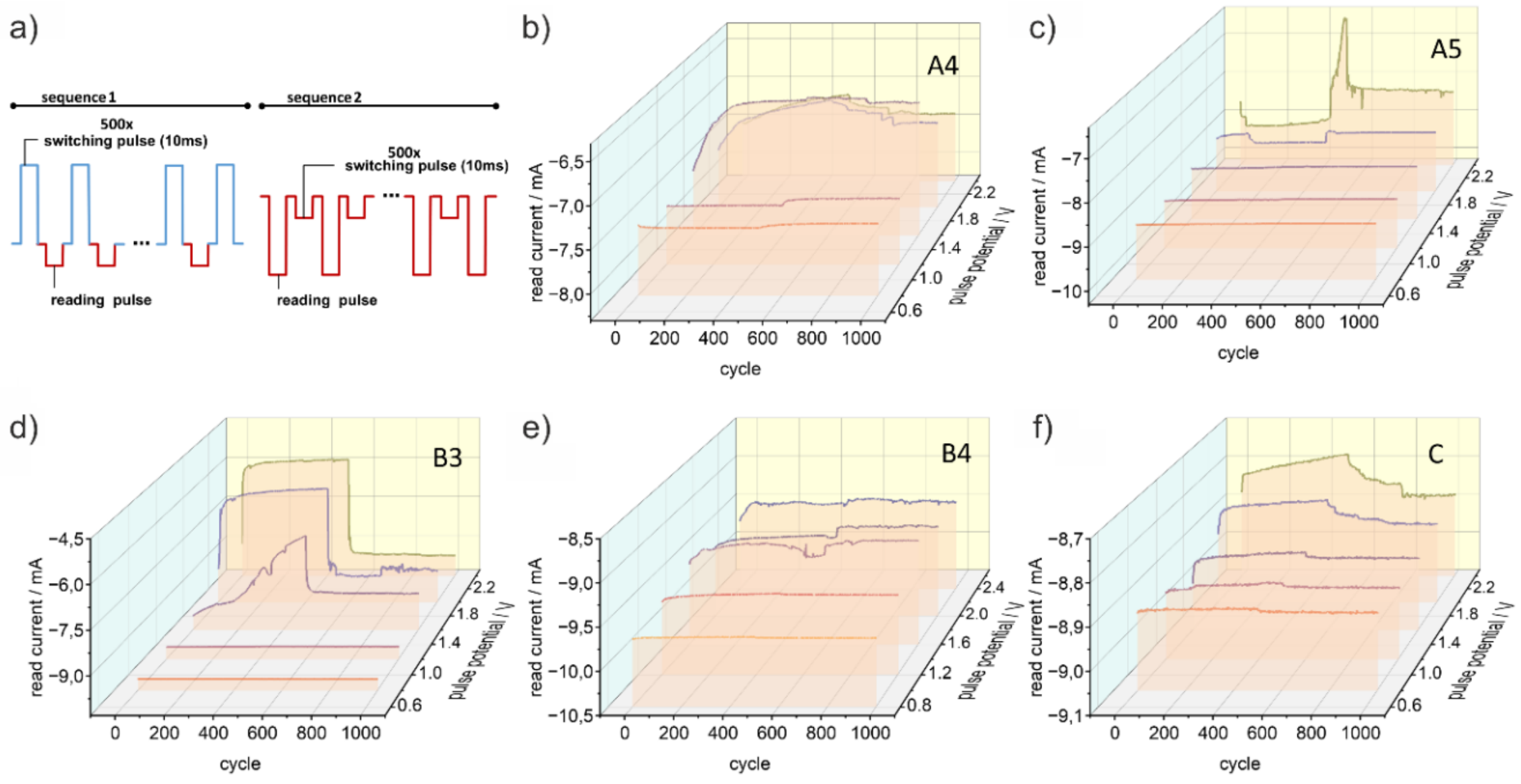

***Figure 16.*** *The potentiation depression measurements for the selected TzTzs; basic plasticity features for various potential pulses: a) pulse sequence for the measurements; b)* ***A4****; c)* ***A5****; d)* ***B3****; e)* ***B4****; f)* ***C****.*

Compound **B3** exhibited dynamic switching responses from 1.6 V with steep but gradual changes, indicating its suitability for spiking neural network simulations, **B4** can operate on pulses higher or equal to 1.2 V, however poor retention of the states causes instability of the responses. Compound **C**, with dynamic responses initiated at lower voltages (1.2 V pulses), shows promise for low-power neuromorphic applications, though its limited endurance suggests constraints in high-voltage regimes.

Overall, the presented behaviours across these compounds highlight their adaptability for different memristive and neuromorphic applications. Compounds like **B4** and **C** provide opportunities for volatile memory applications and energy-efficient synaptic emulation, while the increased stability of **A5** makes it a strong candidate for non-volatile, high-endurance memory systems. This diversity in performance underlines the potential of TzTz derivatives as versatile materials for next-generation neuromorphic and memristive technologies.

**Conclusions**

Presented extensive research on diverse thiazolothiazole (TzTz) derivatives has been driven by their significant potential in the semiconductor industry, particularly for the development of optoelectronic and sensing devices, most likely on flexible substrates. These compounds exhibit well-defined structural motifs and notable optical properties, positioning them as promising candidates for advancements in optoelectronic and sensing technologies. At the same time their syntheses are simple and proceed is a single step with easily available substrates. Some of them can also serve as ligands for complexes with transition metals, which may further modify their properties. Key interactions, including π-π stacking, hydrogen-bonds and specific chalcogen and halogen interactions, are the cause of their stability influencing also spectroscopic fingerprints.

Thiazolothiazoles can be synthesized with the whole range of substituents (both electron donors and electron acceptors) in high yields in one-pot reaction. Their optical properties demonstrate significant tunability, including shifts in spectral behaviour and variability in bandgap energies, making them suitable for specific and targeted applications. These changes are a consequence of electronic interactions within individual molecules as well as rich intermolecular interactions in solid phase. Three distinct packing modes have been identified: herringbone (usually found in various organic semiconductors), parallel stacking and grid-like structures. Furthermore, some intermolecular electronic interactions have been identified, which may further affect electrical properties of these materials. Surprisingly, the structural diversity of thiazolothiazoles is not reflected in their XAS spectra, where only molecule-level scattering pathways are responsible for absorption profiles, however, changes in white line intensity vary from compound to compound (lowest for **A1** and highest for **A4**), indirectly indicating the role of delocalization of antibonding orbitals.

Determination of electrical properties of thiazolothiazoles is partially limited by their solubility. Due to high planarity and resulting high lattice stabilization energies many derivatives are hardly soluble in any solvents. The most soluble **A4**, **A5**, **B3**, **B4** and **C** can be used for fabrication of two-electrode planar devices. All the them show relatively good, temperature independent conductivity and significant hysteresis suitable for reproduction of neuromimetic plasticity behaviour. Symmetry of hysteresis loops suggest bulk virtual filamentary mechanism of resistive switching. Despite symmetric and reversible hysteresis loops, prolonged polling with positive voltage pulses leads to gradual degradation of the materials, except for **A5** and **B3**. Out of these only **A5** is stable under repetitive potentiation-depression cycles and **B3** shows fully reversible switching. It can be concluded that only these two materials have sufficiently high redox stability for further applications in thin layer organic electronic devices.

Future characterization efforts will focus on usage of the coordination sites of thiazolothiazole derivatives to further enhance their properties. Investigations will prioritize the incorporation of several metal ions, to explore their impact on the material's functional characteristics.

Overall, the simplicity of the synthetic procedures and the ease of selecting derivatives with diverse properties make thiazolothiazole systems highly versatile. It can be treated as an alternative strategy that involves fine-tuning of the intrinsic properties of the compounds *via* targeted substitution.

## Experimental

### *Synthetic procedures*

All thiazolothiazoles were synthesized following a uniform procedure. The compounds were obtained by mixing the substrates in a molar ratio of 1 : 2.3 (dithiooxamide : aldehyde), starting with 1 g of dithiooxamide. The mixture was dissolved in 30 mL of DMF and heated at the appropriate temperature (see below).

**2,5-bis(phenyl)-1,3-thiazolo[5,4-d]-1,3-thiazole (A1):** The reaction mixture was heated at 140–145°C for 1.5 h**.** Yellowish, crystalline product was obtained in the form of blocks; yield 69%, m.p. 209-213°C. Elemental analysis: calcd C 65.27, H 3.42, N 9.52; found C 65.25, H 3.40, N 9.53.
*FT-IR ATR:* 3061(w), 3014(w), 1953(w), 1881(w), 1824(w), 1750(w), 1679(w), 1600(w), 1501(m), 1456(vs), 1430(s), 1336(w), 1309(s), 1219(m), 1199(w), 1164(w), 1100(w), 1071(w), 1031(w), 1006(m), 996(m), 910(w), 883(w), 754(s), 680(s), 637(w), 611(s), 483(w) $cm^{-1}$.
*NMR:* $^{1}$H NMR (400 MHz, $CDCl_3$-$d_1$): δ = 8.02-7.97 (m, 4H, *ortho*-C-**H** (**H$_o$**)), 7.51-7.44 (m, overlapped, 6H, *meta*-C-**H** and *para*-C-**H** (**H$_m$** and **H$_p$**)); $^{13}$C{$^{1}$H} NMR (100 MHz, $CDCl_3$-$d_1$): δ = 169.21 (s, S**C**N **(C2)**), 150.86 (s, **C**=**C** **(C1)**), 133.94 (s, **C$_i$** **(C3)**), 130.71 (s, *meta*-**C**H **(C6)**), 129.16 (s, *para*-**C**H **(C5** and **C7)**), 126.45 (s, *ortho*-**C**H **(C4** and **C8)**)

**2,5-bis(4-trifluoromethylphenyl)-1,3-thiazolo[5,4-d]-1,3-thiazole (A2)**: The reaction mixture was heated at reflux for °C for 18 h. Greenish-yellow crystalline product was obtained and purified in 30 mL of dichloromethane, yield 35%, m.p. 269.9-264.9 °C. Elemental analysis: calcd C 49.99, H 2.33, N 6.48,; found C 49.98, H 2.35, N 6.50.

*FT-IR ATR:* 3083(w), 3055(w), 1941(w), 1889(w), 1590(w), 1570(w), 1496(m), 1436(s), 1397(s), 1319(), 1291(w), 1222(m), 1199(w), 1168(w), 1103(w), 1085(vs), 1014(m), 1006(s), 974(w), 881(m), 839(s), 809(s), 729(w), 711(w), 659(m), 621(m), 613(m), 516(m), 497(m), 459(w) $cm^{-1}$.
*NMR:* $^{1}$H NMR (400 MHz, $CDCl_3$-*d*$_1$): δ = 8.13 (d, $^{3}J_{HH}$ = 8, 2H, *ortho*-C-**H (H$_o$)**), 7.58 (d, $^{3}J_{HH}$ = 8, 2H, *meta*-C-**H (H$_m$)**); $^{13}$C{$^{1}$H} NMR (100 MHz, $CDCl_3$-*d*$_1$): δ = 167.9 (s, S**C**N **(C2)**), 151.8 (s, , **C**=**C (C1)**), 136.9 (s, **C*i* (C3)**), 132.4 (m*, $^{2}J_{CF}$=33, **C**$CF_3$ **(C6)**) 126.7 (br s, *ortho*-**C**H (**C4** and **C8**)), 126.2 (q, $^{3}J_{CF}$ = 4, *meta*-**C**H (**C5** and **C7**)), 123.7 (m*, $^{1}J_{CF}$ = 272, **C**$F_3$); *m – Expected to be a quartet, but due to the low solubility of the compound, it is not clearly visible. $^{19}$F NMR (376 MHz, $CDCl_3$-*d*$_1$): δ = - 62.9 (s, C**F$_3$**).

**2,5-bis{bis(3,5-trifluoromethyl)phenyl}-1,3-thiazolo[5,4-d]-1,3-thiazole (A3):** The reaction mixture was heated at reflux for 18 h. Greenish-yellow crystalline product was obtained and and purified in 30 mL of dichloromethane, yield 42 %, m.p. 263.5-264.5°C Elemental analysis: calcd C 41.67, H 2.80, N 4.86; found C 41.70, H 2.81, N 4.83.
*FT-IR ATR:* 3113(w), 1619(w), 1484(w), 1436(w), 1360(s), 1325(w), 1283(s), 1217(m), 1164(s), 1109(vs), 1041(s), 918(m), 897(s), 846(m), 776(w), 706(w), 698(m), 684(m), 654(w), 448(w) $cm^{-1}$.
*NMR:* $^{1}$H NMR (400 MHz, $CDCl_3$-*d*$_1$): δ = 8.46 (s, 4H, *ortho*-C**H (H$_o$)**), 7.99 (s, 2H, *para*-C**H (H$_p$)**); $^{13}$C{$^{1}$H} NMR (100 MHz, $CDCl_3$-*d*$_1$): δ = 166.5 (s, S**C**N **(C2)**), 152.0 (s, **C**=**C (C1)**), 135.5 (s, **C*i* (C3)**), 132.9 (q, $^{2}J_{CF}$ =34, **C**$CF_3$ (**C5** and **C7**)), 126.4 (br s, *ortho*-**C**H (**C4** and **C8**)), 124.1 (m, *para*-**C**H **(C6)**), 122.9 (m*, $^{1}J_{CF}$ = 273, **C**$F_3$), *m – Expected to be a quartet, but due to the low solubility of the compound, it is not clearly visible. $^{19}$F NMR (376 MHz, $CDCl_3$-*d*$_1$): δ = - 63.0 (s, C**F$_3$**);

**2,5-bis(2-hydroxy-3-methoxyphenyl)-1,3-thiazolo[5,4-d]-1,3-thiazole (A4):** The reaction mixture was heated at 120–125 °C for 4–6 h. A yellow crystalline product in the form of thin needles was obtained and purified in 30 mL of ethanol; yield 48,19% (based on dithiooxamide); m.p. 288 °C. Elemental analysis: calcd C 55.94, H 3.65, N 7.25, S 16.60; found C 55.95, H 3.65, N 7.33, S 16.49.
*FT-IR ATR:* 3006(w), 2968(w), 2835(w), 1584(w), 1502(w), 1561(s), 1443(s), 1415(s), 1248(vs), 1171(m), 1115(m), 1094(m), 1021(s), 924(w), 870(w), 832(w), 777(s), 718(s), 678(m), 626(w) $cm^{-1}$.
*NMR:* $^{1}$H NMR (400 MHz, DMSO-*d*$_6$): δ = 10.68 (bs, 2H, O**H**), 7.84 (d, overlapped, $^{3}J_{HH}$= 8, 2H, *ortho*-C-**H (H$_o$)**), 7.12 (d, overlapped, $^{3}J_{HH}$= 8, 2H, *para*-C-**H (H$_p$)**), 6.97 (t, $^{3}J_{HH}$= 8, 2H, *meta*-C-**H (H$_m$)**), 3.89 (s, 6H, C**H$_3$**);$^{13}$C{$^{1}$H} NMR (100 MHz, DMSO-*d*$_6$): δ = 164.34 (s, S**C**N **(C2)**), 151.14 (s, **C=C (C1)**), 148.91 (s, **C**-$OCH_3$ **(C5)**), 145.31 (s, **C**-OH **(C4)**), 120.82 (s, **C*i* (C3)**), 119.97 (s, **C**$H_m$ **(C7)**), 119.44 (s, **C**$H_o$ **(C8)**), 113.68 (s, **C**$H_p$ **(C6)**), 56.8 (s, **C**$H_3$).

**2,5-bis(4-hydroxy-3-methoxyphenyl)-1,3-thiazolo[5,4-d]-1,3-thiazole (A5):** The reaction mixture was heated at 120–125 °C for 4–6 h. An orange powder was obtained and purified in 30 mL of ethanol; yield 33,41% (based on dithiooxamide); m.p. 275-280 °C. Elemental analysis: calcd C 55.95, H 3.65, N 7.25, S 16.49; found C 55.80, H 4.01, N 7.19, S 16.42.
*FT-IR ATR:* 3525(m), 3396(s), 2990(w), 2943(w), 2879(w), 2834(w), 1667(w), 1596(m), 1520(vs), 1466(s), 1445(w), 1432(s), 1376(w), 1290(m), 1264(s), 1246(s), 1217(s), 1191(s), 1137(m), 1112(m), 1024(m), 857(w), 839(m), 820(), 795() $cm^{-1}$.
*NMR:* $^1$H NMR (400 MHz, DMSO-*d*$_6$,): δ = 9.82 (bs, 2H, O**H**), 7.54 (s, overlapped, 2H, *ortho*-C**H (H1)**), 7.45 (m, overlapped, 2H, *ortho*-C**H (H2)**), 6.92 (d, $^3J_{HH}$= 8 Hz, 2H, *meta*-C**H (H3)**), 3.84 (s, 6H, OC**H$_3$**); $^{13}$C{$^1$H} NMR (100 MHz, DMSO-*d*$_6$): δ = 168.75 (s, S**C**N **(C2)**), 150.17 (s, **C**=**C (C1)**), 149.46 (s, **C**-OH **(C6)**), 148.65 (s, **C**-OCH$_3$ **(C5)**), 125.23 (s, **C***$_i$* **(C3)**), 120.34 (s, *ortho*-**C**H **(C8)**), 116.45 (s, *meta*-**C**H **(C7)**), 109.78 (s, *ortho*-**C**H **(C4)**), 56.20 (s, **C**H$_3$).

**2,5-bis(2-furyl)-1,3-thiazolo[5,4-d]-1,3-thiazole (B1):** Beige crystalline product was obtained in the form of blocks, yield % 70% (based on dithiooxamide), m.p. 240-244°C. Elemental analysis: calcd C 52.54, H 2.20, N 10.21; found C 52.52, H 2.21, N 10.18.
*FT-IR ATR:* 3145(s), 3111(vs), 3067(w), 2872(w), 2824(w), 2706(w), 2653(w), 2566(w), 2142(w), 1750(w), 1702(w), 1654(w), 1583(m), 1495(s), 1436(m), 1311(s), 1257(m), 1239(s), 1213(s), 1147(m), 1076(w), 1030(m), 1012(s), 926(m), 878(m), 844(s), 828(s), 752(m), 739(s), 660(m), 592(m) $cm^{-1}$.
*NMR:* $^1$H NMR (400 MHz, DMSO-*d*$_6$): δ = 7.99 (m, 2H, **H$_c$**), 7.30 (m, 2H, **H$_a$**), 6.80 (m, 2H, **H$_b$**); $^{13}$C{$^1$H} NMR (100 MHz, DMSO-*d*$_6$): δ = 158.3 (s, S**C**N, **(C2)**), 150.2 (s, **C=C** (**C1**)), 148.1 (s, O**C**C, **(C3)**) 146.2 (s, **C**-H$_c$ **(C6)**), 113.7 (s, **C**-H$_b$ **(C5)**), 111.6 (s, **C**-H$_a$ **(C4)**).

**2,5-bis(2-thiophyl) -1,3-thiazolo[5,4-d]-1,3-thiazole (B2):** The reaction mixture was heated at reflux in microwave for 1 h after which it was allowed to cool down to room temperature. THF (*ca.* 5 mL) and chloranil (8.3 mmol) were added to the reaction mixture, which was reflux for 10 min. Reaction was quenched by addition of cold methanol (ca. 10 mL) and subsequent cooling to 0 °C. A yellow crystalline powder was obtained, yield 30 % (based on dithiooxamide); m.p. 236-238 °C. Elemental analysis of: calcd C 38.94, H 1.31, N 18.17; found C 38.90, H 1.29, N 18.19.
*FTIR ATR:* 3116(w), 3071(w), 1539(w), 1451(m), 1409(s), 1339(w), 1308(w), 1241(w), 1221(m), 1208(w), 1075(w), 1054(m), 969(s), 822(s), 707(vs), 687(m), 614(w), 560(w), 483(w).
*NMR:* $^1$H NMR (400 MHz, CDCl$_3$-*d*$_1$): δ = 7.58 (m, 2H, **H$_c$**), 7.46 (m, 2H, **H$_a$**), 7.12 (m, 2H, **H$_b$**); $^{13}$C{$^1$H} NMR (100 MHz, CDCl$_3$-*d*$_1$): δ = 162.6 (s, S**C**N, **(C2)**), 149.7 (s, **C=C** (**C1**)), 137.5 (s, S**C**C, **(C3)**), 128. 7 (s, **C**-H$_a$ **(C4)**), 128.1 (s, **C**-H$_b$ **(C5)**), 126.9 (s, **C**-H$_c$ **(C6)**).

**2,5-bis(2-thiazolyl)-1,3-thiazolo[5,4-d]-1,3-thiazole (B3):** The reaction mixture was heated at 120–125°C for approximately 3 h. A yellow crystalline product in the form of blocks was obtained and purified in 30 mL of ethanol; yield 68.46% (based on dithiooxamide); m.p. 292.5-294.1 °C. Elemental analysis: calcd C 38.94, H 1.31, N 18.17; found C 38.90, H 1.29, N 18.19.

*FTIR ATR:* 3000(m), 3078(s), 3063(s), 1446(s), 1370(vs), 1324(w), 1241(w), 1158(w), 1149(w), 1068(w), 1058(w), 980(vs), 913(w), 845(vs), 758(w), 742(vs), 731(m), 698(w), 635(w), 615(w), 591(m), 499(w), 479(w) $cm^{-1}$.
*NMR:* $^1$H NMR (400 MHz, $CDCl_3$-$d_1$): δ = 7.90 (d, $^3J_{HH}$ = 3, 2H, **$H_b$**), 7.48 (d, $^3J_{HH}$ = 3, 2H, **$H_a$**); $^{13}$C{$^1$H} NMR (100 MHz, $CDCl_3$-$d_1$): δ = 171.15 (s, S**C**N, **(C2)**), 163.45 (s, S**C**N, **(C3')**), 161.25 (s, S**C**N, **(C3)**),152. 11 (s, **C=C** (**C1**)), 144.33(s, **C**-$H_b$), 122.12 (s, **C**-$H_a$).

**2,5-bis(2-imidazolyl)-1,3-thiazolo[5,4-d]-1,3-thiazole (B4):** The reaction mixture was heated at 120–125°C for approximately 3 h. A brownish product in the form of crystalline powder was obtained; yield 75,44 % (based on dithiooxamide); m.p. 323-338°C. Elemental analysis: calcd C 43.48, H 2.20, N 30.64; found C 43.45, H 2.18, N 30.62. To obtain single crystals for XRD measurement, the compound was recrystallized from N,N-dimethylformamide.
*FTIR ATR:* 3123(w), 3059(w), 3013(m), 2925(m), 2860(m), 2806(m), 2766(m), 2713(s), 2680(s), 2590(m), 2535(m), 1653(w), 1484(m), 1448(m), 1431(w), 1363(vs), 1297(w), 1157(w), 1118(s), 1050(m), 964(m), 926(w), 911(w), 870(m), 739(s), 710(m) $cm^{-1}$.
*NMR:* $^1$H NMR (400 MHz, DMSO-$d_6$): δ = 13.43 (bs, 2H, N**H**), 7.43 (bs, 2H, **$H_a$**), 7.17 (bs, 2H, **$H_b$**); $^{13}$C{$^1$H} NMR (100 MHz, DMSO-$d_6$): δ = 160.5 (s, S**C**N **(C2)**), 150.0 (s, **C**=**C** **(C1)**), 141.0 (s, N**C**NH **(C3)**), 131.0 (s, **C**$H_b$ **(C5)**), 121.0 (s, **C**$H_a$ **(C4)**).

**2,5-bis((2E)-3-fenyloprop-2-enyl))-1,3-thiazolo[5,4-d]-1,3-thiazole (C):** The reaction mixture was heated 130–140 °C for 3–3.5 h. A light orange crystalline product was obtained; yield 70.53 % (based on dithiooxamide); m.p. 239-242 °C. Elemental analysis: calcd C 69.33, H 4.07, N 8.09; found C 69.30, H 4.10, N 8.03.
*FTIR ATR:* 3103(w), 3080(w), 3052(w) 3032(m), 3011(w), 2993(m), 2805(w), 2635(w), 1947(w), 1875(w), 1779(w), 1745(w), 1698(w), 1662(w), 1624(m), 1595(w), 1576(w), 1497(m), 1453(w), 1427(vs), 1332(w), 1301(m), 1254(m), 1220(s), 1187(s), 1172(s), 1100(w), 1071(m), 1029(w), 999(w), 973(w), 943(vs), 847(w),743(s), 682(vs), 650(m), 603(w), 567(), 491(m) $cm^{-1}$.
*NMR:* $^1$H NMR (400 MHz, $CDCl_3$-$d_1$): δ = 7.58-7.54 (m, 4 H, *ortho*-C-**H** **($H_o$)**), 7.50 (d, $^3J_{HH}$ = 16, 2 H, **H**C=CH, **($H_a$)**), 7.44-7.35 (m, 6 H, *meta-* and *para*-C-**H**, **($H_m$** and **$H_p$)**), 7.30 (d, $^3J_{HH}$ = 16, 2 H, HC=C**H**, **($H_b$)**); $^{13}$C{$^1$H} NMR (100 MHz, $CDCl_3$-$d_1$): δ =168.2 (s, S**C**N **(C2)**), 150.3 (s, **C**=**C** **(C1)**), 135.4 (s, **$C_i$** **(C5)**), 135.1 (s, H**C**=CH **(C3)**), 129.4 (s, *meta*-**C**H **(C7** and **C9)**), 129.0 (s, *para*-**C**H **(C8)**), 127.3 (s, *ortho*-**C**H **(C6** and **C10)**), 122.12 (s, HC=**C**H **(C4)**).

### *Crystal Structures*

Crystal data, data collection and structure refinement details are summarized in Table S1. The crystal structure data for **A3-A5, B1, C**, were collected on an IPDS 2T dual beam diffractometer (STOE & Cie GmbH, Darmstadt, Germany) at 120.0(2) K with $MoK_\alpha$ radiation of a microfocus X-ray source – for **A4**, **A5**, **B1** and **C** (GeniX 3D Mo High Flux, Xenocs, Sassenage, France) and $CuK_\alpha$ radiation for **A3**. Small crystal of **B4** was measured on Eulerian 4-circle diffractometer STOE Stadivari (Mo source, EIGER2 1M CdTe detector). Crystals were cooled using a Cryostream 800 open flow nitrogen cryostat (Oxford Cryosystems). The structures were solved using intrinsic phasing procedure implemented in SHELXT and all non-hydrogen atoms were refined with anisotropic displacement parameters by full-matrix least squares procedure based on F2 using the SHELX–2014 program package.[65] The Olex2[66] and Wingx[67] program suites were used to prepare the final version of CIF files. Olex2[67] and Mercury[68] were used to prepare the figures.

Hydrogen atoms were refined using isotropic model with Uiso(H) values fixed to be 1.2 or 1.5 times Ueq of the carbon atoms to which they were attached. Hydrogen atoms bonded to the electronegative oxygen or nitrogen atoms were located in the electron density maps and refined without constraints with the exception of **B4**.

***DFT modelling***

Single molecule calculations were performed using Gaussian 16 Rev. C.01 software package[69] using B3LYP hybrid functional[70] and the TZVP basis set[71] under tight convergence criteria. Results have been processed and visualized using GaussView 5.08 software package.[72]

The geometry relaxation and the electronic structures (band structure and the density of states) calculations of the crystals were calculated using the plane-wave basis set and pseudopotentials implemented on the CASTEP (CAmbridge Serial Total Energy package) code. The PBE-GGA exchange-correlation function was applied and the ion-electron interactions were defined using the projected augmented wave formalism. The non-covalent interaction corrections were applied in the calculation DFT-MBD calculations. The periodic boundary conditions convergence tolerance such as cut-off energy, energy, maximum force, stress, and displacements were set to be 410 eV, $5x10^{-6}$ eV/atom, 0.01 eV/Å, 0.02 GPa, and $5x10^{-4}$ Å. A Monkhorst-Pack of k-points was applied to be 4x3x2 for all the calculations. All the conditions were uniformly applied to all the crystal structures and molecules.

***Spectroscopic measurements***

**NMR Spectroscopy**

($^1$H, $^{13}$C{$^1$H}, $^{19}$F) NMR spectra were recorded on a Bruker AV400 MHz spectrometer (external standard TMS for $^1$H and $^{13}$C and $CFCl_3$ for $^{19}$F)) at ambient temperature. $^1$H, $^{13}$C{$^1$H}, and $^{19}$F chemical shifts (δ) are reported in ppm relative to the residual solvent signals at 2.50 and 39.5 ppm (DMSO-$d_6$) and 7.26 and 77.16 ppm ($CDCl_3$-$d_1$). Coupling constants (J) are given in hertz (Hz). Multiplicities are abbreviated as singlet (s), doublet (d), triplet (t), quartet (q), multiplet (m) and broad (br). In order to improve the solubility of **B3** and enhance the peaks, a small amount of ethanol was added to the NMR tube.

**FTIR ATR Spectroscopy**

FT-IR ATR spectra were recorded for the crystalline compounds using Nicolet iS50 equipped with Specac Quest diamond ATR device; the spectra were collected and formatted by OMNIC software

**UV-Vis Spectroscopy**

Absorbance spectra for acetonitrile solution were measure on Agilent 8453 UV-vis Spectrometer in standard 1 cm quartz cells in the range 200 – 1200 nm with resolution 1 nm.

Diffuse reflectance spectra for solid state samples were measured on LAMBDA 750 UV/vis/NIR spectrophotometer equipped with a 100 mm-integrating sphere (PerkinElmer Inc., USA). Spectra were recorded for powder samples prepared by grinding the mixture of powder samples with $BaSO_4$ in an agate mortar. $BaSO_4$ was used as a reference sample.

Photoluminescence spectra have been recorded on FS5 (Edinburgh Instruments) spectrophotometer with high pressure xenon lamp as the excitation source and single-monochromator photomultiplier detector. The emission spectra were recorded in the range of 390-800 nm with 375 nm excitation wavelength and resolution of 1 nm. Measurements were conducted in standard 1 cm quartz cells for solutions or in solid state sample holder for powder.

Time-resolved photoluminescence kinetic traces for all samples were recorded at room temperature using FS5 spectrofluorometer (Edinburg Instruments Company) in time-correlated single photon counting (TCSPC) mode. The EPL pulsed diode laser of 375 nm with a 50 ps pulse and repetition rate of 2 MHz (for solution) or 10 MHz (for powder) was used as the excitation source. The spectra were collected in 20 ns time window with a time resolution of 2 ps. The instrument response function (IRF) has been recorded for light scattering ($\lambda_{ex} = \lambda_{em}$ = 375 nm) independently for each sample.

The lifetime of fluorescence was determined by using Fluoracle software with the Reconvolution Fit Analysis function including the IRF according to the fitting formula:

$$I(t) = B_1 \cdot e^{-\frac{t}{\tau_1}} + B_2 \cdot e^{-\frac{t}{\tau_2}} \qquad \text{(Equation 7)}$$

where $\tau_1$ – represents the lifetime of the radiative energy decay process, $\tau_2$ – represents the lifetime of the nonradiative energy decay process, $B_1$, $B_2$ – fitting amplitudes corresponding to $\tau_1$ and $\tau_2$, respectively.

***Synchrotron measurements* – *XAS at the S K-edge***

Radial distribution functions were calculated for every S – S pair for the DFT-optimized[73].

XAS were measured at the SOLARIS National Synchrotron Radiation of Poland[74]. The spectra at the S K-edge were collected at the bending magnet ASTRA beamline. A thin layers of powder samples were applied on sulfur-free a Kapton tape and the excess was shaken off and removed with a cotton swab. The measurements were performed in transmission mode using an incident photon beam provided by a modified Lemmonnier-type double-crystal monochromator equipped with InSb (111) crystals. For the monochromator energy calibration, we used $ZnSO_4$ reference placed in the reference chamber, with white line at 2481.4 eV. Positions of the absorption edges were determined based on a maximum of the first derivative of the spectrum. The final spectra were merged from at least three consecutive scans. All spectra were processed using the Athena program from the Demeter software package[75].

XAS spectra were modelled using density functional theory (DFT) calculations performed with FDMNES software[64] using local spin density approximation. The structure for the calculations was previously DFT-optimized using CASTEP. The finite difference method was used for X-ray absorption fine structure[76] with dipole ($\Delta l=\pm1$) and 6 Å cluster radius. It was verified that the quadrupole transitions ($\Delta l=0, \pm2$) do not contribute to the spectrum, therefore they were omitted. Relativistic and spin-orbit coupling effects were neglected. Lorentzian convoluted spectra were presented.

***Electric measurements***

For thin film preparations only representatives: **A4**, **A5**, **B3**, **B4** and **C** were chosen, based on their better solubility in DMF in comparison to the remaining compounds. Thin film samples were prepared according to following procedural steps. Firstly, all compounds were dissolved in DMF (typically 50mg of compound in 1mL of DMF) and mixed on magnetic stirrer at elevated temperatures (100 °C). Substrates, ITO glass (Ossila, The Netherlands) were washed (water, isopropyl alcohol), dried and cleaned with $O_2$ plasma, then heated to 100 °C. Thin films were deposited on hot substrates via spincoating technique, typically 3000 rpm for 45s and post-baked in 100 °C on a hotplate for 10 min. Copper electrodes were thermally deposited afterwards, through shadow mask (Ossila, The Netherlands) with electrode dimensions 1.3 x 1.5 mm.

All of the I-V responses, state retention measurements, endurance tests and potentiation-depression tests were registered on SP-300 potentiostat (BioLogic, France). Temperature-dependent I-V responses measurements were conducted on Instec TP102V Thermoelectric Probe Station. The system was designed as two-terminal device, with working electrode (WE) connected to Cu electrodes and counter (CE) and reference (RE) electrodes were connected the ITO substrate.

State switching stability was evaluated for both high-resistance state (HRS) and low-resistance state (LRS). After applying a DC bias of either -3 V or +3 V for 1 s, the device state was measured at a reading voltage of +500 mV every 5 minutes over a period of 2 hours. The endurance test (on-off switching) was conducted by subjecting the devices to alternating extreme potential values (-3 V and +3 V) for 100 ms. This switching sequence was repeated 750 times, with the device state measured at a reading voltage of +500 mV after each cycle.

**Acknowledgements**

The authors acknowledge the financial support from the Polish National Science Center within the OPUS programme (grant agreement No. 2022/47/B/ST4/00728). This research was partly supported by program "Excellence initiative–research university" for the AGH University of Science and Technology and Gdańsk University of Technology: SILICIUM SUPPORTING CORE R&D FACILITIES DEC-2/2021/IDUB/V.6/Si (crystallographic measurements). The authors gratefully acknowledge Polish high-performance computing infrastructure PLGrid (HPC Center: ACK Cyfronet AGH) for providing computer facilities and support within computational grant no. PLG/2024/017405. The XANES measurements at ASTRA beamline were made under the provision of the Polish Ministry of Education and Science project: 'Support for research and development with the use of research infrastructure of the National Synchrotron Radiation Centre SOLARIS' under contract nr 1/SOL/2021/2. The further development of the ASTRA beamline for measuring at low photon energies was supported within the EU Horizon2020 programme (952148-Sylinda).

**References**

[1] (a) A. Khasbaatar, Z. Xu, J.-H. Lee, G. Campillo-Alvarado, C. Hwang, B. N. Onusaitis, Y. Diao, *Chem. Rev.* **2023**, *123*, 8395-8487; (b) L. Ding, Z.-D. Yu, X.-Y. Wang, Z.-F. Yao, Y. Lu, C.-Y. Yang, J.-Y. Wang, J. Pei, *Chem. Rev.* **2023**, *123*, 7421-7497; (c) G. Zhang, F. R. Lin, F. Qi, T. Heumüller, A. Distler, H.-J. Egelhaaf, N. Li, P. C. Y. Chow, C. J. Brabec, A. K. Y. Jen, H.-L. Yip, *Chem. Rev.* **2022**, *122*, 14180-14274.

[2] (a) F. Yang, S. Cheng, X. Zhang, X. Ren, R. Li, H. Dong, W. Hu, *Adv. Mater.* **2018**, *30*, 1702415; (b) N. A. Kukhta, M. R. Bryce, *Mater. Horiz.* **2021**, *8*, 33-55; (c) F. Gao, E. Reichmanis, *Chem. Rev.* **2023**, *123*, 10835-10837; (d) J. W. Lim, in *Materials, Vol. 17*, **2024**.

[3] (a) H. Yao, J. Wang, Y. Xu, S. Zhang, J. Hou, *Acc. Chem. Res.* **2020**, *53*, 822-832; (b) D. Meng, R. Zheng, Y. Zhao, E. Zhang, L. Dou, Y. Yang, *Adv. Mater.* **2022**, *34*, 2107330; (c) B. Fan, F. Lin, X. Wu, Z. Zhu, A. K. Y. Jen, *Acc. Chem. Res.* **2021**, *54*, 3906-3916; (d) M. D. M. Faure, B. H. Lessard, *J. Mater. Chem. C* **2021**, *9*, 14-40; (e) E. K. Solak, E. Irmak, *RSC Adv.* **2023**, *13*, 12244-12269.

[4] (a) A. Calabrese, P. Battistoni, S. Ceylan, L. Zeni, A. Capo, A. Varriale, S. D'Auria, M. Staiano, in *Biosensors, Vol. 13*, **2023**; (b) D. Deng, Y. Chang, W. Liu, M. Ren, N. Xia, Y. Hao, in *Biosensors, Vol. 13*, **2023**; (c) J. B. Kaushal, P.

Raut, S. Kumar, in *Biosensors, Vol. 13*, **2023**; (d) J. Ma, T. Shu, Y. Sun, X. Zhou, C. Ren, L. Su, X. Zhang, *Small* **2022**, *18*, 2103516; (e) C. Sun, X. Wang, M. A. Auwalu, S. Cheng, W. Hu, *EcoMat* **2021**, *3*, e12094.

[5] (a) R. Iftikhar, F. Z. Khan, N. Naeem, *Mol. Divers.* **2024**, *28*, 271-307; (b) R. J. Pandhare, P. M. Badani, R. M. Kamble, *J. Mol. Struct.* **2024**, *1298*, 137080; (c) K. Upendranath, T. Venkatesh, T. N. Lohith, M. A. Sridhar, *J. Mol. Struct.* **2022**, *1264*, 133231; (d) J. V. Vaghasiya, C. C. Mayorga-Martinez, M. Pumera, *Chem. Soc. Rev.* **2020**, *49*, 7819-7844; (e) Y. Zhang, J. Song, J. Qu, P.-C. Qian, W.-Y. Wong, *Science China Chemistry* **2021**, *64*, 341-357.

[6] (a) D. Bevk, L. Marin, L. Lutsen, D. Vanderzande, W. Maes, *RSC Adv.* **2013**, *3*, 11418-11431; (b) P. Maienfisch, A. J. F. Edmunds, in *Adv. Heterocycl. Chem., Vol. 121* (Eds.: E. F. V. Scriven, C. A. Ramsden), Academic Press, **2017**, pp. 35-88; (c) M. Saito, I. Osaka, Y. Suzuki, K. Takimiya, T. Okabe, S. Ikeda, T. Asano, *Sci. Rep.* **2015**, *5*, 14202; (d) P. Zahradník, P. Magdolen, P. Zahradník, *Tetrahedron Lett.* **2010**, *51*, 5819-5821.

[7] J. Y. Jung, S. J. Han, J. Chun, C. Lee, J. Yoon, *Dyes Pigments* **2012**, *94*, 423-426.

[8] (a) J. R. Johnson, R. Ketcham, *J. Am. Chem. Soc.* **1960**, *82*, 2719-2724; (b) J. R. Johnson, D. H. Rotenberg, R. Ketcham, *J. Am. Chem. Soc.* **1970**, *92*, 4046-4050.

[9] R. Ketcham, S. Mah, *J. Med. Chem.* **1971**, *14*, 743-747.

[10] X. Huang, W. Lu, Q. Liu, M. Wu, *Comput. Theor. Chem.* **2023**, *1230*, 114387.

[11] (a) A. Shibu, S. Jones, P. L. Tolley, D. Diaz, C. O. Kwiatkowski, D. S. Jones, J. M. Shivas, J. J. Foley, T. A. Schmedake, M. G. Walter, *Materials Advances* **2023**, *4*, 6321-6332; (b) A. Thorat, S. Behera, A. A. Boopathi, C. Kulkarni, *Angew. Chem. Int. Ed.* **2024**, *63*, e202409725; (c) W. Zhou, D. Wu, H. Xiao, J. Song, L. Qu, L. Wang, X. Zhou, Z.-X. Xu, H. Xiang, *Dyes Pigments* **2022**, *197*, 109906.

[12] (a) A. R. Brotherton, A. Shibu, J. C. Meadows, N. A. Sayresmith, C. E. Brown, A. M. Ledezma, T. A. Schmedake, M. G. Walter, *Adv. Sci.* **2023**, *10*, 2205729; (b) Y. Liu, Z. Liang, Z. Li, K. Zhao, Y. Sun, X. Zhang, R. Yang, L. Qu, *Microchem. J.* **2020**, *154*, 104640; (c) S. Shah, N. Naithani, S. C. Sahoo, P. P. Neelakandan, N. Tyagi, *J. Mater. Chem. C* **2024**, *12*, 13088-13095.

[13] (a) A. F. R. Cerqueira, M. E. Pérez, N. S. Gsponer, M. G. P. M. S. Neves, A. Jorge Parola, E. N. Durantini, A. C. Tomé, *J. Photochem. Photobiol., A* **2024**, *456*, 115849; (b) A. F. R. Cerqueira, N. M. M. Moura, M. G. P. M. S. Neves, A. Jorge Parola, A. C. Tomé, *J. Photochem. Photobiol., A* **2024**, *451*, 115490; (c) V. Kumar, S. Sony, N. Kaur, S. M. Mobin, P. Kaur, K. Singh, *Anal. Chim. Acta* **2022**, *1206*, 339776.

[14] X. Li, C. Huang, Y. Fan, Z. Bai, B.-L. An, J. Xu, W. Zheng, Y.-L. Bai, *ACS Appl. Mater. Interfaces* **2023**, *15*, 46022-46030.

[15] (a) R. Li, X. Yang, *J. Phys. Chem. A* **2019**, *123*, 10102-10108; (b) A. N. Woodward, J. M. Kolesar, S. R. Hall, N.-A. Saleh, D. S. Jones, M. G. Walter, *J. Am. Chem. Soc.* **2017**, *139*, 8467-8473.
[16] (a) S. Ando, J.-i. Nishida, Y. Inoue, S. Tokito, Y. Yamashita, *J. Mater. Chem.* **2004**, *14*, 1787-1790; (b) S. Van Mierloo, S. Chambon, A. E. Boyukbayram, P. Adriaensens, L. Lutsen, T. J. Cleij, D. Vanderzande, *Magn. Reson. Chem.* **2010**, *48*, 362-369; (c) S. Ando, J. Nishida, E. Fujiwara, H. Tada, Y. Inoue, S. Tokito, Y. Yamashita, *Synth. Met.* **2006**, *156*, 327-331; (d) J.-M. Jiang, M.-C. Yuan, K. Dinakaran, A. Hariharan, K.-H. Wei, *J. Mater. Chem. A* **2013**, *1*, 4415-4422; (e) I. Osaka, G. Sauvé, R. Zhang, T. Kowalewski, R. D. McCullough, *Adv. Mater.* **2007**, *19*, 4160-4165.
[17] (a) T. Naito, Y. Kita, T. Shimazaki, M. Tachikawa, *RSC Adv.* **2022**, *12*, 34685-34693; (b) V. Malytskyi, J.-J. Simon, L. Patrone, J.-M. Raimundo, *RSC Adv.* **2015**, *5*, 354-397; (c) A. Dessì, M. Calamante, A. Mordini, M. Peruzzini, A. Sinicropi, R. Basosi, F. Fabrizi de Biani, M. Taddei, D. Colonna, A. Di Carlo, G. Reginato, L. Zani, *Chem. Commun.* **2014**, *50*, 13952-13955; (d) T. Shimazaki, M. Tachikawa, *Phys. Chem. Chem. Phys.* **2021**, *23*, 21078-21086.
[18] M. Mamada, J.-i. Nishida, D. Kumaki, S. Tokito, Y. Yamashita, *Chem. Mater.* **2007**, *19*, 5404-5409.
[19] A. Sławek, L. Alluhaibi, E. Kowalewska, G. Abdi, T. Mazur, A. Podborska, K. Mech, M. Marciszko-Wiąckowska, A. Maximenko, K. Szaciłowski, *Adv. Electron. Mater.* **2024**, 2300818.
[20] A. Hantzsch, *Berichte der deutschen chemischen Gesellschaft* **1881**, *14*, 1637-1638.
[21] J. Ephraim, *Berichte der deutschen chemischen Gesellschaft* **1891**, *24*, 1026-1031.
[22] (a) K. Wang, H. Zhang, S. Chen, G. Yang, J. Zhang, W. Tian, Z. Su, Y. Wang, *Adv. Mater.* **2014**, *26*, 6168-6173; (b) M. Ghora, P. Majumdar, M. Anas, S. Varghese, *Chem.: Eur. J.* **2020**, *26*, 14488-14495.
[23] A. Bolognesi, M. Catellani, S. Destri, W. Porzio, *Acta Crystallogr. C* **1987**, *43*, 2106-2108.
[24] J. E. Campbell, J. Yang, G. M. Day, *J. Mater. Chem. C* **2017**, *5*, 7574-7584.
[25] L. D. Costa, S. Guieu, M. d. A. F. Faustino, A. C. Tomé, *New J. Chem.* **2022**, *46*, 3602-3615.
[26] D. Li, Z. Zhang, S. Zhao, Y. Wang, H. Zhang, *Dalton Trans.* **2011**, *40*, 1279-1285.
[27] B. P. Biswal, D. Becker, N. Chandrasekhar, J. S. Seenath, S. Paasch, S. Machill, F. Hennersdorf, E. Brunner, J. J. Weigand, R. Berger, X. Feng, *Chem.: Eur. J.* **2018**, *24*, 10868-10875.
[28] S. Ando, J.-i. Nishida, H. Tada, Y. Inoue, S. Tokito, Y. Yamashita, *J. Am. Chem. Soc.* **2005**, *127*, 5336-5337.
[29] P. Wagner, M. Kubicki, *Acta Crystallogr. C* **2003**, *59*, o91-o92.

[30] T. Tao, J. Geng, L. Hong, W. Huang, H. Tanaka, D. Tanaka, T. Ogawa, *J. Phys. Chem. C* **2013**, *117*, 25325-25333.
[31] R. J. Baker, P. E. Colavita, D. M. Murphy, J. A. Platts, J. D. Wallis, *J. Phys. Chem. A* **2012**, *116*, 1435-1444.
[32] P. R. Spackman, M. J. Turner, J. J. McKinnon, S. K. Wolff, D. J. Grimwood, D. Jayatilaka, M. A. Spackman, *J. Appl. Crystallogr.* **2021**, *54*, 1006-1011.
[33] (a) X. Yu, L. Zheng, J. Li, P. Yu, Z. Liu, C. Li, Y. Zou, X. Zhang, W. Hu, *Org. Electron.* **2020**, *87*, 105941; (b) J.-I. Park, J. W. Chung, J.-Y. Kim, J. Lee, J. Y. Jung, B. Koo, B.-L. Lee, S. W. Lee, Y. W. Jin, S. Y. Lee, *J. Am. Chem. Soc.* **2015**, *137*, 12175-12178; (c) J. Podlesný, F. Bureš, in *Organics, Vol. 3*, **2022**, pp. 446-469.
[34] (a) Y. A. Berlin, G. R. Hutchison, P. Rempala, M. A. Ratner, J. Michl, *J. Phys. Chem. A* **2003**, *107*, 3970-3980; (b) S. T. Bromley, M. Mas-Torrent, P. Hadley, C. Rovira, *J. Am. Chem. Soc.* **2004**, *126*, 6544-6545; (c) R. A. Marcus, *Rev. Mod. Phys.* **1993**, *65*, 599-610; (d) R. A. Marcus, *J. Chem. Phys.* **2004**, *26*, 867-871; (e) R. A. Marcus, *J. Chem. Phys.* **2004**, *26*, 872-877; (f) R. A. Marcus, *J. Chem. Phys.* **2004**, *24*, 966-978.
[35] (a) O. López-Estrada, H. G. Laguna, C. Barrueta-Flores, C. Amador-Bedolla, *ACS Omega* **2018**, *3*, 2130-2140; (b) P. K. Behara, M. Dupuis, *Phys. Chem. Chem. Phys.* **2020**, *22*, 10609-10623.
[36] A. Datta, S. Mohakud, S. K. Pati, *J. Mater. Chem.* **2007**, *17*, 1933-1938.
[37] E. Cechosz, L. Alluhaibi, T. Mazur, A. Sławek, N. Pandurangan, K. Szaciłowski, *Adv. Electron. Mater.* **2024**, *n/a*, 2400654.
[38] S. Slassi, M. Aarjane, A. Amine, *J. Mol. Struct.* **2023**, *1276*, 134788.
[39] (a) V. S. Naik, P. S. Patil, Q. A. Wong, C. K. Quah, N. B. Gummagol, H. S. Jayanna, *J. Mol. Struct.* **2020**, *1222*, 128901; (b) V. S. Naik, A. Pragasam, H. S. Jayanna, G. Vinitha, *Chem. Phys. Lett.* **2020**, *754*, 137680.
[40] (a) K. Pyta, P. Przybylski, A. Huczyński, A. Hoser, K. Woźniak, W. Schilf, B. Kamieński, E. Grech, B. Brzezinski, *J. Mol. Struct.* **2010**, *970*, 147-154; (b) A. Mielcarek, A. Wiśniewska, A. Dołęga, *Struct. Chem.* **2018**, *29*, 1189-1200.
[41] (a) P. Rajkumar, S. Selvaraj, P. Anthoniammal, A. Ram Kumar, K. Kasthuri, S. Kumaresan, *Chemical Physics Impact* **2023**, *7*, 100257; (b) R. P. C. L. Sousa, R. B. Figueira, B. R. Gomes, S. Sousa, R. C. M. Ferreira, S. P. G. Costa, M. M. M. Raposo, in *Nanomaterials, Vol. 11*, **2021**.
[42] B. Brzezinski, G. Zundel, *The Journal of Physical Chemistry* **1982**, *86*, 5133-5135.
[43] M. Ventura, J. R. Silva, T. Catunda, L. H. C. Andrade, S. M. Lima, *J. Mol. Liq.* **2021**, *328*, 115414.
[44] (a) O. Unsalan, H. Arı, C. Altunayar-Unsalan, K. Bolelli, M. Boyukata, I. Yalcin, *J. Mol. Struct.* **2020**, *1218*, 128454; (b) T.-X. Luan, P. Zhang, Q. Wang, X. Xiao, Y. Feng, S. Yuan, P.-Z. Li, Q. Xu, *Nano Lett.* **2024**, *24*, 5075-5084.

[45] (a) P. Sowmya, S. Prakash, A. Joseph, *J. Solid State Chem.* **2023**, *320*, 123836; (b) R. Balogh, A. Eckstein, K. Tokár, M. Danko, *J. Photochem. Photobiol., A* **2023**, *434*, 114217; (c) X. Lv, Q. Hu, T. Miao, Y. Li, B. Cui, Y. Fang, *Anal. Bioanal. Chem.* **2022**, *414*, 4837-4847.
[46] U. Olgun, Z. Dikmen, H. Çetin, F. Arıcan, M. Gülfen, *J. Mol. Struct.* **2022**, *1250*, 131816.
[47] (a) A. Rocheteau, L. Lemeur, M. Cordier, M. Paris, J.-Y. Mevellec, C. Latouche, H. Serier-Brault, S. Perruchas, *Adv.Opt. Mater.* **2024**, *n/a*, 2402040; (b) B. Liu, Q. Zhou, Y. Li, Y. Chen, D. He, D. Ma, X. Han, R. Li, K. Yang, Y. Yang, S. Lu, X. Ren, Z. Zhang, L. Ding, J. Feng, J. Yi, J. Chen, *Angew. Chem. Int. Ed.* **2024**, *63*, e202317185; (c) T. G. Paiva, M. Klem, S. L. Silvestre, J. Coelho, N. Alves, E. Fortunato, E. J. Cabrita, M. C. Corvo, *ChemSusChem* **2024**, *n/a*, e202401710; (d) R. A. Yadav, M. Kumar, R. Singh, P. Singh, S. Jaiswal, G. Srivastav, R. L. Prasad, *Spectrochim. Acta, Part A* **2008**, *71*, 1565-1570.
[48] R. Berger, G. Resnati, P. Metrangolo, E. Weber, J. Hulliger, *Chem. Soc. Rev.* **2011**, *40*, 3496-3508.
[49] M. Asemani, A. R. Rabbani, *Journal of Petroleum Science and Engineering* **2020**, *185*, 106618.
[50] X. Li, Y. Ju, Q. Hou, H. Lin, *Science China Earth Sciences* **2012**, *55*, 1269-1279.
[51] A. K. Sahoo, C. Yadav, J. N. Moorthy, *Appl. Catal., A: Gen.* **2024**, *671*, 119557.
[52] (a) M. D. Smith, B. L. Watson, R. H. Dauskardt, H. I. Karunadasa, *Chem. Mater.* **2017**, *29*, 7083-7087; (b) C. H. Cheng, H. Y. Huang, M. J. Talite, W. C. Chou, J. M. Yeh, C. T. Yuan, *J. Colloid Interface Sci.* **2017**, *508*, 105-111.
[53] (a) S. Kumar, M. Singh, P. Gaur, J.-H. Jou, S. Ghosh, *ACS Omega* **2017**, *2*, 5348-5356; (b) H. Wu, S. Wang, J. Ding, R. Wang, Y. Zhang, *Dyes Pigments* **2020**, *182*, 108665.
[54] S. J. Cassidy, I. Brettell-Adams, L. E. McNamara, M. F. Smith, M. Bautista, H. Cao, M. Vasiliu, D. L. Gerlach, F. Qu, N. I. Hammer, D. A. Dixon, P. A. Rupar, *Organometallics* **2018**, *37*, 3732-3741.
[55] (a) E. Lippert, **1955**, *10*, 541-545; (b) N. Mataga, Y. Kaifu, M. Koizumi, *Bull. Chem. Soc. Jpn.* **1955**, *28*, 690-691.
[56] (a) M. R. Pinto, Y. Takahata, T. D. Z. Atvars, *J. Photochem. Photobiol., A* **2001**, *143*, 119-127; (b) A. Brillante, B. Samorì, C. Stremmenos, P. Zanirato, *Mol. Cryst. Liq. Cryst.* **1983**, *100*, 263-274.
[57] S. Landi, I. R. Segundo, E. Freitas, M. Vasilevskiy, J. Carneiro, C. J. Tavares, *Solid State Commun.* **2022**, *341*, 114573.
[58] J. I. Pankove, *Optical Processes in Semiconductors*, Dover Publications, Inc., Mineola NY, **1975**.

[59] J. Tauc, in *Amorphous and Liquid Semiconductors* (Ed.: J. Tauc), Plenum Press, London, **1974**.

[60] (a) S. Kasap, C. Koughia, J. Singh, H. Ruda, S. O'Leary, in *Springer Handbook of Electronic and Photonic Materials* (Eds.: S. Kasap, P. Capper), Springer US, Boston, MA, **2007**, pp. 47-77; (b) K. Morigaki, C. Ogihara, in *Springer Handbook of Electronic and Photonic Materials* (Eds.: S. Kasap, P. Capper), Springer US, Boston, MA, **2007**, pp. 565-580.

[61] E. I. Solomon, B. Hedman, K. O. Hodgson, A. Dey, R. K. Szilagyi, *Coordination Chemistry Reviews* **2005**, *249*, 97-129.

[62] G. Almkvist, K. Boye, I. Persson, *Journal of Synchrotron Radiation* **2010**, *17*, 683-688.

[63] (a) R. Alonso Mori, E. Paris, G. Giuli, S. G. Eeckhout, M. Kavčič, M. Žitnik, K. Bučar, L. G. M. Pettersson, P. Glatzel, *Anal. Chem.* **2009**, *81*, 6516-6525; (b) G. Almkvist, K. Boye, I. Persson, *Journal of Synchrotron Radiation* **2010**, *17*, 683-688.

[64] O. Bunău, Y. Joly, *J. Phys.: Condens. Matter* **2009**, *21*, 345501.

[65] G. Sheldrick, *Acta Cryst. C* **2015**, *71*, 3-8.

[66] O. V. Dolomanov, L. J. Bourhis, R. J. Gildea, J. A. K. Howard, H. Puschmann, *J. Appl. Crystallogr.* **2009**, *42*, 339-341.

[67] L. Farrugia, *J. Appl. Crystallogr.* **2012**, *45*, 849-854.

[68] C. F. Macrae, P. R. Edgington, P. McCabe, E. Pidcock, G. P. Shields, R. Taylor, M. Towler, J. van de Streek, *J. Appl. Crystallogr.* **2006**, *39*, 453-457.

[69] M. J. Frisch, G. W. Trucks, H. B. Schlegel, G. E. Scuseria, M. A. Robb, J. R. Cheeseman, G. Scalmani, V. Barone, G. A. Petersson, H. Nakatsuji, X. Li, M. Caricato, A. V. Marenich, J. Bloino, B. G. Janesko, R. Gomperts, B. Mennucci, H. P. Hratchian, J. V. Ortiz, A. F. Izmaylov, J. L. Sonnenberg, Williams, F. Ding, F. Lipparini, F. Egidi, J. Goings, B. Peng, A. Petrone, T. Henderson, D. Ranasinghe, V. G. Zakrzewski, J. Gao, N. Rega, G. Zheng, W. Liang, M. Hada, M. Ehara, K. Toyota, R. Fukuda, J. Hasegawa, M. Ishida, T. Nakajima, Y. Honda, O. Kitao, H. Nakai, T. Vreven, K. Throssell, J. A. Montgomery Jr., J. E. Peralta, F. Ogliaro, M. J. Bearpark, J. J. Heyd, E. N. Brothers, K. N. Kudin, V. N. Staroverov, T. A. Keith, R. Kobayashi, J. Normand, K. Raghavachari, A. P. Rendell, J. C. Burant, S. S. Iyengar, J. Tomasi, M. Cossi, J. M. Millam, M. Klene, C. Adamo, R. Cammi, J. W. Ochterski, R. L. Martin, K. Morokuma, O. Farkas, J. B. Foresman, D. J. Fox, *Gaussian 16, Rev. C.01*, Gaussian, Inc., Wallingford CT, **2019**.

[70] A. D. Becke, *J. Chem. Phys.* **1993**, *98*, 5648-5652.

[71] (a) A. Schäfer, C. Huber, R. Ahlrichs, *J. Chem. Phys.* **1994**, *100*, 5829-5835; (b) A. Schäfer, H. Horn, R. Ahlrichs, *J. Chem. Phys.* **1992**, *97*, 2571-2577.

[72] R. T. Dennington, T. Keith, J. Millam, Semichem Inc. Shawnee Mission KS, **2009**.

[73] A. Stukowski, *Modelling and Simulation in Materials Science and Engineering* **2010**, *18*, 015012.
[74] J. Szlachetko, J. Szade, E. Beyer, W. Błachucki, P. Ciochoń, P. Dumas, K. Freindl, G. Gazdowicz, S. Glatt, K. Guła, J. Hormes, P. Indyka, A. Klonecka, J. Kołodziej, T. Kołodziej, J. Korecki, P. Korecki, F. Kosiorowski, K. Kosowska, G. Kowalski, M. Kozak, P. Kozioł, W. Kwiatek, D. Liberda, H. Lichtenberg, E. Madej, A. Mandziak, A. Marendziak, K. Matlak, A. Maximenko, P. Nita, N. Olszowska, R. Panaś, E. Partyka-Jankowska, M. Piszak, A. Prange, M. Rawski, M. Roman, M. Rosmus, M. Sikora, J. Sławek, T. Sobol, K. Sowa, N. Spiridis, J. Stępień, M. Szczepanik, M. Ślęzak, T. Ślęzak, T. Tyliszczak, G. Ważny, J. Wiechecki, D. Wilgocka-Ślęzak, B. Wolanin, P. Wróbel, T. Wróbel, M. Zając, A. Wawrzyniak, M. Stankiewicz, *The European Physical Journal Plus* **2023**, *138*, 10.
[75] B. Ravel, M. Newville, *Journal of Synchrotron Radiation* **2005**, *12*, 537-541.
[76] J. D. Bourke, C. T. Chantler, Y. Joly, *Journal of Synchrotron Radiation* **2016**, *23*, 551-559.

**Supporting Information section**

# Structural Insights and Advanced Spectroscopic Characterization of Thiazolothiazoles: Unveiling Potential for Optoelectronic and Sensing Applications

Karolina Gutmańska[a]*, Agnieszka Podborska[b], Andrzej Sławek[b], Ramesh Sivasamy[b], Lulu Alluhaibi[c], Alexey Maximenko[c], Anna Ordyszewska[a], Konrad Szaciłowski[b,d]*, Anna Dołęga[a], Tomasz Mazur[b]*

[a] *Gdansk University of Technology, Chemical Faculty, Department of Inorganic Chemistry, Narutowicza 11/12, 80-233 Gdańsk, Poland*

[b] *AGH University of Krakow, Academic Centre of Materials and Technology, Mickiewicza 30, 30-059 Kraków, Poland*

[c] *National Synchrotron Radiation Centre SOLARIS, Jagiellonian University, Czerwone Maki 98, Kraków 30-392, Poland*

[d] *University of the West of England, Unconventional Computing Lab, Bristol BS16 1QY, United Kingdom*

*Corresponding authors:* karolina.gutmanska@pg.edu.pl, szacilow@agh.edu.pl, tmazur@agh.edu.pl

## ***Table S1*** *Crystal data and structure refinement for compounds* ***A3-A5, B1, B4, C***

| Compound | A3 | A4 | A5 | B1 | B4 | C |
|---|---|---|---|---|---|---|
| Empirical formula | $C_{20}H_6F_{12}N_2S_2$ | $C_{18}H_{14}N_2O_4S_2$ | $C_{18}H_{14}N_2O_4S_2$ | $C_{12}H_6N_2O_2S_2$ | $C_{10}H_6N_6S_2$ | $C_{20}H_{14}N_2S_2$ |
| Formula weight (g mol$^{-1}$) | 566.39 | 386.43 | 386.43 | 274.31 | 274.33 | 346.45 |
| Wavelength (Å) | 1.54186 | 0.71073 | 0.71073 | 0.71073 | 0.71073 | 0.71073 |
| Temperature (K) | 120(2) | 120(2) | 120(2) | 120(2) | 110(2) | 120(2) |
| Crystal system | orthorhombic | monoclinic | monoclinic | monoclinic | monoclinic | monoclinic |
| Space group | *Ibam* | $P2_1/n$ | $P2_1/n$ | $P2_1/n$ | $P2_1/c$ | $P2_1/c$ |
| *a* (Å) | 15.4405(9) | 6.2324(13) | 10.2538(10) | 5.6092(8) | 3.7442(3) | 5.8842(4) |
| *b* (Å) | 17.4483(8) | 20.207(4) | 3.9636(2) | 5.8044(7) | 14.7018(13) | 4.7578(3) |
| *c* (Å) | 7.4365(4) | 6.673(2) | 20.4696(18) | 16.727(2) | 9.9993(6) | 28.4784(16) |
| *α* (°) | 90 | 90 | 90 | 90 | 90 | 90 |
| *β* (°) | 90 | 106.84(2) | 98.554(7) | 95.631(11) | 92.063(5) | 92.464(5) |
| *γ* (°) | 90 | 90 | 90 | 90 | 90 | 90 |
| Volume (Å$^3$) | 2003.47(18) | 804.4(4) | 822.67(12) | 541.98(12) | 550.07(7) | 796.54(9) |
| *Z* | 4 | 2 | 2 | 2 | 2 | 2 |
| Calculated density (g cm$^{-3}$) | 1.878 | 1.595 | 1.560 | 1.681 | 1.656 | 1.444 |
| Crystal size (mm) | 0.238x0.109x 0.037 | 0.426x0.167x0 0.031 | 0.147x 0.075x 0.027 | 0.245x 0.111x 0.041 | - | 0.397x 0.155x 0.033 |
| Absorption coefficient (mm$^{-1}$) | 3.607 | 0.360 | 0.352 | 0.483 | 0.473 | 0.337 |
| *F*(000) | 1120 | 400 | 400 | 280 | 280 | 360 |
| θ range (°) | 3.823 to 67.532 | 3.345 to 29.246 | 2.377 to 29.248 | 2.447 to 29.215 | 2.464 to 29.983 | 2.864 to 29.339 |
| Limiting indices | -17≤h≤17 | -8≤h≤7 | -14≤h≤14 | -7≤h≤7 | -4≤h≤5 | -8≤h≤7 |
| | -20≤k≤18 | -27≤k≤26 | -5≤k≤5 | -7≤k≤7 | -20≤k≤19 | -6≤k≤6 |
| | -8≤l≤8 | -9≤l≤9 | -25≤l≤28 | -22≤l≤22 | -13≤l≤13 | -38≤l≤38 |
| Reflections collected / unique/unique [$I>2\sigma(I)$] | 5735/974/738 | 10068/ 2161/ 1355 | 6230/2218/ 1641 | 4703/1454/ 1073 | 4955/ 1384/ 853 | 7692/ 2162/ 1824 |
| $R_{int}$ | 0.0341 | 0.0805 | 0.0519 | 0.0571 | 0.0835 | 0.0608 |
| Completeness to $\theta_{max}$ (%) | 98.5 | 98.5 | 99.0 | 99.2 | 86.6 | 98.4 |
| Data / restraints / parameters | 974/0/103 | 2161/0/123 | 2218/0/123 | 1454/0/82 | 1308/0/82 | 2162/0/109 |
| Goodness-of-fit on $F^2$ | 1.072 | 0.971 | 1.133 | 1.106 | 1.039 | 1.143 |
| Final R indices [$I>2\sigma(I)$] | $R_1$ = 0.0396 | $R_1$ = 0.0488 | $R_1$ = 0.0649 | $R_1$ = 0.0602 | $R_1$ = 0.0718 | $R_1$ = 0.0769 |
| | $wR_2$ = 0.104 | $wR_2$ = 0.1193 | $wR_2$ = 0.1269 | $wR_2$ = 0.1232 | $wR_2$ = 0.171 | $wR_2$ = 0.1486 |
| R indices (all data) | $R_1$ = 0.0537 | $R_1$ =0.0876 | $R_1$ = 0.1012 | $R_1$ = 0.094 | $R_1$ = 0.1308 | $R_1$ =0.061 |
| | $wR_2$ = 0.1136 | $wR_2$ = 0.1399 | $wR_2$ = 0.1481 | $wR_2$ = 0.1437 | $wR_2$ = 0.2024 | $wR_2$ = 0.1658 |
| Largest diff. peak and hole (e Å$^{-3}$) | 0.308, -0.363 | 0.41, -0.502 | 0.494, -0.429 | 0.51, -0.567 | 0.53, −0.665 | 0.515, -0.517 |
| CCDC deposition number | 2414396 | 2414397 | 2414398 | 2414399 | 2414400 | 2414401 |

## ***Table S2*** *Selected bond lengths for the atoms of the TzTz system.*

| Compound/ Bond [Å] | A1[a] | A1[b] | A2[c] | A3 | A4 | A5 | B1 | B2 [d] | B3 [e] | B4 | C |
|---|---|---|---|---|---|---|---|---|---|---|---|
| C1/C6/C8–C1$^{ii/iv/v/vii/viii/ix}$/C6B$^{vi}$/C1B$^{iii/ix}$/C8B$^{i}$ | 1.371(3) | 1.382(2) | 1.379(5) | 1.365(7) | 1.363(4) | 1.378(5) | 1.387(5) | 1.382(2) | 1.377(3) | 1.372(6) | 1.360(3) |
| C1/C1$^{vii}$/C6B$^{vi}$/C1B$^{ix}$/C8B$^{i}$–N1 | 1.376(2) | 1.321(2) | 1.358(5) | 1.368(5) | 1.363(3) | 1.363(4) | 1.358(5) | 1.362(2) | 1.362(2) | 1.357(5) | 1.366(3) |
| C1/C1B$^{iii}$/C6/C8– S1/S2 | 1.711(2) | 1.721(1) | 1.725(4) | 1.723(4) | 1.725(2) | 1.724(3) | 1.723(3) | 1.733(1) | 1.723(2) | 1.726(4) | 1.726(3) |
| C2/C2B$^{ii}$/C5/C7– S1/S1$^{ii/iv/v/viii/v}$/S2 | 1.759(2) | 1.762(2) | 1.757(4) | 1.751(4) | 1.754(3) | 1.765(3) | 1.757(3) | 1.761(1) | 1.747(2) | 1.750(4) | 1.765(2) |
| C2/C5/C7– N1/N1$^{ix}$ | 1.330(2) | 1.321(2) | 1.312(5) | 1.315(5) | 1.330(3) | 1.310(4) | 1.317(4) | 1.323(2) | 1.313(3) | 1.321(5) | 1.320(3) |
| C3/C1/C6– C2/C5/C7 | 1.471(2) | 1.465(2) | 1.467(5) | 1.468(5) | 1.463(3) | 1.460(4) | 1.434(5) | 1.444(2) | 1.451(2) | 1.444(6) | 1.441(4) |

$^{i}$:2-x,2-y,2-z, $^{ii}$:1-x,2-y,1-z, $^{iii}$:-x,-y,-z, $^{iv}$: 1-x,1-y,z, $^{v}$:1-x,1-y,1-z,$^{vi}$:1-x,1-y,-z,$^{vii}$:2-x,-y,1-z , $^{viii}$:2-x,1-y,1-z, $^{ix}$:1-x,-y,-z .

## ***Table S3*** *Selected angles for the atoms of the TzTz system.*

| Compound/ Angles [°] | A1[a] | A1[b] | A2[c] | A3 | A4 | A5 | B1 | B2[d] | B3[e] | B4 | C |
|---|---|---|---|---|---|---|---|---|---|---|---|
| C1/C1$^{ii/viii}$/C6/C8– S1/S1$^{ii/viii}$/S1B $^{iii}$ /S2–C2/C2 $^{ii/iv/v}$/C5/C2B $^{ii/iii}$/C7 | 88.46(8) | 88.57(7) | 88.3(2) | 88.24(17 | 88.3(1) | 88.4(2) | 88.0(2) | 87.9(2) | 87.76(8) | 88.38(6) | 88.1(1) |
| C1/C1$^{vii}$ /C6B$^{vi}$ /C8B $^{i}$ – N1–C2/C2B$^{ix}$ /C5/C7 | 107.1(1) | 108.1(1) | 108.5(3) | 108.4(3) | 108.6(2) | 109.1(3) | 108.0(3) | 107.7(3) | 107.9(2) | 108.4(1) | 108.2(2) |
| N1/N1B$^{ix}$– C2/C5/C7–S1/S1 $^{v/viii}$/S2/S1B $^{ii}$ | 116.3(1) | 116.0(1) | 116.1(2) | 116.1(3) | 115.5(2) | 115.6(2) | 116.7(3) | 116.7(3) | 117.1(1) | 116.0(1) | 115.7(2) |
| N1/N1B$^{ix}$ – C2/C5/C7–C3/C1/C6 | 123.4(1) | 123.5(1) | 123.2(3) | 121.9(3) | 123.7(2) | 123.4(3) | 122.9(3) | 123.8(4) | 123.0(2) | 123.6(1) | 122.2(2) |
| N1/N1$^{vii}$ /N1B/N1B$^{i}$– C1/C6/C8–S1/S1B $^{iii}$ /S2 | 131.9(1) | 132.6(1) | 132.9(3) | 132.7(3) | 132.4(2) | 133.1(2) | 132.7(3) | 132.3(3) | 132.7(1) | 132.8(1) | 132.0(2) |
| S1/S2– C1/C6/C1B $^{iii}$ /C8–C1 $^{ii/iv/v/vii/viii/ix}$/C1B $^{ii}$ /C6B$^{vi}$ /C8B | 109.7(1) | 108.9(1) | 108.9(2) | 109.6(3) | 109.9(2) | 109.1(2) | 109.2(3) | 109.2(3) | 109.6(1) | 108.9(1) | 109.5(2) |
| S1/S1 $^{ii/iv/v/viii}$/S1B $^{ii}$ /S2– C2/C5/C7–C3/C1/C6 | 120.3(1) | 120.5(1) | 123.2(3) | 122.0(3) | 120.8(2) | 121.1(2) | 120.3(3) | 119.5(3) | 119.9(1) | 120.3(1) | 122.1(2) |
| C1/C1$^{ii/iv/v/viii/ix}$ / C1B $^{ii/iii}$ /C6/C8– C1/C1$^{vii}$/C6B$^{vi}$/C8B$^{i}$–N1 | 118.4(1) | 118.4(1) | 118.2(3) | 117.8(4) | 117.8(2) | 117.8(3) | 118.1(3) | 118.5(4) | 117.7(2) | 118.3(1) | 118.5(2) |

$^{i}$:2-x,2-y,2-z, $^{ii}$:1-x,2-y,1-z, $^{iii}$:-x,-y,-z, $^{iv}$: 1-x,1-y,z, $^{v}$:1-x,1-y,1-z,$^{vi}$:1-x,1-y,-z,$^{vii}$:2-x,-y,1-z , $^{viii}$:2-x,1-y,1-z, $^{ix}$:1-x,-y,-z .

[a] D. Li, Z. Zhang, S. Zhao, Y. Wang, H. Zhang, *Dalton Trans.* 2011, **40**, 1279.

[b] B.P. Biswal, D. Becker, N. Chandrasekhar, J.S. Seenath, S. Paasch, S. Machill, F.Hennersdorf, E.Brunner, J.J. Weigand, R. Berger, X. Feng, *Chem.-Eur. J.* 2018, **24**, 10868.

[c] S. Ando, J. Nishida, H. Tada, Y. Inoue, S. Tokito, Y. Yamashita, *J. Am. Chem. Soc.* 2005, **127**, 5336.

[d] P. Wagner, M. Kubicki, *Acta Crystallogr., Sect. C: Cryst. Struct. Commun.* 2003, **59**, o91.

[e] T. Tao, J. Geng, L. Hong, W. Huang, H. Tanaka, D. Tanaka, T. Ogawa, *J. Phys. Chem. C* 2013, **117**, 25325.

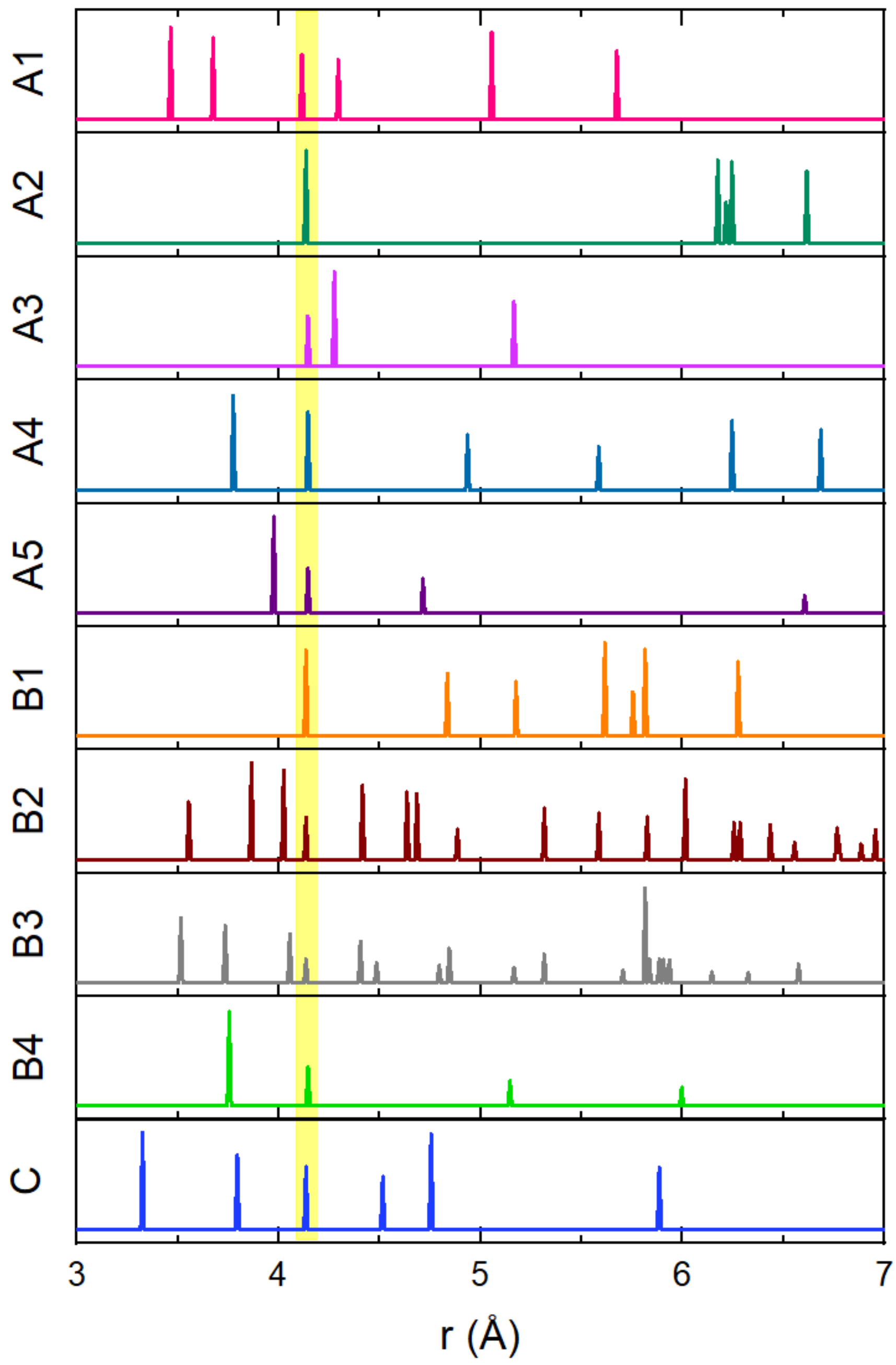

***Figure S1***. *Plots of radial distribution of sulfur atoms in fully relaxed crystal lattices of thiazolothiazols. Sulfur-sulfur distance within single thiazolothiazole system highlighted in yellow.*

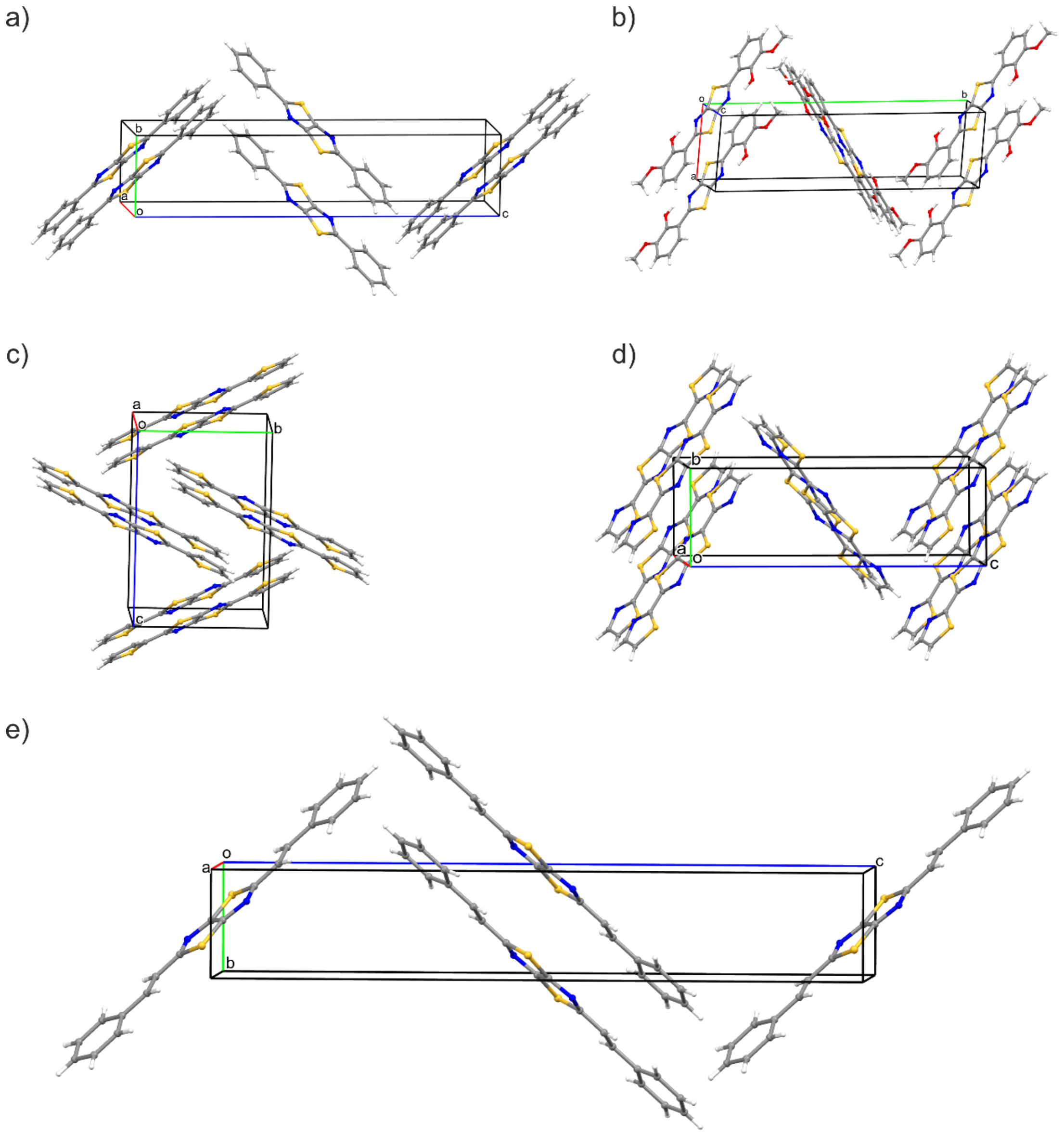

**Figure S2.** Herringbone-type crystal packing in a) **A1**; b) **A4**; c) **B2**; d) **B3**; e) **C**

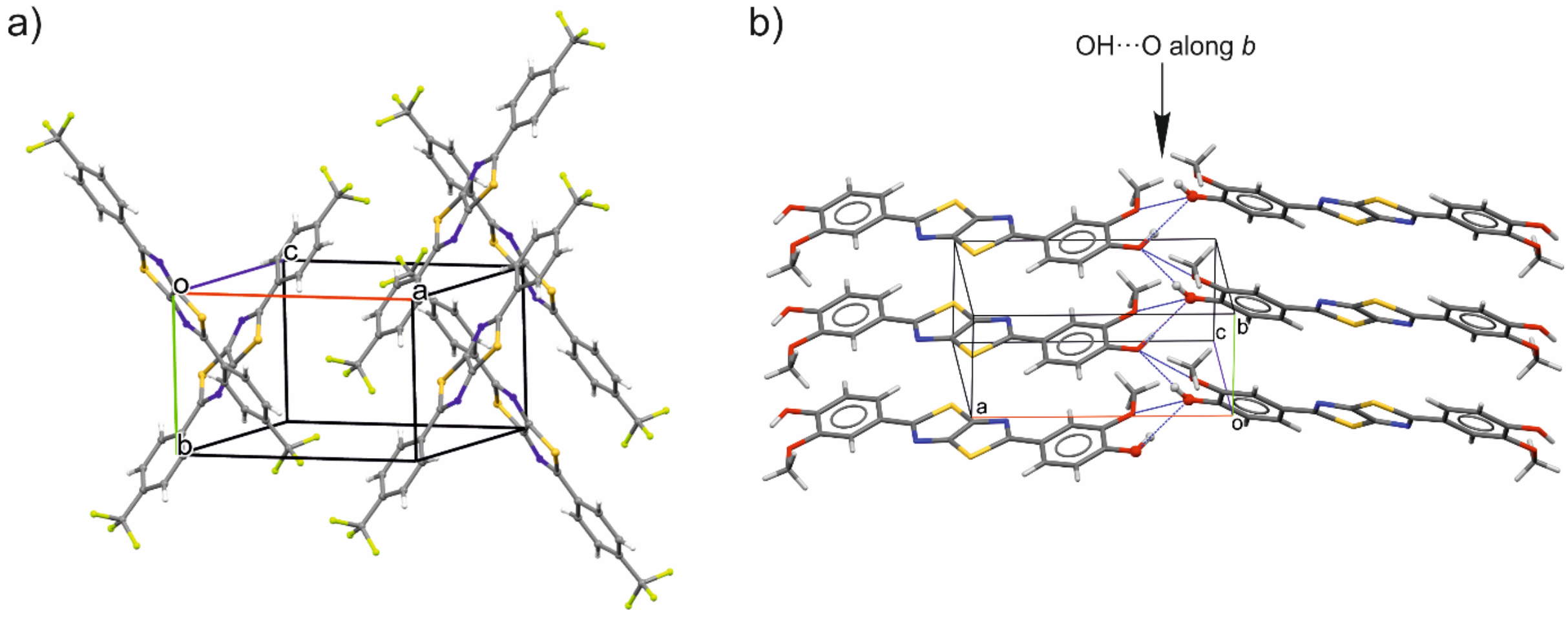

**Figure S3.** Grid-type crystal packing in a) **A2**; b) **A5**

***Table S4.*** *Theoretical crystal lattice parameters and optical bandgap values for optimized A-type TzTz structures*

| | **A1** | **A2** | **A3** | **A4** | **A5** |
|---|---|---|---|---|---|
| substituent | 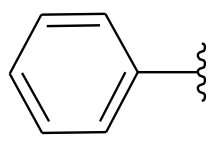 | 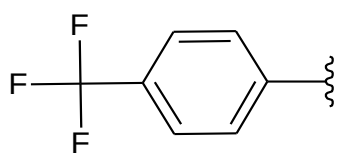 | 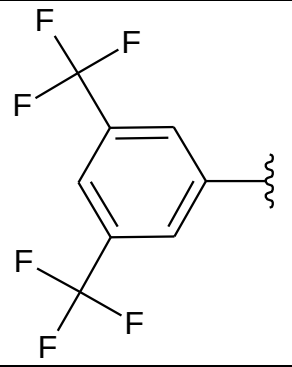 | 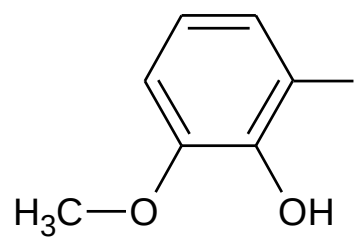 | 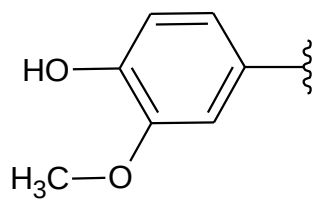 |
| Crystal system, space group | Monoclinic; $P2_1/c$ (14) | Monoclinic; $P2_1/c$ (14) | Orthorhombic *Ibam* (72) | Monoclinic; $P2_1/c$ (14) | Monoclinic; $P2_1/c$ (14) |
| Lattice parameters: a,b,c, [Å] α,β,γ [°] | *a* = 5.6757<br>*b* = 5.0599<br>*c* = 22.503<br>*α* = 90.000<br>*β* = 92.796<br>*γ* = 90.000 | *a* = 10.650<br>*b* = 6.600<br>*c* = 11.900<br>*α* = 90.000<br>*β* = 91.940<br>*γ* = 90.000 | *a* = 15.4405<br>*b* =17.4483<br>*c* = 7.4365<br>*α* = 90.000<br>*β* = 1<br>*γ* = 90.000 | *a* = 6.2324<br>*b* = 20.207<br>*c* = 6.673<br>*α* = 90.000<br>*β* = 106.84<br>*γ* = 90.000 | *a* = 10.2538<br>*b* = 3.9636<br>*c* = 20.4696<br>*α* = 90.000<br>*β* = 98.554<br>*γ* = 90.000 |
| V [Å$^3$] | 645.50 | 835.97 | 2003.47 | 804.35 | 822.67 |
| Bangap [eV] | 2.21 | 2.13 | 2.17 | 2.04 | 1.92 |
| Bandgap type | indirect | direct | direct | direct | indirect |

***Table S5.*** *Theoretical crystal lattice parameters and optical bandgap values for optimized B- and C-type TzTz structures*

| | **B1** | **B2** | **B3** | **B4** | **C** |
|---|---|---|---|---|---|
| substituent | 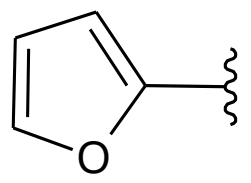 | 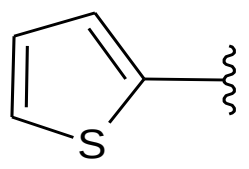 | 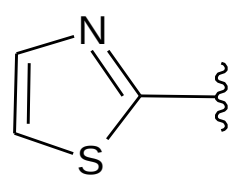 | 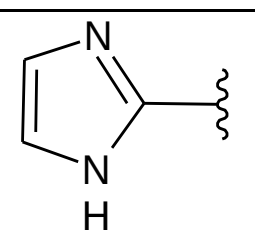 | 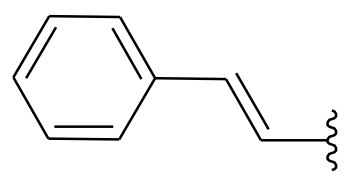 |
| Crystal system, space group | Monoclinic; $P2_1/c$ (14) | Monoclinic; $P2_1/c$ (14) | Monoclinic; $P2_1/c$ (14) | Monoclinic; $P2_1/c$ (14) | Monoclinic; $P2_1/c$ (14) |
| Lattice parameters: a,b,c, [Å] α,β,γ [°] | *a* = 5.6092<br>*b* = 5.8044<br>*c* = 16.727<br>*α* = 90.000<br>*β* = 95.631<br>*γ* = 90.000 | *a* = 6.004<br>*b* = 8.358<br>*c* = 12.270<br>*α* = 90.000<br>*β* = 92.720<br>*γ* = 90.000 | *a* = 5.8139<br>*b* = 5.8101<br>*c* = 17.366<br>*α* = 90.000<br>*β* = 101.60<br>*γ* = 90.000 | *a* = 3.7442<br>*b* = 14.7018<br>*c* = 9.9993<br>*α* = 90.000<br>*β* = 92.063<br>*γ* = 90.000 | *a* = 5.8842<br>*b* = 4.7578<br>*c* = 28.4784<br>*α* = 90.000<br>*β* = 92.464<br>*γ* = 90.000 |
| V [Å$^3$] | 541.97 | 615.03 | 574.62 | 550 | 796.54 |
| Bangap [eV] | 1.77 | 1.90 | 1.85 | 1.98 | 1.83 |
| Bandgap type | indirect | indirect | direct | indirect | indirect |

***Table S6.*** *DFT predicted absorption maxima (in vacuo, B3LYP/TZVP) and band gaps (solid state, periodic GGA-PWA) of studied thiazolothiazoles.*

| **compound** | **$\lambda_{max}$ [nm]** | **$E_g$ [eV]** | **band gap type** |
|---|---|---|---|
| **A1** | 374 | 2.21 | indirect |
| **A2** | 379 | 2.13 | direct |
| **A3** | 374 | 2.17 | indirect |
| **A4** | 392 | 2.04 | direct |
| **A5** | 398 | 1.92 | indirect |
| **B1** | 390 | 1.92 | indirect |
| **B2** | 406 | 1.92 | indirect |
| **B3** | 400 | 1.85 | direct |
| **B4** | 377 | 1.98 | indirect |
| **C** | 448 | 1.83 | indirect |

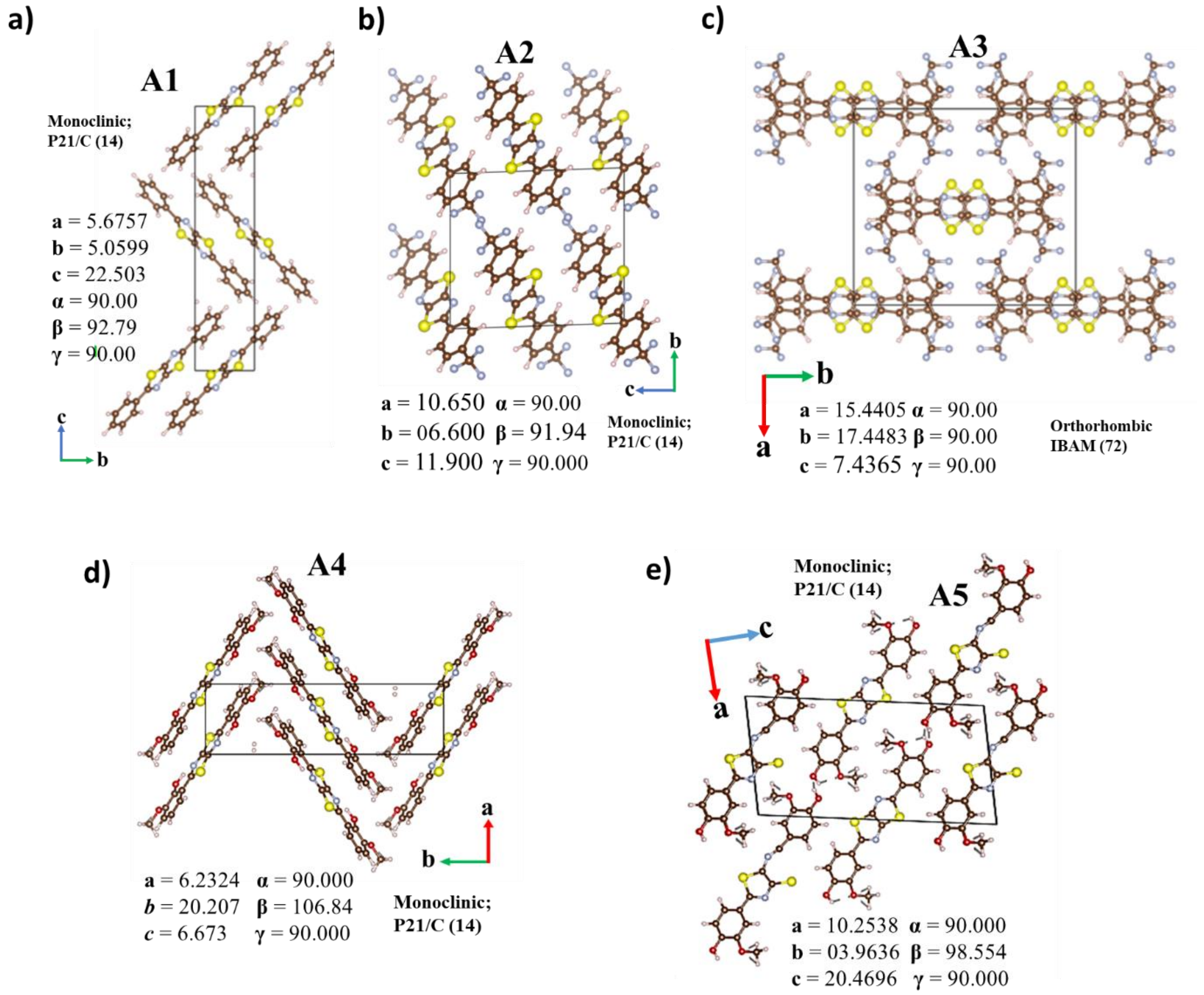

***Figure S4.*** *Optimized geometries for the a)-e) A-type TzTz derivatives.*

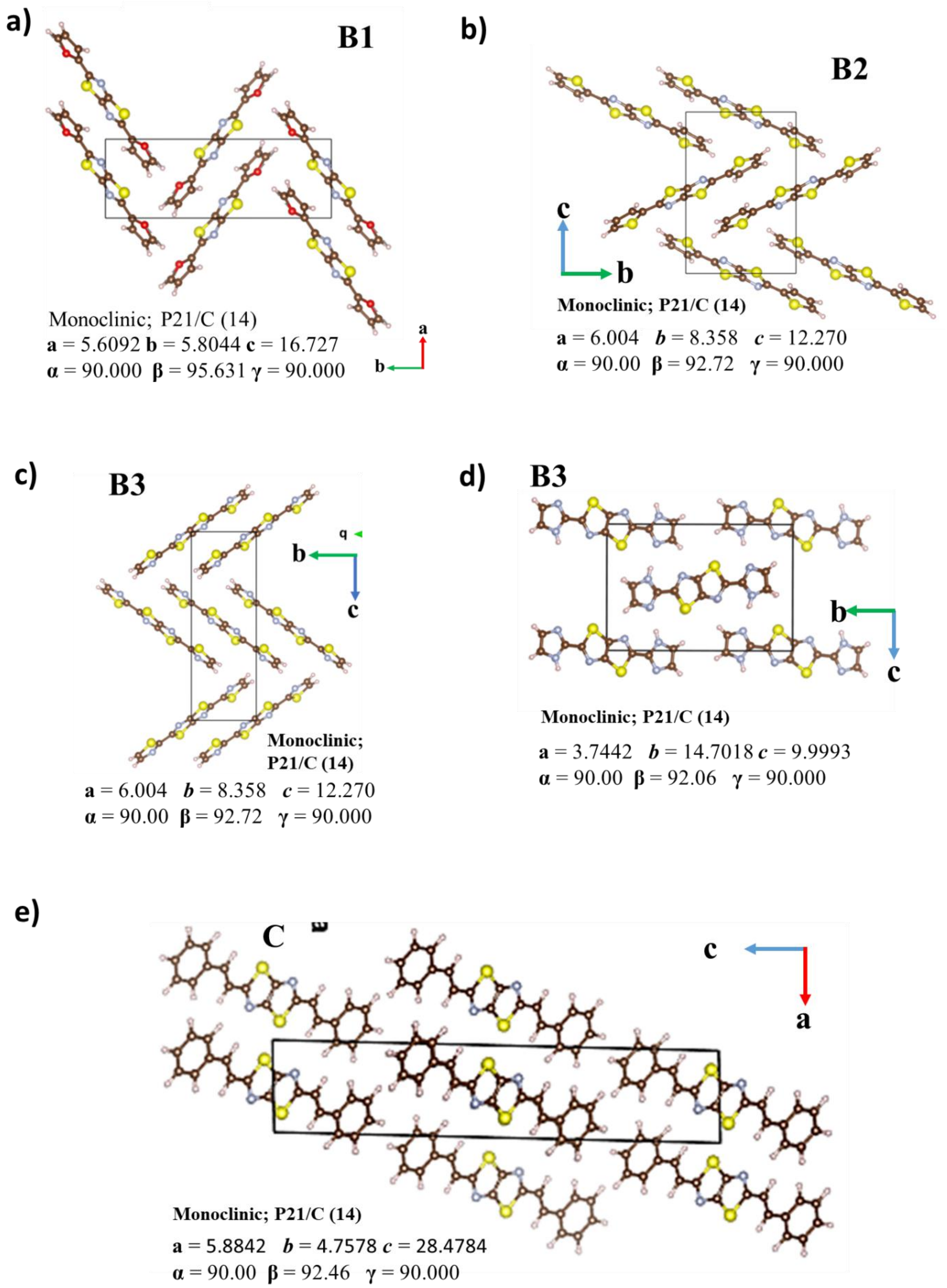

***Figure S5.*** *Optimized geometries for the a)-d) B-type TzTz derivatives and e) C-type TzTz derivatives.*

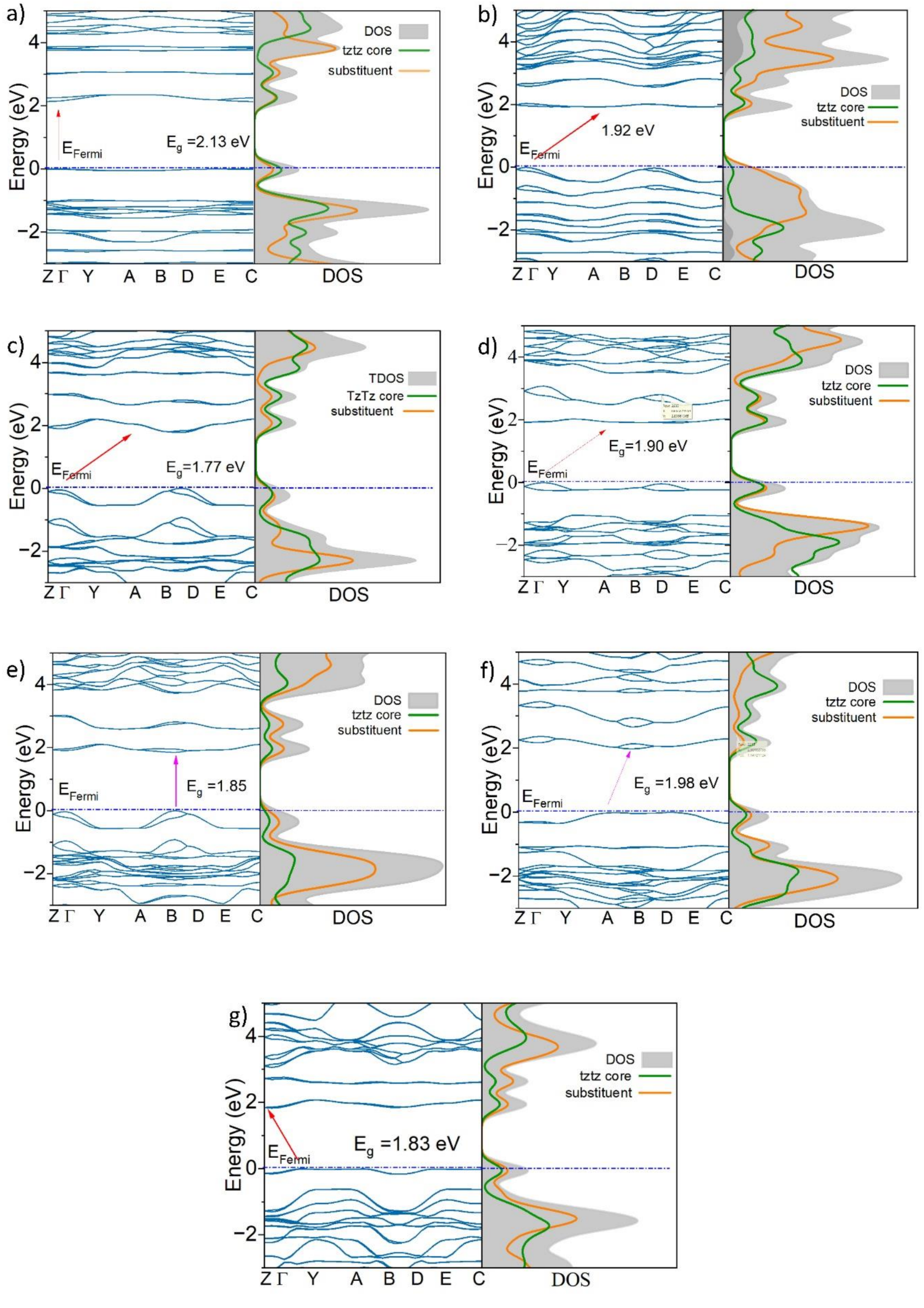

***Figure S6.*** *Theoretical band structure and partial density of states (pDOS), division between TzTz core and substituent for: (a)* ***A2****, (b)* ***A5****, (c)* ***B1****, (d)* ***B2****, (e)* ***B3****, (f)* ***B4*** *and (g)* ***C****.*

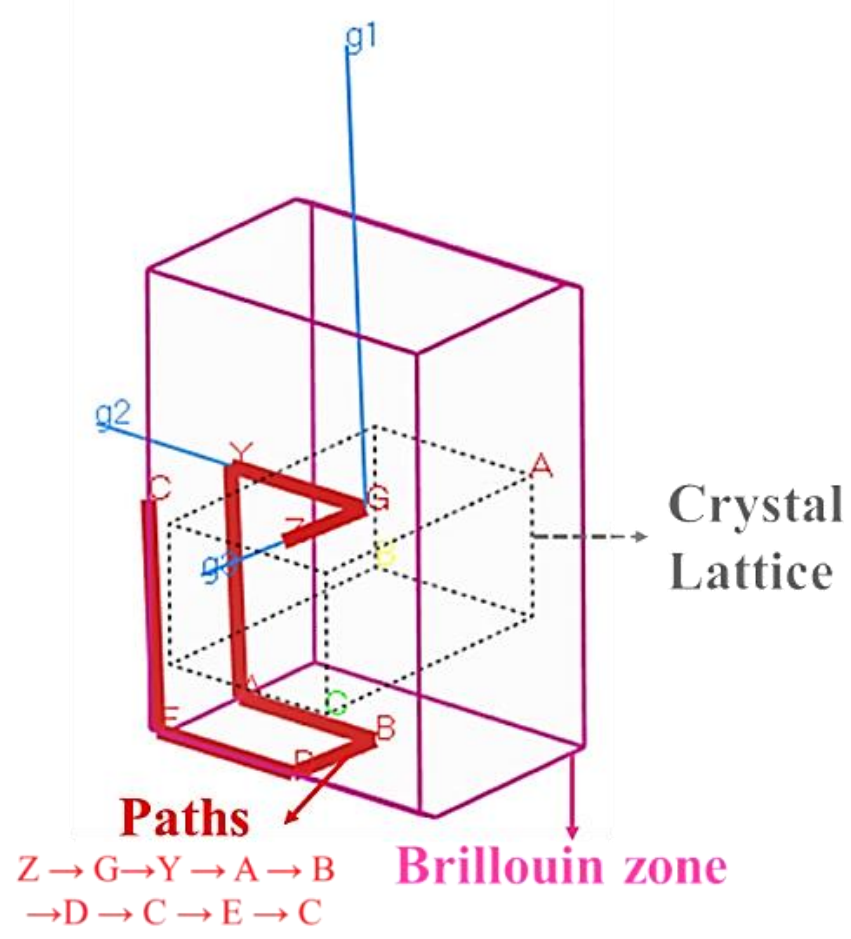

***Figure S7.*** *Outline of the paths through the Brillouin zone for the crystal structure calculation.*

***Table S7.*** *NMR spectra of studied thiazolothiazoles. Abbreviations:* ***s*** *- deuterated solvent (residual signal),* ***w*** *– water,* ******* *- impurities,* ***e*** *- ethanol*

NMR Spectra for **A1**

$^{1}$H spectrum ($CDCl_3$-*d*$_1$) of **A1:**

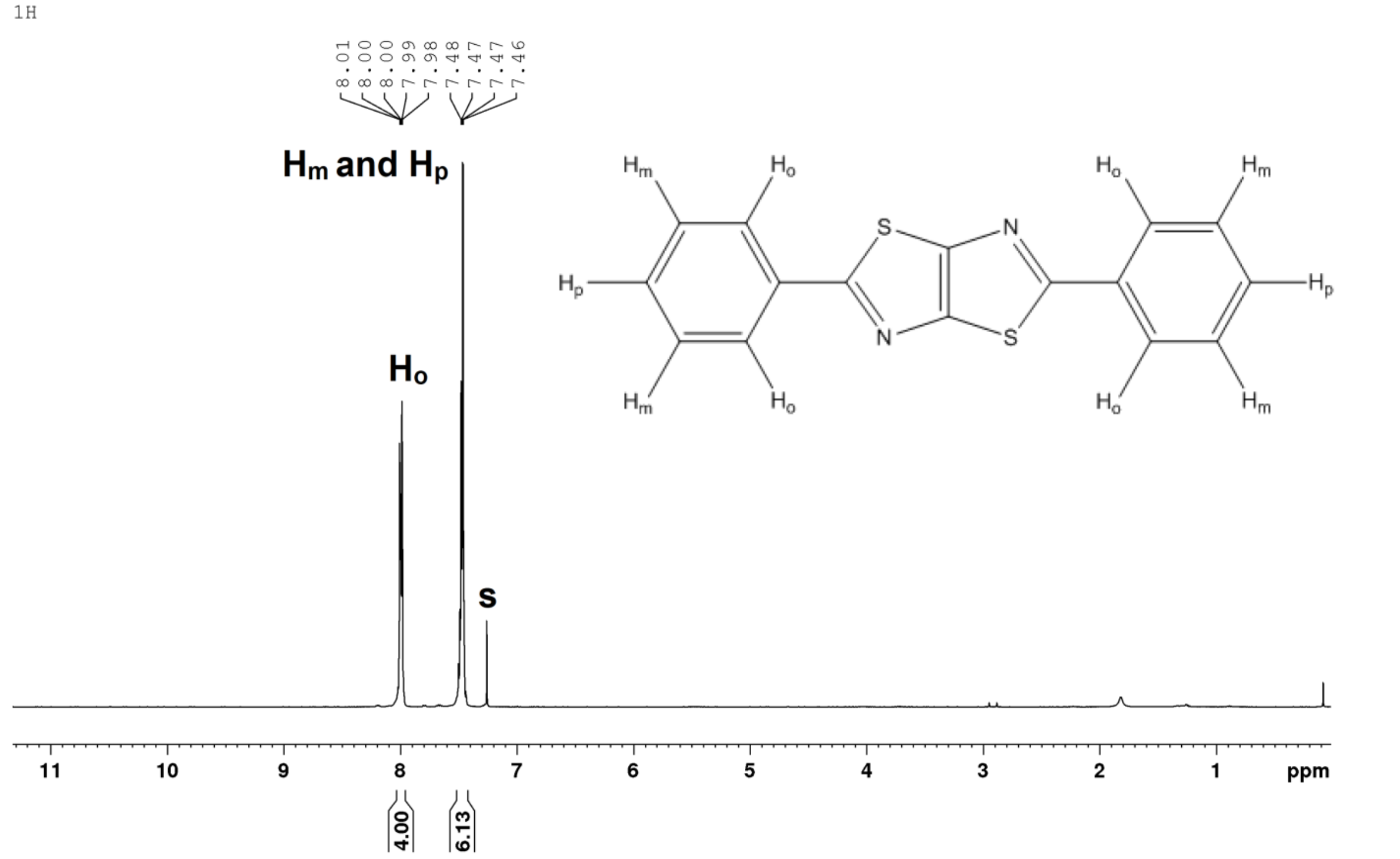

$^{13}$C{$^{1}$H} spectrum ($CDCl_3$-*d*$_1$) of **A1:**

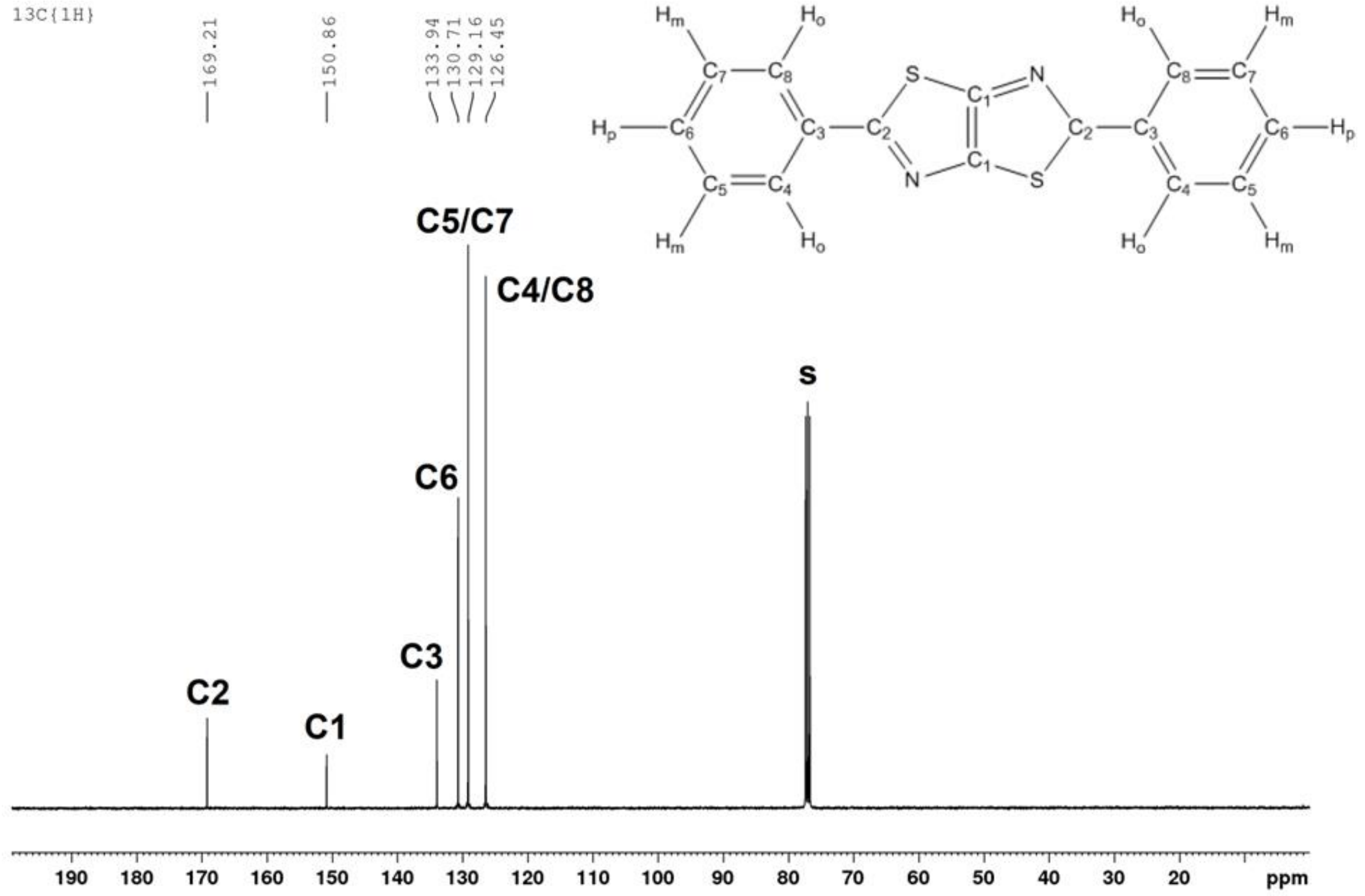

NMR Spectra for **A2**

$^{1}H$ spectrum ($CDCl_3$-$d_1$) of **A2:**

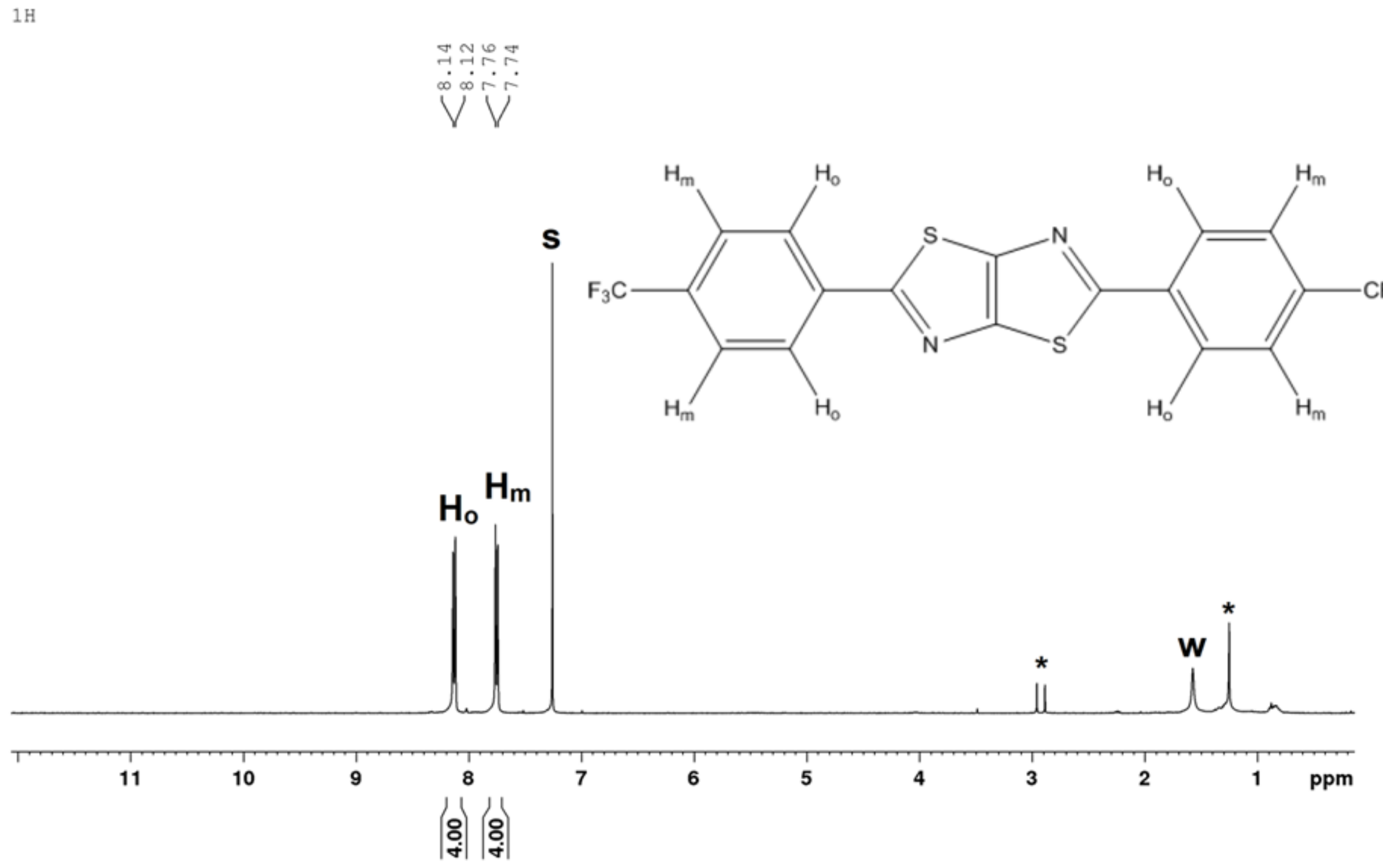

$^{13}C\{^{1}H\}$ spectrum ($CDCl_3$-$d_1$) of **A2**

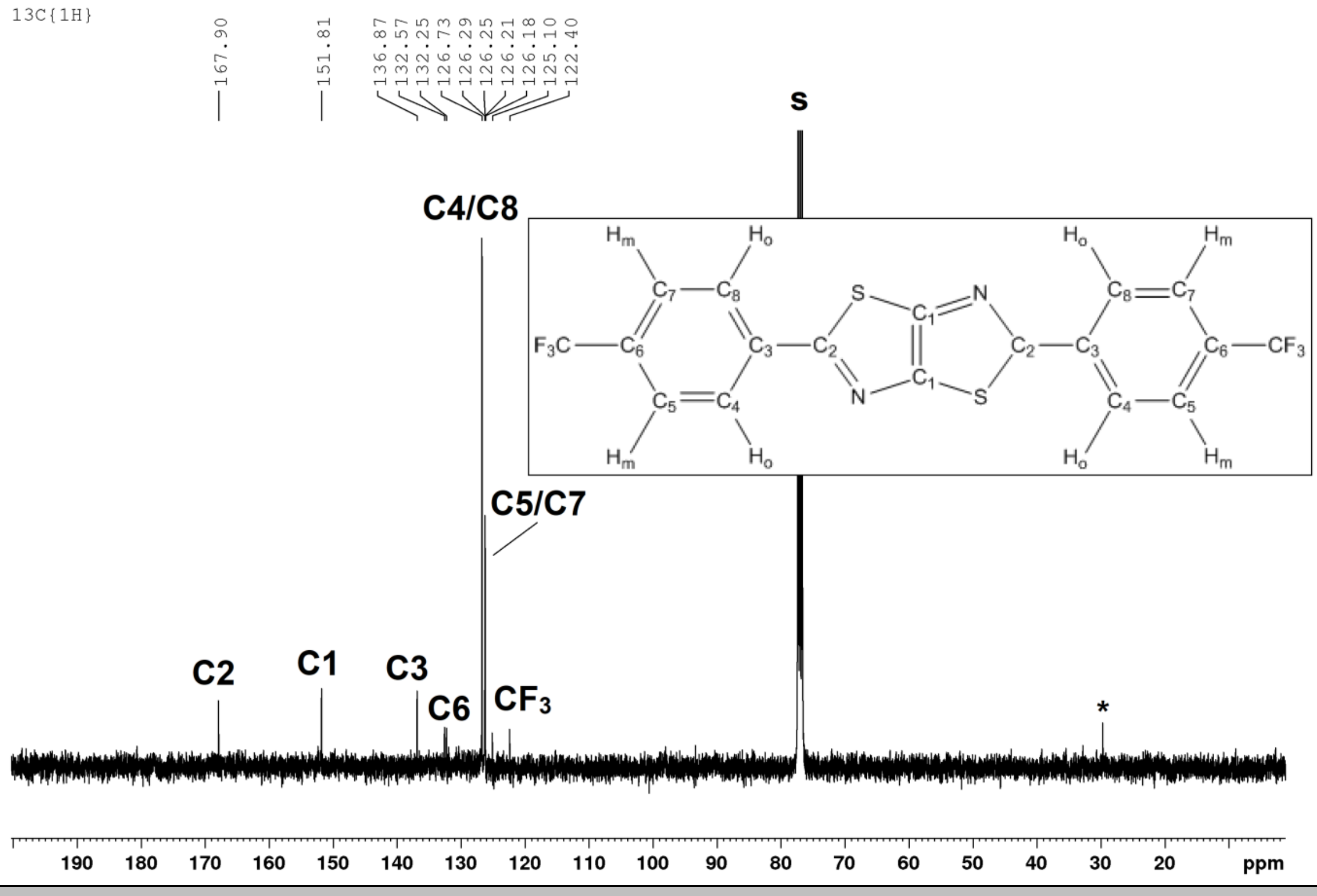

$^{19}$F spectrum ($CDCl_3$-$d_1$) of **A2:**

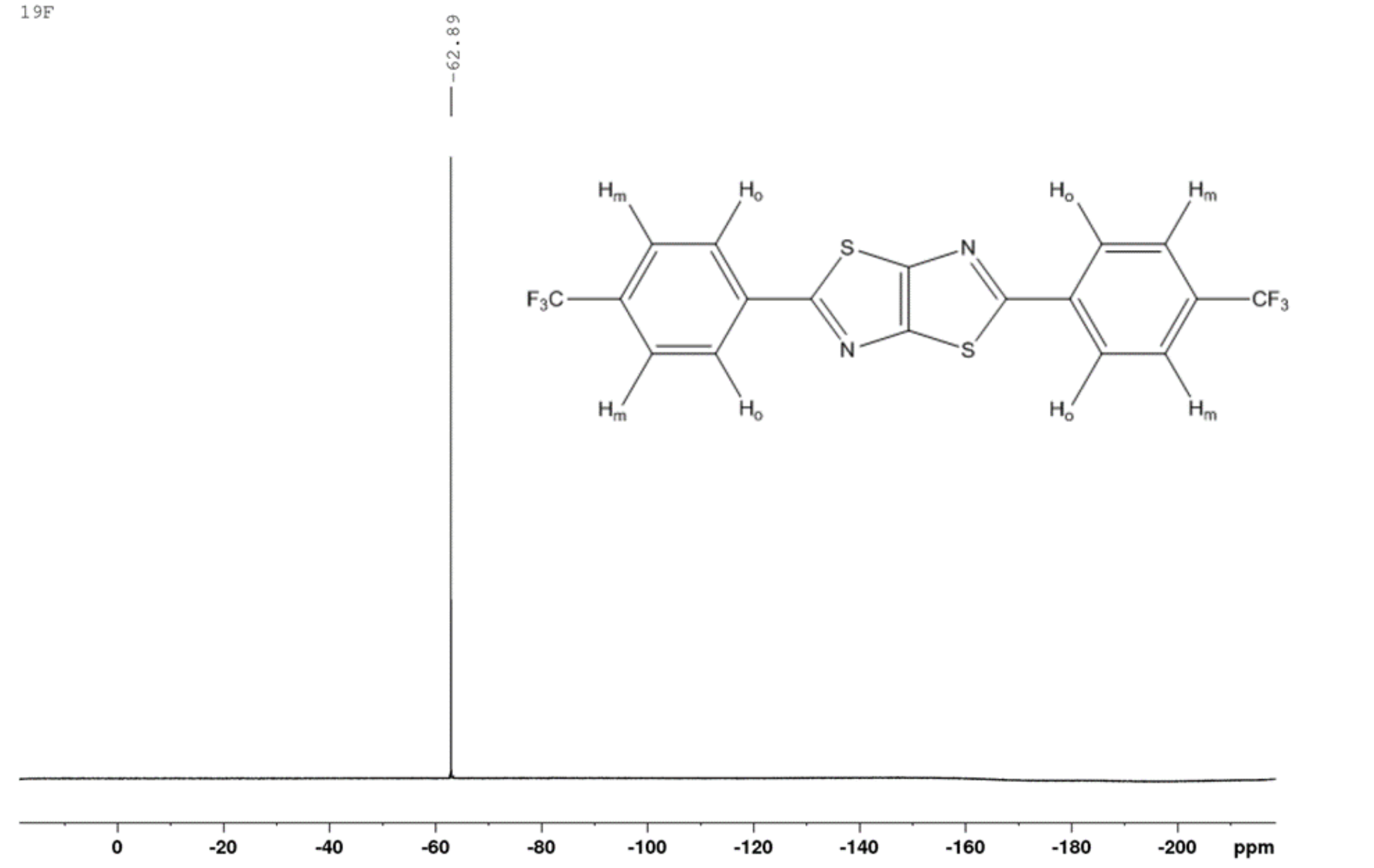

NMR Spectra for **A3**

$^{1}$H spectrum ($CDCl_3$-$d_1$) of **A3:**

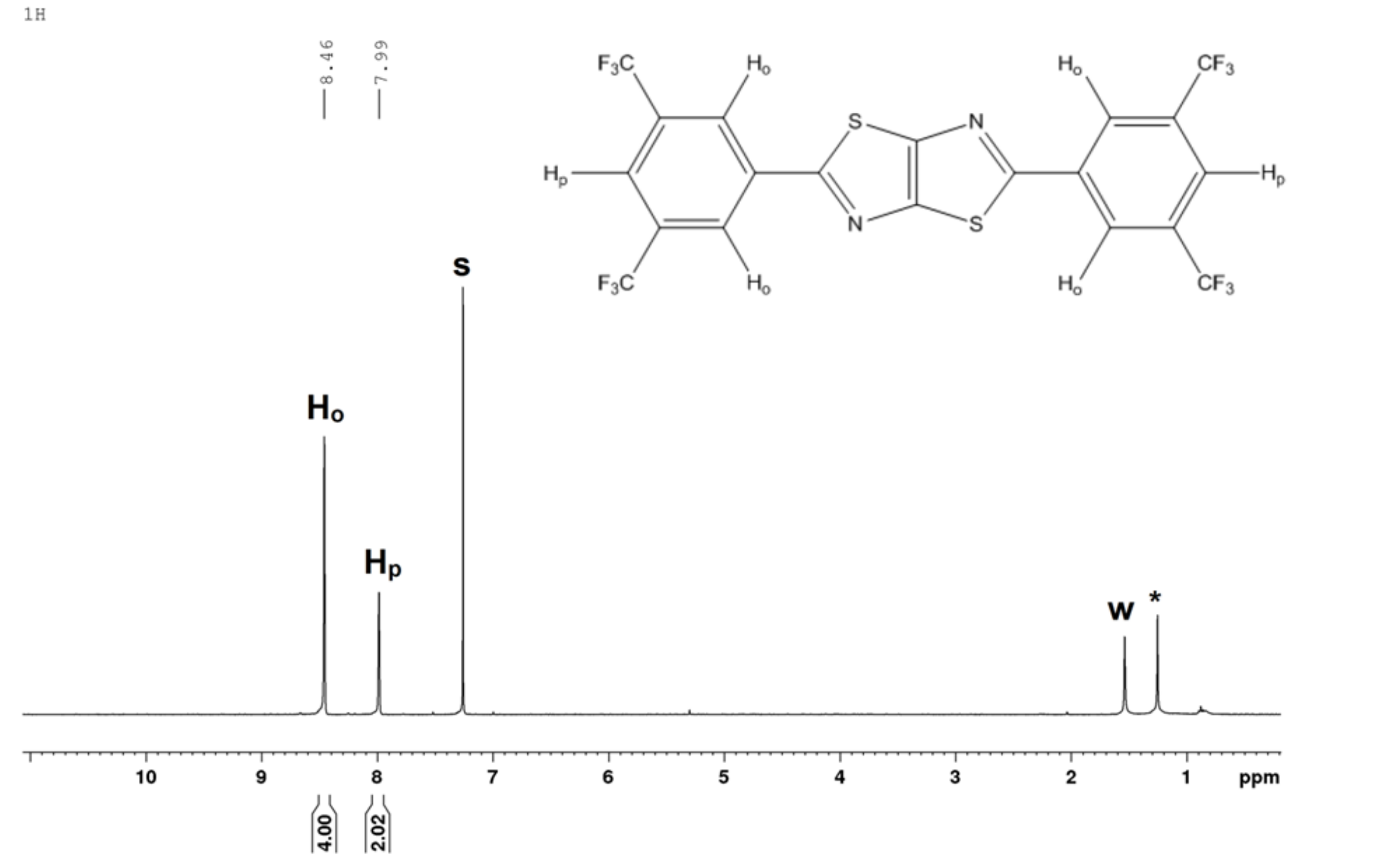

$^{13}C\{^{1}H\}$ spectrum ($CDCl_3$-$d_1$) of **A3:**

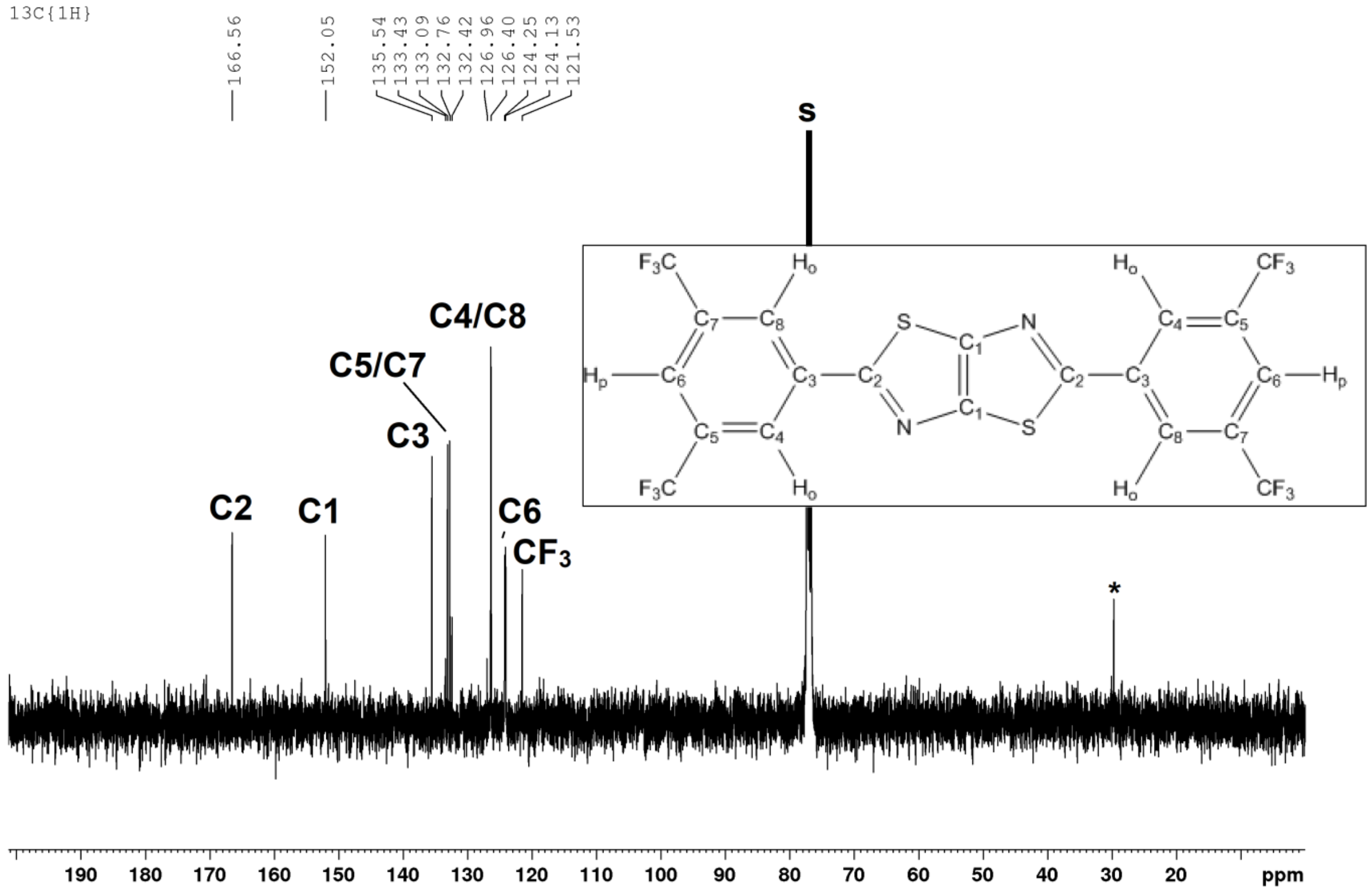

$^{19}F$ spectrum ($CDCl_3$-$d_1$) of **A3**:

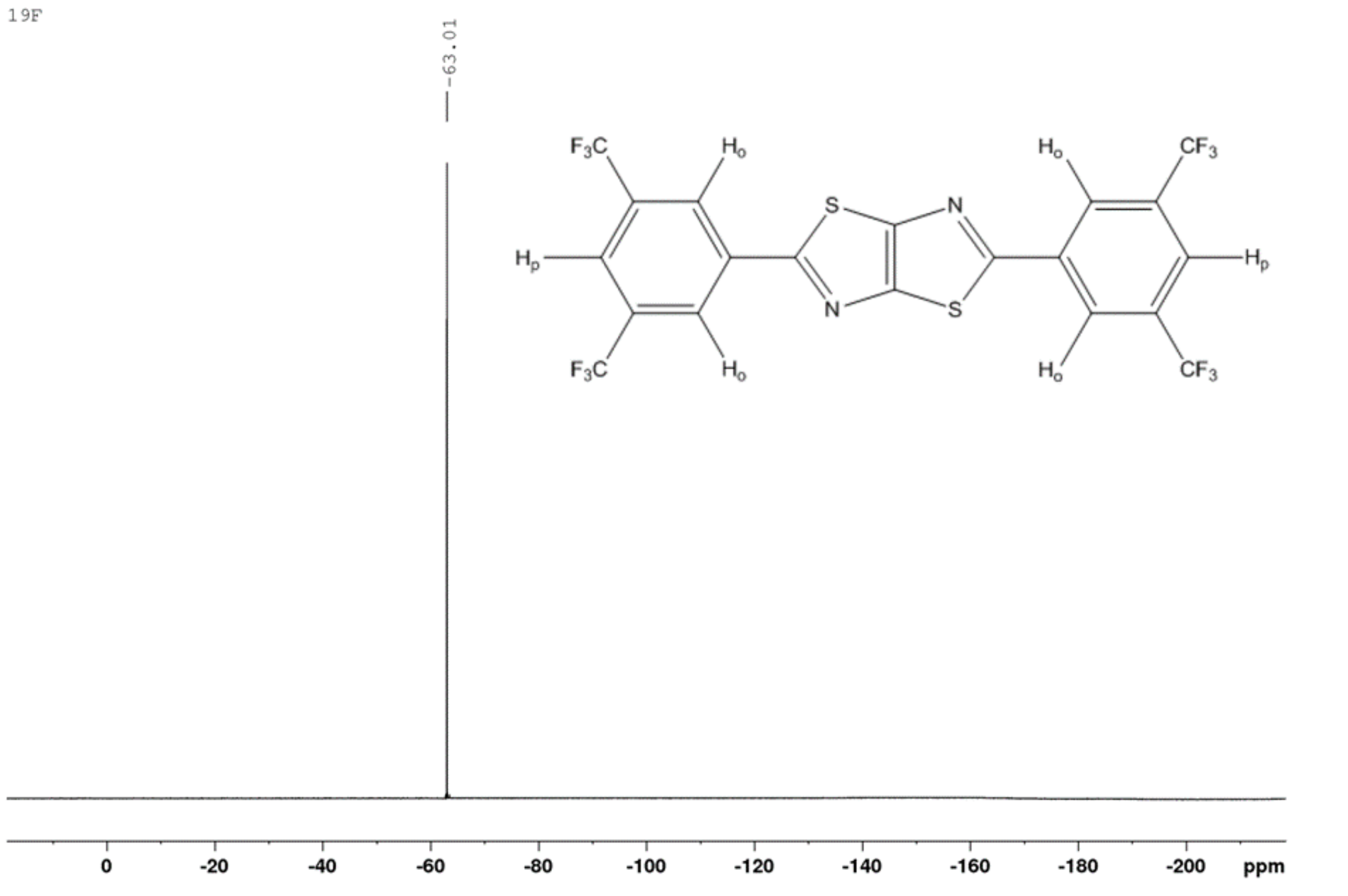

NMR Spectra for **A4**

$^{1}H$ spectrum (DMSO-$d_6$) of **A4:**

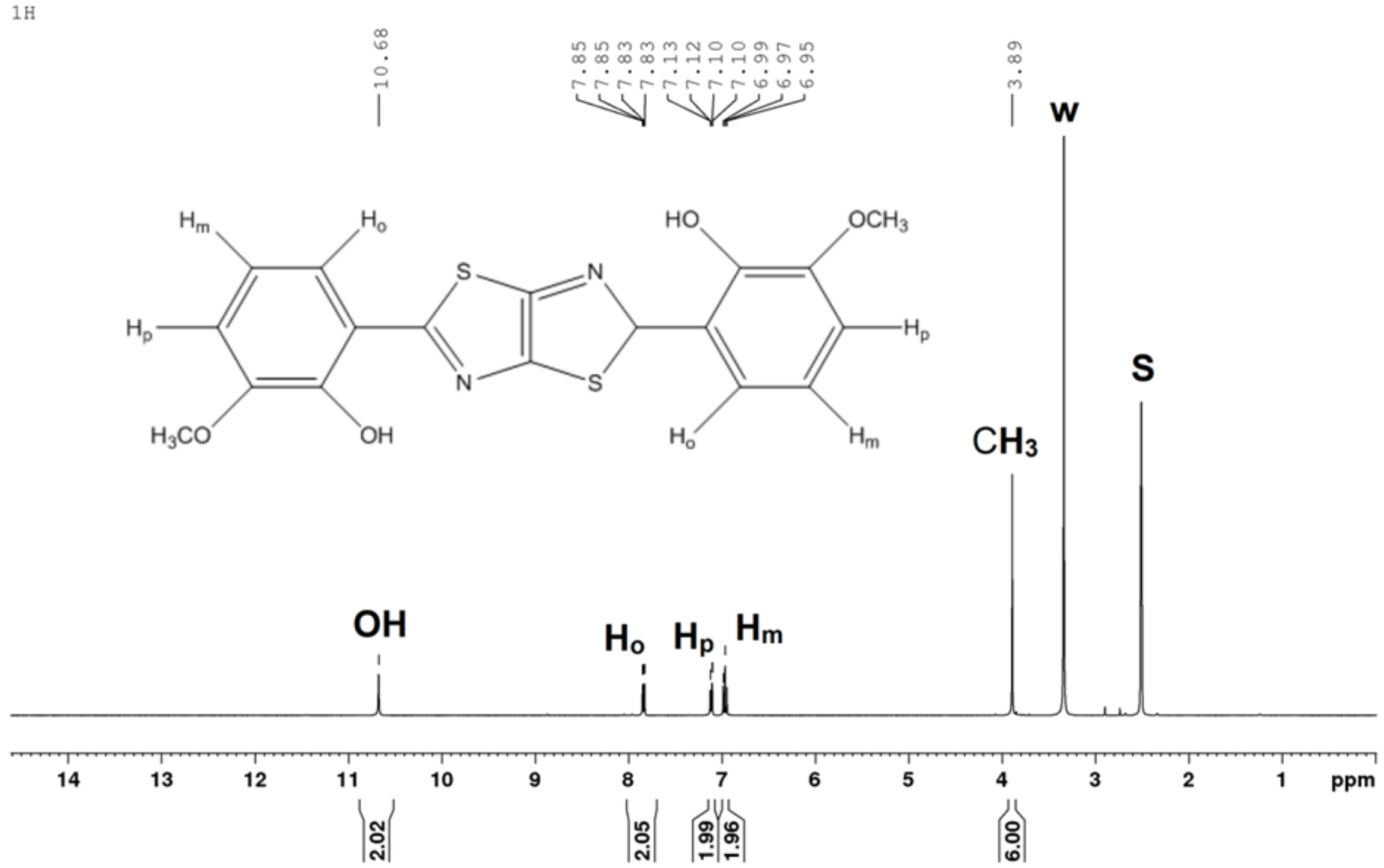

$^{13}C\{^{1}H\}$ spectrum (DMSO-$d_6$) of **A4:**

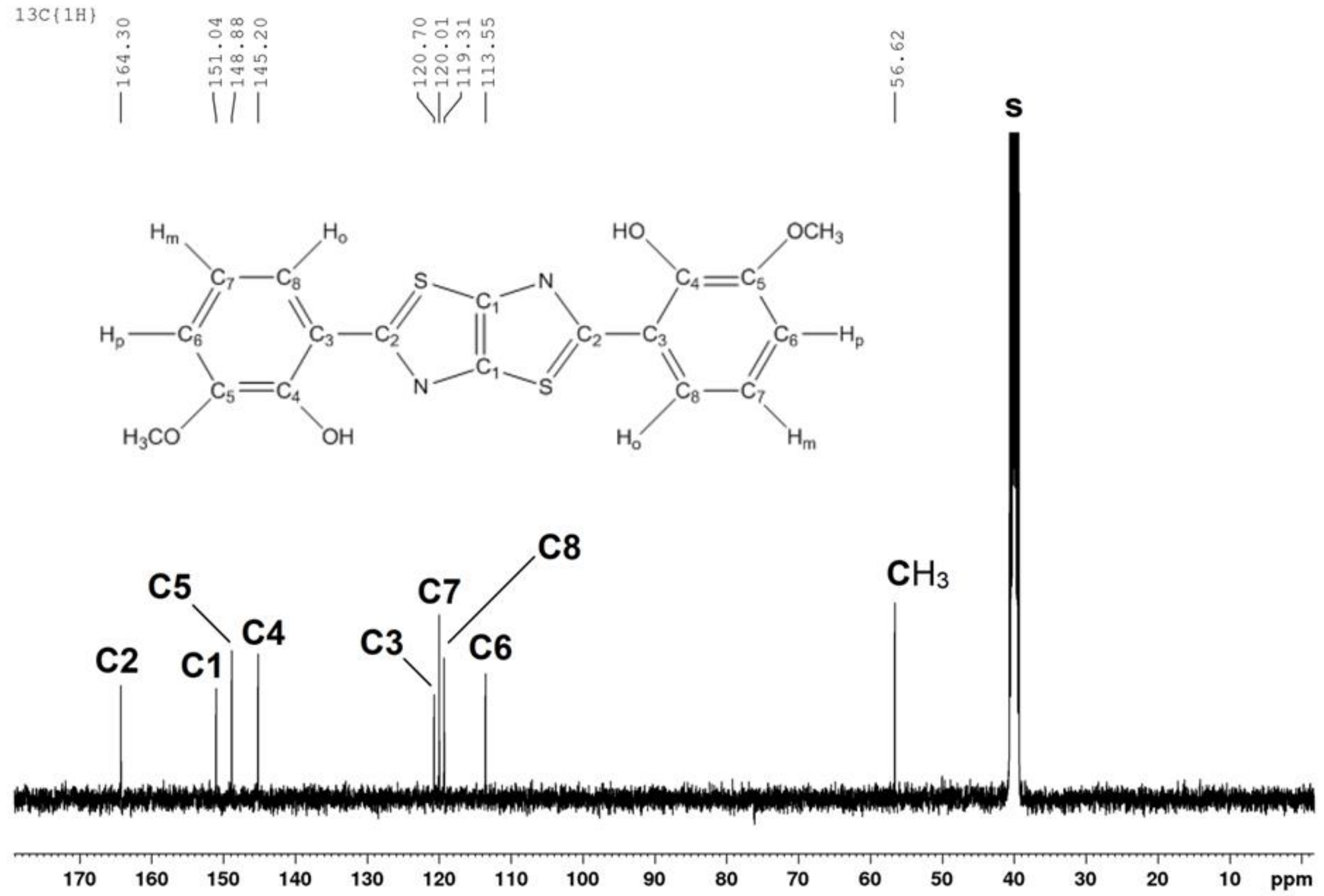

NMR Spectra for **A5**

$^{1}H$ spectrum (DMSO-$d_6$) of **A5:**

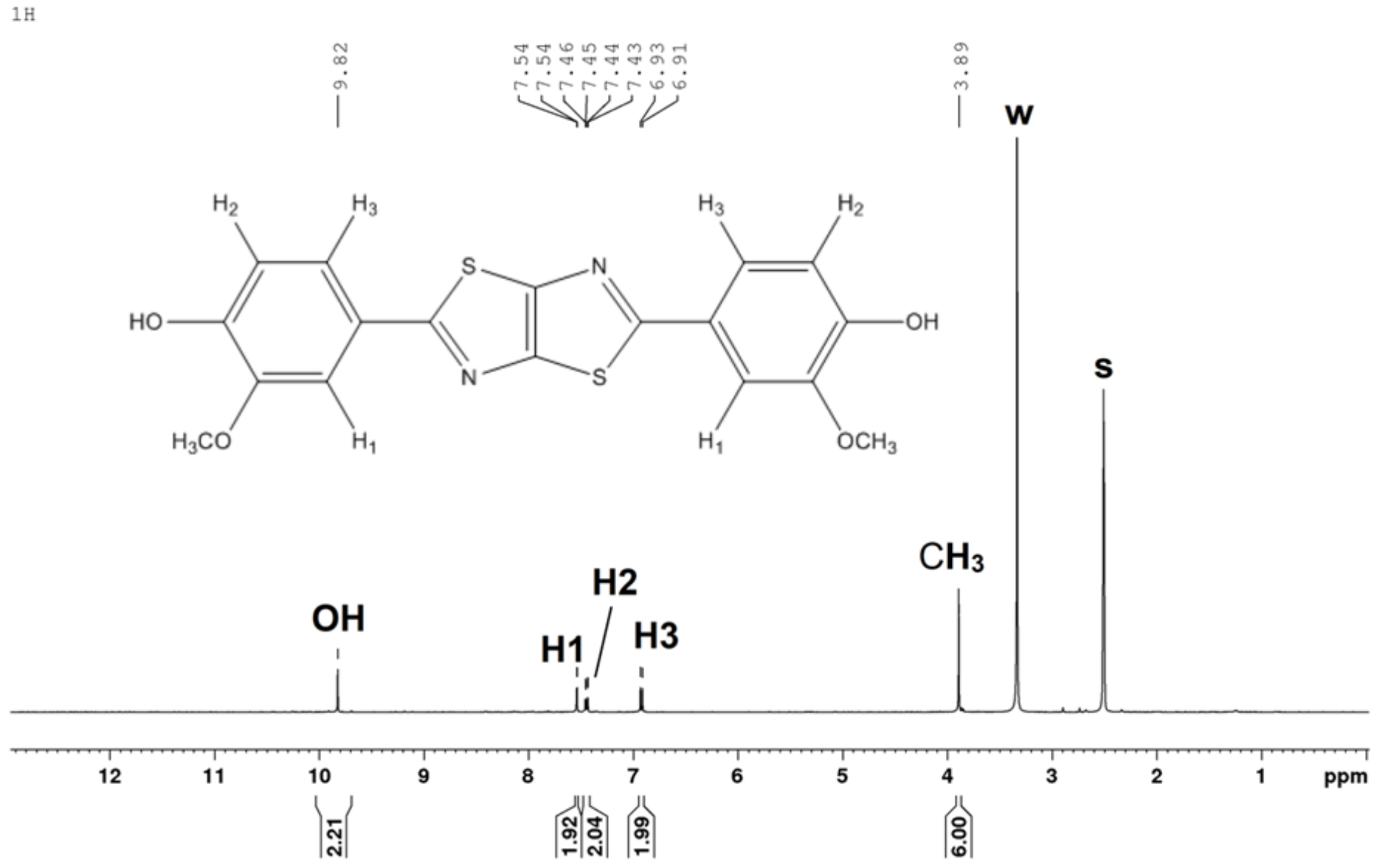

$^{13}C\{^{1}H\}$ spectrum (DMSO-$d_6$) of **A5:**

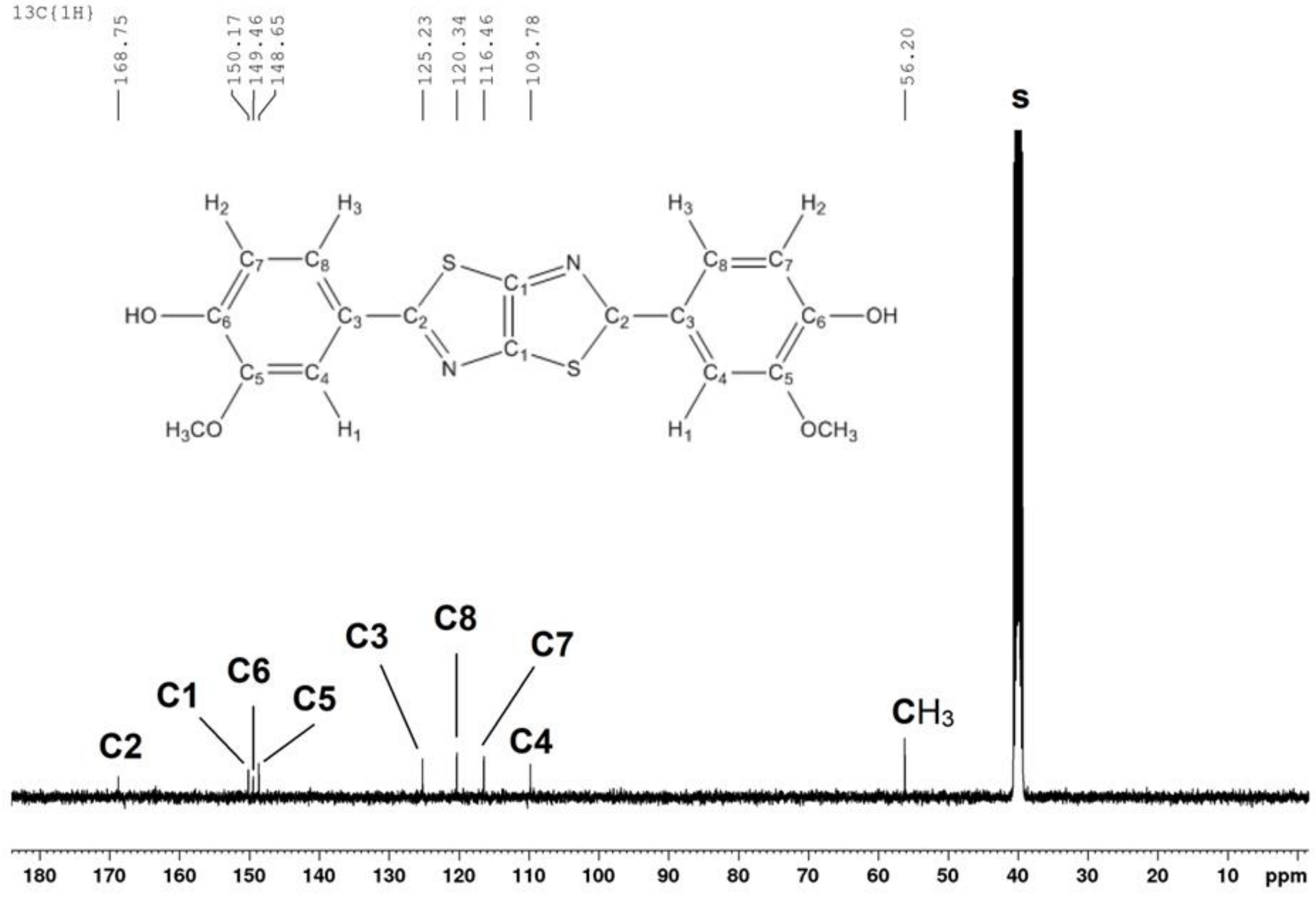

NMR Spectra for **B1**

$^{1}H$ spectrum (DMSO-$d_6$) of **B1:**

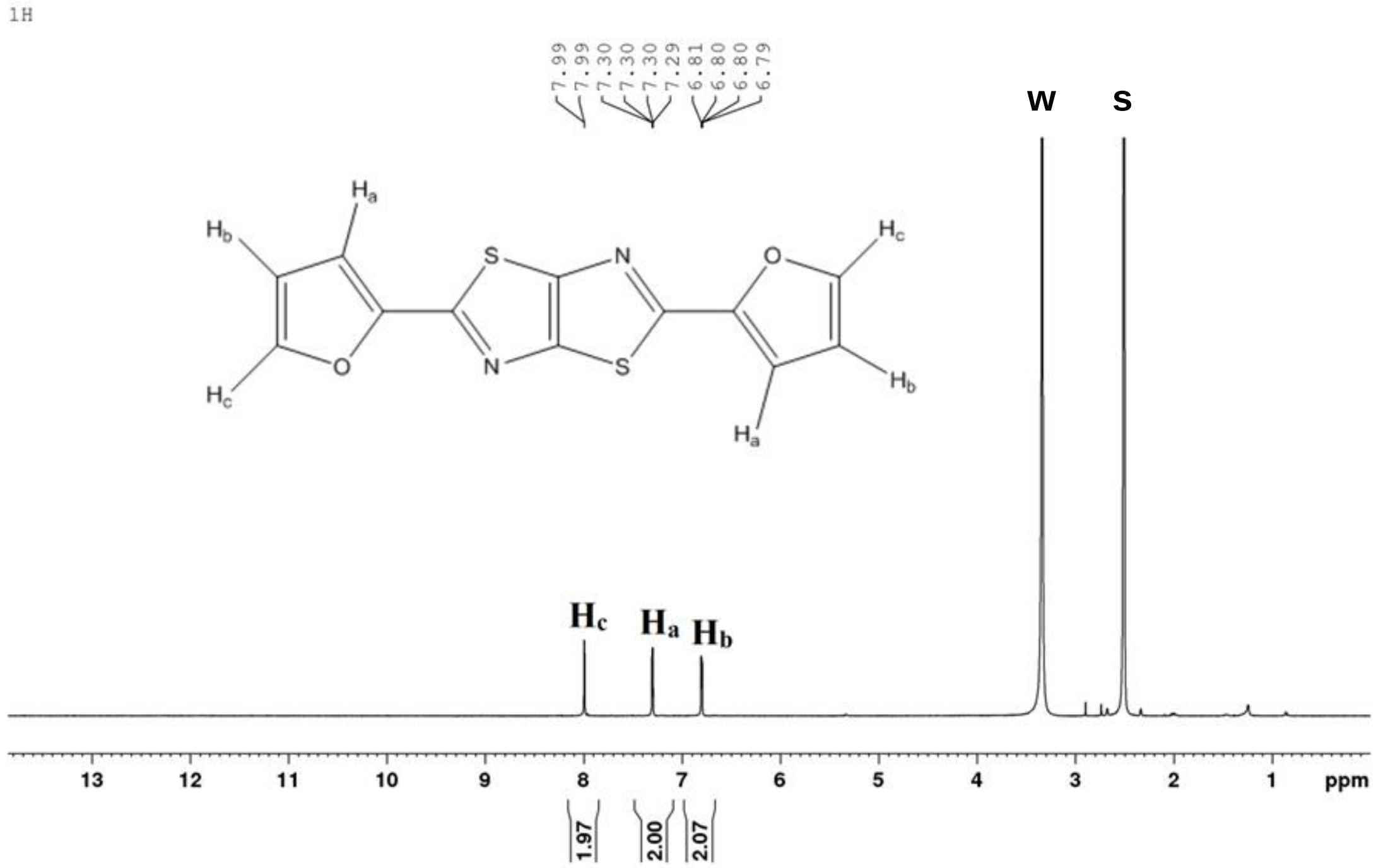

$^{13}C\{^{1}H\}$ spectrum (DMSO-$d_6$) of **B1:**

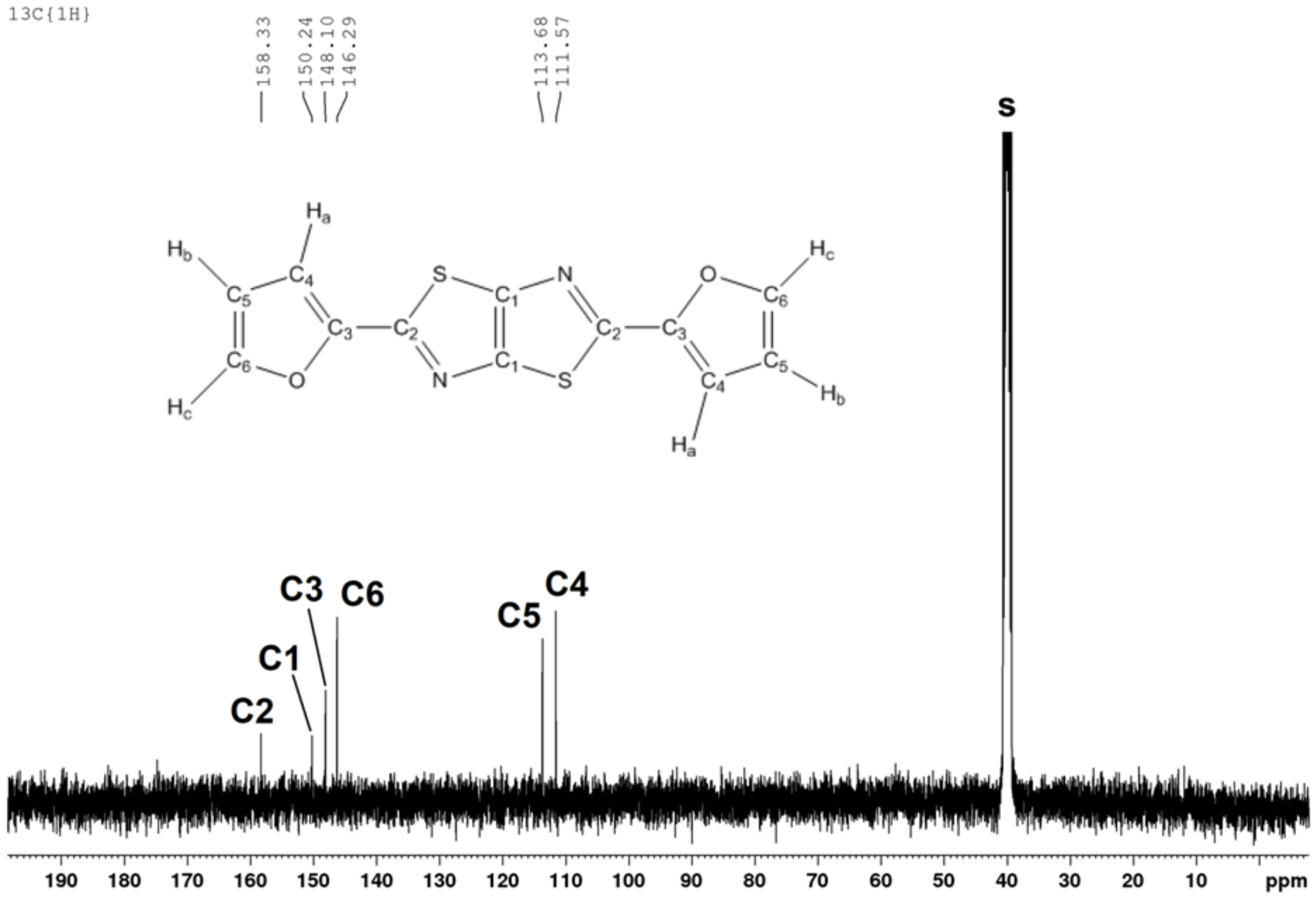

NMR Spectra for **B2**

$^1$H spectrum ($CDCl_3$-$d_1$) of **B2:**

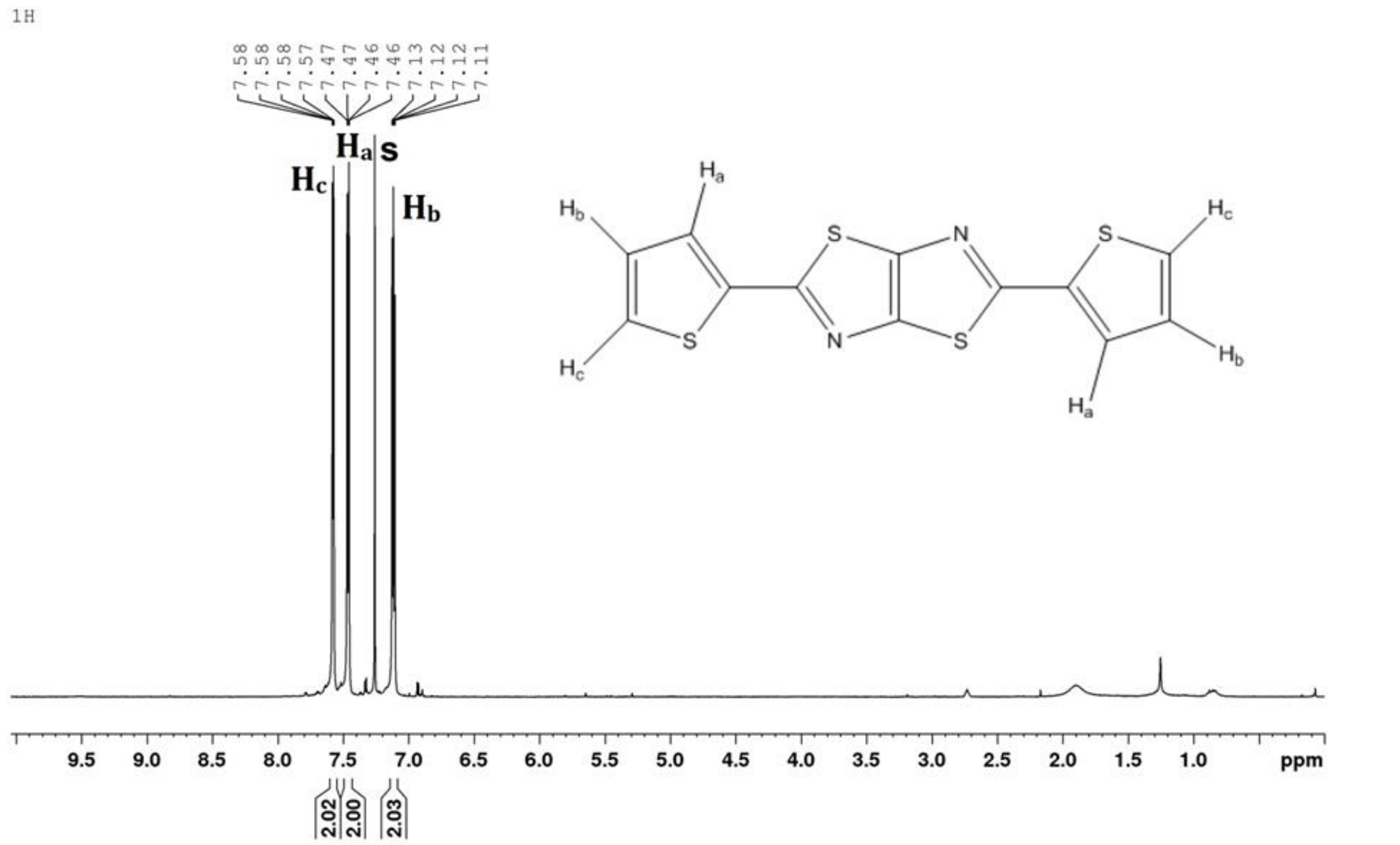

$^{13}$C{$^1$H} spectrum ($CDCl_3$-$d_1$) of **B2:**

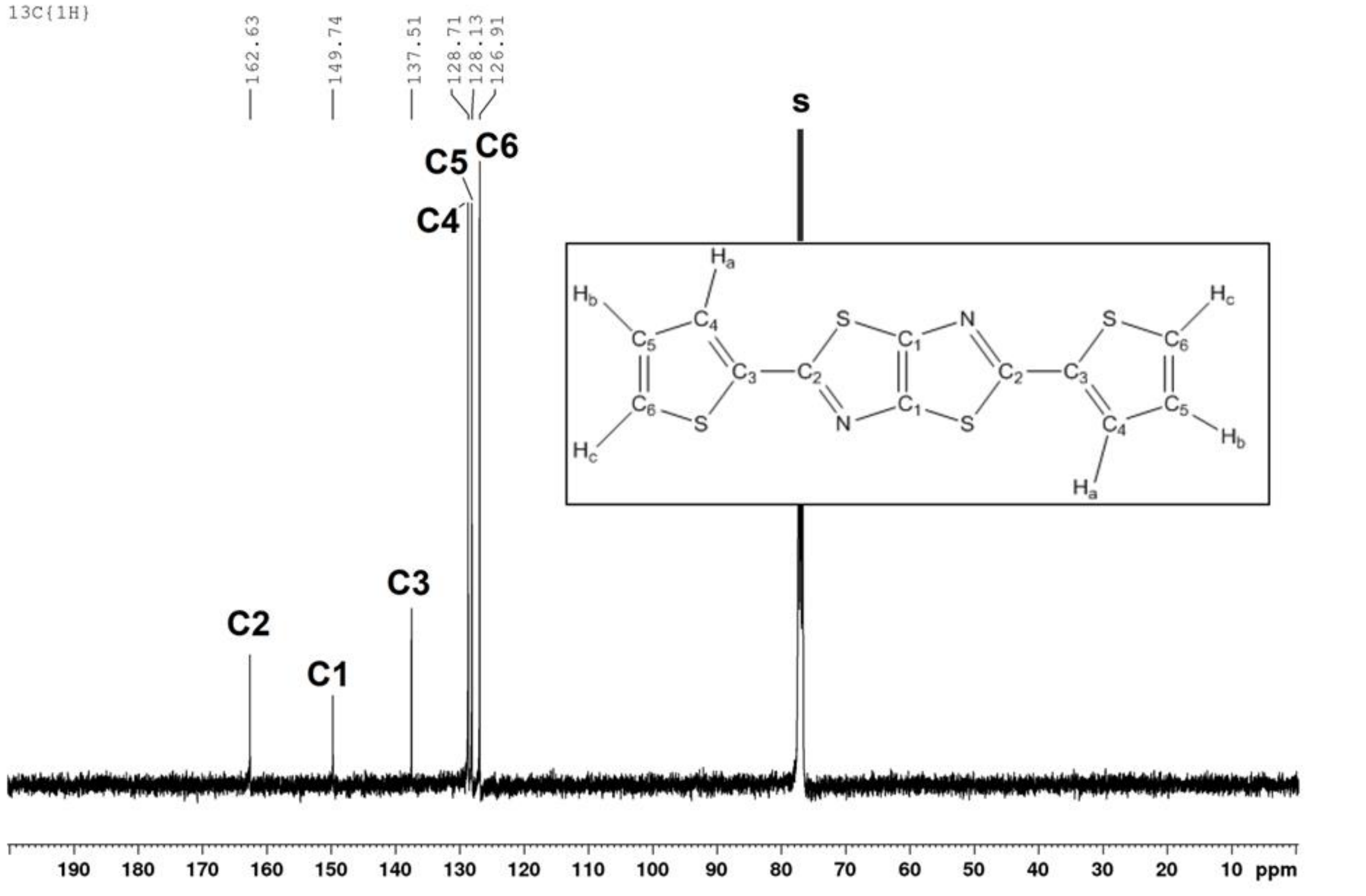

NMR Spectra for **B3**

$^{1}H$ spectrum ($CDCl_3$-$d_1$) of **B3:**

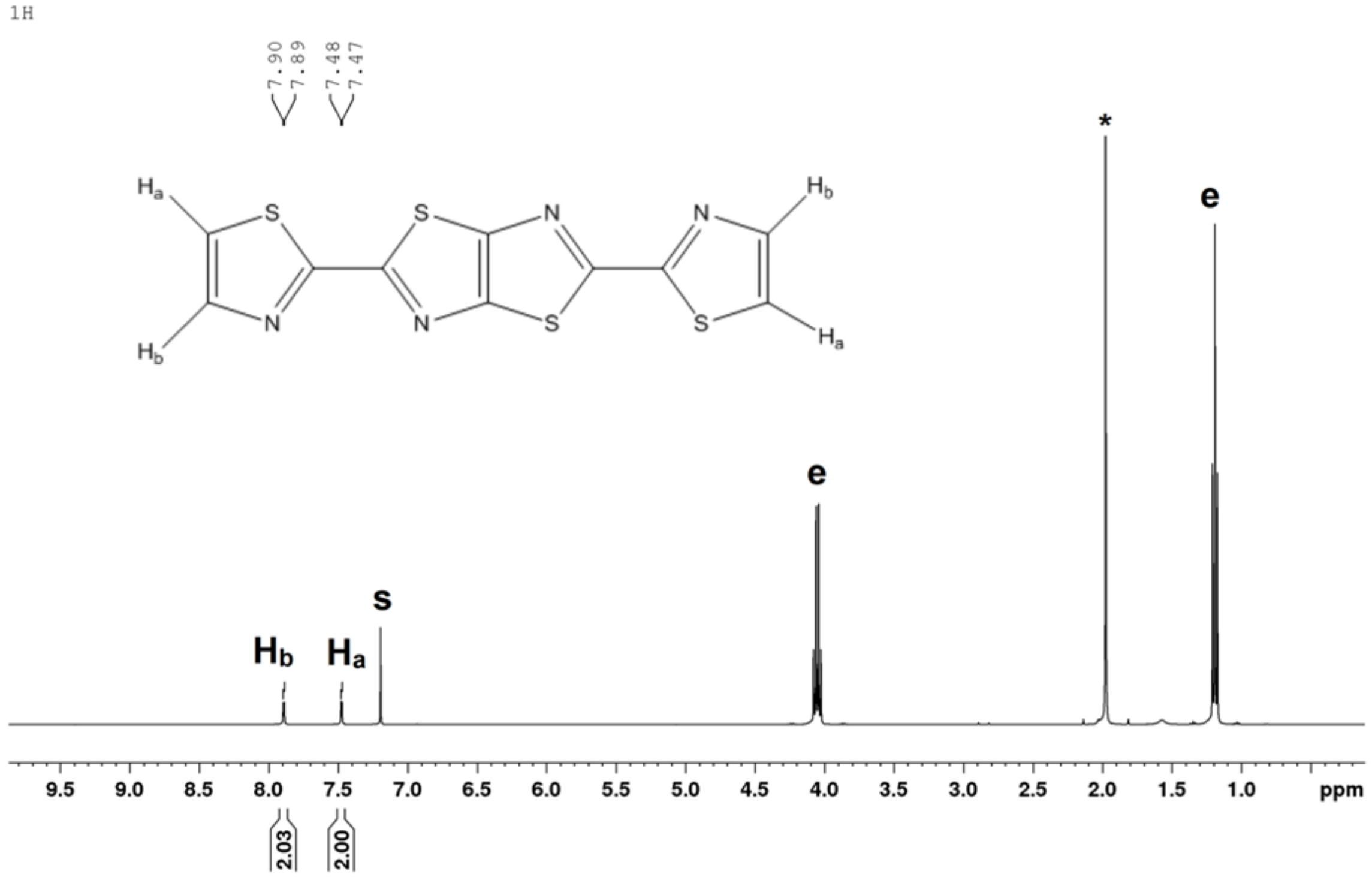

$^{13}C\{^{1}H\}$ spectrum ($CDCl_3$-$d_1$) of **B3:**

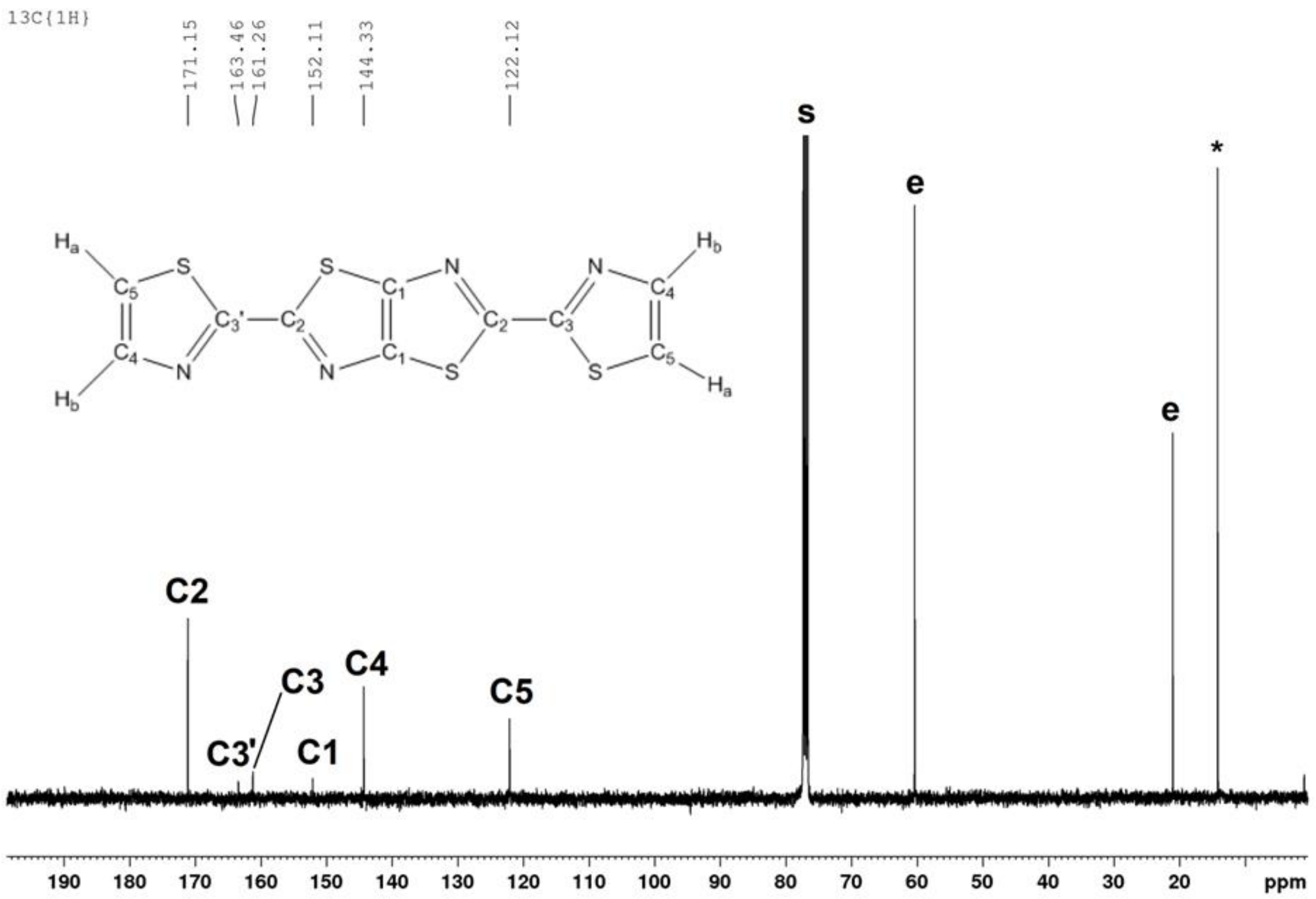

NMR Spectra for **B4**

$^{1}H$ spectrum (DMSO-$d_6$) of **B4:**

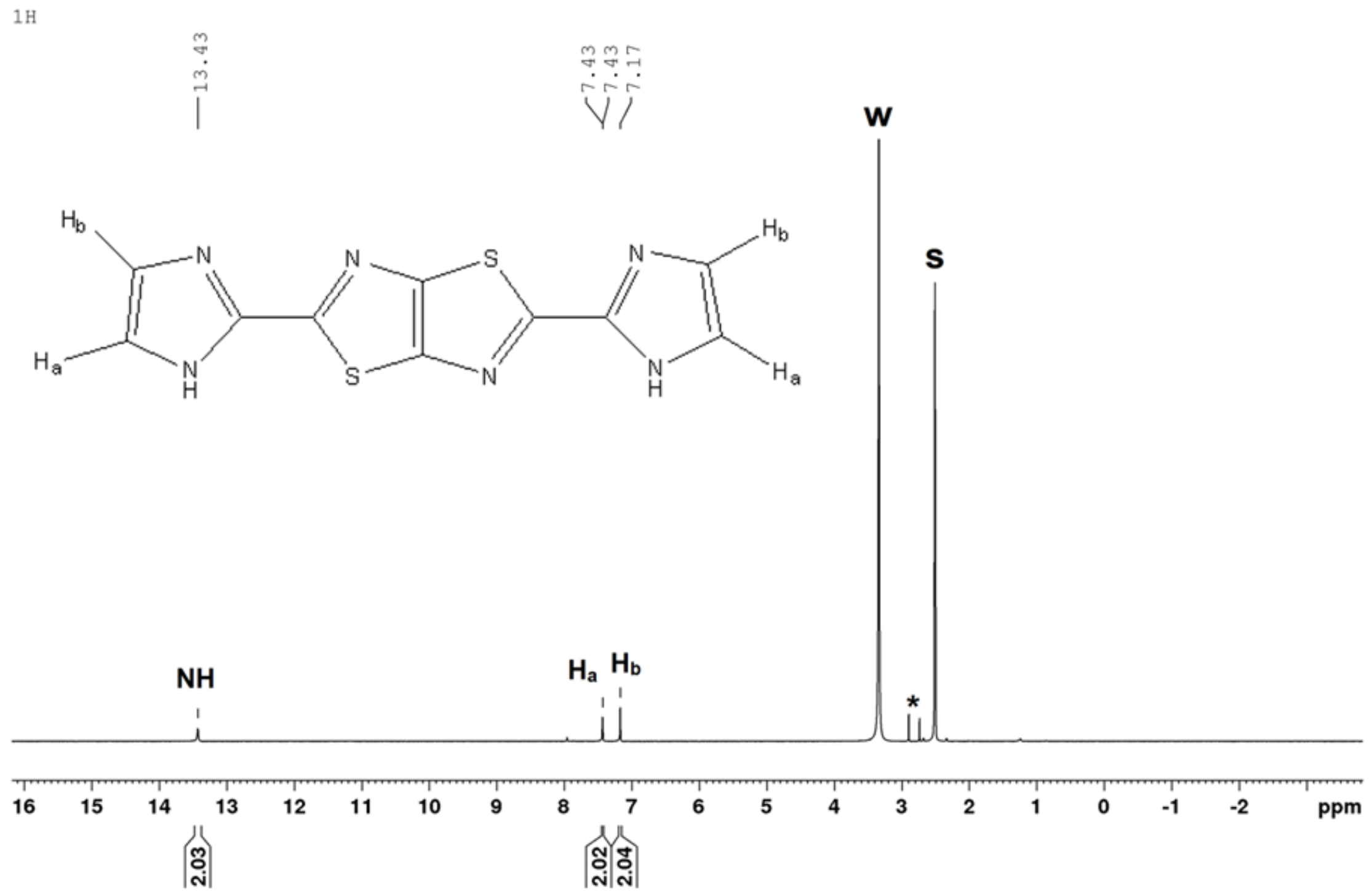

$^{13}C\{^{1}H\}$ spectrum (DMSO-$d_6$) of **B4:**

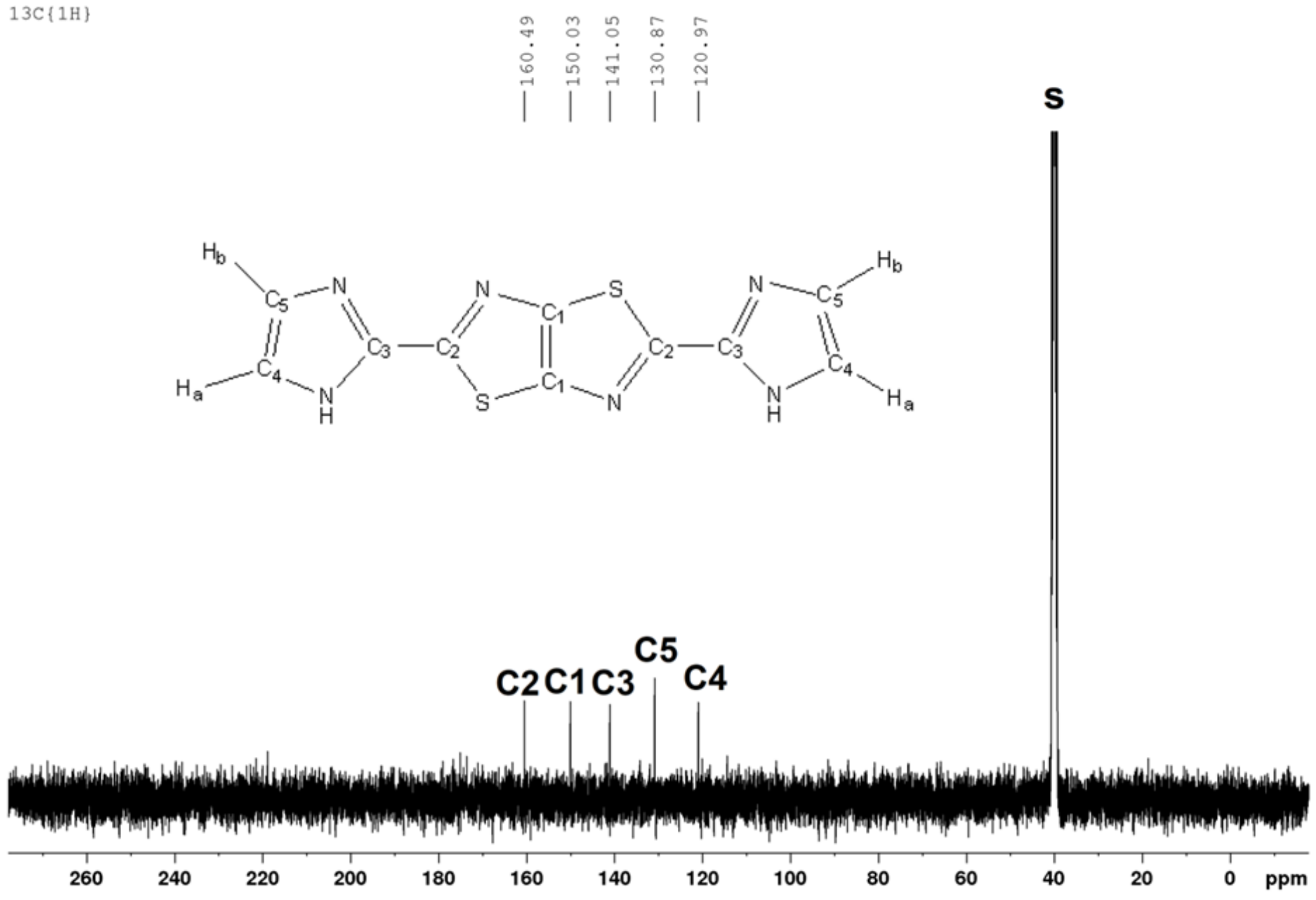

NMR Spectra for **C**

$^{1}H$ spectrum ($CDCl_3$-$d_1$) of **C:**

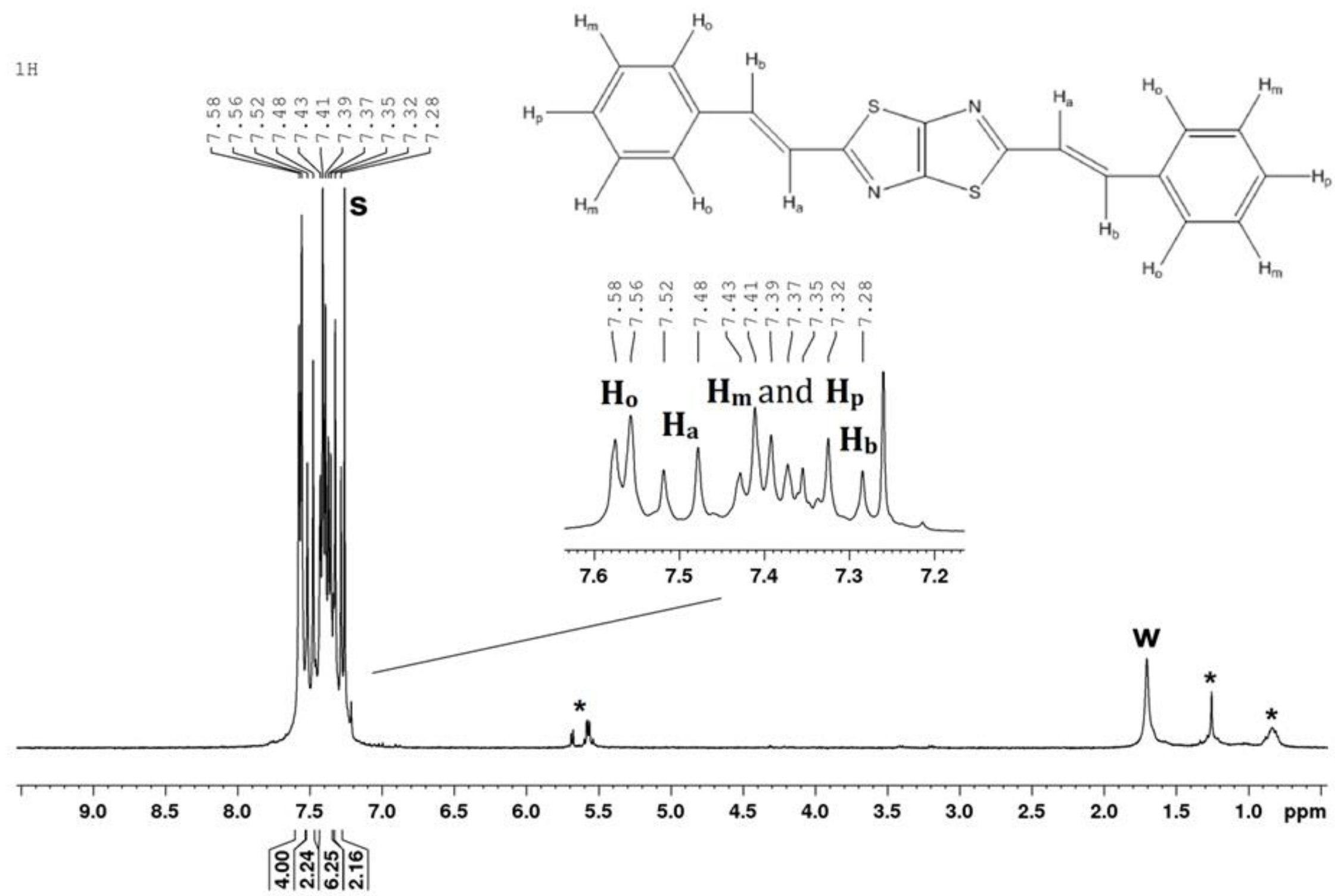

$^{13}C\{^{1}H\}$ spectrum($CDCl_3$-$d_1$) of **C:**

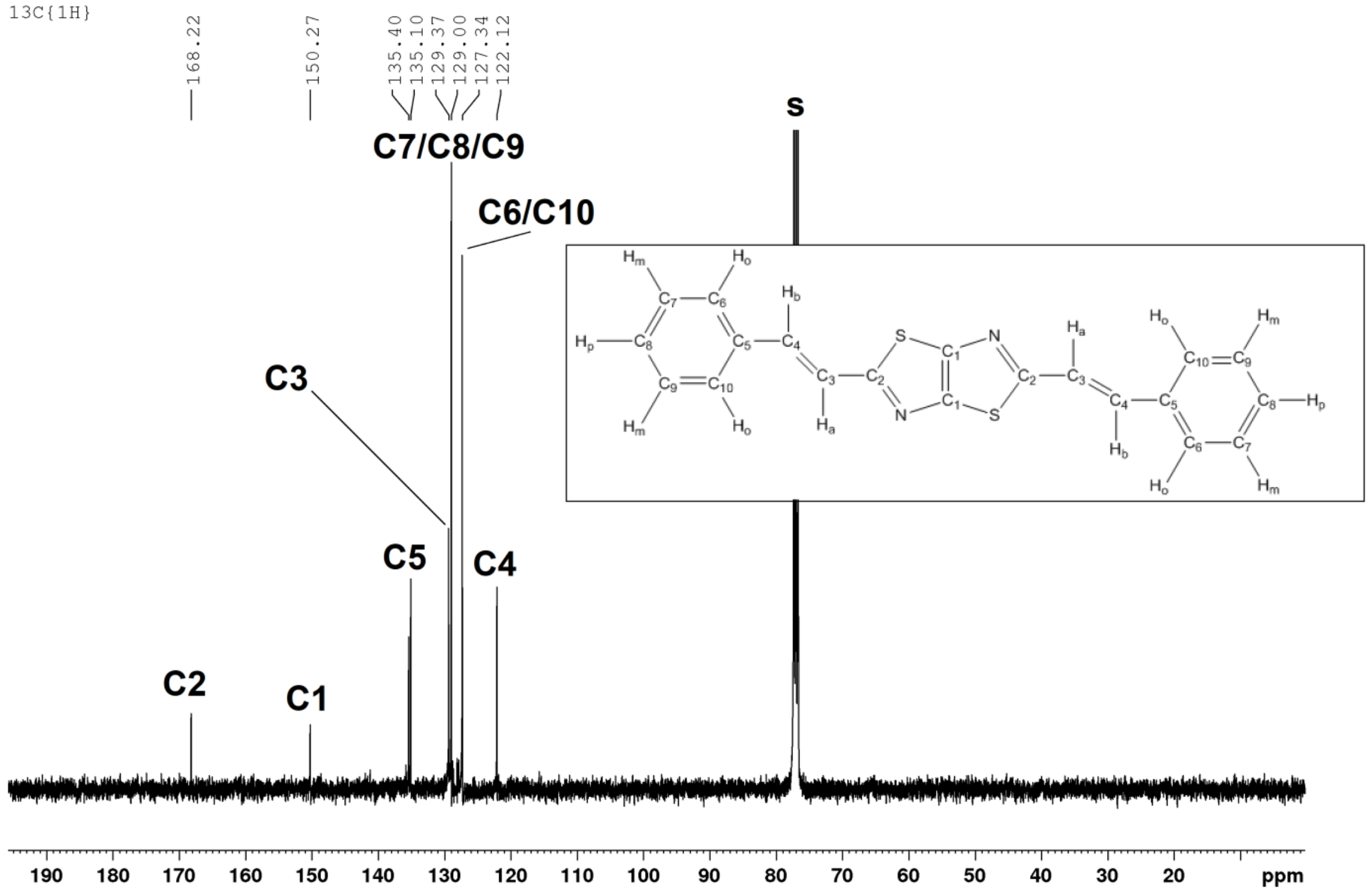

***Table S8.** Comparison of experimental and calculated (metoda!!!) FTIR spectra of studied thiazolothiazoles.*

FTIR Spectra for **A1**

Experimental:

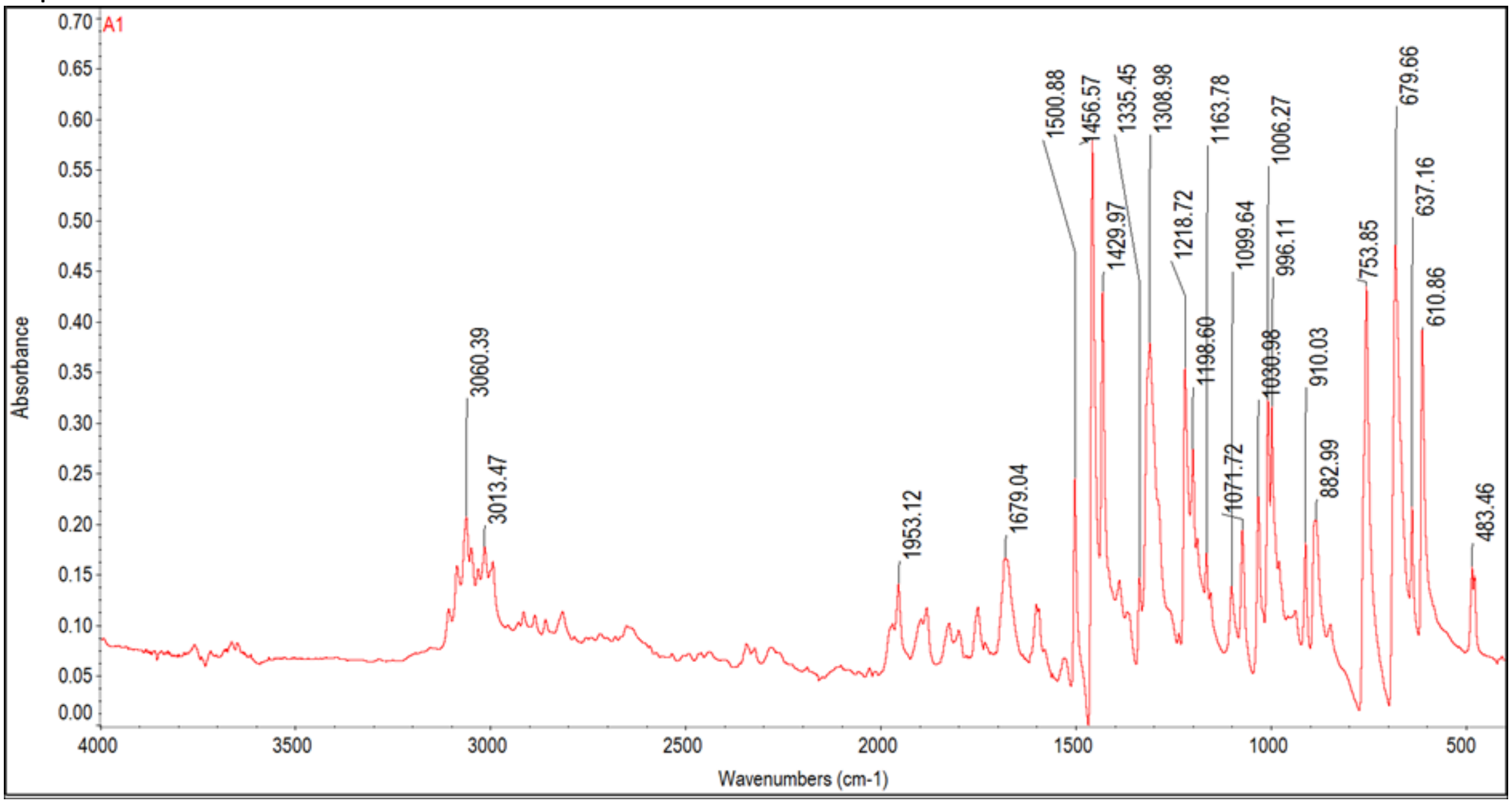

Theoretical (calculated):

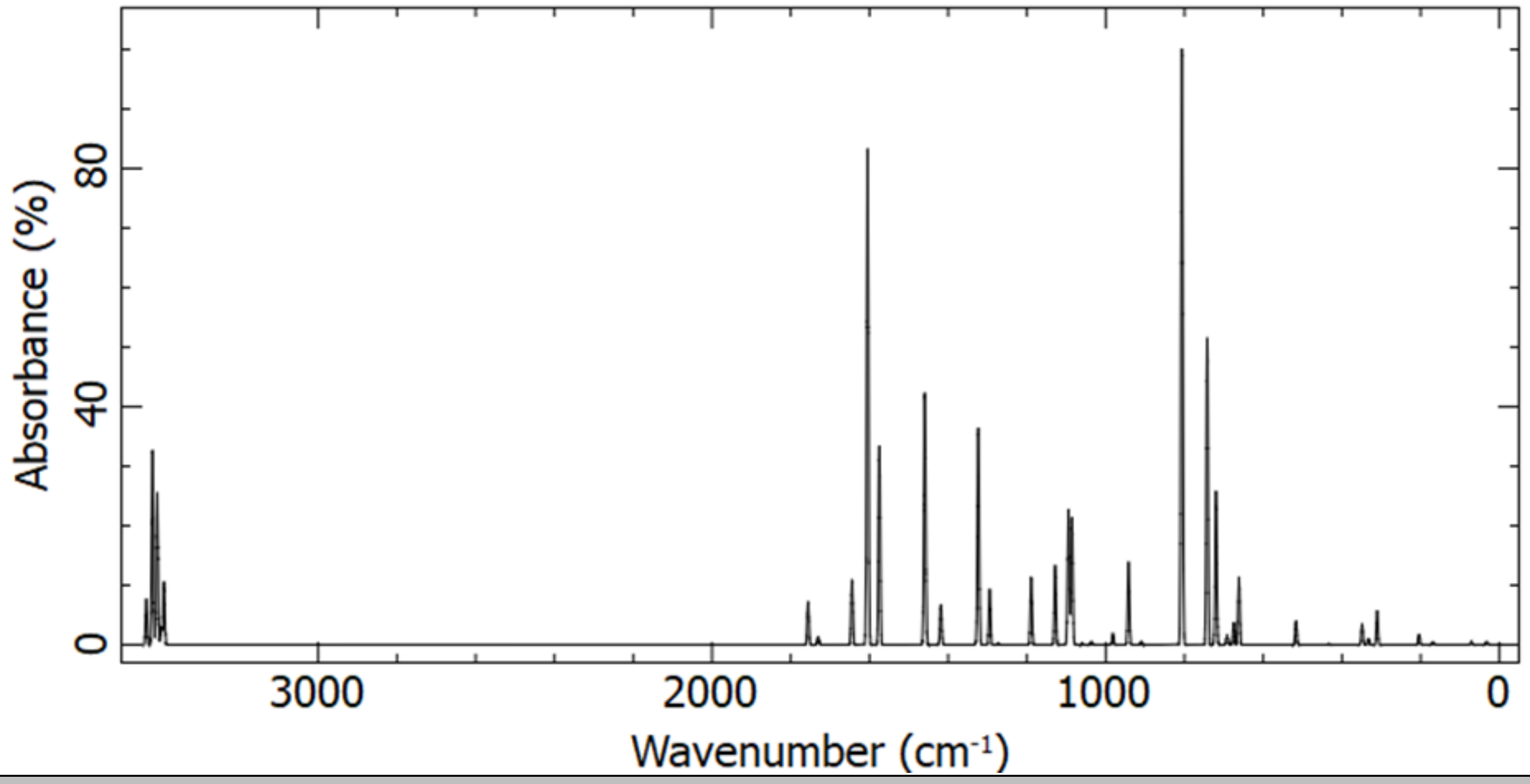

FTIR Spectra for **A2**

Experimental:

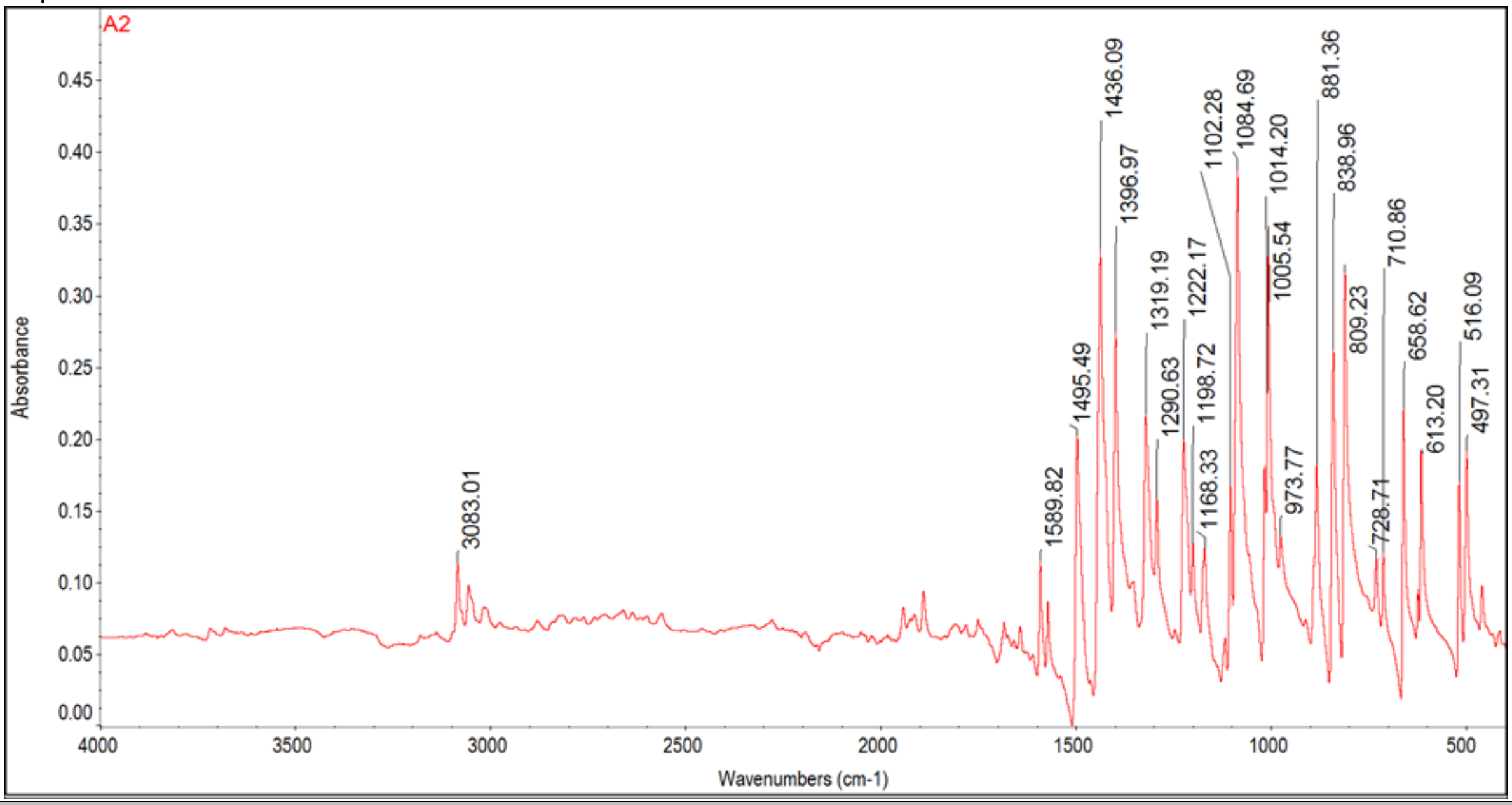

Theoretical (calculated):

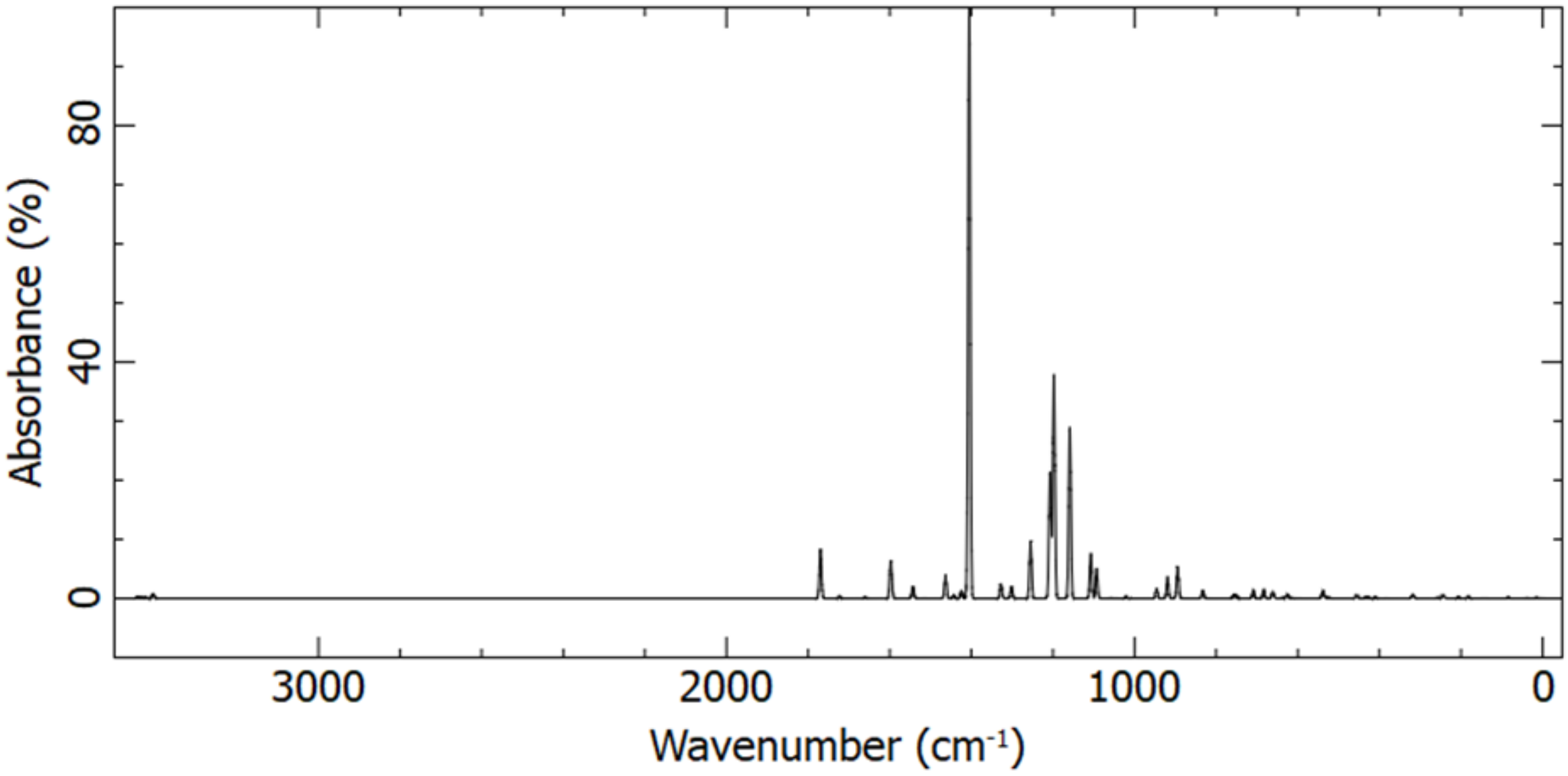

FTIR Spectra for **A3**

Experimental:

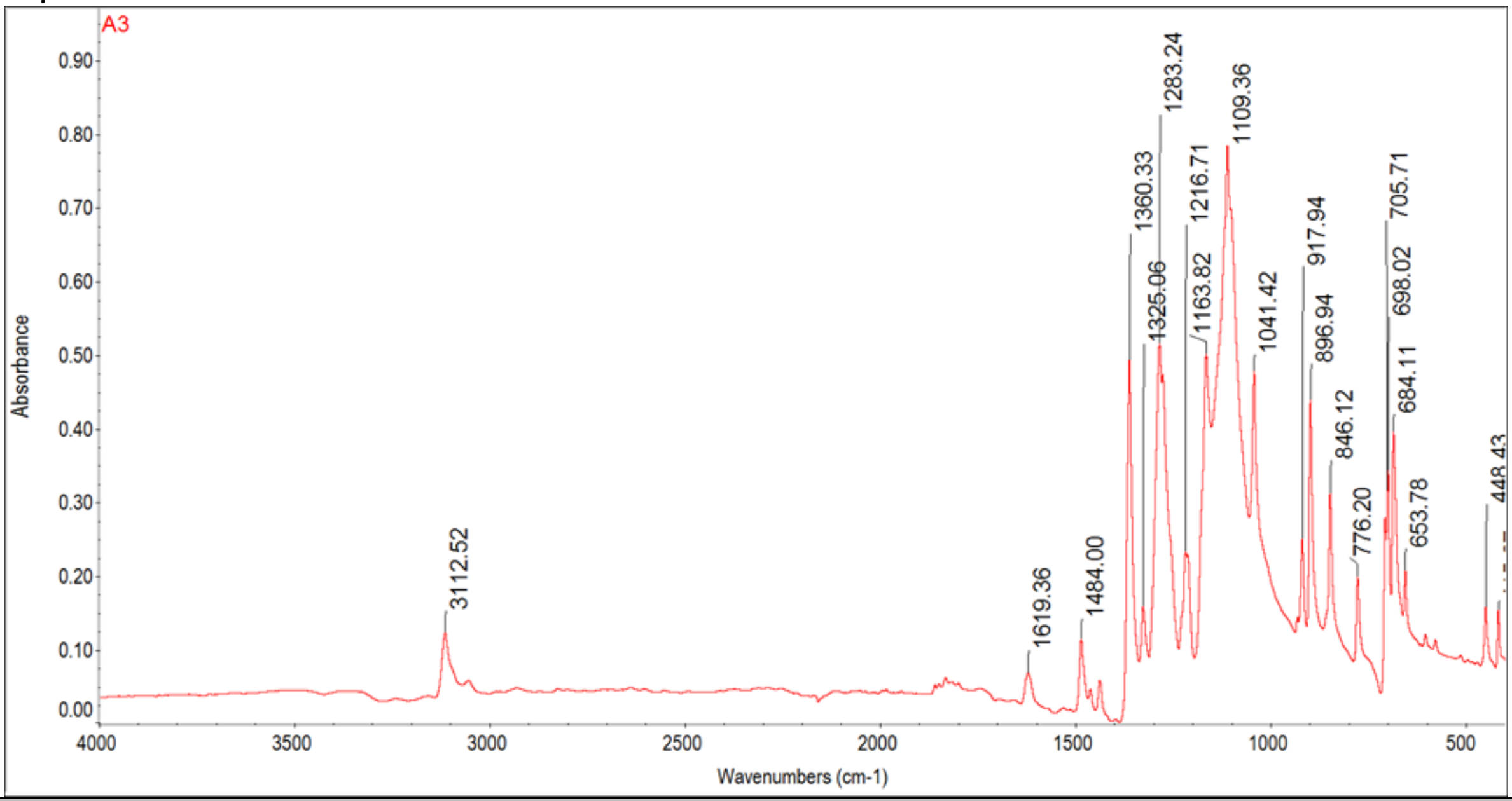

Theoretical (calculated):

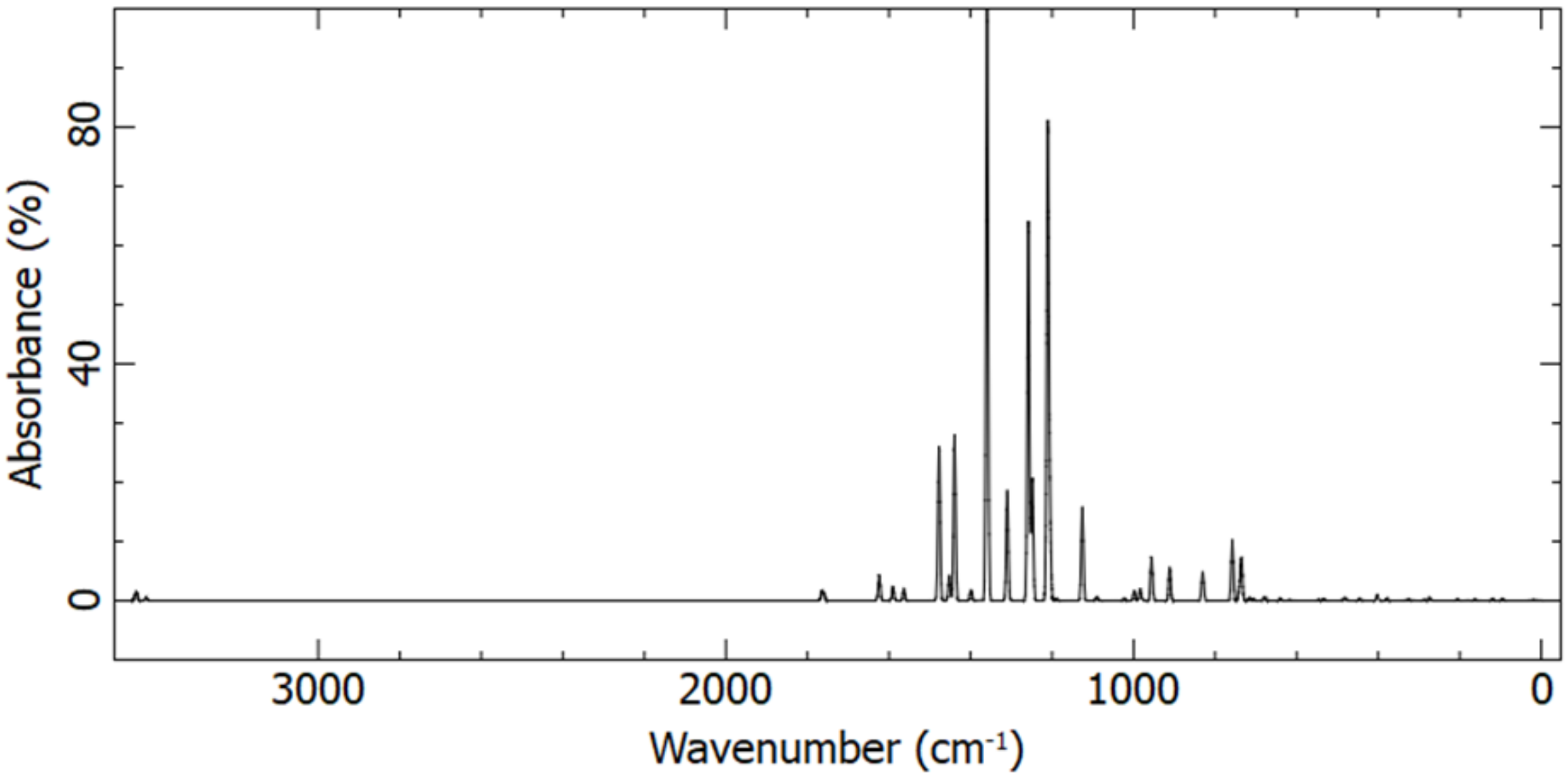

FTIR Spectra for **A4**

Experimental:

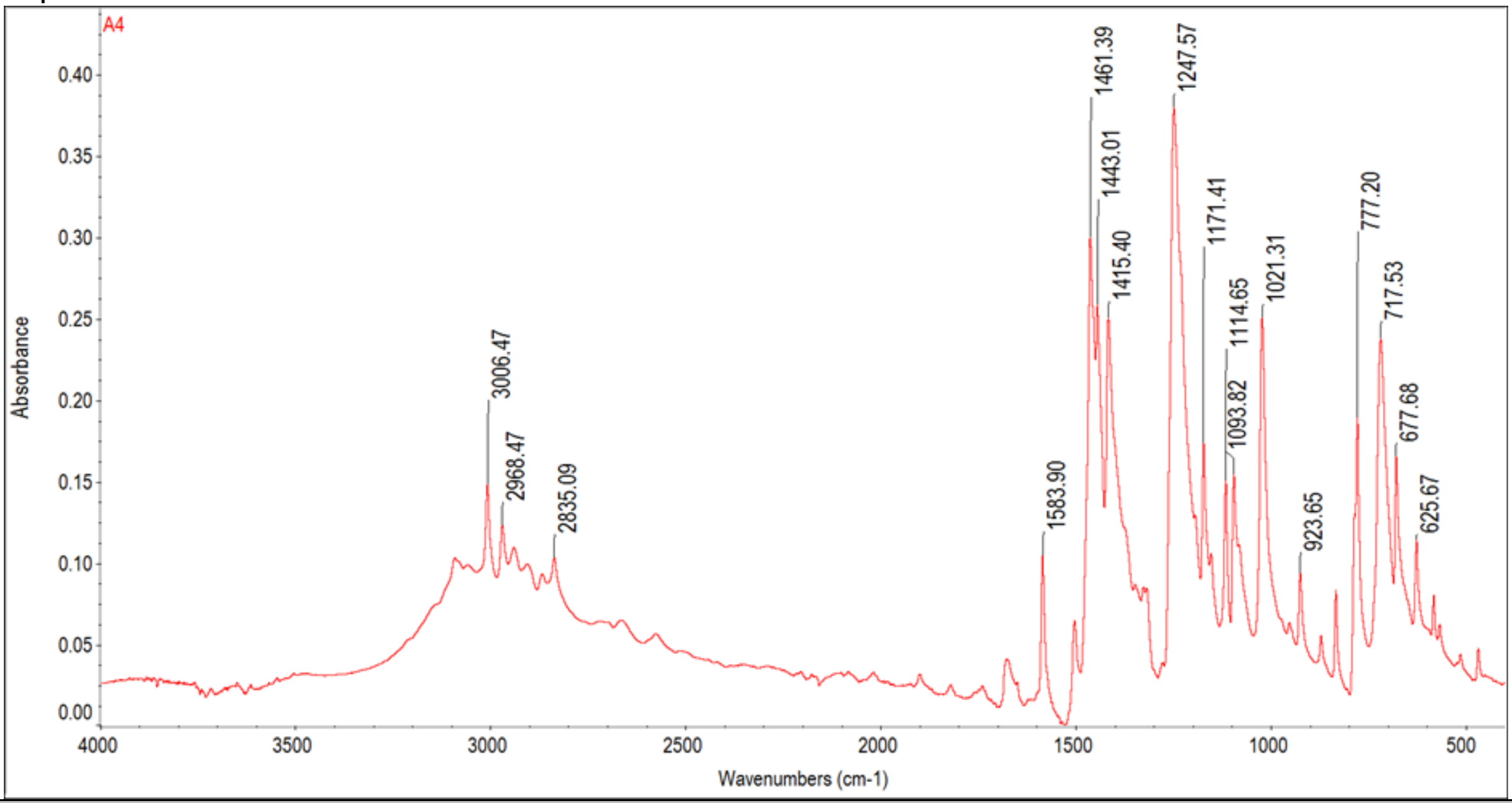

Theoretical (calculated):

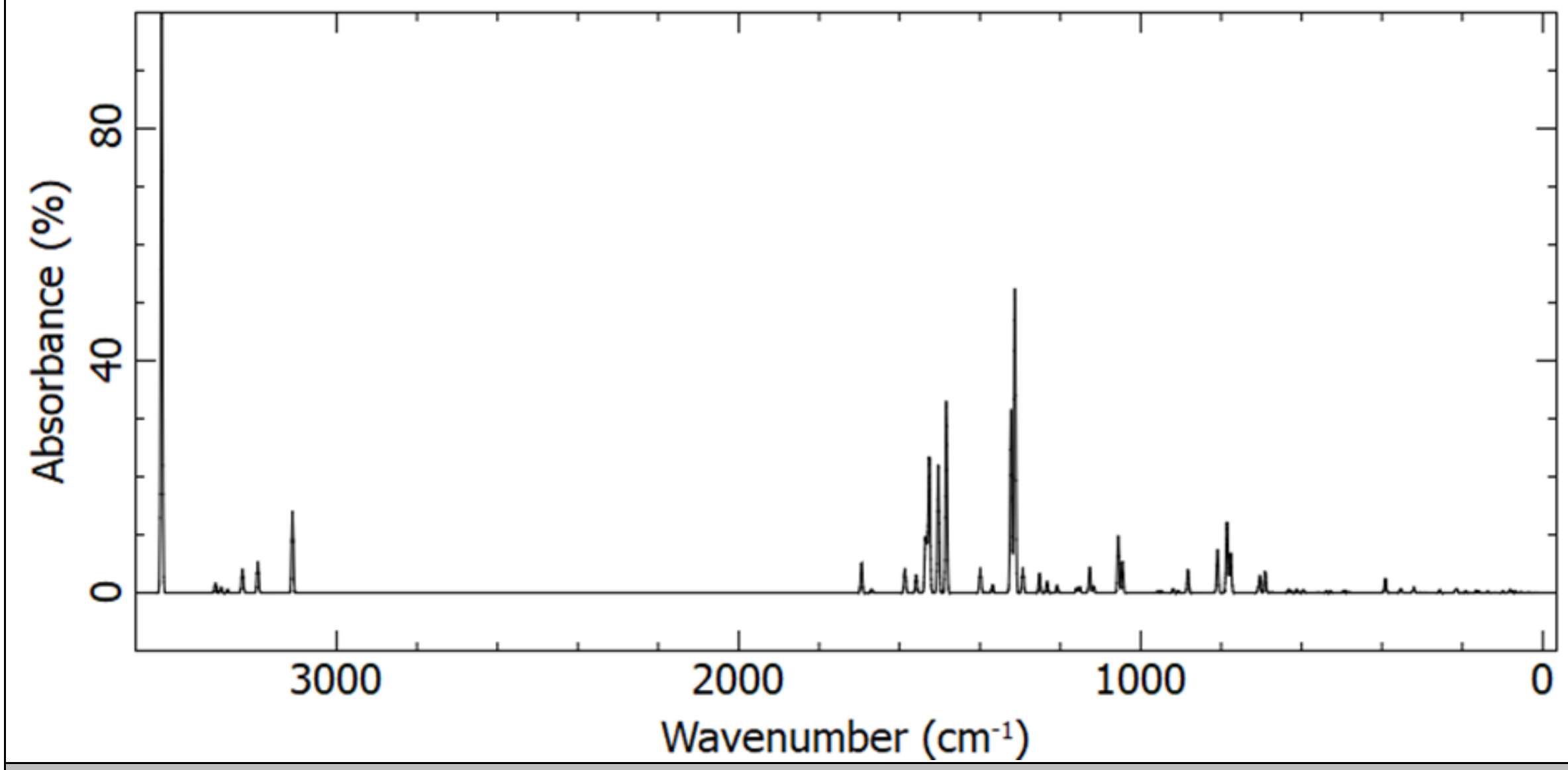

FTIR Spectra for **A5**

Experimental:

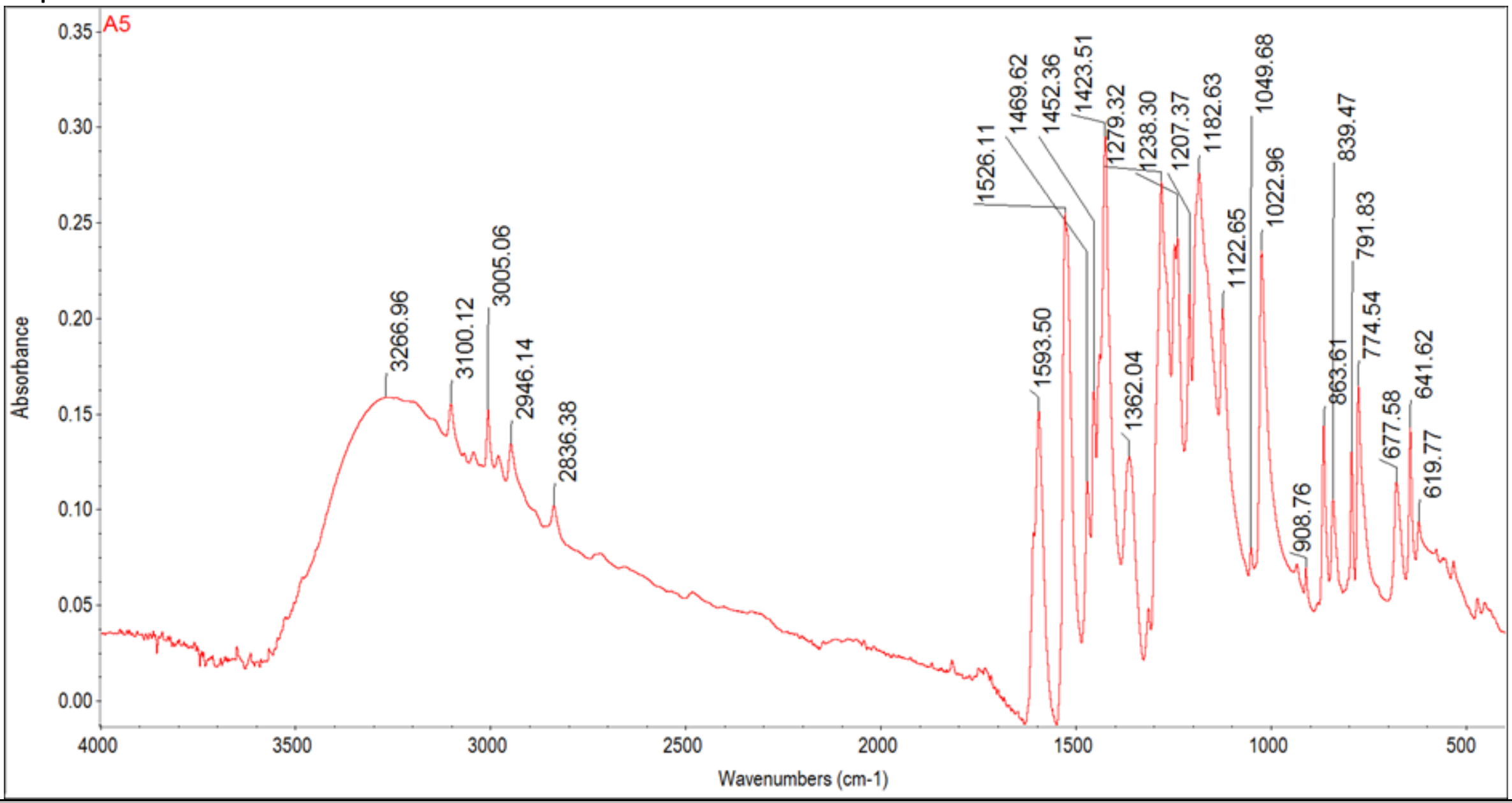

Theoretical (calculated):

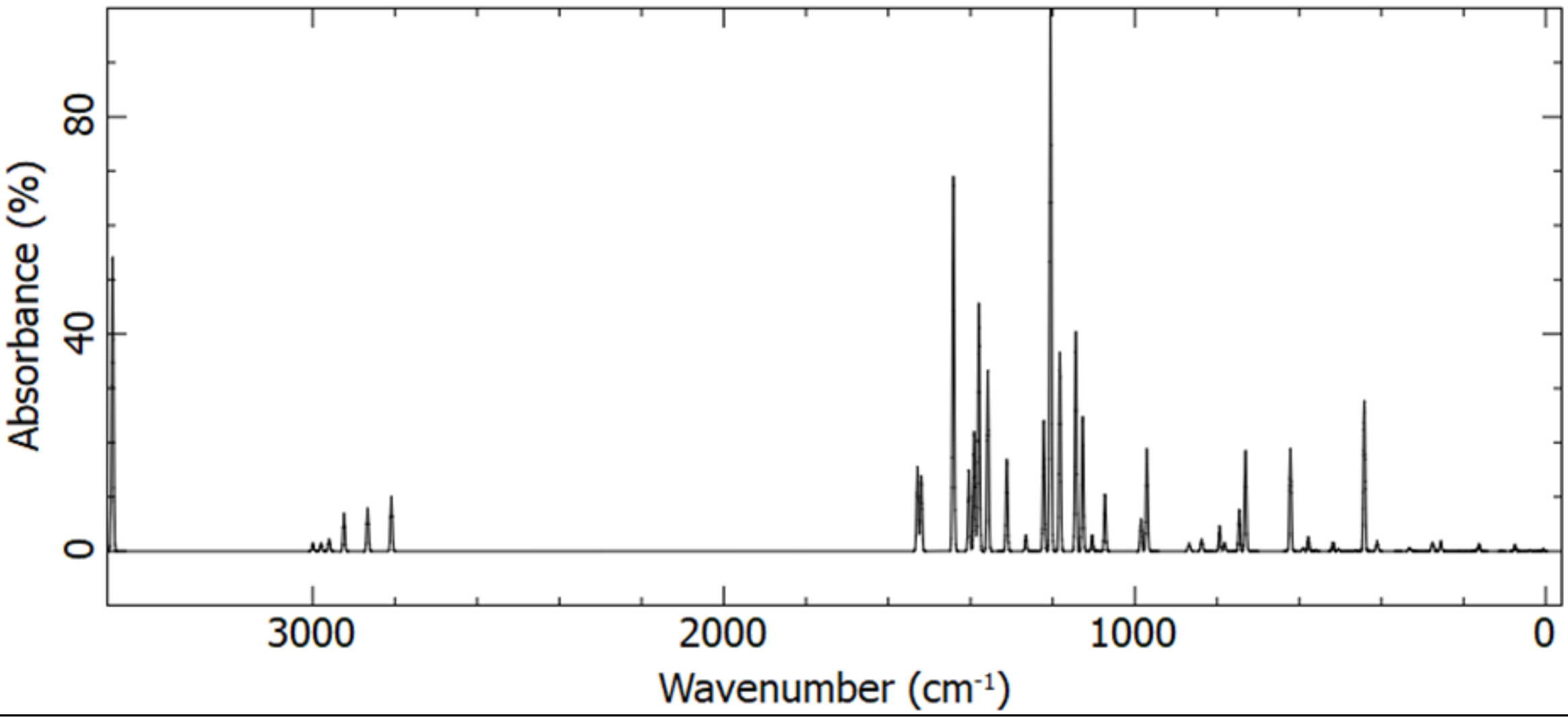

FTIR Spectra for **B1**

Experimental:

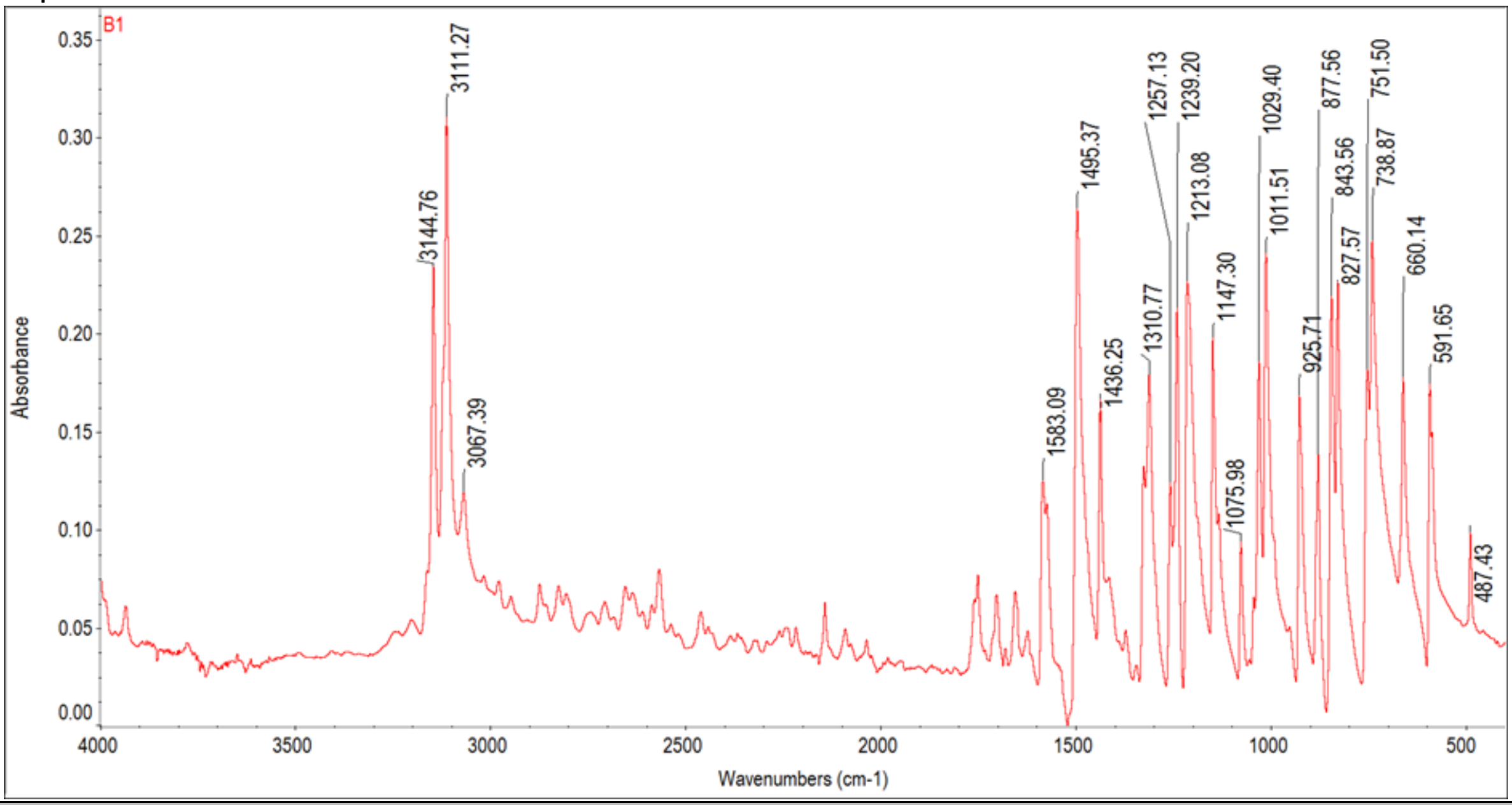

Theoretical (calculated):

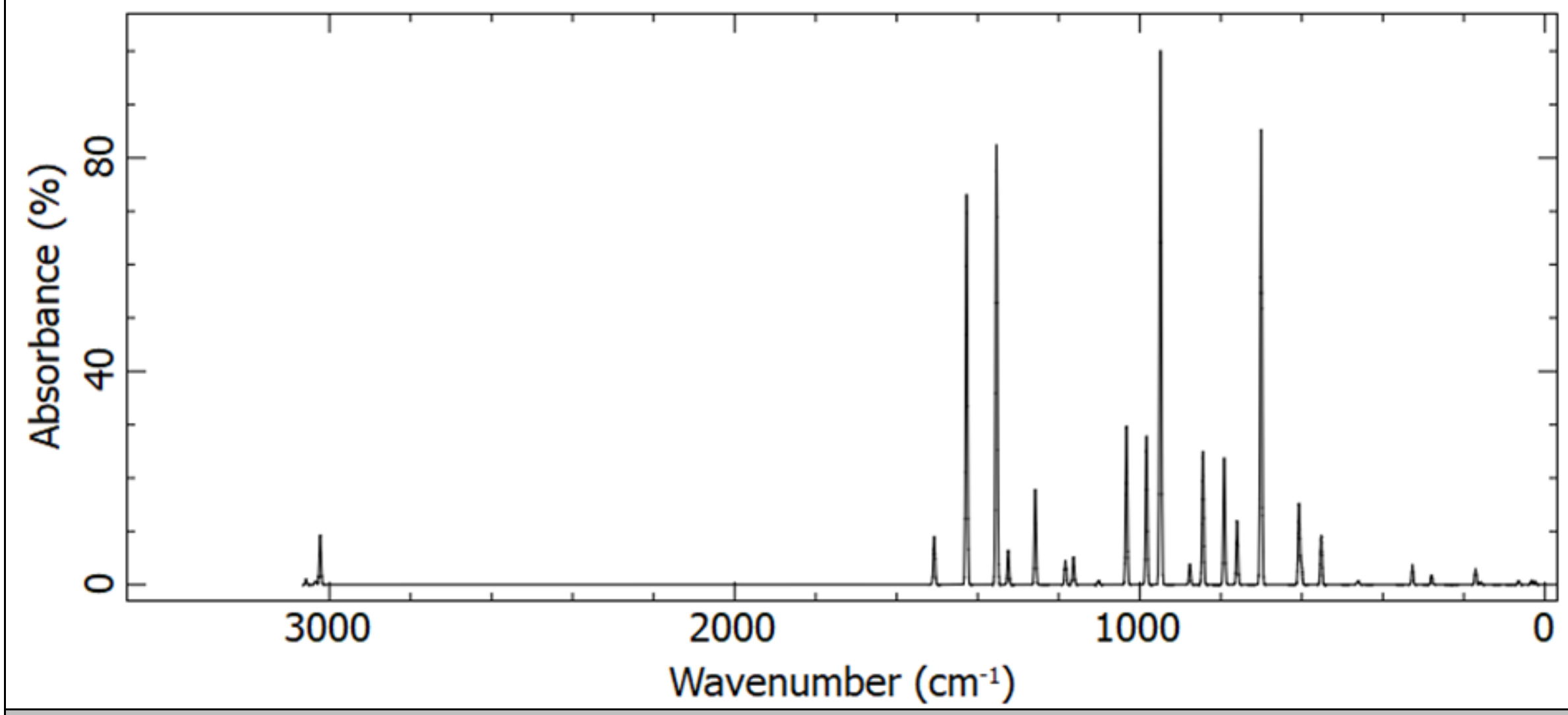

FTIR Spectra for **B2**

Experimental:

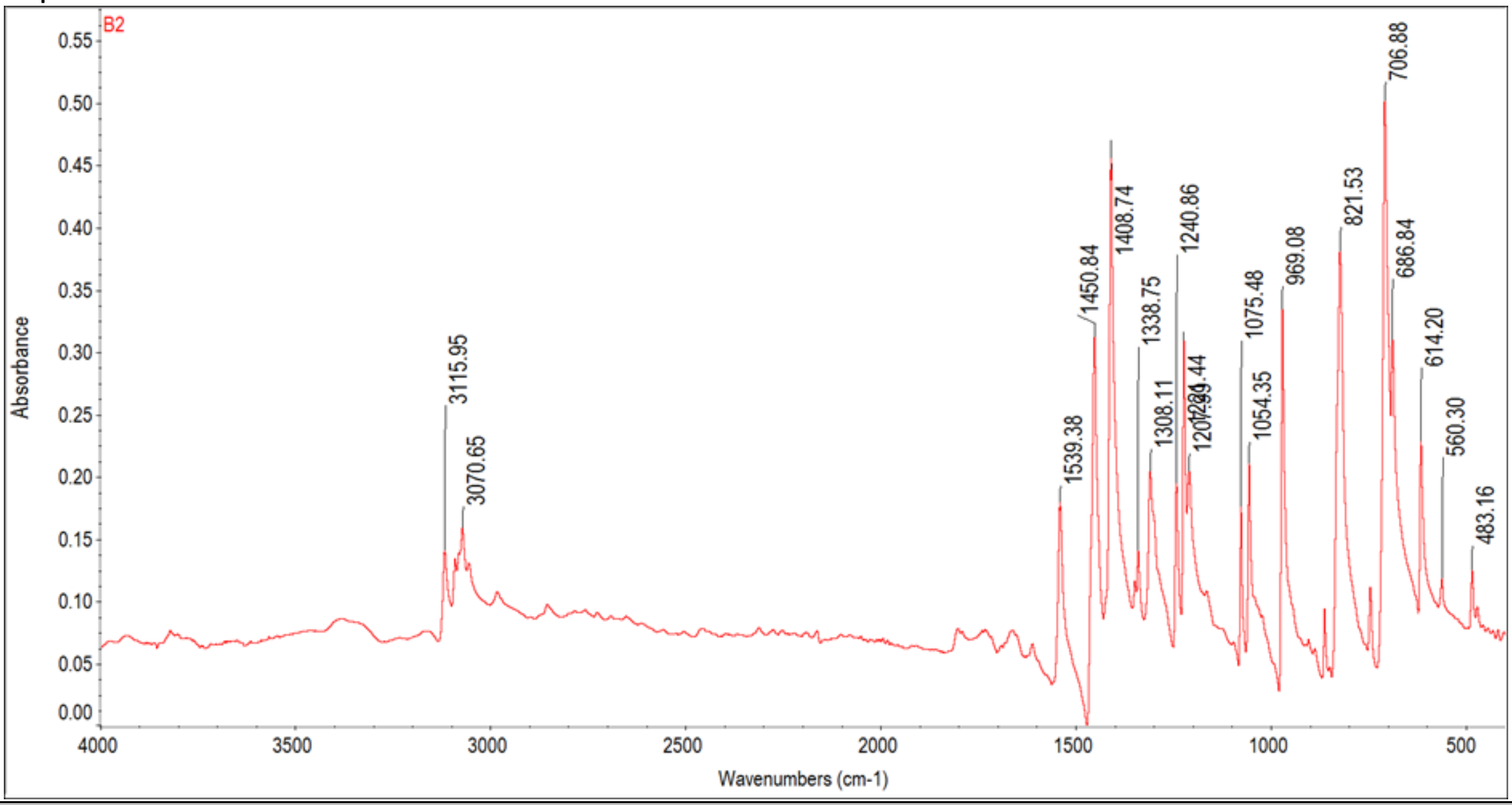

Theoretical (calculated):

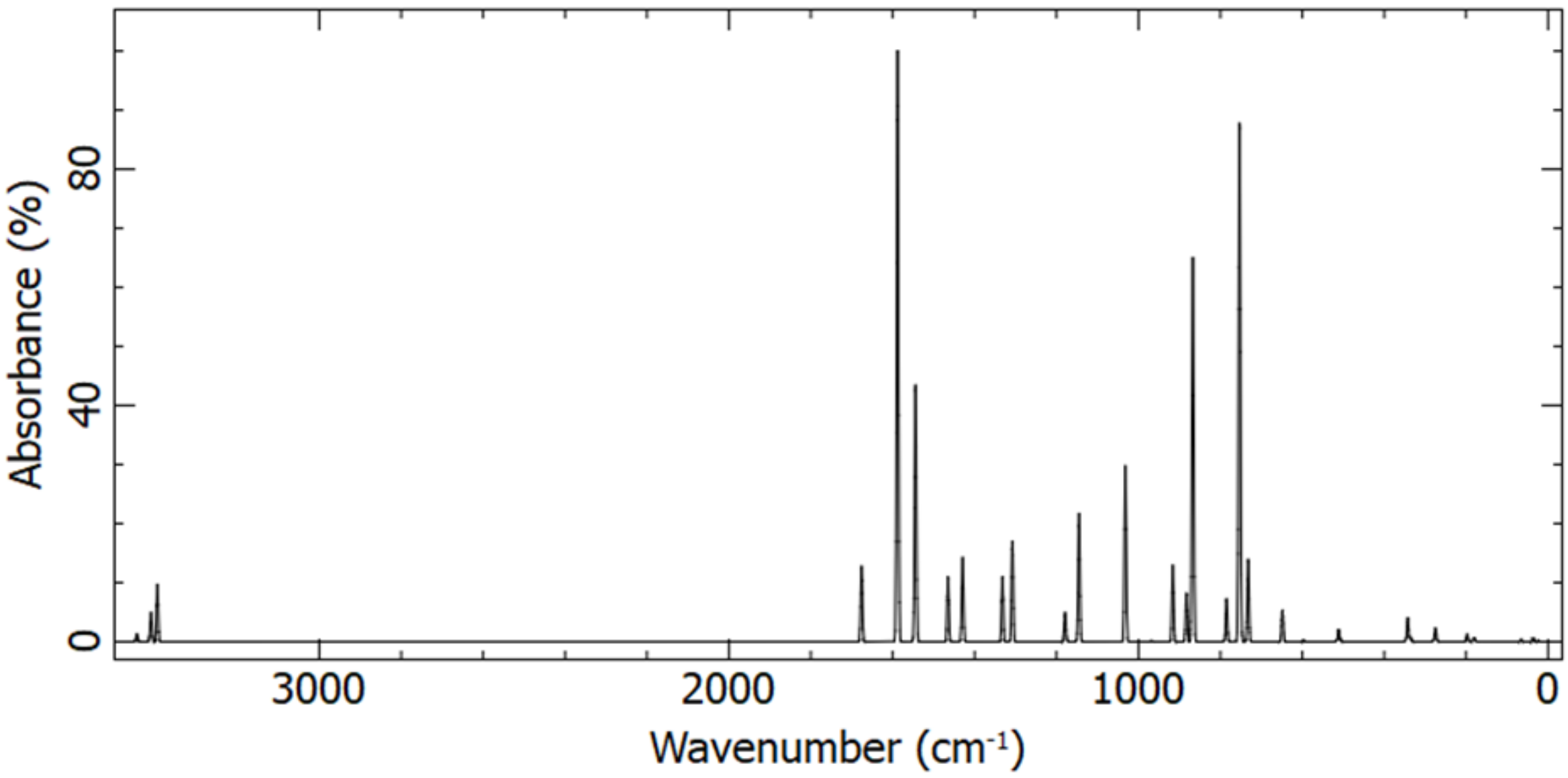

FTIR Spectra for **B3**

Experimental**:**

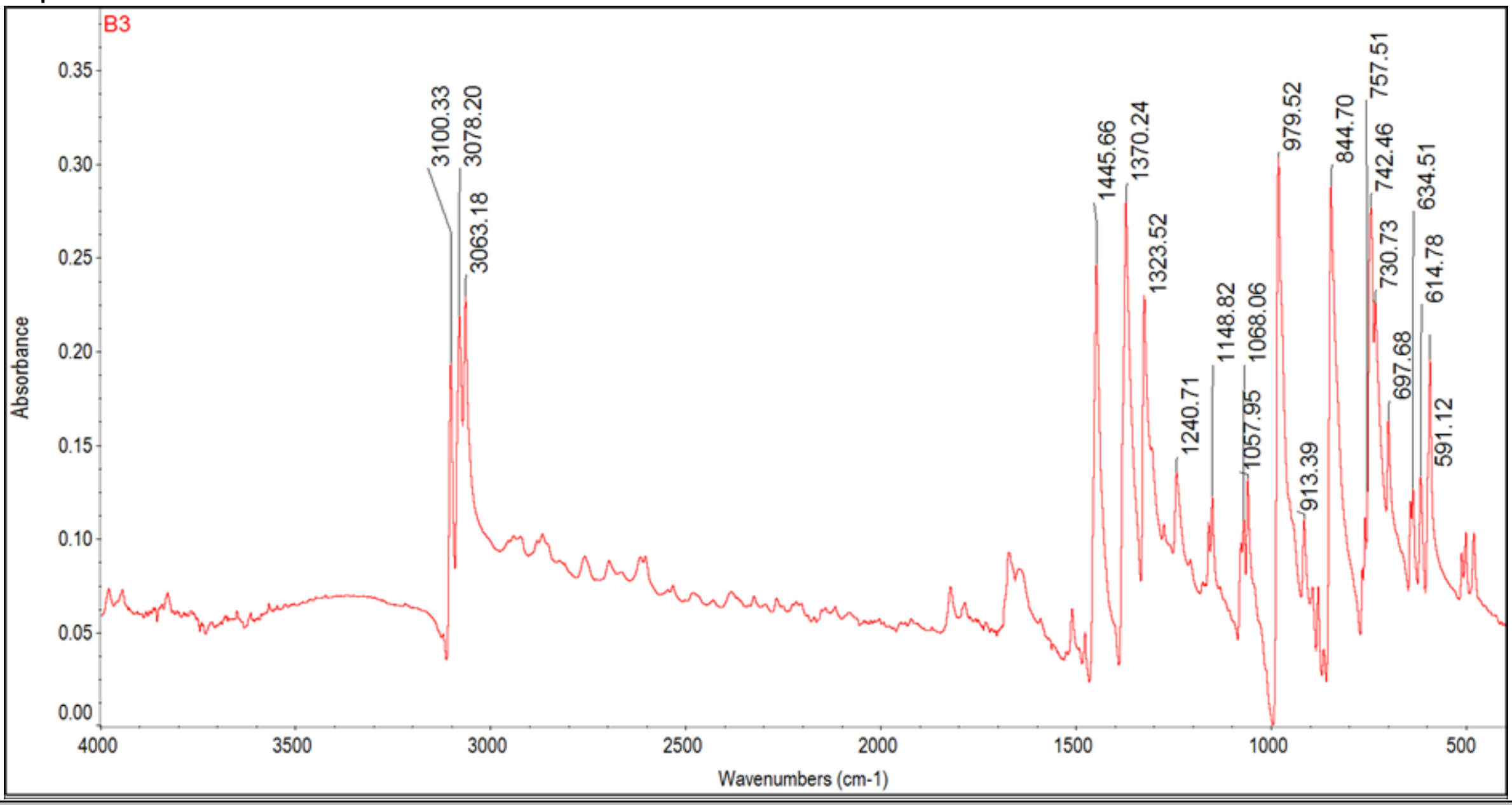

Theoretical (calculated)**:**

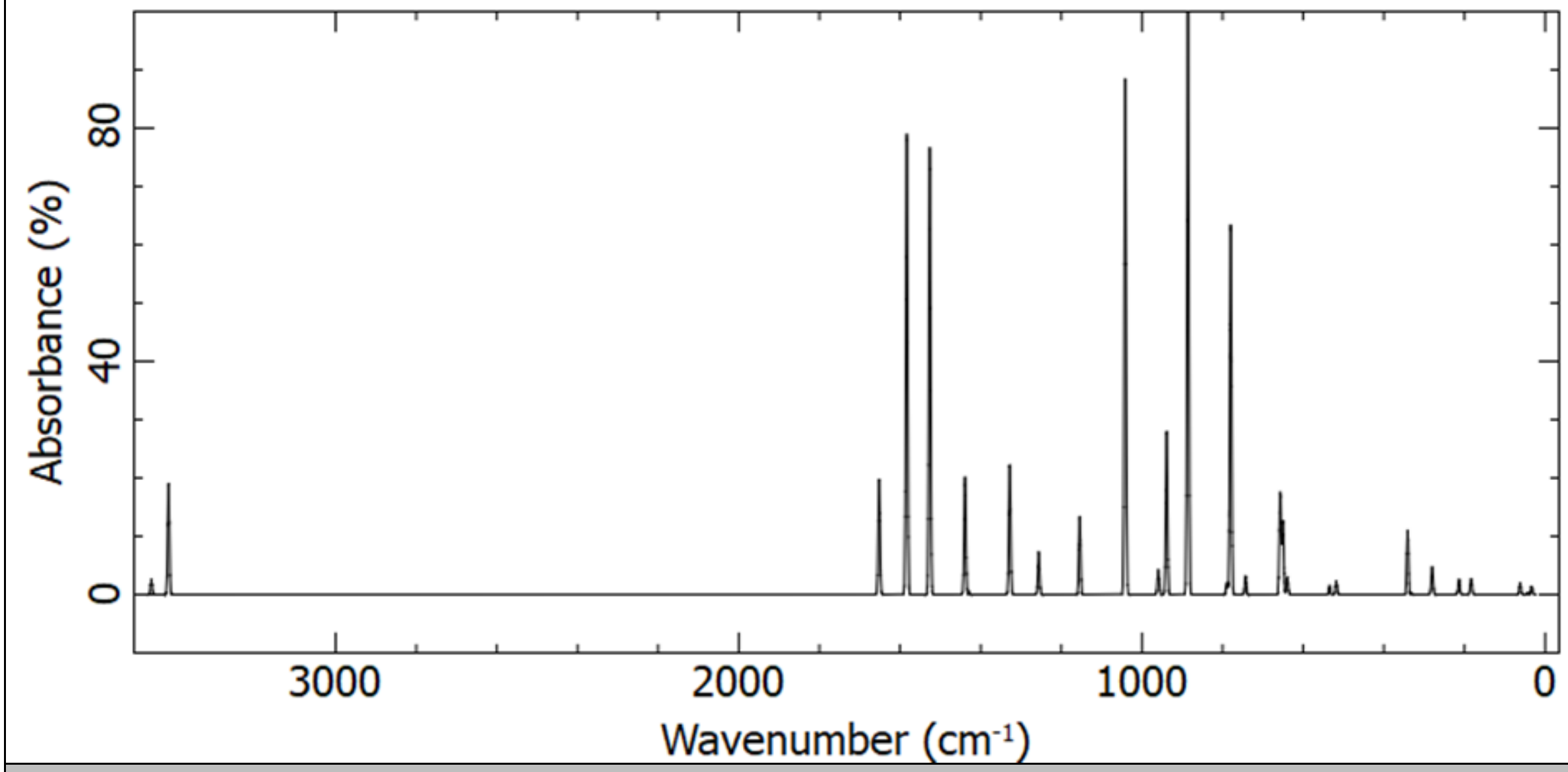

FTIR Spectra for **B4**

Experimental:

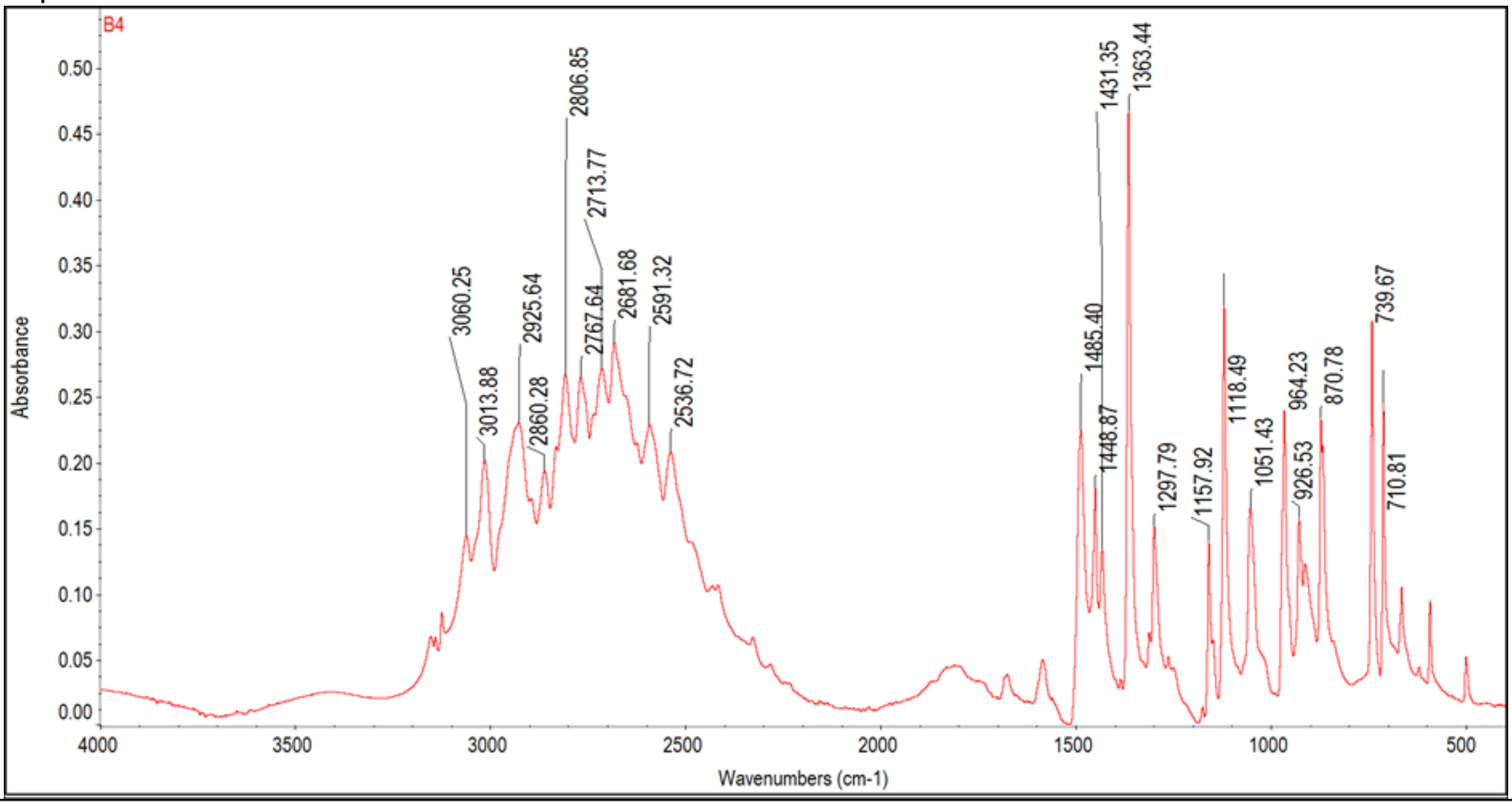

Theoretical(calculated):

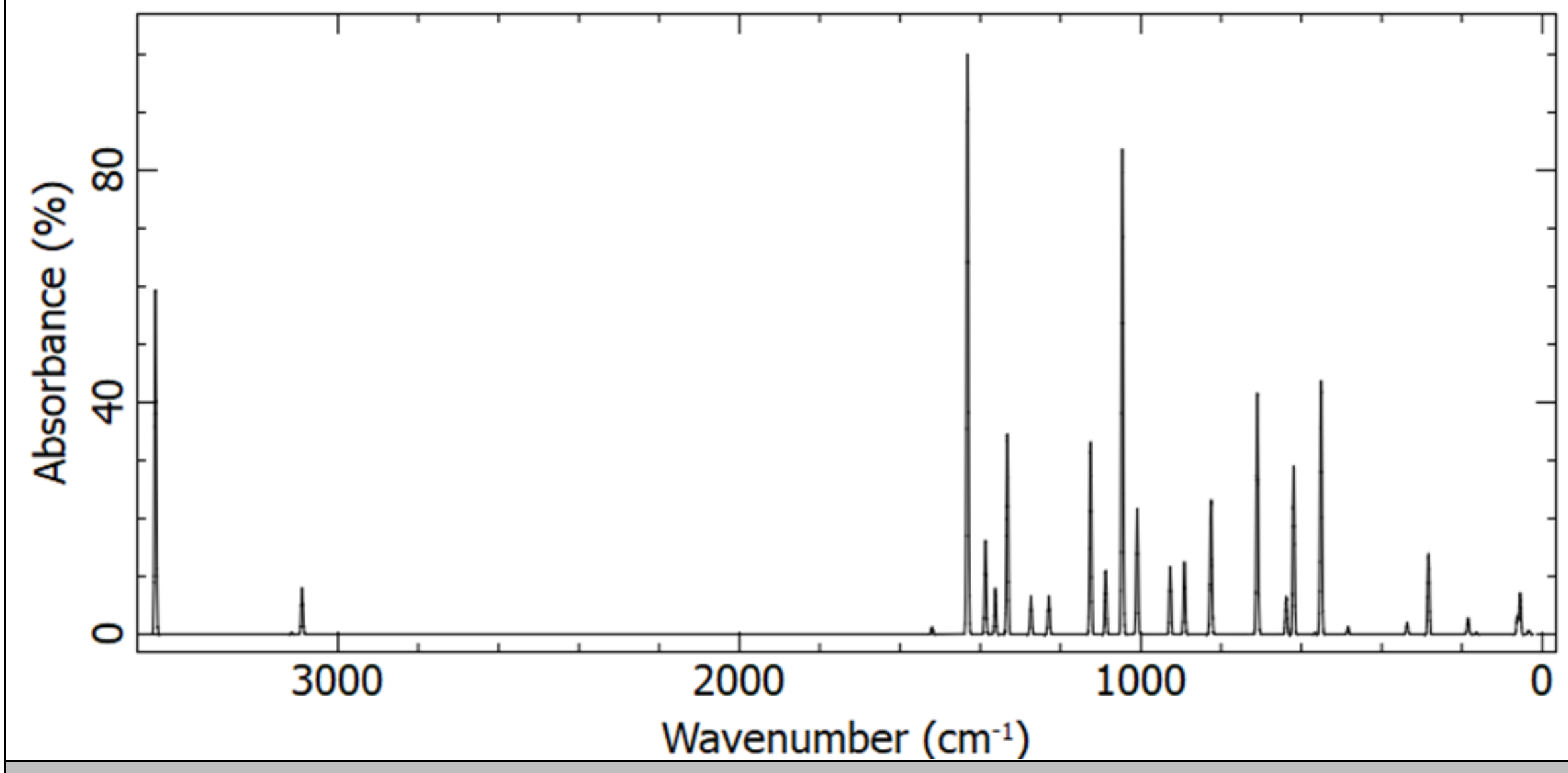

FTIR Spectra for **C**

Experimental:

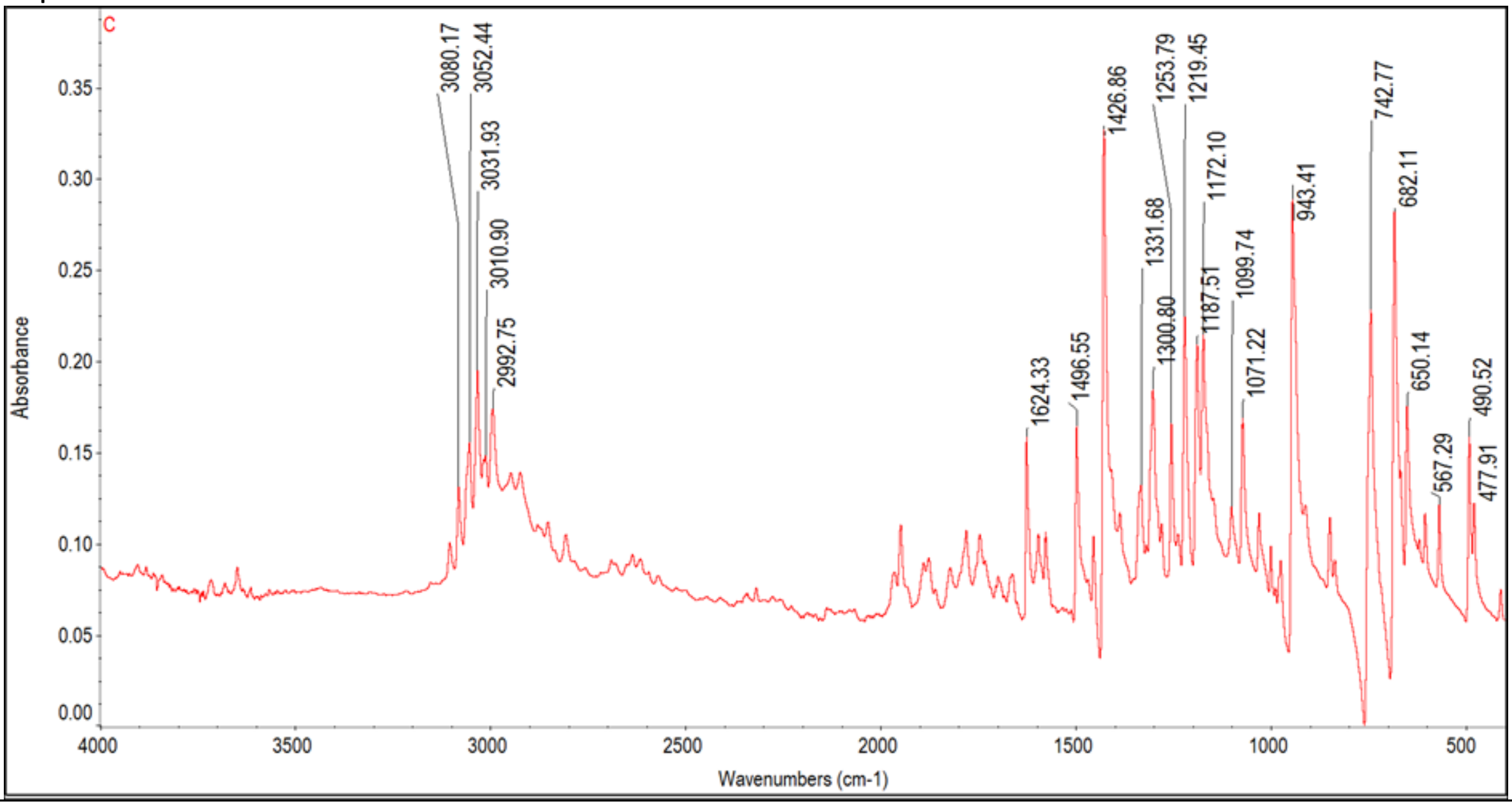

Theoretical (calculated):

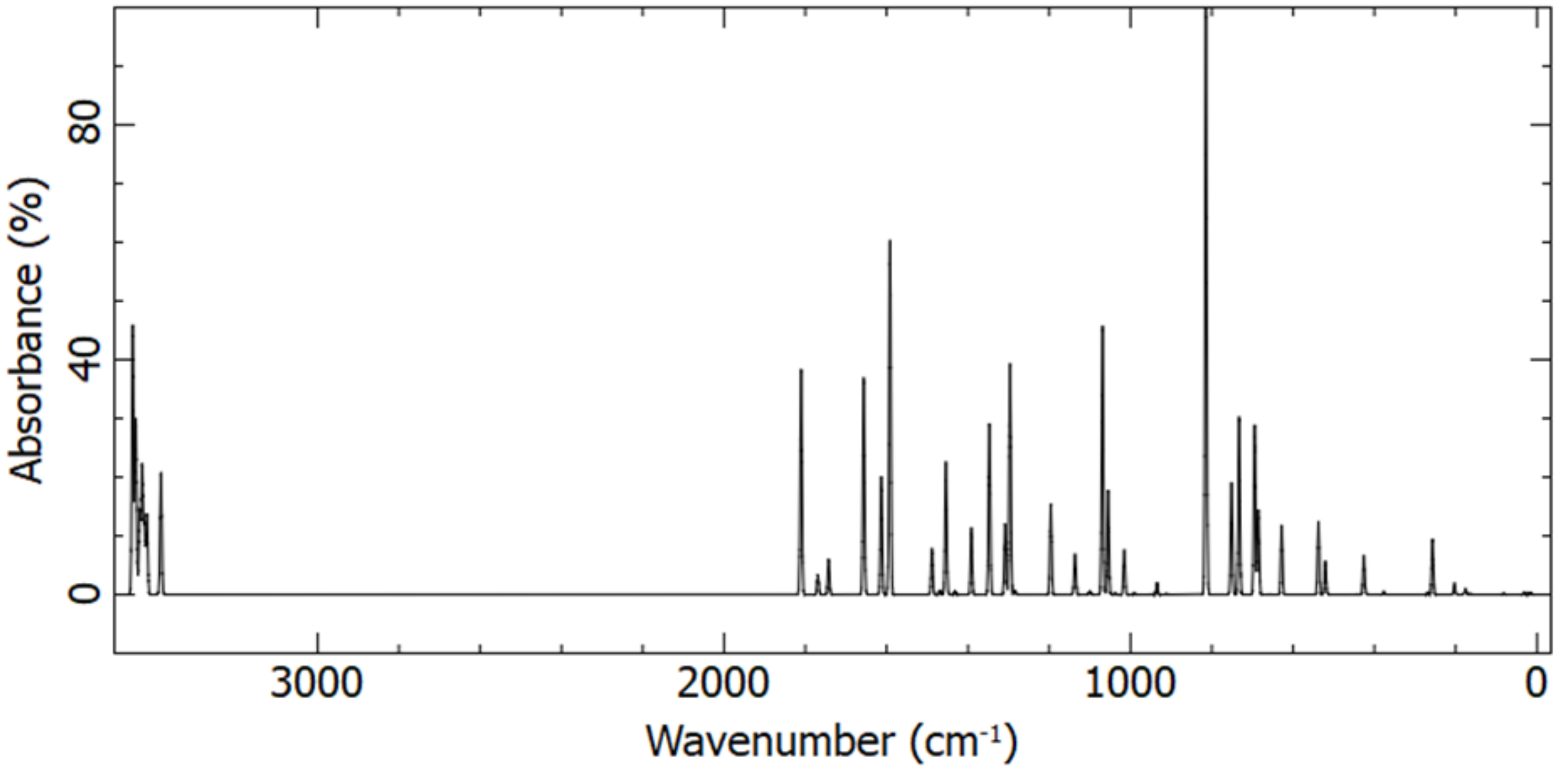

***Table S9*** *Selected IR active modes [cm$^{-1}$] and their assignment from DFT calculations – type A compounds.*

| A1 | | A2 | | A3 | | A4 | | A5 | | Band Assignment<br>ip – in-plane, oop – out-of-plane |
|---|---|---|---|---|---|---|---|---|---|---|
| Experimental | Calculated | Experiment | Calculated | Experimenta | Calculated | Experimental | Calculated | Experimental | Calculated | |
| - | - | - | - | - | - | - | 3335(vs) | 3267(m) | 3749(vs) | O–H stretching |
| 3030(w) | 3212(w),<br>3197(m) | 3083(vw) | - | - | - | - | 3205(m) | 3100(m) | 3204(w) | C–H stretching, sym. (from phenyl group, ip) |
| 3013(w) | 3186(m),<br>3170(w) | 3055(vw) | 3182(w) | 3112(w) | - | 3006(w) | - | 3005(m) | 3183(w) | C–H stretching, asym. (from phenyl group, ip) |
| - | - | - | - | - | - | 2968(w) | 3140(m),<br>3103(m) | 2946(m) | 3144(m),<br>3082(m) | C–H stretching, asym. (from –$CH_3$) |
| - | - | - | - | - | - | 2835(w) | 3019(m) | 2836(m) | 3020(m) | C–H stretching, sym. (from –$CH_3$) |
| 1953(w) | - | - | - | - | - | - | - | - | - | aromatic C–H bending overtones |
| 1679(w) | 1641(w) | 1590(vw) | 1654(vs) | 1619(w) | 1649(w),<br>1644 (w) | - | - | - | - | C=C stretching (from phenyl group, ip) |
| - | - | - | - | - | - | 1584(w) | 1645(m),<br>1540(m) | 1593(m) | 1644(s),<br>1634(s) | C=C stretching (from phenyl group,ip) + C–O–H stretching (ip) |
| 1501(w) | 1536(w) | 1495(m) | 1493(s) | - | 1517(m) | - | - | 1526(s) | 1550(vs) | C=N stretching (from TzTz system + C=C stretching (from phenyl group), ip) |
| - | - | - | - | - | - | - | 1512.88(m) | - | 1509(s),<br>1493(s) | C–H (from –$CH_3$) bending, scissoring (ip) |
| 1457(vs),<br>1430(s) | 1499(vs),<br>1471(m) | 1436(s),<br>1397(m) | 1442(m) | - | 1486(w)<br>1461 (w) | - | - | - | - | C=N stretching (from TzTz system, ip) + C=C stretching (from phenyl group, ip) |
| - | - | - | - | - | - | 1461(s)<br>1443(m) | 1491(s),<br>1486(m),<br>1481(s),<br>1460(s),<br>1440(vs), | 1470(m) | 1496(s),<br>1483(vs),<br>1460(s),<br>1410(m) | C=N stretching (from TzTz system, ip) + C=C stretching (from phenyl group, ip) + C–H (from –$CH_3$, ip) bending, scissoring + C–O–H bending (ip) |
| - | - | - | - | - | 1357(w) | 1415(m) | - | - | 1452(m),<br>1424(vs) | C=C stretching (from phenyl group, ip) |
| 1335(w),<br>1309(m) | 1364(m),<br>1326(w) | 1319(m) | 1367(s),<br>1348(w),<br>1331(w) | 1360(m),<br>1325(w) | 1381(m),<br>1357(w),<br>1344(m) | - | 1358(m) | - | - | C–N stretching (from TzTz system, ip) + C=C stretching (from phenyl group, ip) |
| 1219(m) | 1236.95(m) | - | - | - | - | - | - | 1362(m) | 1360(w) | C=N stretching (from TzTz system) + C=C stretching (from phenyl group), C–H (from phenyl ring), bending, rocking (ip) |
| - | - | - | - | - | - | - | 1328(w) | | 1313(m) | C–N stretching (from TzTz system, ip) + C=C stretching (from phenyl group, ip) + C=N stretching (from TzTz system, ip) + C–O–H bending (ip) |

| A1 | | A2 | | A3 | | A4 | | A5 | | Band Assignment<br>ip – in-plane, oop – out-of-plane |
|---|---|---|---|---|---|---|---|---|---|---|
| Experimental | Calculated | Experiment | Calculated | Experimenta | Calculated | Experimental | Calculated | Experimental | Calculated | |
| - | - | 1291(w) | 1313(vs) | 1283(m) | 1307 (w)<br>1271(vs) | - | - | - | - | C=C stretching (from phenyl group, ip) + C–C stretching (from $CF_3$, ip) |
| | | 1222(w) | 1241(m) | 1216.71(w) | 1224(w) | - | 1284(m) | - | - | C=N stretching (from TzTz system, ip) + C=C stretching (from phenyl group, ip) |
| - | - | - | - | - | - | 1248(vs) | 1275(m) | 1279(vs) | 1296(vs), | C=C stretching (from phenyl group, ip) + C–O stretching (ip) |
| - | - | - | - | - | - | - | - | 1238(s),<br>1207(m) | 1271(s)<br>1230(s) | C–O stretching, O–H bending (ip), C–H ) bending , scissoring (ip) |
| - | - | - | - | - | - | - | - | 1183(vs)<br>1123(s)<br>1023(s) | 1212(s),<br>1188(w),<br>1153(m),<br>1059(w), | C–H bending, scissoring (ip) +C–O–H bending (ip) + C–O stretching (ip) |
| - | - | - | - | - | - | - | 1255(m) | - | | C–H (from phenyl ring, ip) bending , scissoring (ip) + C=N stretching (from TzTz system, ip) + C=C stretching (from phenyl group, ip) |
| - | - | - | - | - | - | 1115(w),<br>1094(w) | 1215(m),<br>1197(w),<br>1094(w),<br>1175(w) | - | - | C–H (-$CH_3$ and phenyl ring) bending , scissoring (ip) |
| - | - | - | - | - | - | 1021(m) | 1025(s)<br>1015(m) | - | 1044(s)<br>933(w) | C=C ring bending (from phenyl group) + C–O stretching + C=N ring bending (from TzTz system) |
| 1199(m)<br>1164(w)<br>1100(w)<br>1072(w) | 1210(w)<br>1111(w)<br>1055(w) | 1199(w) | 1216(w) | - | - | - | - | - | - | C–H (from –$CH_3$) bending , scissoring (ip) |
| - | - | 1168(vw)<br>1102(w) | 1173(w)<br>1128(m) | - | - | - | - | - | - | C–H (from –$CH_3$) bending , scissoring (ip), C–F stretching (from $CF_3$,ip) |
| - | - | 1085(vs) | 1119(m) | 1164(m),<br>1109(vs) | 1176(s),<br>1167(w),<br>1131(s),<br>1130(w)<br>1124(s) | 1171(w) | - | - | - | C=C stretching (from phenyl group, ip) + C–F stretching (from $CF_3$,ip) |
| - | - | - | 1083(w),<br>1083(w),<br>1035(w) | - | - | - | - | - | - | C=C –H bending (ip) (from phenyl ring and TzTz system), C–F stretching (from $CF_3$) |
| 1031(w),<br>1007(m),<br>996(m) | 1024(m),<br>1015(m) | 1014(w),<br>1006(s),<br>974(vw) | 1021(w) | 1041(w) | 1052(m), | - | - | - | - | C=C ring bending (from phenyl ring and TzTz system, ip) |
| 910(w) | - | - | - | 918(w) | 933(w) | - | - | - | 900(w),<br>853(w) | C-H bending, twisting (oop) |

| A1 | | A2 | | A3 | | A4 | | A5 | | Band Assignment<br>ip – in-plane, oop – out-of-plane |
|---|---|---|---|---|---|---|---|---|---|---|
| Experimental | Calculated | Experiment | Calculated | Experimenta | Calculated | Experimental | Calculated | Experimental | Calculated | |
| 883(w) | 880(w) | 881(w) | 884 | - | 919(w) | - | - | - | - | C-S stretching, C=C ring bending (from phenyl ring and TzTz system) |
| - | - | 839(m),<br>809(s) | 859<br>836 | 897(w) | 893(w) | - | - | 864(m) | 802(m) | C-H bending, wagging (oop) |
| | | - | - | 846.12(w) | 852(w) | - | 857(vw) | - | | C=C ring bending (from phenyl ring and TzTz system, oop) |
| - | - | - | - | 776(w),<br>706(w),<br>684(m),<br>654(w) | 776(w),<br>708(w),<br>687(w) | - | - | 839(w) | 840(w),<br>785(m) | C-S stretching (ip), C=C ring bending (from phenyl ring and TzTz system, oop) |
| - | - | - | - | - | | 777(m) | 763(w) | | | C–O–H bending (oop) |
| - | - | 659(m),<br>624(vw) | 778(w),<br>663(w),<br>639(w) | - | - | - | - | - | - | C=C ring bending (from phenyl ring and TzTz system, oop), C-S stretching |
| - | - | 728.(vw),<br>710.86(vw) | - | - | - | - | - | - | - | C=C ring bending (from phenyl ring, oop) |
| 680(s) | 694(s) | - | - | - | - | - | - | - | | C-S stretching i C=C ring bending (from phenyl ring and TzTz system, oop) |
| 637(w) | 672(m) | - | - | - | - | - | - | - | - | C-H bending, twisting (oop) |
| 611(s) | 618(w) | 613(m) | 618(w) | - | - | - | - | - | - | C=C ring bending (from phenyl ring and TzTz system, oop) |
| - | - | 516(m) | 503(w) | - | - | - | - | 678(w),<br>642(w) | 668(m) | C-S stretching |
| 483(w) | - | 497(m) | - | - | - | 626(w) | - | 620(w) | 621(w) | C-H bending, wagging (oop), C=N (from TzTz system) + C=C ring bending (from phenyl group, oop) |

***Table S10*** *Selected IR active modes [cm$^{-1}$] and their assignment from DFT calculations – type B and C compounds.*

| B1 | | B2 | | B3 | | B4 | | C | | Band Assignment<br>ip – in-plane, opp – out-of-plane |
|---|---|---|---|---|---|---|---|---|---|---|
| Experimental | Calculated | Experimental | Calculated | Experimental | Calculated | Experimental | Calculated | Experimental | Calculated | |
| - | - | - | - | - | - | - | 3637(m) | - | - | N–H stretching, sym. (ip) |
| | - | 3116(w) | - | - | - | - | - | - | - | C–H stretching, sym. (ip) |

| B1 | | B2 | | B3 | | B4 | | C | | Band Assignment |
|---|---|---|---|---|---|---|---|---|---|---|
| - | - | - | - | - | - | - | - | - | 3199(m) | C–H stretching, sym. (from phenyl ring+ CH=C–H, ip) |
| 3111(vs), 3067(w) | 3251(w) | 3071(w) | 3203(w) | 3100(m) | 3221(w) | 3060(m) | 3252.60(w) | 3080(w), 3052(w), 3032(m), 3011(w), 2993(w) | 3192(w), 3183(w), 3178(m) 3172(w) 3167(w) 3135(w) | C–H stretching, asym. (from phenyl ring+ CH=C–H, ip) |
| - | - | - | - | 3078(m), 3063(s) | - | 3014(m) | - | - | - | C–H stretching, sym. (ip) |
| - | - | - | - | - | - | 2926(m), 2860(m), 2807(m), 2768(m), 2714(m), 2682(m), 2591(m), 2538(m) | - | - | - | aromatic C–H bending overtones + aromatic N–H bending overtones |
| - | - | - | - | - | - | - | - | 1624(w) | 1676(m) | C=C stretching (from phenyl group,ip) |
| 1583(m), 1495(vs), 1436(m) | 1621(w), 1535(m), 1455(m), 1424(w) | 1539(m), 1451(s), 1409(vs) | 1581(w), 1498(m), 1457(w) | 1446(s) | 1558(w), 1494(m), 1440(m) | 1485(m), 1449(w), 1431(w) | 1507(s), 1460(w), 1434(w), 1402(w), | 1497(w), 1427(vs) | 1534(m), 1494(w), 1474(s) | C=N stretching (from TzTz system, ip) + C=C stretching (ip) |
| 1311(m) | 1353(w) | 1339(w) | 1382(w), 1348(w) | 1370(vs), 1323(s) | 1357(w) | 1363(vs) | 1341(w), 1294(w) | 1332(w) | 1347(w) | C–N stretching (from TzTz system, ip) + C=C stretching (ip) |
| 1257(w) 1239(s) | - | 1308(m) 1241(m) | - | - | - | - | - | - | - | C=C stretching (ip) |
| 1213(s) | - | 1221(s) | - | 1240(w) | 1252(w) | - | - | 1301(w) 1254(w) | 1288(vw) 1247(w) | C=N stretching (from TzTz system, ip) + C=C stretching (ip) |
| - | - | - | - | - | - | - | 1184(w) | - | - | C–N stretching (from TzTz system, ip ) |
| - | - | 1208(m) | 1241(w) | - | - | 1298(w), 1062(w) | | 1219(m), 1188(w) | 1200(w) | C=N stretching (from TzTz system, ip) + C=C stretching +C–H bending , scissoring (ip) |
| - | - | - | - | 1158(w) | - | 1119(s) | | - | - | C–N stretching (from imidazole ring) + C–H/N–H bending, scissoring (ip) |
| - | - | - | 1234(w) | - | - | 1158(w) | - | - | - | C=C stretching (ip) |
| 1147(m) | 1110(w) | - | - | - | - | - | - | - | - | C–O stretching (ip) |
| 1076(w) | 1057(w) | 1075(w) | 1080(w) | 1149(w), 1068(w), | 1089(w) | - | 1144(w) 1101(s) | 1172(w), 1100(w), | 1211(w), 1108(w) | C–H bending , scissoring (ip) |

| B1 | | B2 | | B3 | | B4 | | C | | Band Assignment |
|---|---|---|---|---|---|---|---|---|---|---|
| | | | | 1058(w) | | | 1062(w) | 1071(w) | | |
| 1029(m) | 1023(m) | - | - | - | - | - | - | - | - | C=N (from TzTz system, ip) + C=C ring bending (from TzTz system, ip) +C–H bending, scissoring (ip) |
| - | - | 1054(m) | - | - | - | 1051(w) | - | - | - | C=C ring bending (from phenyl group, ip) |
| - | - | - | - | - | - | - | - | 943(s) | 990(m), 977(w) | C–H bending, twisting (oop) |
| 1012(s) | - | 969(s) | 974w) 864(w) | 980(vs), 913(w) | | 964(m) 927(w) | 975(w), 938(w) | - | 940(w) | C=N (from TzTz system, ip) + C=C ring bending (from phenyl group, ip) |
| 926(m) | - | 822(s) | - | - | - | - | - | - | - | C=N (from TzTz system) + C=C ring bending (oop) |
| - | - | 707(vs) | 832.80(w) | 845(vs) | - | 871(m) | 867(w) | 743(m) | - | C–H bending, twisting (oop) |
| - | - | - | - | - | - | 740(s), 711(m) | - | - | - | C=C bending (ip) |
| 878(m), 844(s) | 851(w) | 687(m) | 818(m), 740(w) | - | 983(m), 886(m), 836(w) | - | 868(w), 671(w) | - | - | C–S stretching (ip) |
| 828(s) | 817(w) | - | - | 742(vs) | | - | 748(w) | - | - | C–H bending, twisting (oop) |
| 751.50(m), 738.87(s) | 753(m) | 614(m) | 710(w) | 758(w), 731(s), 698(m) | 736(m) | | 652(w) | 682(s) | 754(vs) 679(m) | C–H/N–H bending, wagging (oop) |
| - | - | - | - | - | - | - | - | 696(w), 650(w) | - | C=N (from TzTz system, ip) + C=C ring bending (ip) |
| - | - | - | - | - | 620(w) | - | - | - | - | C=C ring bending + C–H bending, wagging (oop) |
| 660.17(m) | 653(w) | 560(w) | 690(w) | 635(w) | 614(w) | - | - | - | - | C–S stretching (ip) |
| 591.65(m) | 594(w) | 483(w) | - | 615(w) | - | - | - | 567(w) | - | C=C ring bending, (oop) |
| 487.43(w) | - | - | - | 591(m) | - | - | - | 491(w) | - | C–H bending, wagging (oop) |
| - | - | - | - | - | | - | - | 478(w) | 643(m), 635(w), 582(w) | C–S stretching i C=C ring bending, (oop) |

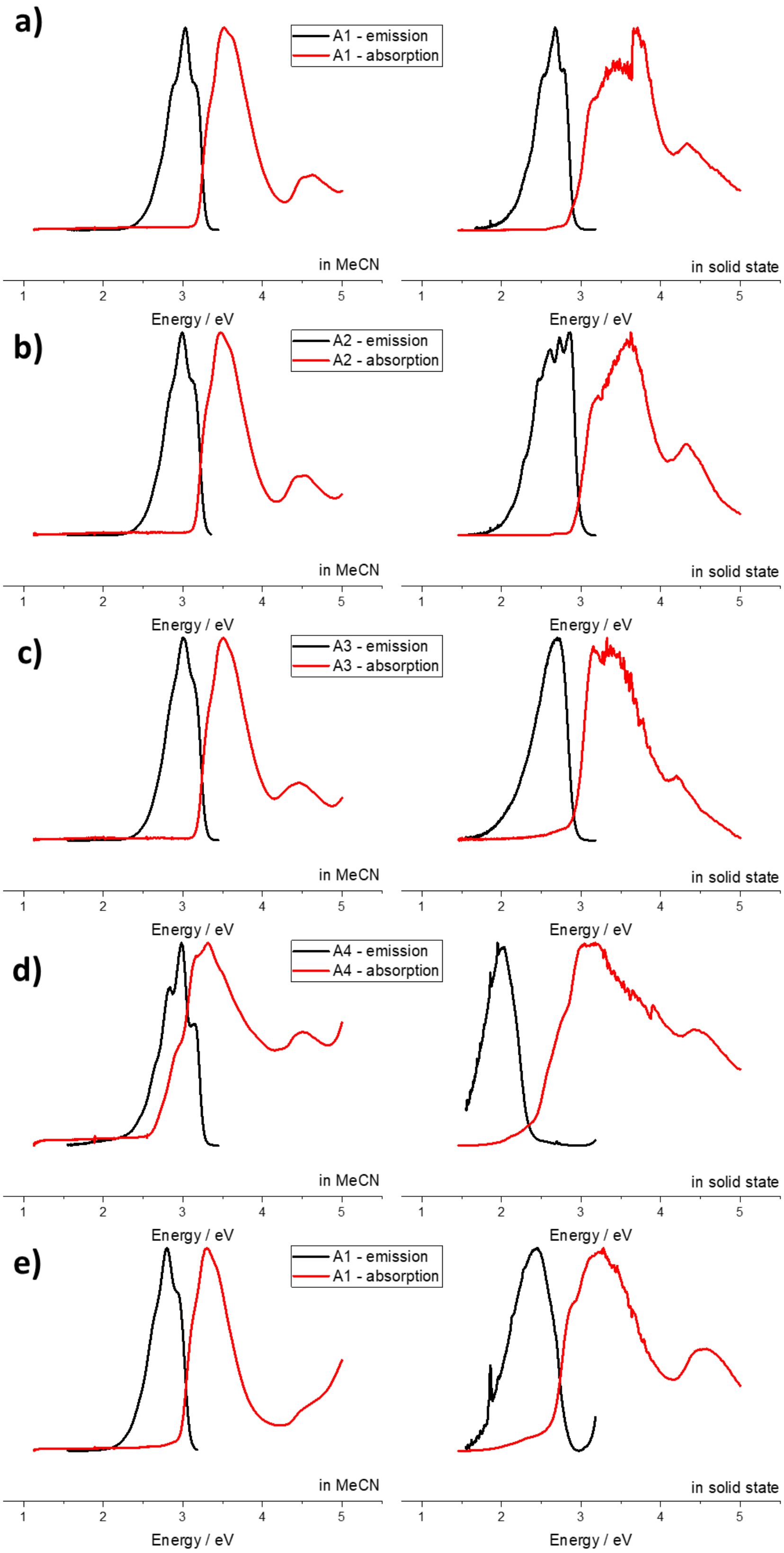

***Figure S8.*** *Absorption and emission spectra in acetonitrile solution (left column) and in solid state (right column) of the remaining A-type compounds: a)* ***A1****, b)* ***A3****, c)* ***A4****, d)* ***A5****.*

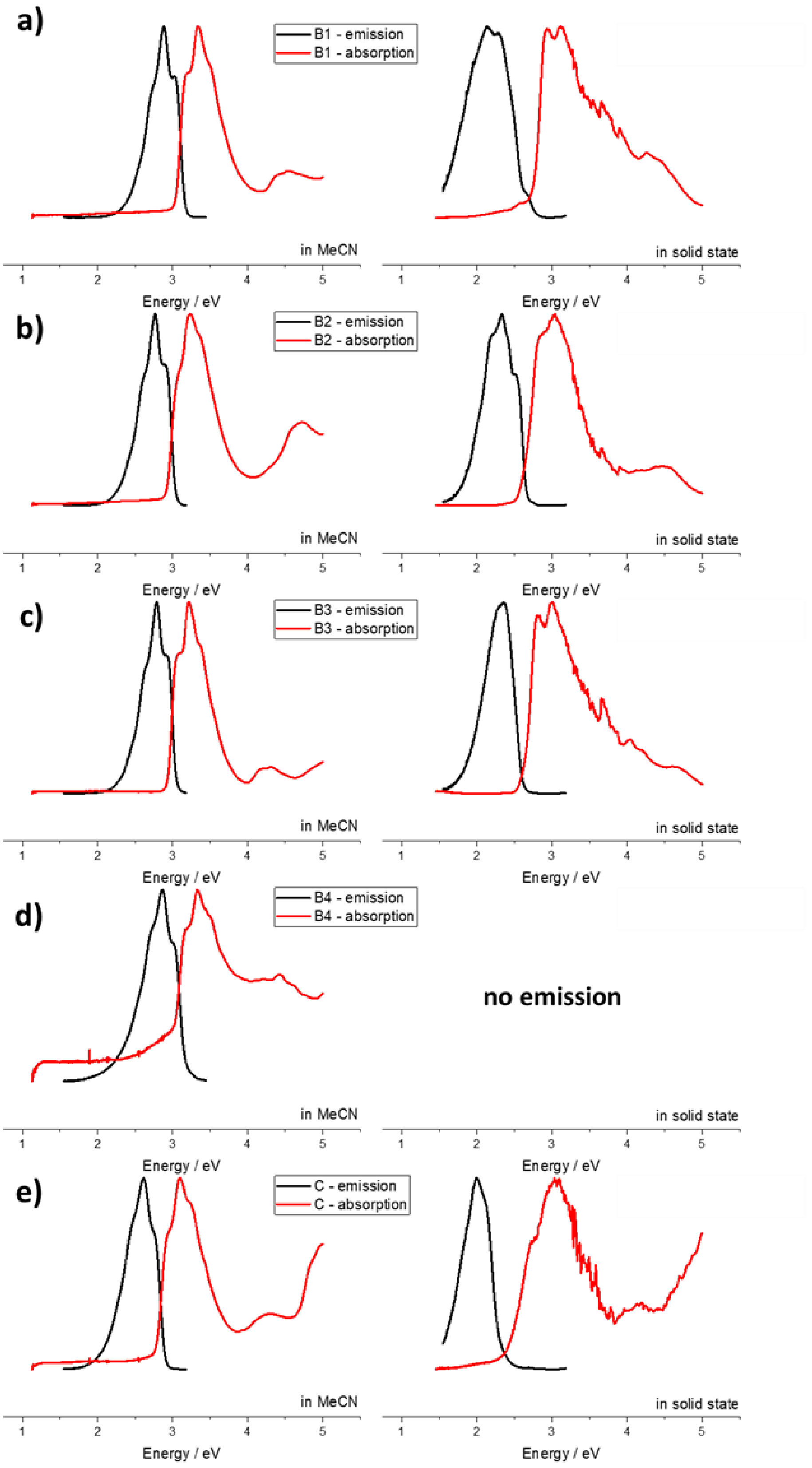

***Figure S9.*** *Absorption and emission spectra in acetonitrile solution (left column) and in solid state (right column) of B- and C-type compounds: a)* ***B1****, b)* ***B2****, c)* ***B3****, d)* ***B4****, e)* ***C****.*

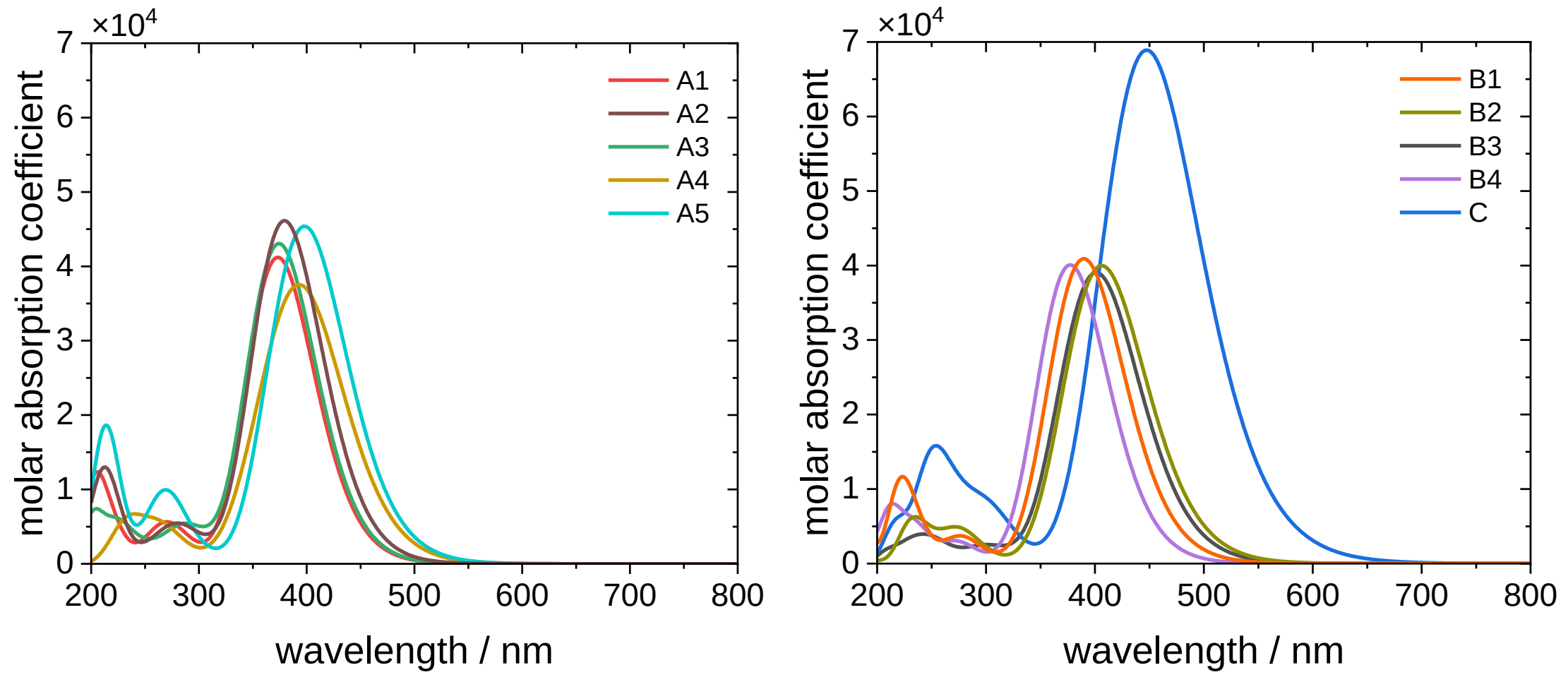

***Figure S10***. *DFT prediction (B3LYP/TZVP) of absorption spectra of studied thiazolothiazoles.*

***Table S11.*** *DFT predicted absorption maxima (in vacuo, B3LYP/TZVP) and band gaps (solid state, periodic GGA-PWA) of studied thiazolothiazoles.*

| compound | $\lambda_{max}$ / nm | $E_g$ / eV | band gap type |
|---|---|---|---|
| A1 | 374 | 2.21 | indirect |
| A2 | 379 | 2.13 | direct |
| A3 | 374 | 2.17 | indirect |
| A4 | 392 | 2.04 | direct |
| A5 | 398 | 1.92 | indirect |
| B1 | 390 | 1.92 | indirect |
| B2 | 406 | 1.92 | indirect |
| B3 | 400 | 1.85 | direct |
| B4 | 377 | 1.98 | indirect |
| C | 448 | 1.83 | indirect |

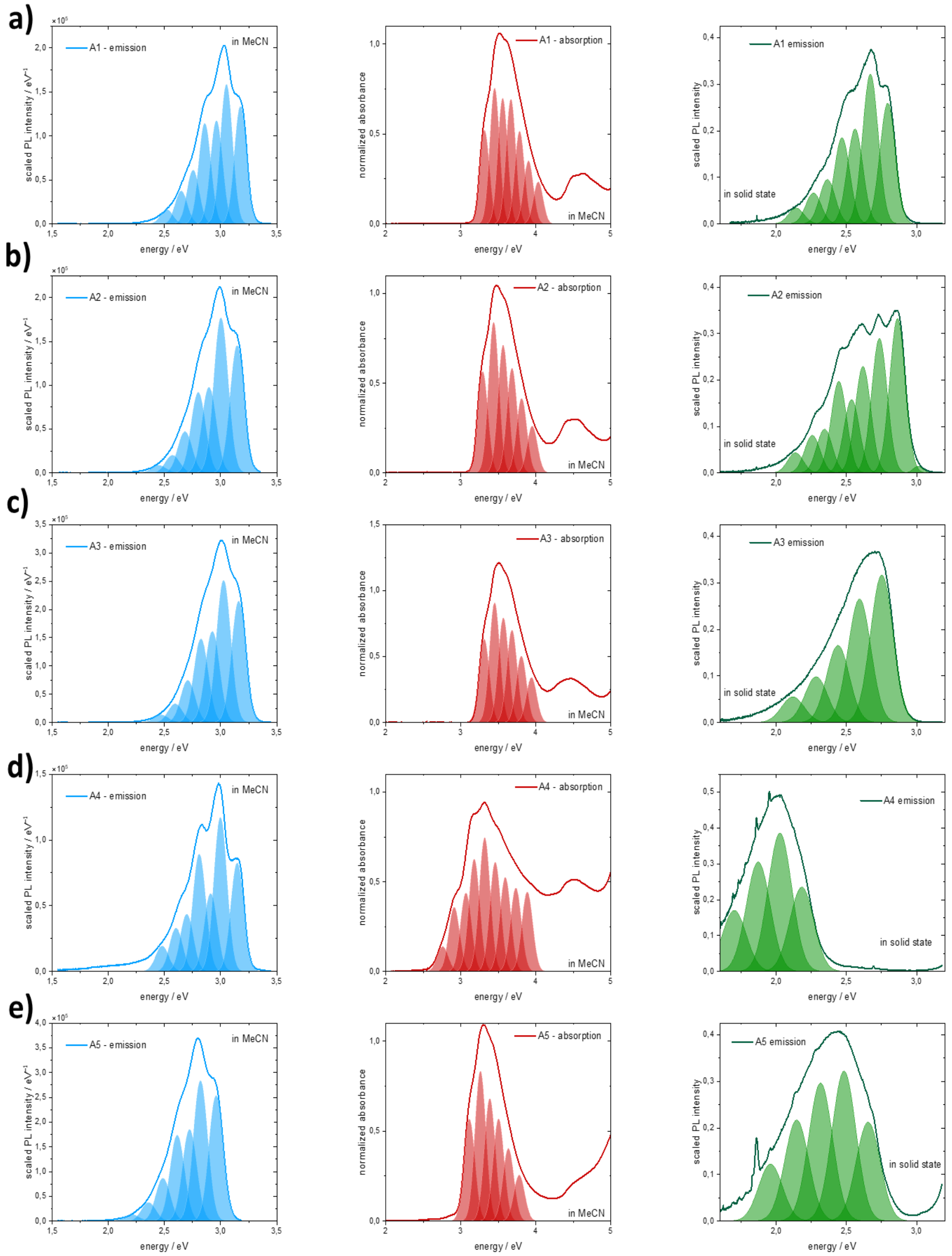

***Figure S11.*** *Emission spectra in acetonitrile (left column), Absorption spectra in acetonitrile (middle column) and emission spectra in solid state (right column) with deconvoluted peaks of spectral broadening for A-type compounds: a)* ***A1****, b)* ***A2****, c)* ***A3****, d)* ***A4*** *and e)* ***A5****.*

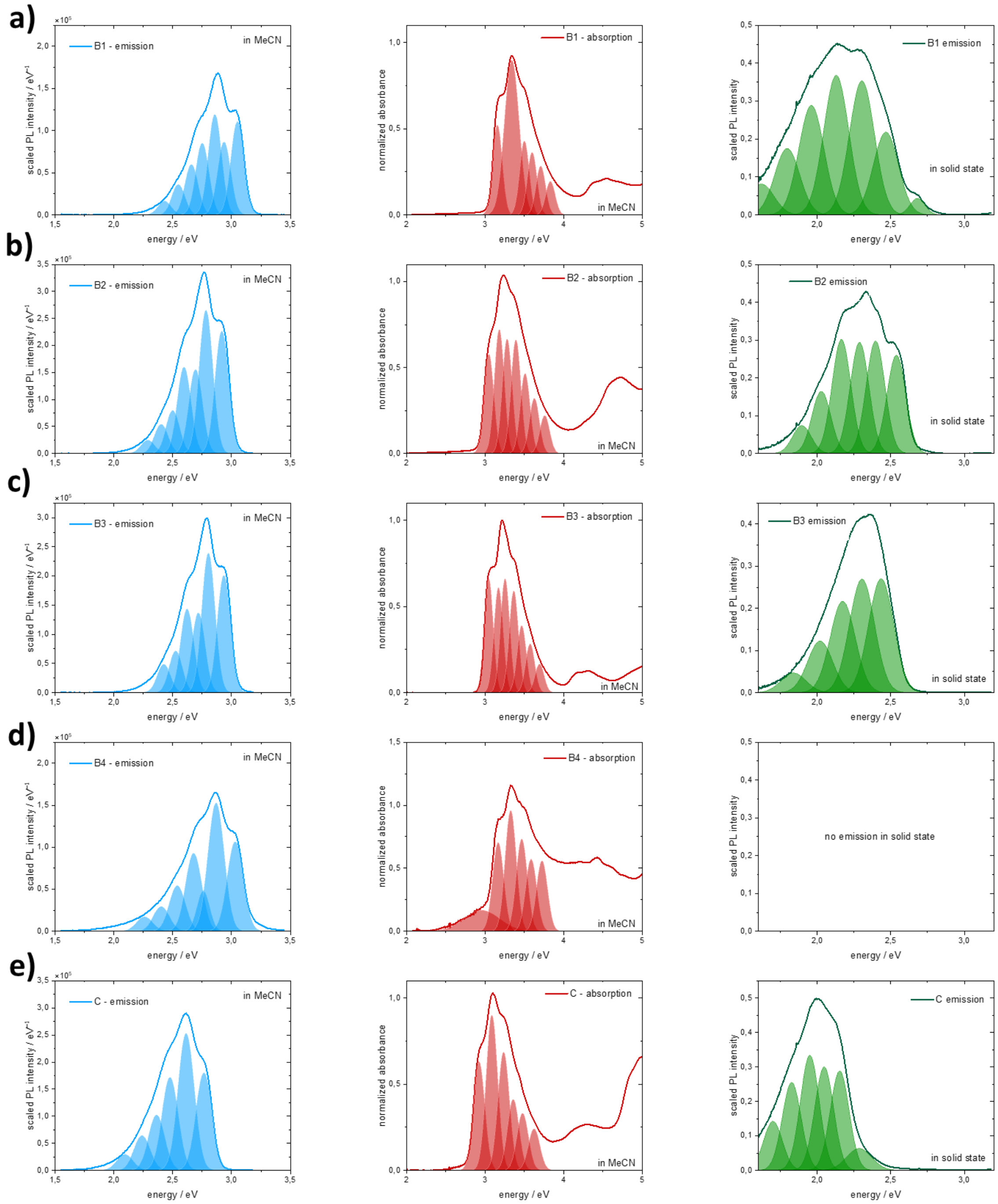

**Figure S12.** Emission spectra in acetonitrile (left column), Absorption spectra in acetonitrile (middle column) and emission spectra in solid state (right column) with deconvoluted peaks of spectral broadening for A-type compounds: a) **B1**, b) **B2**, c) **B3**, d) **B4** and e) **C**.

***Table S12.*** *Bandgap determination: fitted linear function parameters for A,B,C-type compounds*

| code | **A1** | **A2** | **A3** | **A4** | **A5** |
|---|---|---|---|---|---|

| substituent | 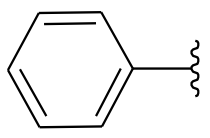 | 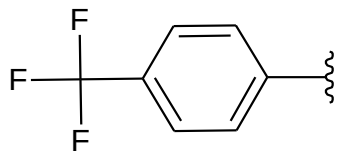 | 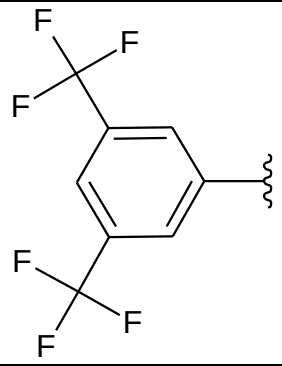 | 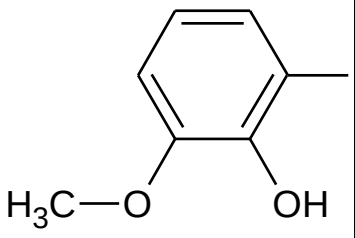 | 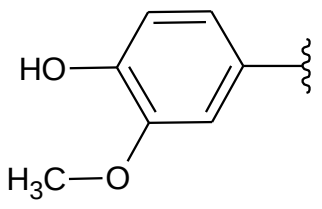 |
|---|---|---|---|---|---|
| intercept | -36.29 ± 0.92 | -32.79 ± 0.79 | -43.16 ± 1.01 | -11.62 ± 0.10 | -26.31 ± 0.41 |
| slope | 12.55 ± 0.31 | 11.32 ± 0.26 | 14.83 ± 0.34 | 4.96 ± 0.04 | 10.08 ± 0.15 |
| R-square | 0.9872 | 0.9844 | 0.9888 | 0.9967 | 0.9933 |
| code | **B1** | **B2** | **B3** | **B4** | **C** |
| substituent | 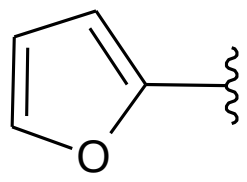 | 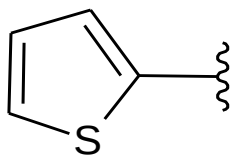 | 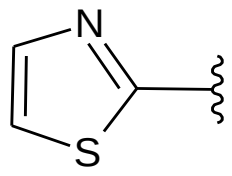 | 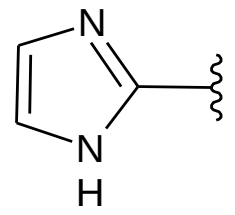 | 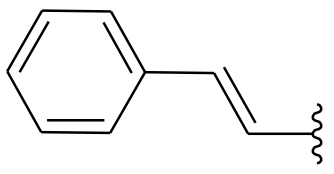 |
| intercept | -43.21 ± 1.61 | -29.77 ± 0.89 | -39.35 ± 1.43 | -28.01 ± 0.36 | -16.03 ± 0.19 |
| slope | 15.83 ± 0.57 | 11.56 ± 0.33 | 15.05 ± 0.53 | 10.69 ± 0.13 | 6.69 ± 0.07 |
| R-square | 0.9718 | 0.9743 | 0.9734 | 0.9964 | 0.9949 |

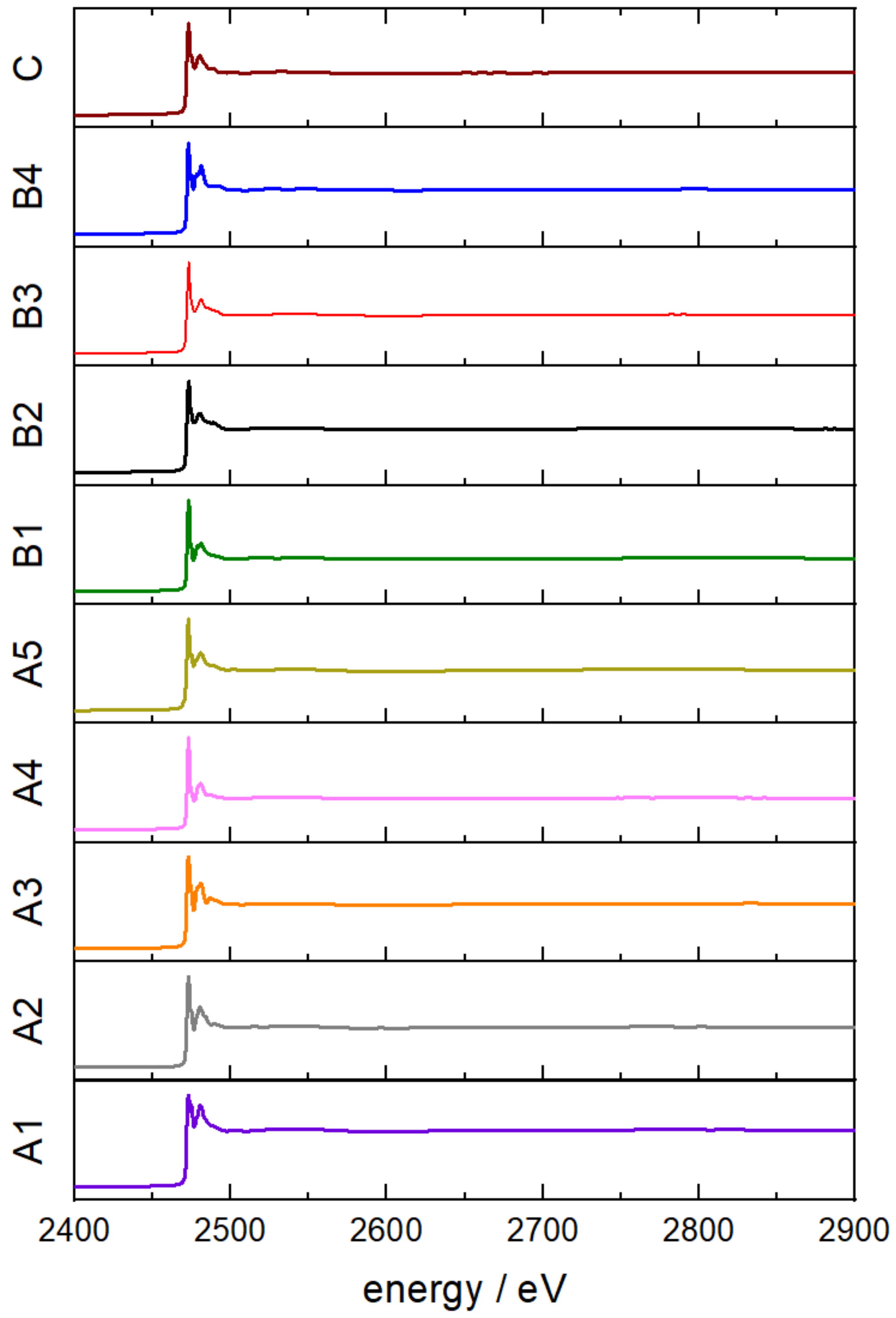

***Figure S13***. *Normalized full-scale XAS spectra of the studied thiazolothiazoles.*

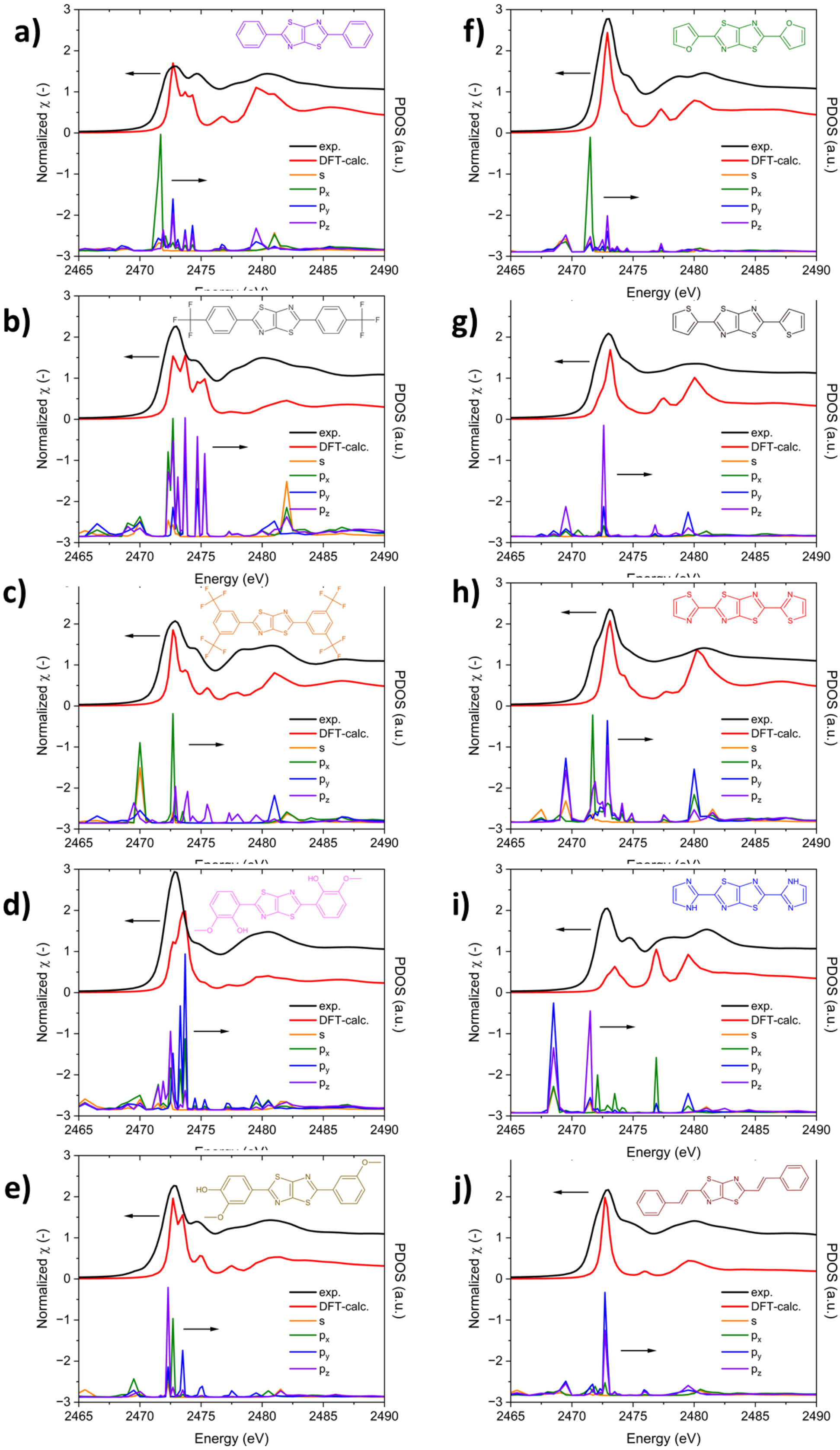

***Figure S14.*** *XANES spectra analysis for a) A1, b) A2, c) A3, d) A4, e) A5, f) B1, g) B2, h) B3 i) B4 and j) C compounds. Description for each panel: left axis - experimental (black) and DFT-calculated (red) S K-edge XANES spectra for A, B, C-type compounds measured in the transmission mode. Right axis: PDOS corresponding to the calculated spectrum.*

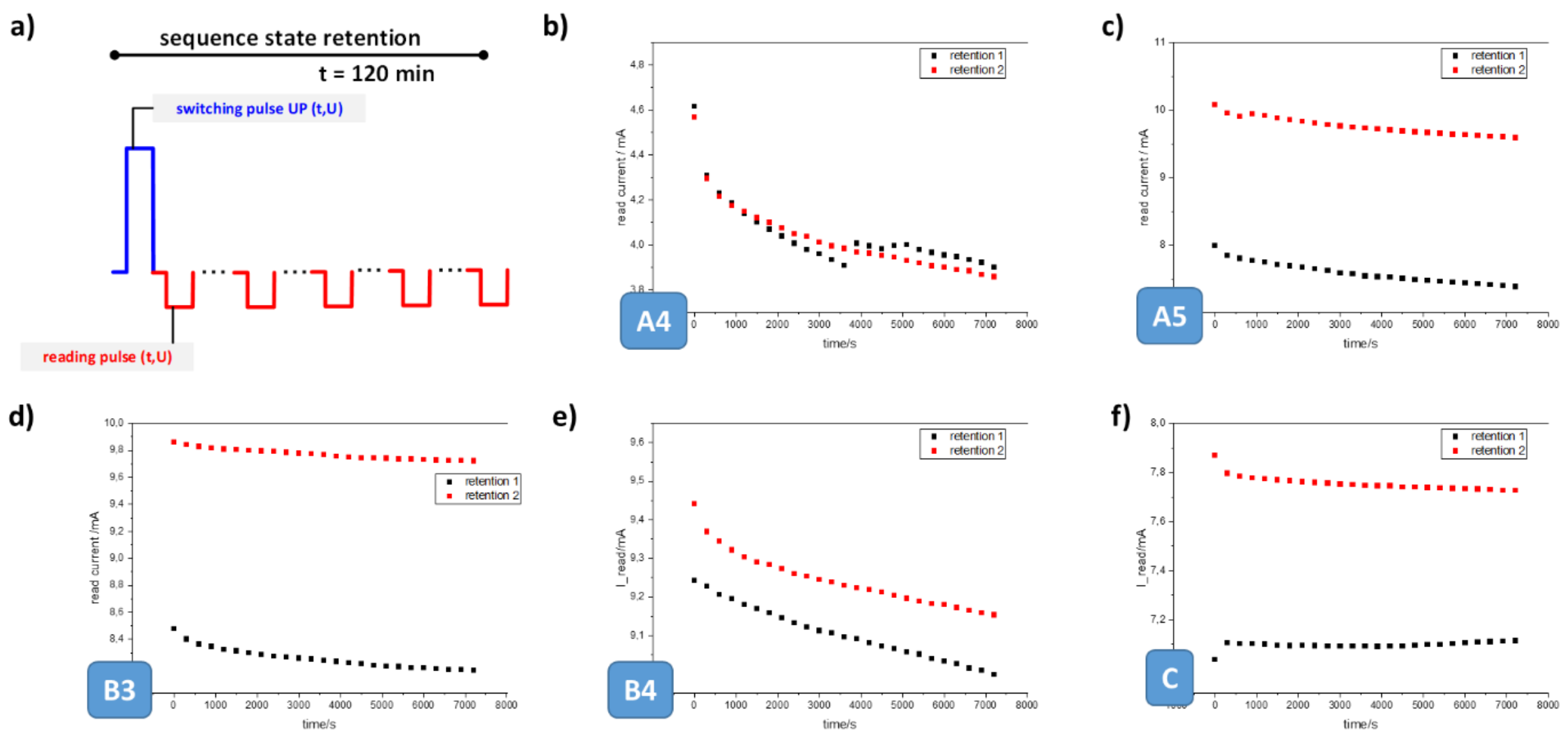

***Figure S15.*** *The state retention test for the thin film layer device of TzTzs: a) electric pulse sequence for the test. Switching pulse potential was -3/+3V with length of 5s, reading pulse potential -500mV with the length equal to 100ms. Results of the test b)* ***A4****, c)* ***A5****, d)* ***B3****, e)* ***B4****, f)* ***C****.*

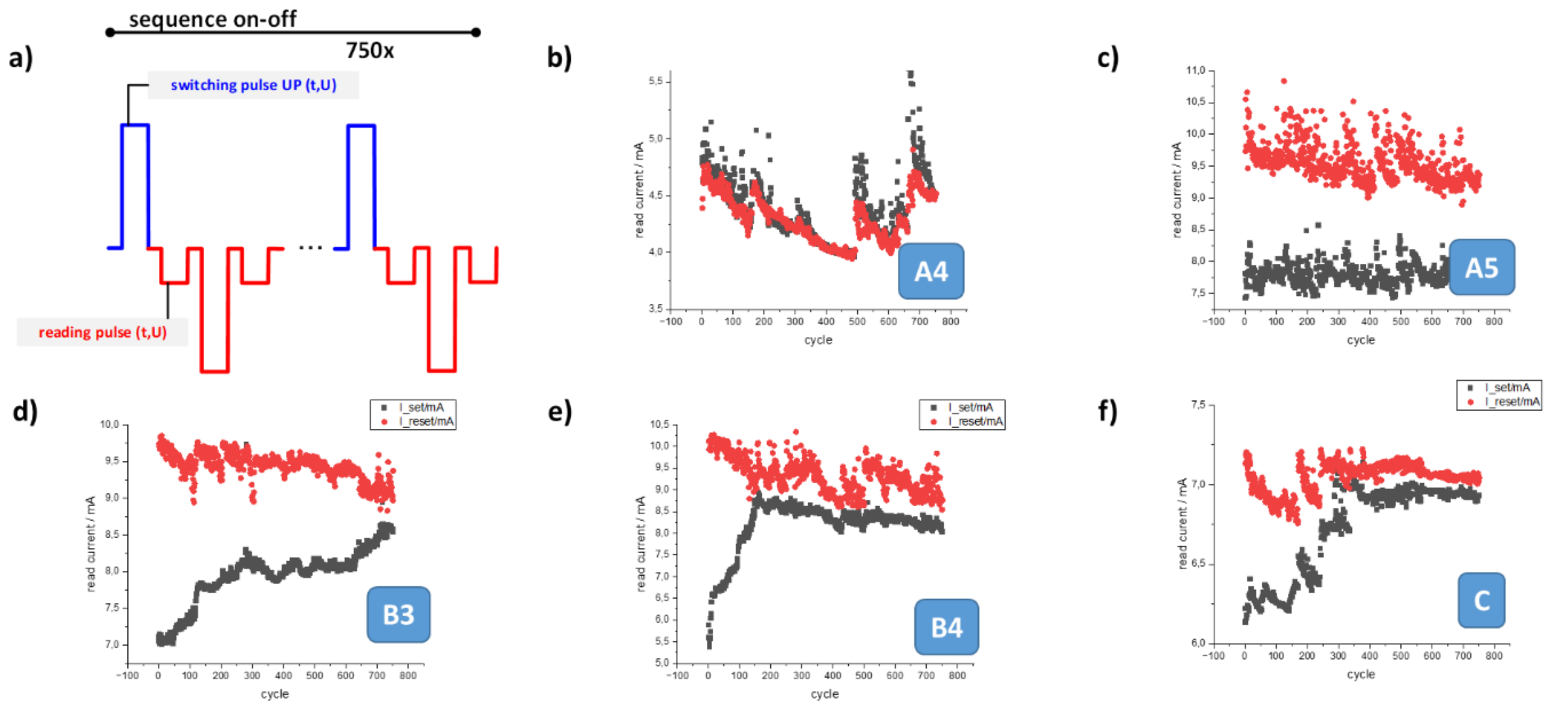

***Figure S16.*** *Results of the on-off ratio test: a) electric pulse sequence; switching pulse potential was -3/+3V, reading pulse potential -500mV, pulses length equal 100ms. Results of the test for the thin film layer devices of b)* ***A4****, c)* ***A5****, d)* ***B3****, e)* ***B4****, f)* ***C****.*